\documentclass[referee,a4wide]{aa}

\usepackage{natbib,graphicx}
\bibpunct{(}{)}{;}{a}{}{,}

\newcommand{\eg}{e.g. }
\newcommand{\ie}{i.e. }

\begin{document}

\thesaurus{02.08.1;    
             11.01.2;  
             11.09.1;  
             11.10.1;  
             13.18.1}  

     \title{
Complex shock structure in the western hot-spot of Pictor~A
}


     \author{C.J. Saxton \inst{1,2}
            R.S. Sutherland \inst{2}
             G.V. Bicknell \inst{1,2}
             G. Blanchet \inst{1,2,3}
             S.J. Wagner \inst{4}
            \and
             M.V. Metchnik \inst{2}
            }

     \offprints{C.J. Saxton}

     \institute{
Department of Physics \& Theoretical Physics, Australian National 
University, Canberra, Australia
\and
Research School of Astronomy \& Astrophysics, Australian
National University, Canberra, Australia
\\
email: saxton@mso.anu.edu.au,
mvm@mso.anu.edu.au,
ralph@mso.anu.edu.au,
Geoff.Bicknell@mso.anu.edu.au
\and
INSA, 135, Avenue de Rangueil, 31077 Toulouse Cedex 4, France
\and
Landessternwarte Heidelberg-Konigstuhl, Konigstuhl, D-69117
Heidelberg, Germany  \\
email: swagner@lsw.uni-heidelberg.de}

\date{}

\authorrunning{Saxton et al.}
\titlerunning{Shock structure in western hot-spot of Pictor~A}

\maketitle

\begin{abstract}

We have carried out simulations of supersonic light jets in order
to model the features observed in
optical and radio images of the western hot-spot
in the radio galaxy Pictor~A.
We have considered jets with density ratios
$\eta=10^{-2} - 10^{-4}$,
and Mach numbers ranging between $5$ and $50$.
 From each simulation,
we have generated ray-traced maps of radio surface brightness
at a variety of jet inclinations,
in order to study the appearance of time-dependent luminous structures
in the vicinity of the western hot-spot.
We compare these rendered images with observed features of Pictor~A.
A remarkable feature of Pictor~A observations
is a bar-shaped ``filament'' inclined almost at right angles
to the inferred jet
direction and extending $24''$ ($10.8 h^{-1} \rm kpc$)
along its longest axis.
The constraints of reproducing the appearance of this structure
in simulations indicate that the jet of Pictor~A
lies nearly in the plane of the sky.
The results of the simulation are also consistent
with other features found in the radio image of Pictor~A.
This filament arises
from the surging behaviour of the jet near the hot-spot; the surging is
provoked by alternate compression and decompression of the jet by the
turbulent backflow in the cocoon. 
We also examine the arguments for the jet
in Pictor A being at a more acute angle to the line of sight and find that
our preferred orientation is just consistent with the limits on the
brightness ratio of the X-ray jet and counter-jet. 
We determine from our simulations, 
the structure function of hot-spot brightness and also
the cumulative distribution of 
the ratio of intrinsic hot-spot brightnesses.
The latter may be used to quantify the use of hot-spot ratios for the
estimation of relativistic effects.


\end{abstract}

%

\section{Introduction}

The FR II radio galaxy Pictor~A ($z=0.0342$)
consists of two lobes  dominated by hot-spots separated by
$4'$ ($210 h^{-1} \> \rm kpc$) from the flat spectrum nucleus.
Near the western hot-spot, there is a faint, $24''$
($10.8 h^{-1} \> \rm kpc$) long  radio and optical filament,
``upstream of the hot-spot'' first detected by
\citet{roser87a}.
\citet{wagner00a}
found the filament to be
highly polarised, confirming that
it radiates via optical synchrotron emission.

The filament has been studied in detail at several radio frequencies by
\citet{perley97a}.
The radio emission at 8415MHz, figure~\ref{f:radio_image},
which bears a striking correspondence to the optical emission,
also reveals a knot further upstream from the filament.
This is of particular interest since
at various timesteps  throughout our simulation such a morphology is
captured.

The discovery of optical synchrotron radiation, emitted by electrons with
radiative life-times of about 100 yr
(for magnetic fields of the order of equipartition),
in a filament which extends several thousand light years
perpendicular to the jet  strongly indicates that
the electrons are accelerated {\em in situ}.
The crossing time from the hot-spot to the filament is much
longer than the cooling time-scale,
and efficient acceleration to optical emitting energies is required.

Our objective here is to ascertain whether purely hydrodynamical 
simulations can
qualitatively reproduce the radio and optical morphology.  Indeed  on the basis
of our simulations we suggest below that the filament is the result of a
quasi-periodic pulsing behaviour of the terminal jet shock.

\section{Summary of the Observations}

Before discussing the simulations, we summarise relevant aspects
of the observations and in the following sections we point out the relevance of
the features in the simulations to the observational data.

\begin{figure}
\centering \leavevmode
\includegraphics[height=5cm]{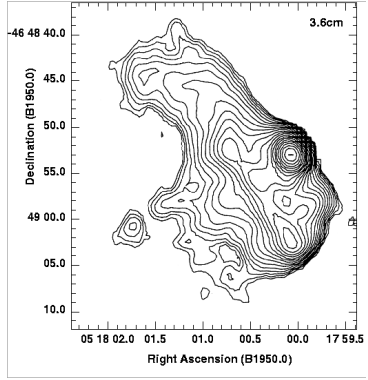}
\caption{Radio image. VLA 8415MHz map of the western lobe,
with contour levels at multiples of $1/\sqrt{2}$
\citep{perley97a}.
}
\label{f:radio_image}
\end{figure}

A total intensity radio map
\citep{perley97a},
displayed in figure
\ref{f:radio_image} shows the high-surface brightness western hot-spot;
a bar-shaped region (the filament) east of the hot-spot,
extends perpendicular to the inferred direction of the jet.
The region between
the hot-spot and the ridge line of the bar shows a ``pedestal'' of lower
surface  brightness.

\begin{figure}
\centering \leavevmode
\includegraphics[height=5cm]{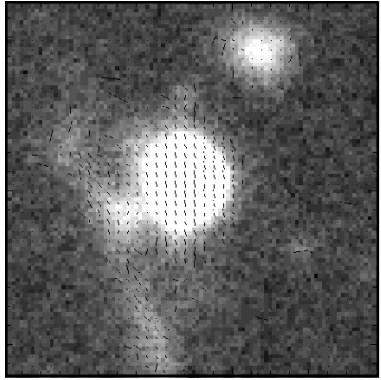}
\caption{Optical image. Total intensity emission at 450 nm with
polarization map overlay.}
\label{f:optical_image}
\end{figure}

High S/N observations were obtained with the VLT (UT1)
during commissioning of FORS1.
Figure~\ref{f:optical_image} shows total intensity emission at 450 nm.
\citet{wagner00a}
presented polarization maps derived from these data,
showing up to $60\%$ polarization in the hot-spot,
and up to $40\%$ polarization in the filament.
Such a high degree of polarization cannot be the result of
scattering and strongly suggests that the filament radiates optical
synchrotron emission.  Moreover, in the optical image, the bar-shaped
filament is quite bright, and spatially coincides  with a similar
feature in the radio images.
The eastern edge of the bar is very pronounced,
but the surface brightness gradient in the western direction
is more gradual.
The filament is brightest at its ends and near the centre,
where it overlaps the inferred jet direction.
In the optical image there are
faint extensions from the hot-spot extending
in the direction of the ends of the filament.


\section{Simulation Methodology}

\subsection{Hydrodynamics}

The earliest simulations of astrophysical jets were published by
\citet{norman82a} and \citet{smith85a}.
Their simulations revealed
hot-spot morphologies that were different from the simple one-shock structure
envisaged by
\citet{blandford74a}.
These papers were seminal in leading to a good physical understanding of
the properties of supersonic extragalactic jets.

In this work we have carried out similar simulations with a view to
addressing in
detail the time variability of hot-spot morphology.
In particular we address
the  question of whether the Pictor~A morphology could arise naturally
in avlight supersonic jet.
Our simulations were conducted using the VH-1 code
\citep{blondin93a} which is  an implementation of the
Piecewise Parabolic Method (PPM) \citep{colella84a}.
An advantage of PPM for this type of simulation is its
excellent resolution of shocks.

The simulations were computed on an axisymmetric grid,
with $600 \times 300$ cells
in the axial and radial directions.
We performed calculations
with nine different cases of the jet parameters,
i.e. jet Mach number $M=5, 10, 50$
and the ratio of jet to background densities,
$\eta =10^{-2}, 10^{-3}, 10^{-4}$.
In each case the jet density and pressure were
normalized by the respective values, $(\rho_0,p_0)$,
in the background medium.
If we take $T =10^7 T_7 \> \rm K$
to be the temperature of the external medium,
then the normalising value of the velocity is
$v_0 = \sqrt{p_0/\rho_0}
= \sqrt{kT/\mu m_p}\approx 3.65 \times 10^7 \,
T_7^{1/2} \> \rm cm \> s^{-1}$.
The unit of length, $x_0$, is defined by the
value of the $\hbox{jet radius} = 0.25$, in numerical units.
If we take the observed western hot-spot
$\hbox{FWHM}
\approx 1.2^{\prime\prime} \approx 2.7 \times 10^{21} \> \rm cm$,
for $H_0 = 70 \> \rm km \> s^{-1} \> Mpc^{-1}$,
as indicative of the jet diameter, then
$x_0 \approx 9.0\times 10^{21} \>\rm cm$
and the unit of time,
$t_0 =x_0 v_0^{-1} \approx 9.4 \times 10^6 \,T_7^{-1/2} \> \rm yr$.
As with simulations of this type,
the exact physical parameters of velocity and density are not recovered.
However, a high Mach number and low
density ratio serve to reproduce,
at least qualitatively, many of the features found in real jets.

Table~\ref{table.simulation.parameters}
lists the jet parameters, time resolution
and left boundary condition of our simulations.
We define the grid to be $10$ units in the $z$ direction
and $5$ units in the radial direction.
Table~\ref{table.jet.properties}
indicates the jet's total mass flux, force and power
for each choice of the basic parameters,
in terms of fiducial values of the temperature, $T_7$,
and the number density of the background medium,
$n_{-3}=n/10^{-3}{\rm cm}^{-3}$.
For a jet with Mach number $M$,
density contrast $\eta$,
adiabatic index $\gamma$ (assumed equal to the ambient medium),
and cross-sectional area $A=\pi r_j^2$,
the initial flow velocity,
mass flux, force and kinetic power are
\begin{equation}
v_{\rm j}=v_0\sqrt{\gamma\over\eta}M
\ ,
\end{equation}
\begin{equation}
\dot{M}_{\rm j}
=A\rho_0 v_0\sqrt{\gamma\eta}M
\end{equation}
\begin{equation}
F_{\rm j}=A(\rho v^2 +p)=
A\rho_0v_0^2 (\gamma M^2+1)
\ , \mbox{ and}
\end{equation}
\begin{equation}
P_{\rm j}=A\left({ {\frac12}\rho v^3+{\gamma\over{\gamma-1}} pv
}\right)
=A\rho_0v_0^3 \left({
{\frac12}M^2+{1\over{\gamma-1}}
}\right) M\sqrt{{\gamma^3}\over\eta}
\ .
\end{equation}
These are subrelativistic expressions,
although the jets of Pictor~A may involve relativistic flows of plasma.
One may relate the $(\eta,M)$ parameters of a classical jet
to an equivalent pair of parameters for
a relativistic jet with equivalent thrust or power,
e.g. \citet{rosen1999}, \citet{carvalho2001a}.
However we are not modelling the system to determine the Lorentz factor
of the Pictor~A jet to great precision.

\begin{table}
\caption{
Simulation physical parameters $(\eta,M)$,
time coverage ($t_\mathrm{init}\leq t\leq t_\mathrm{end}$)
and frame intervals ($\Delta t$).
Times are expressed in terms of the scale $t_0$.
}
\label{table.simulation.parameters}
\begin{center}
\small
$
\begin{array}{ccc}
\begin{array}{lcrcllc}
\mbox{label}
&\begin{array}{c}\mbox{left}\\\mbox{boundary}\end{array}
&\eta&M&t_{\rm init}&t_{\rm end}&\Delta t
\\\hline
\\
{\tt C2V}&\mbox{closed}&10^{-2}&5&0&2.4&4\times10^{-3}
\\
{\tt C3V}&\mbox{closed}&10^{-3}&5&0&2.4&4\times10^{-3}
\\
{\tt C4V}&\mbox{closed}&10^{-4}&5&0&2.4&4\times10^{-3}
\\
{\tt C2X}&\mbox{closed}&10^{-2}&10&0&1.2&2\times10^{-3}
\\
{\tt C3X}&\mbox{closed}&10^{-3}&10&0&1.2&2\times10^{-3}
\\
{\tt C4X}&\mbox{closed}&10^{-4}&10&0&1.2&2\times10^{-3}
\\
{\tt C2L}&\mbox{closed}&10^{-2}&50&0&0.24&4\times10^{-4}
\\
{\tt C3L}&\mbox{closed}&10^{-3}&50&0&0.24&4\times10^{-4}
\\
{\tt C4L}&\mbox{closed}&10^{-4}&50&0&0.24&4\times10^{-4}
\\
\\
{\tt c3V}&\mbox{closed}&10^{-3}&5&1.6&2.4&4\times10^{-4}
\\
{\tt c4V}&\mbox{closed}&10^{-4}&5&0.8&1.28&2\times10^{-4}
\\
{\tt c3X}&\mbox{closed}&10^{-3}&10&0.8&1.2&2\times10^{-4}
\\
{\tt c4X}&\mbox{closed}&10^{-4}&10&0.4&0.64&1\times10^{-4}
\\
{\tt c3L}&\mbox{closed}&10^{-3}&50&0.05&0.2&5\times10^{-5}
\\ \\
\\\hline
\end{array}
&~~~&
\begin{array}{lcrcllc}
\mbox{label}
&\begin{array}{c}\mbox{left}\\\mbox{boundary}\end{array}
&\eta&M&t_{\rm init}&t_{\rm end}&\Delta t
\\\hline
\\
{\tt O2V}&\mbox{open}& 5&10^{-2}&0&2.4&4\times10^{-3}
\\
{\tt O3V}&\mbox{open}& 5&10^{-3}&0&2.4&4\times10^{-3}
\\
{\tt O4V}&\mbox{open}& 5&10^{-4}&0&2.4&4\times10^{-3}
\\
{\tt O2X}&\mbox{open}&10&10^{-2}&0&1.2&2\times10^{-3}
\\
{\tt O3X}&\mbox{open}&10&10^{-3}&0&1.2&2\times10^{-3}
\\
{\tt O4X}&\mbox{open}&10&10^{-4}&0&1.2&2\times10^{-3}
\\
{\tt O2L}&\mbox{open}&50&10^{-2}&0&0.24&4\times10^{-4}
\\
{\tt O3L}&\mbox{open}&50&10^{-3}&0&0.24&4\times10^{-4}
\\
{\tt O4L}&\mbox{open}&50&10^{-4}&0&0.24&4\times10^{-4}
\\
\\
{\tt o3V}&\mbox{open}&5&10^{-3}&1.0&2.4&4\times10^{-4}
\\
{\tt o4V}&\mbox{open}&5&10^{-4}&1.6&2.4&2\times10^{-4}
\\
{\tt o3X}&\mbox{open}&10&10^{-3}&0.2&1.0&2\times10^{-4}
\\
{\tt o4X}&\mbox{open}&10&10^{-4}&0.8&1.2&1\times10^{-4}
\\
{\tt o3L}&\mbox{open}&50&10^{-3}&0.03&0.15&5\times10^{-5}
\\
{\tt o4L}&\mbox{open}&50&10^{-4}&0.02&0.18&5\times10^{-5}
\\
\\\hline
\end{array}
\\
\end{array}
$
\end{center}
\end{table}

\begin{table}
\caption{
Velocity, mass flux,  force  and power of a jet with radius $r_{\rm j}=0.25x_0$
for our choices of the parameters $(\eta,M)$. These are non-relativistic
evaluations, assuming a number density of $10^{-2} \> {\rm cm}^{-2}$ and a
temperature of $10^7 \> {\rm K}$ in the ambient medium.
}
\label{table.jet.properties}
\begin{center}
$\begin{array}{rcclllllll}
M&\eta
&\begin{array}{c}{{v_{\rm j}}\over{T_7^{1/2} }}
\\({\rm cm} \>.\> {\rm s}^{-1})
\end{array}
&\begin{array}{c}{{\dot{M}_{\rm j}}\over{n_{-3} T_7^{1/2} }}
\\(M_\odot \>.\> {\rm yr}^{-1})
\end{array}
&&\begin{array}{c}{{F_{\rm j}}\over{n_{-3} T_7}}
\\({\rm dyn})
\end{array}
&&\begin{array}{c}
{{P_{\rm j}}\over{n_{-3} T_7^{3/2} }}\\({\rm erg} \>.\> {\rm s}^{-1})
\end{array}
\\\hline
\\
5&10^{-2}&2.36\times10^9
&1.39\times10^{-2}&&2.11\times10^{33}&&2.72\times10^{42}
\\
5&10^{-3}&7.45\times10^9
&4.40\times10^{-3}&&2.11\times10^{33}&&8.61\times10^{42}
\\
5&10^{-4}&2.36\times10^{10}
&1.39\times10^{-3}&&2.11\times10^{33}&&2.72\times10^{43}
\\
10&10^{-2}&4.71\times10^9
&2.78\times10^{-2}&&8.31\times10^{33}&&2.00\times10^{43}
\\
10&10^{-3}&1.49\times10^{10}
&8.79\times10^{-3}&&8.31\times10^{33}&&6.34\times10^{43}
\\
10&10^{-4}&4.71\times10^{10}
&2.78\times10^{-3}&&8.31\times10^{33}&&2.00\times10^{44}
\\
50&10^{-2}&2.36\times10^{10}
&1.39\times10^{-1}&&2.06\times10^{35}&&2.43\times10^{45}
\\
50&10^{-3}&7.45\times10^{10}
&4.40\times10^{-2}&&2.06\times10^{35}&&7.70\times10^{45}
\\
50&10^{-4}&2.36\times10^{11}
&1.39\times10^{-2}&&2.06\times10^{35}&&2.43\times10^{46}
\\
\\\hline
\end{array}$
\end{center}
\end{table}

Each simulation is initialised with the high-velocity, low-density jet material
extruding a small distance into the ambient medium.
At every subsequent time step the conditions of the initial jet
are recreated in the cells at the base of the jet on the left (inner $z$)
boundary of the grid,
and thus the jet's injected fluxes of mass, momentum and energy
remain constant in time.
We performed simulations with two different choices for the condition 
applied at
the left boundary outside the radius of the jet.
(1) A reflecting (``closed'') boundary condition
represents the effects of the symmetry plane through the nucleus,
felt by a jet that is close to its origin and counterjet.
(2) An ``open'' boundary allows free outflow  of mass, momentum and energy,
thereby representing systems where the hot-spot and bow shock
are distant from the nucleus and not directly influenced by the opposite lobe.
The outer $z$ and $r$ boundaries are open to outflow.

\subsection{Ray-tracing}
\label{s.raytracing}

In our simulations, a scalar variable $\varphi$, as introduced in
\citet{saxton01a}, is passively advected with the gas.
Cells occupied purely by jet material are initially assigned the value
$\varphi=1$; and the ambient medium initially has $\varphi=0$.
Thus $\varphi$ traces the local fraction of matter originating in the jet.

We adopt the approximation of using the pressure field to trace the emissivity
of the relativistic particles within jet plasma
\citep[e.g.][]{saxton01a}.
This approximation is based on the fact that the synchrotron emissivity,
$j_\nu$ is proportional to $p B^{1+\alpha}$  where $p$ is the pressure,
$B$ is the magnetic field and $\alpha$ is the spectral index.
If the magnetic
pressure, $B^2/8\pi$ tracks the particle pressure,
then $j_\nu\propto p^{(3+\alpha)/2}$,
and $\alpha\approx0.6$ typically.
Even if the magnetic pressure does not track the pressure faithfully,
it is likely that regions of high pressure still correspond to
regions of high emissivity.

We revolve the hydrodynamic data frames of $\varphi$ and $p$
to form 3D cylindrical structures.
Using $\varphi$ to distinguish synchrotron-emitting jet material
from gas originating in the background medium,
the weighted emissivity is then $j_\nu\propto\varphi p^{1.8}$.
By integrating $j_\nu$ along rays through the 3D volume
we generate surface brightness images of the jet and cocoon
at a selected orientation.
The resulting images are compared with observations
in \S\ref{s.results.raytracing}.

\section{Simulation Results}

\subsection{General features}

\begin{figure}[h]
\centering \leavevmode
$\begin{array}{cc}
\begin{array}{c}\includegraphics[width=6cm]{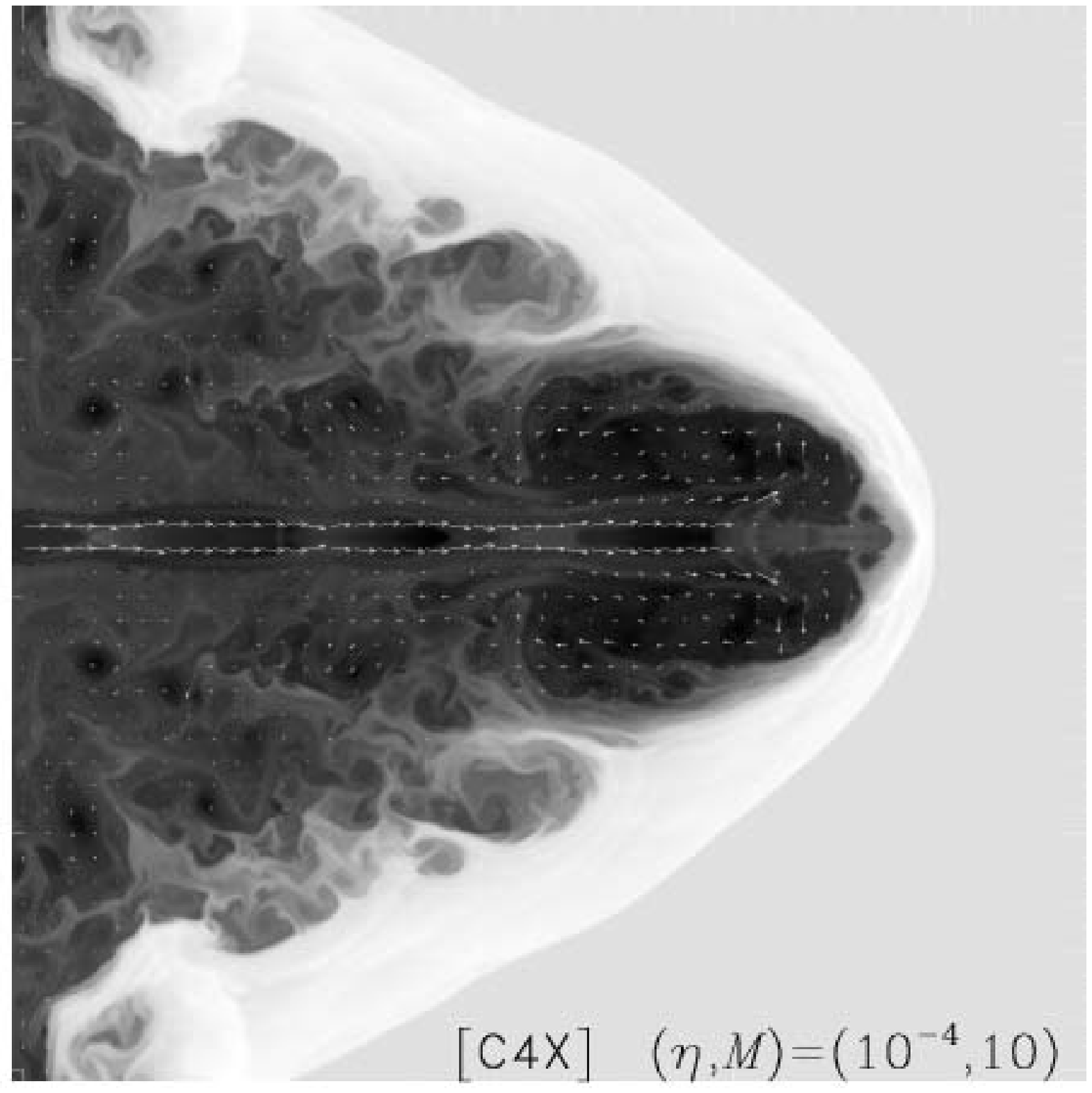}\end{array}
&
\begin{array}{c}\includegraphics[width=6cm]{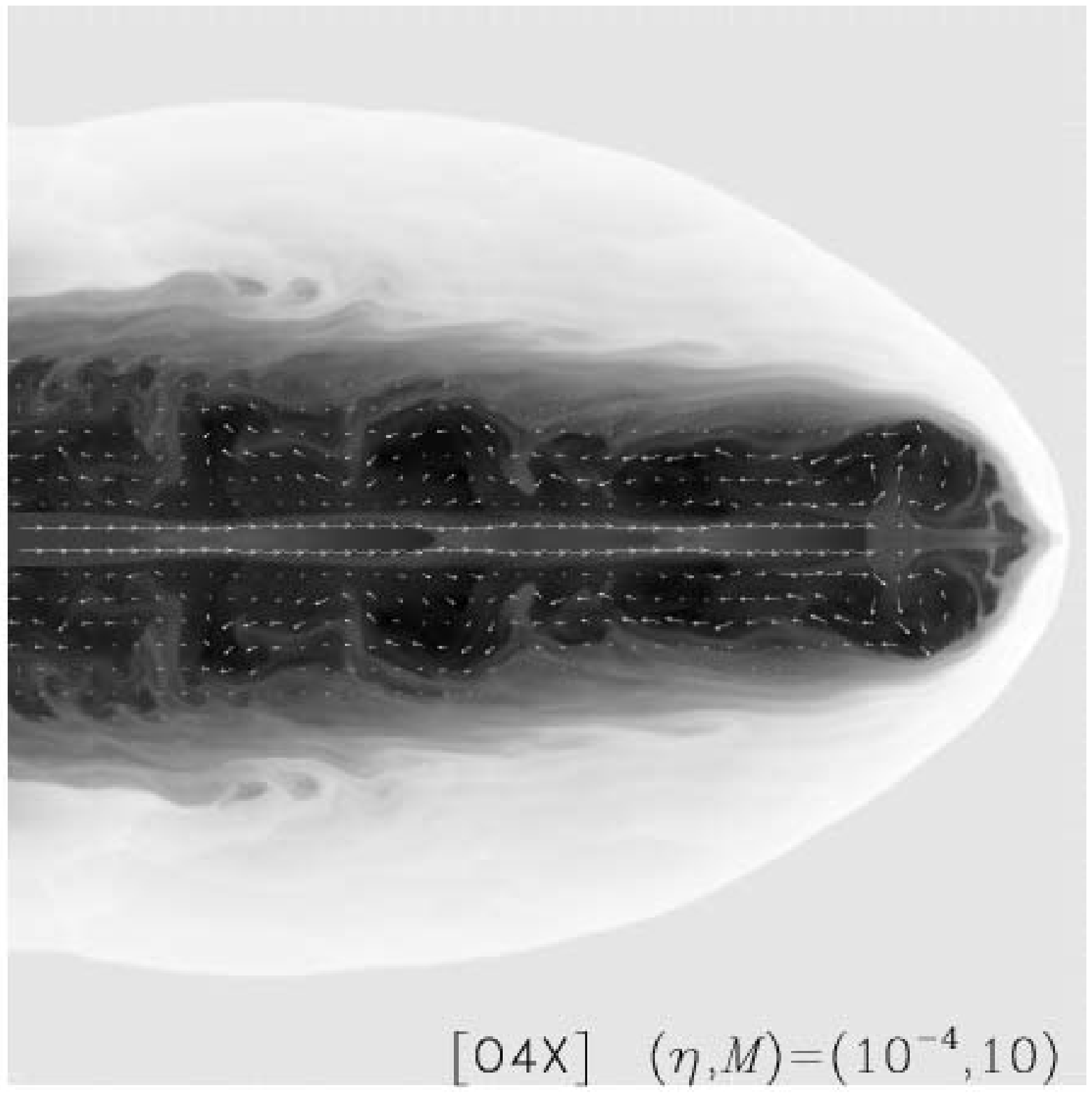}\end{array}
\\
\begin{array}{c}\includegraphics[width=6cm]{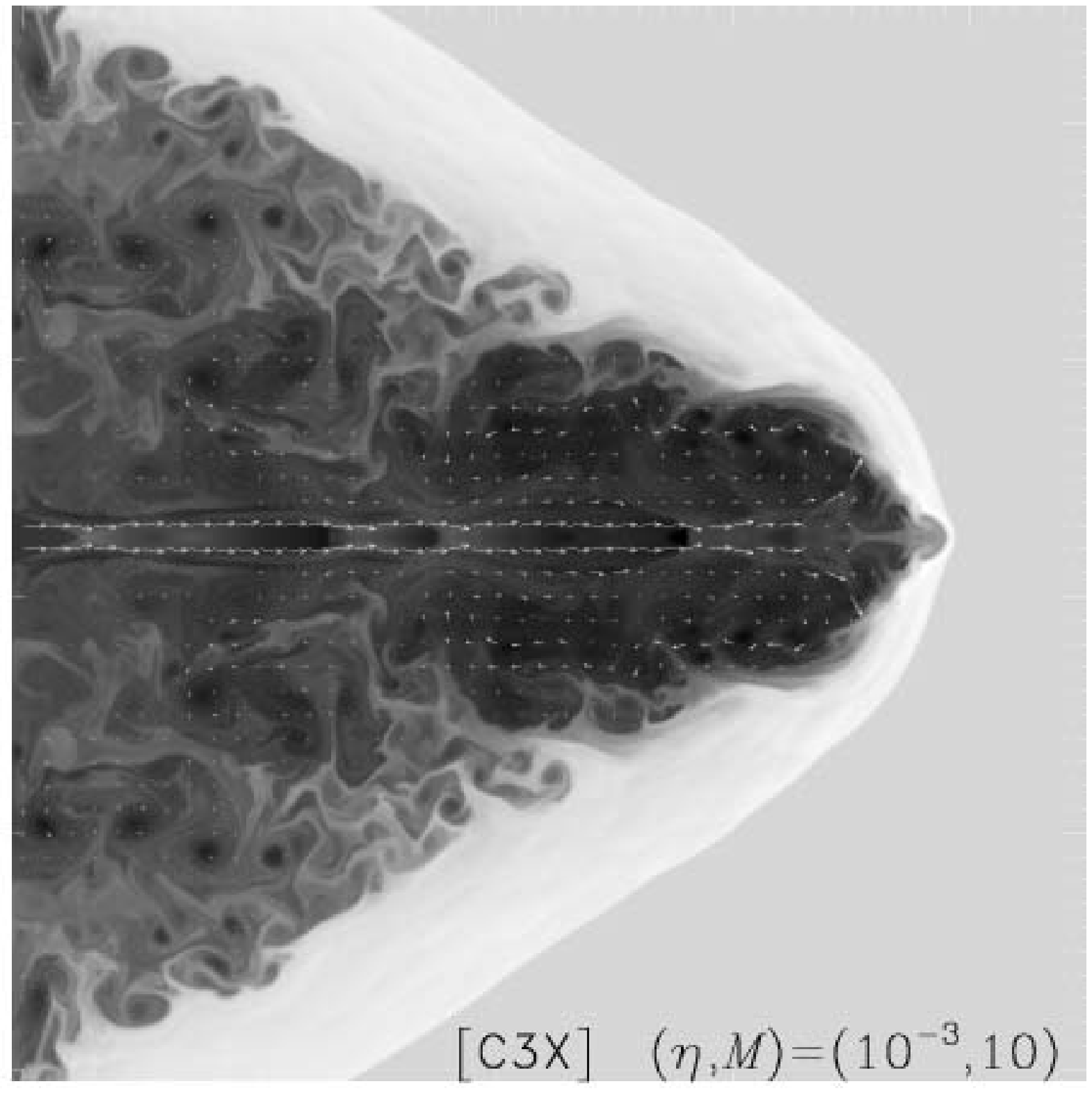}\end{array}
&
\begin{array}{c}\includegraphics[width=6cm]{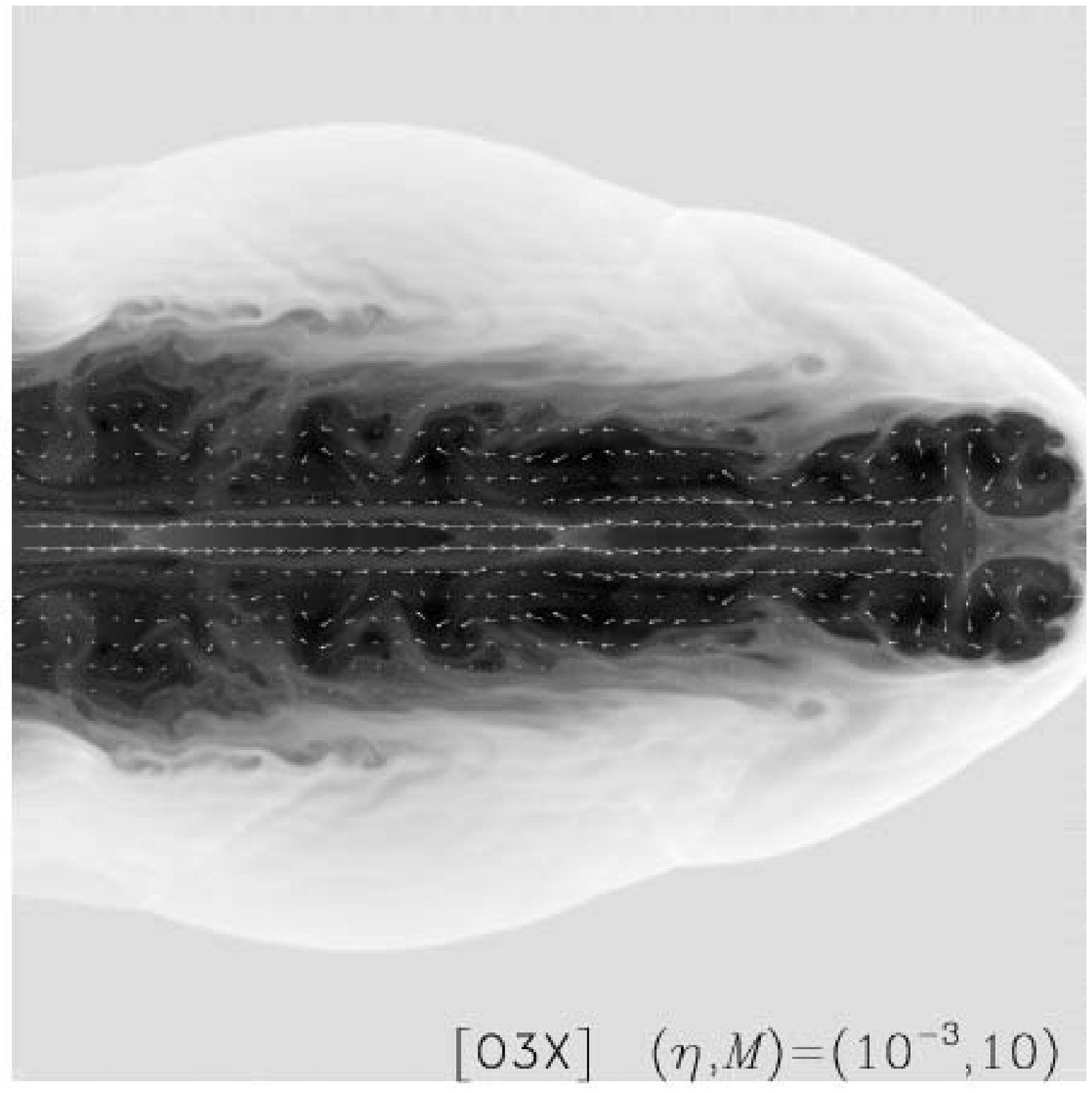}\end{array}
\\
\begin{array}{c}\includegraphics[width=6cm]{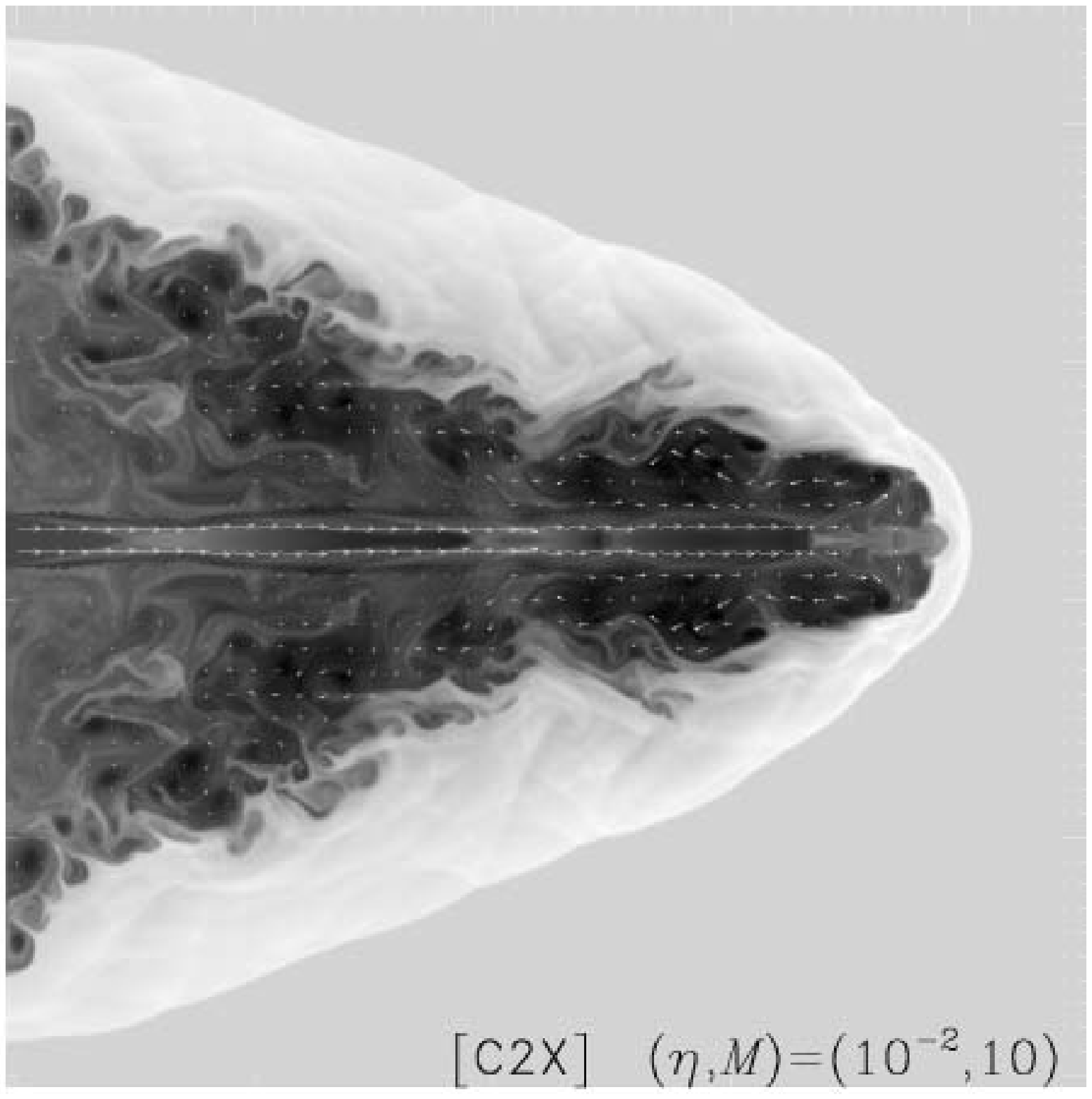}\end{array}
&
\begin{array}{c}\includegraphics[width=6cm]{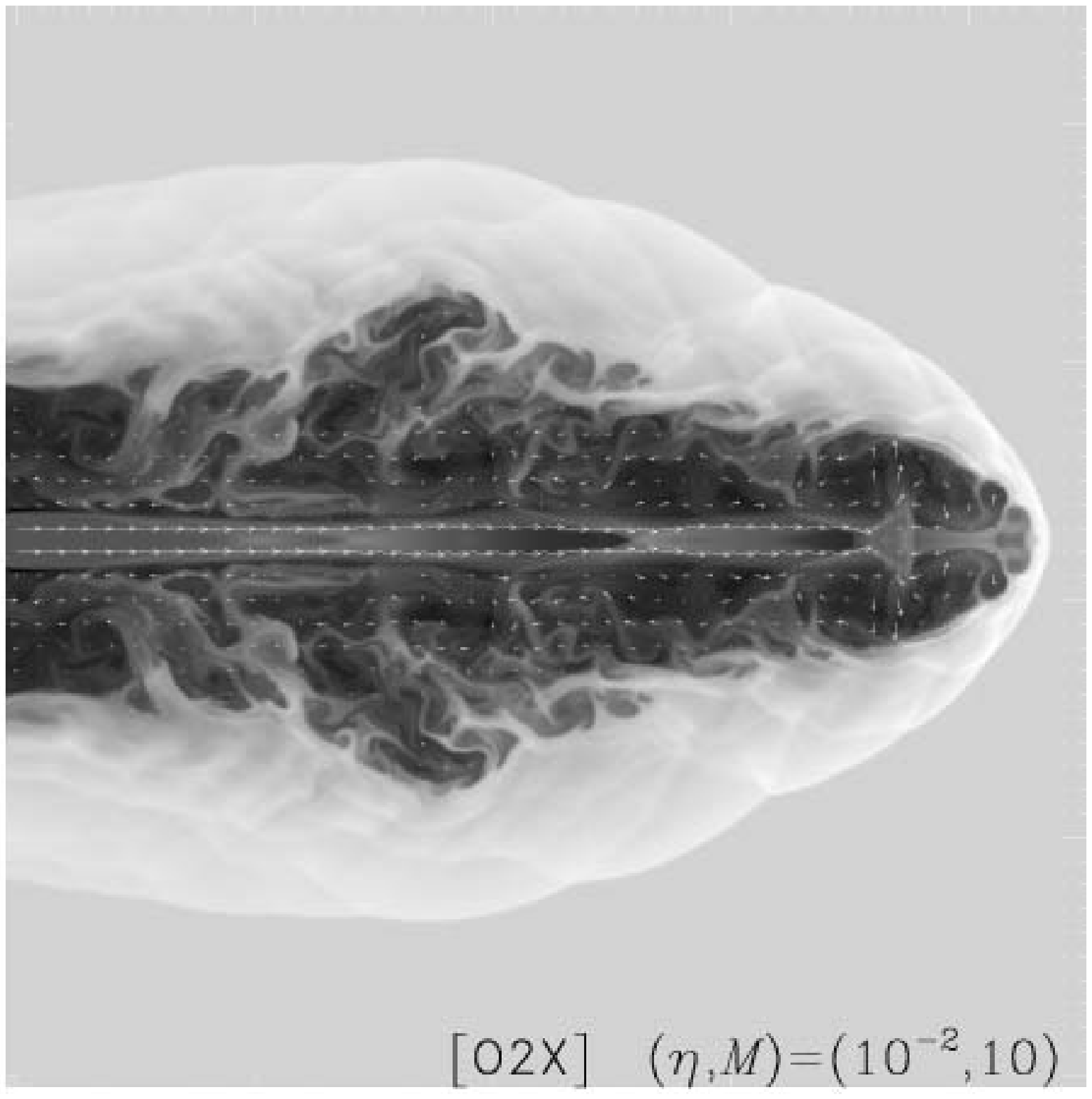}\end{array}
\\
\end{array}$
\caption{
Snapshots of flow velocity vectors superimposed on a $\log\rho$ image,
in a $450 \times 300$ pixel zone
($0<z<7.5x_0$, $r<2.5x_0$),
at comparable stages of the jet's advance across the grid.
Each row presents a different choice of $\eta$;
in all cases $M=10$.
The left and right columns show cases in which
the left boundary is closed and open, respectively.
}
\label{f:backflow.flow}
\end{figure}

\begin{figure}[h]
\centering \leavevmode
$\begin{array}{cc}
\begin{array}{c}\includegraphics[width=6cm]{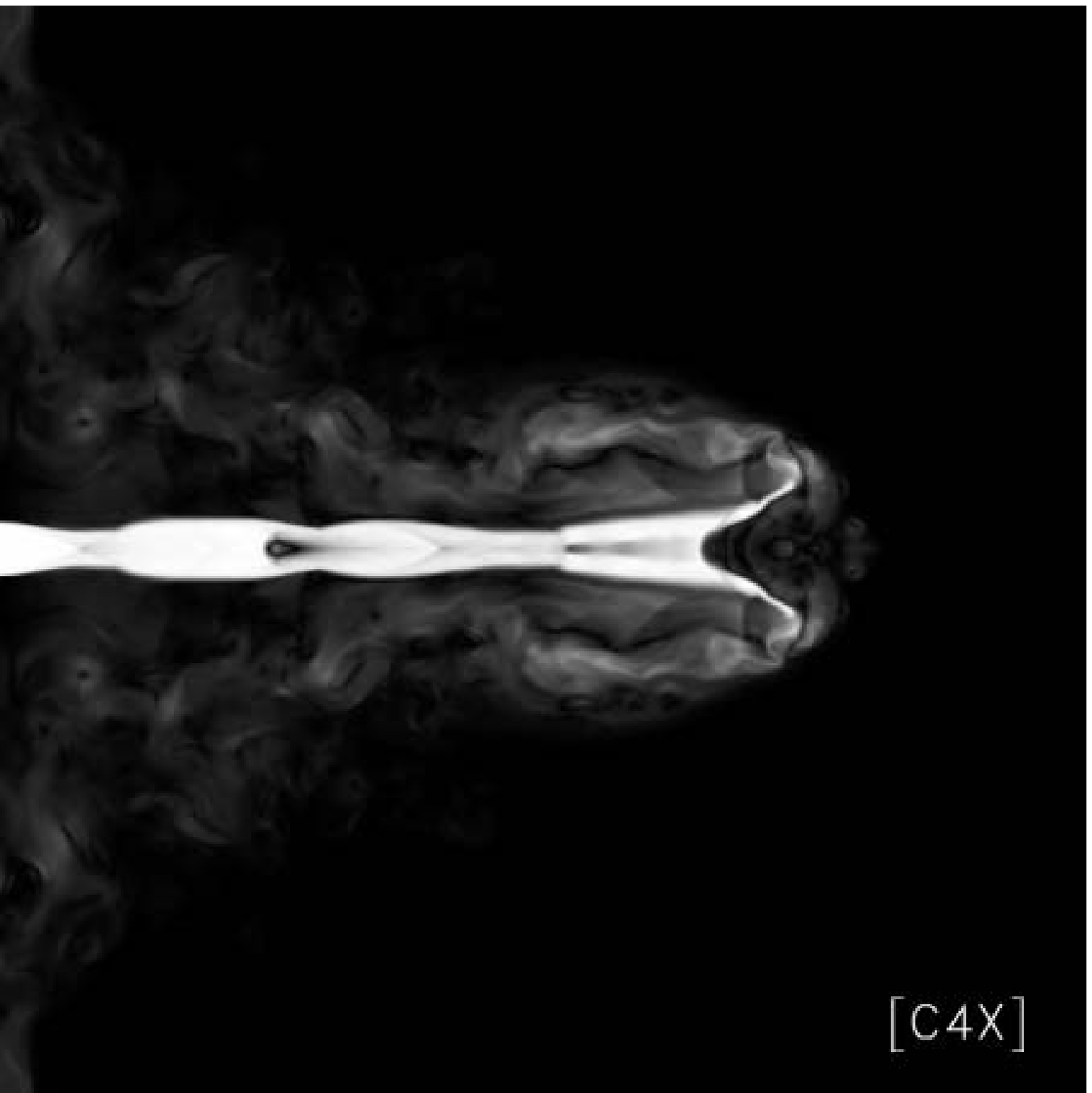}\end{array}
&
\begin{array}{c}\includegraphics[width=6cm]{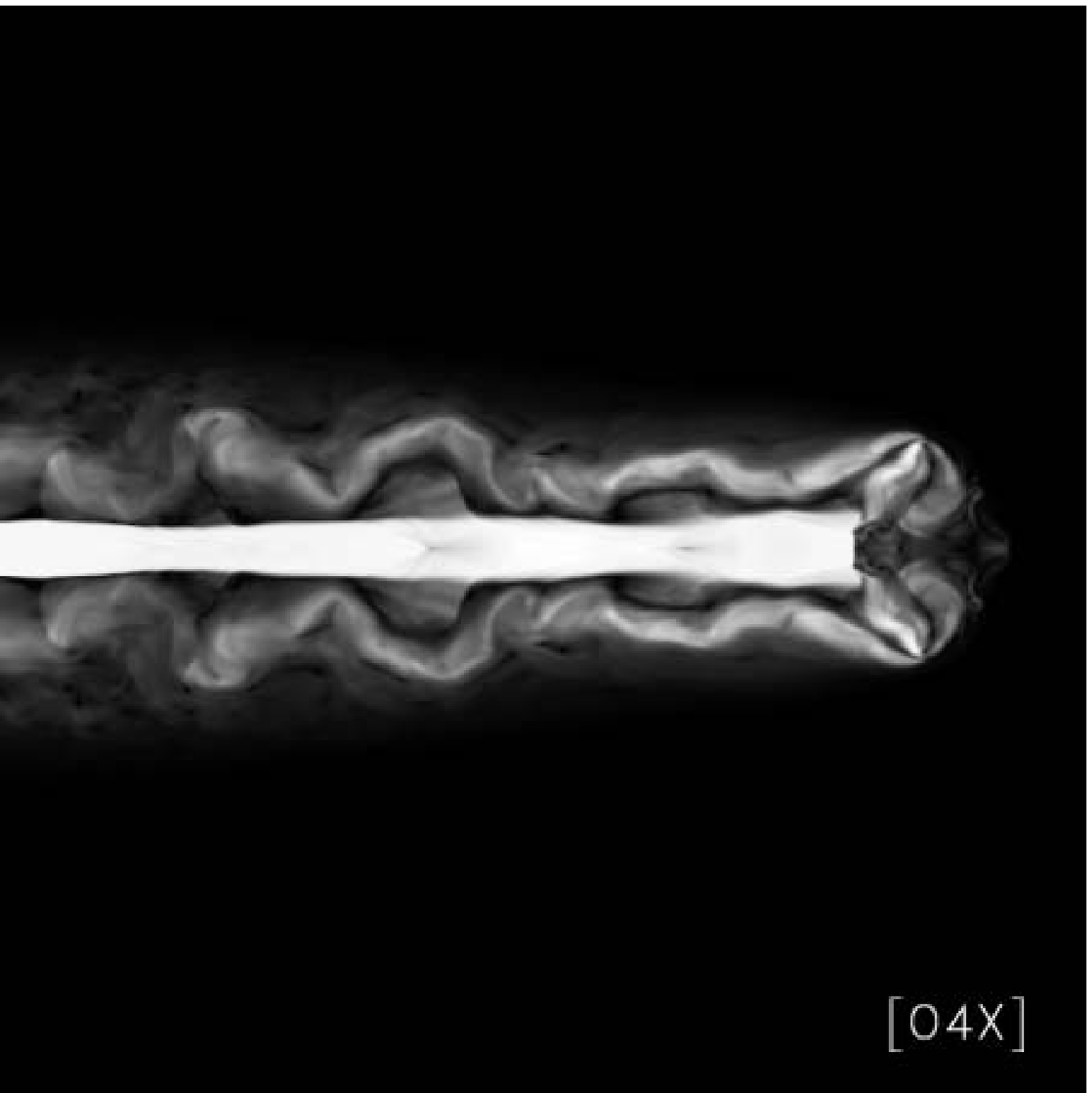}\end{array}
\\
\begin{array}{c}\includegraphics[width=6cm]{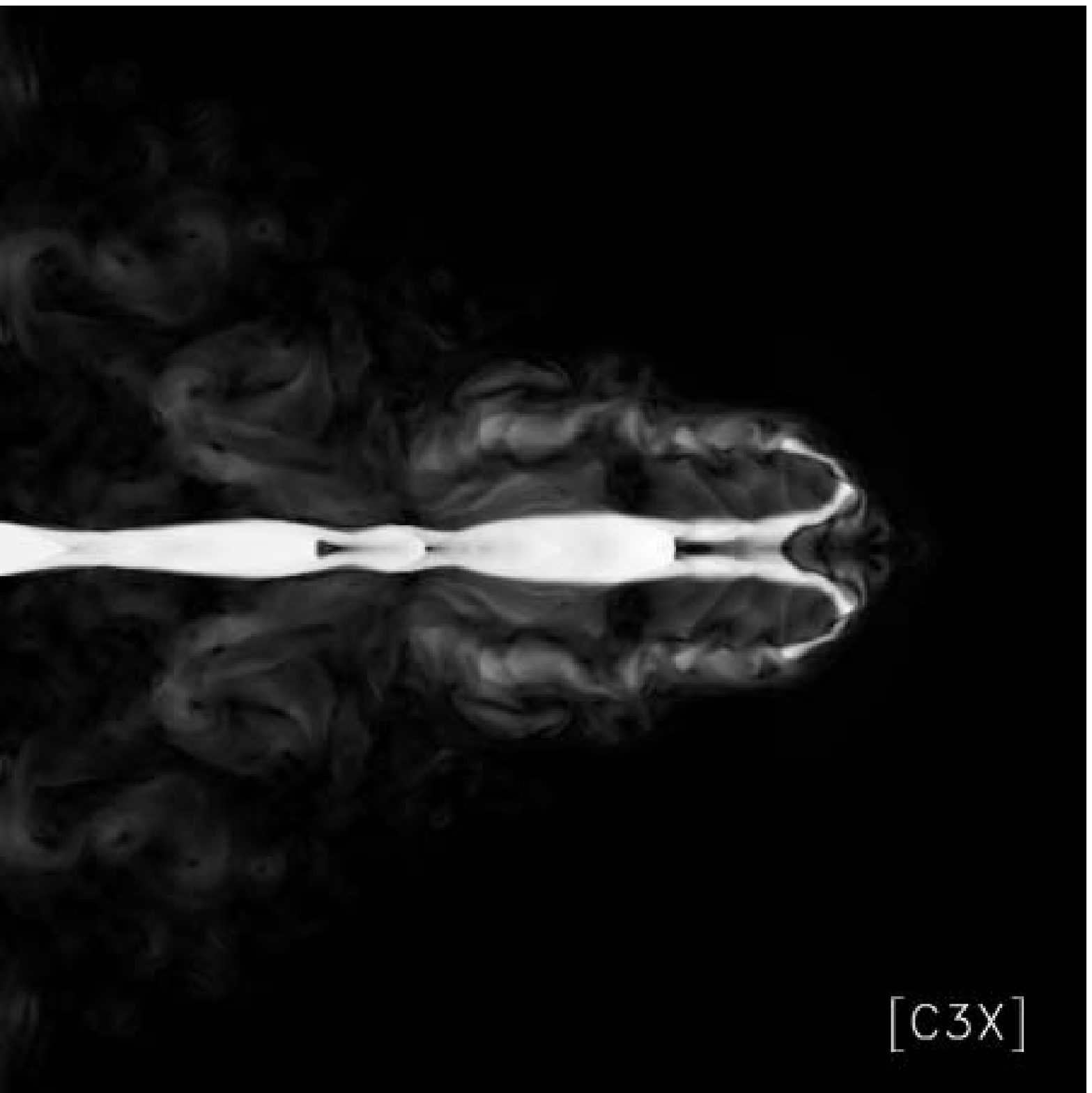}\end{array}
&
\begin{array}{c}\includegraphics[width=6cm]{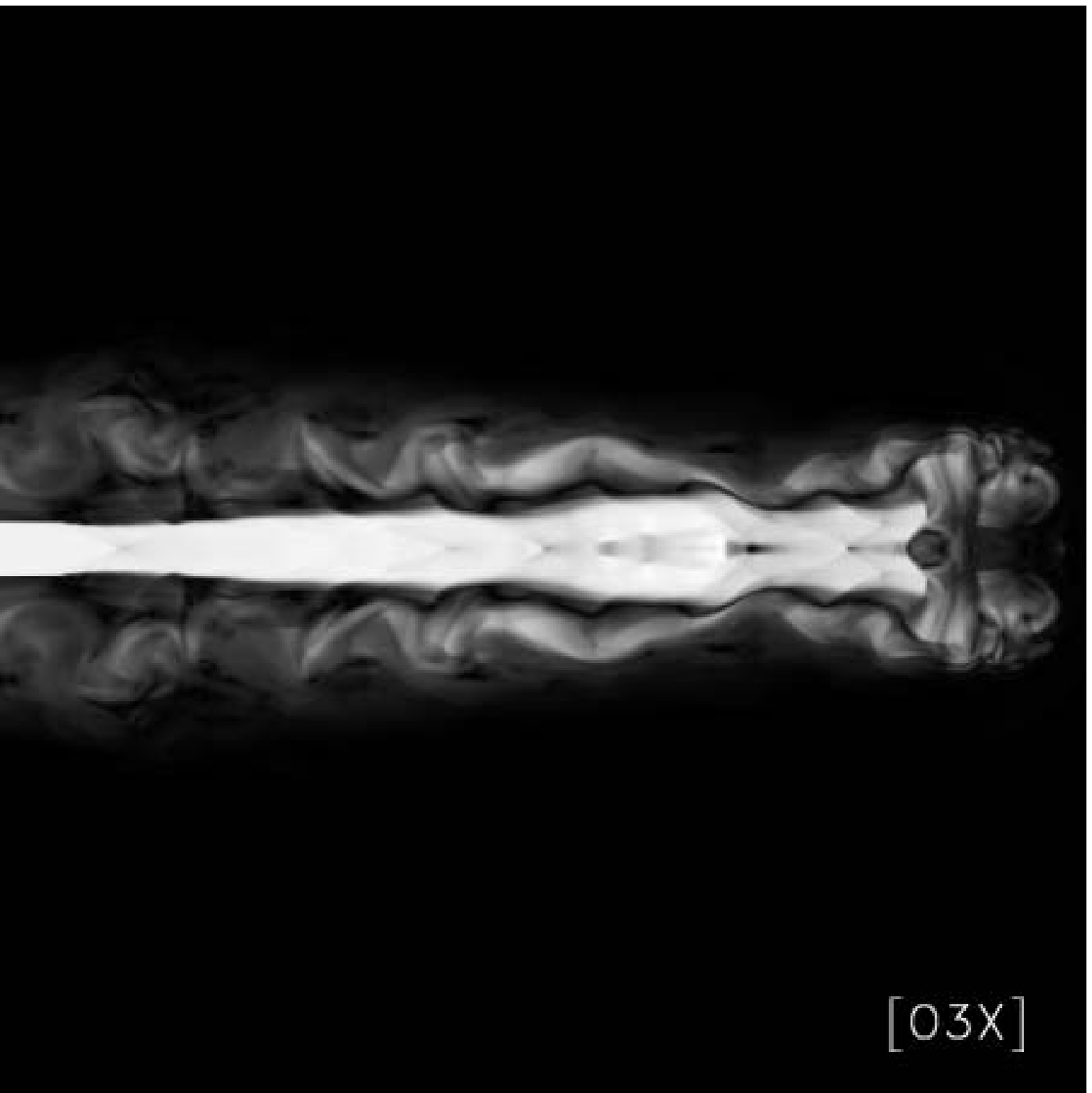}\end{array}
\\
\begin{array}{c}\includegraphics[width=6cm]{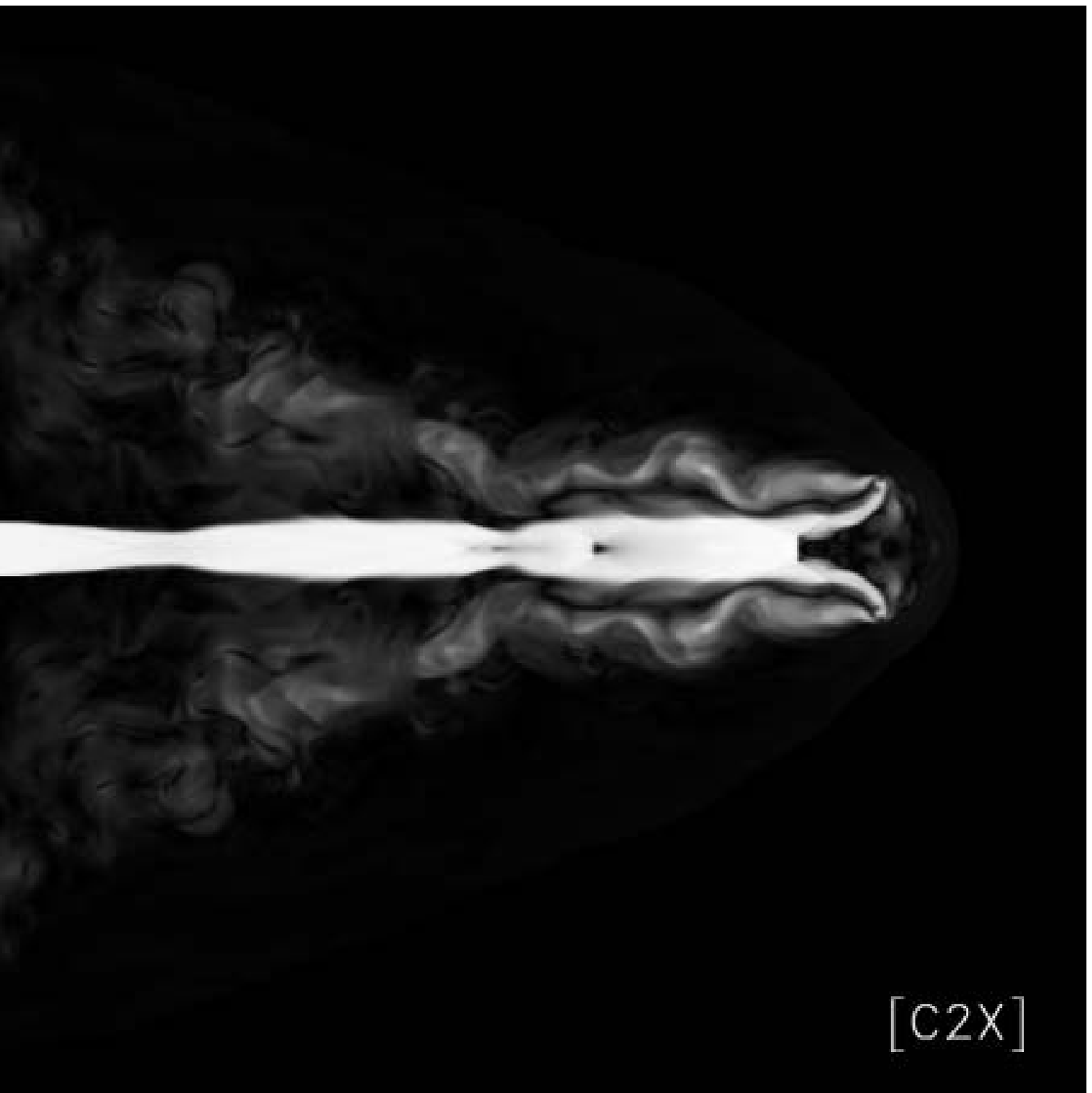}\end{array}
&
\begin{array}{c}\includegraphics[width=6cm]{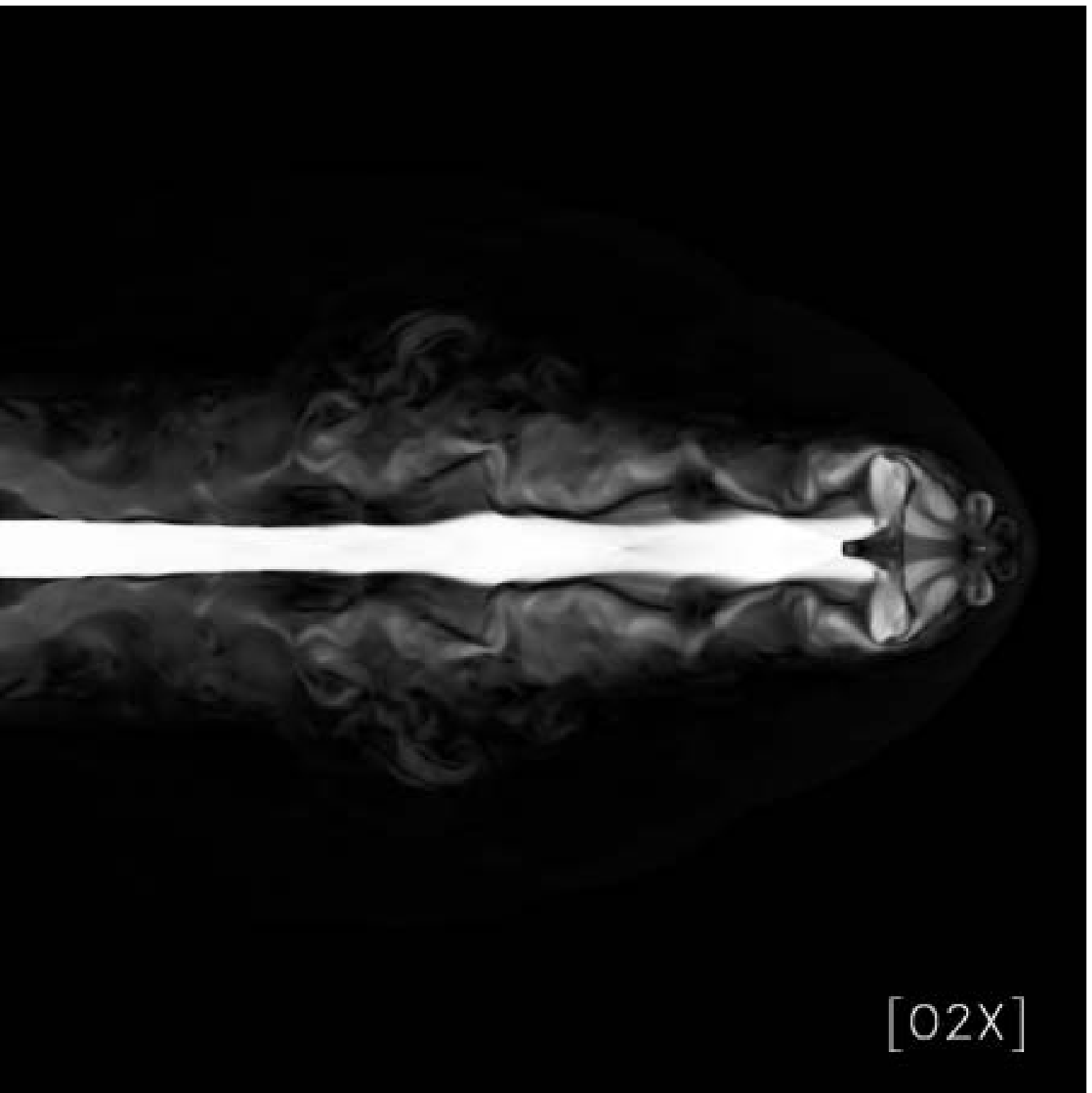}\end{array}
\\
\end{array}$
\caption{
Linearly scaled snapshots of the velocity magnitude,
with the initial jet velocity $v_{\rm j}$ assigned to maximum brightness,
for the same region, time and choice of parameters
as in Figure~\ref{f:backflow.flow}.
}
\label{f:backflow.velocity}
\end{figure}

\begin{figure}[h]
\centering \leavevmode
$\begin{array}{cc}
\begin{array}{c}\includegraphics[width=6cm]{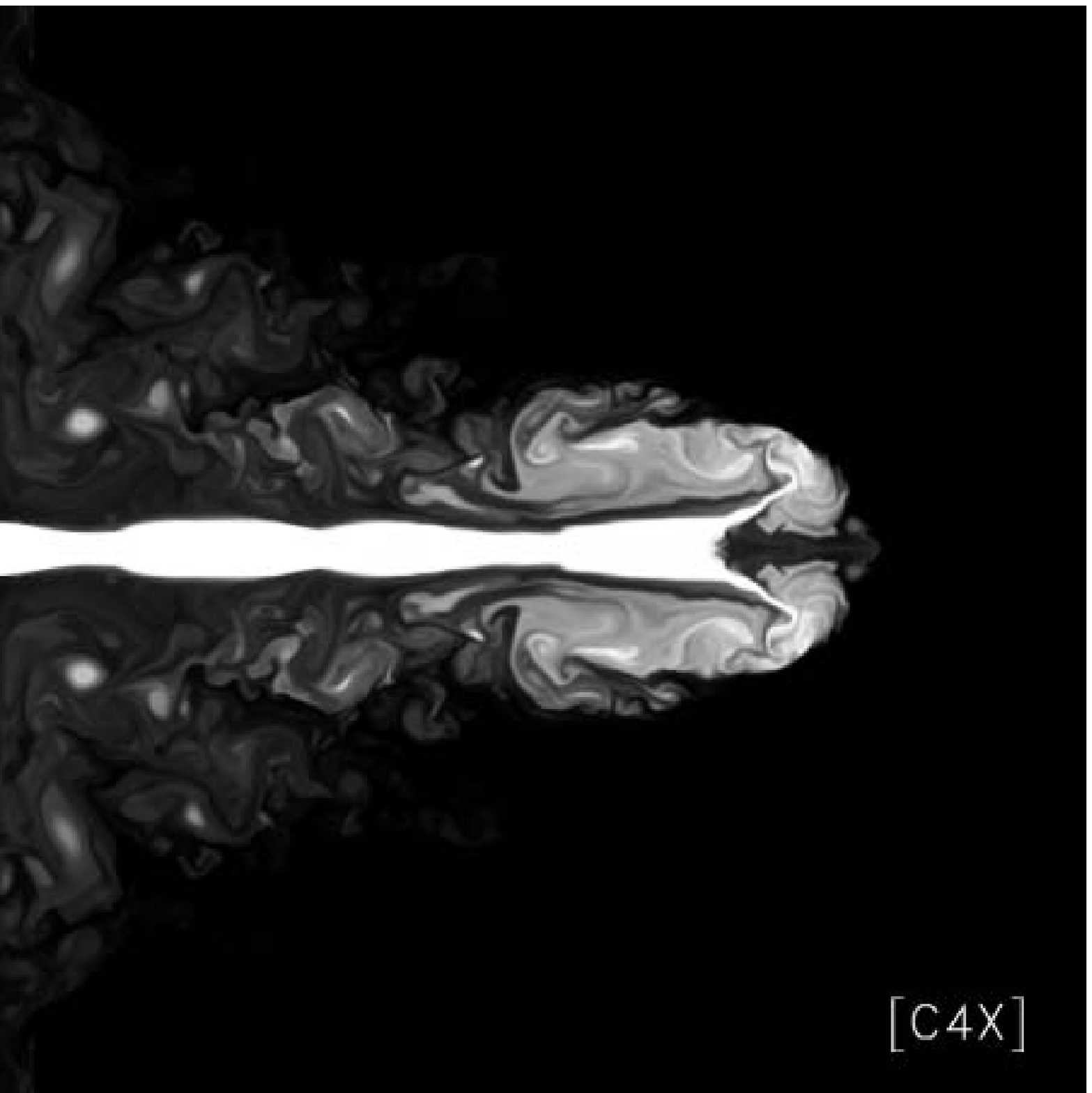}\end{array}
&
\begin{array}{c}\includegraphics[width=6cm]{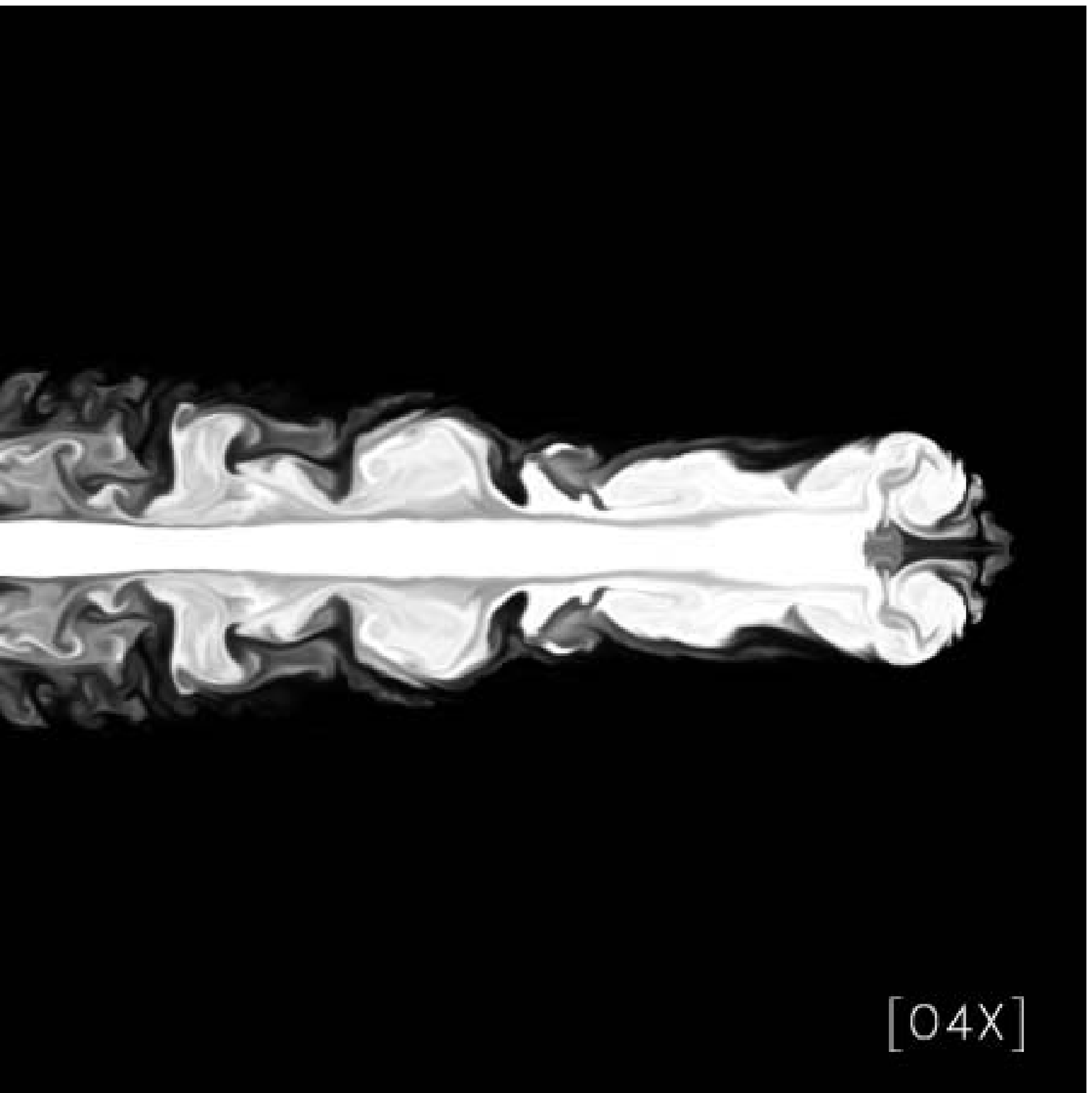}\end{array}
\\
\begin{array}{c}\includegraphics[width=6cm]{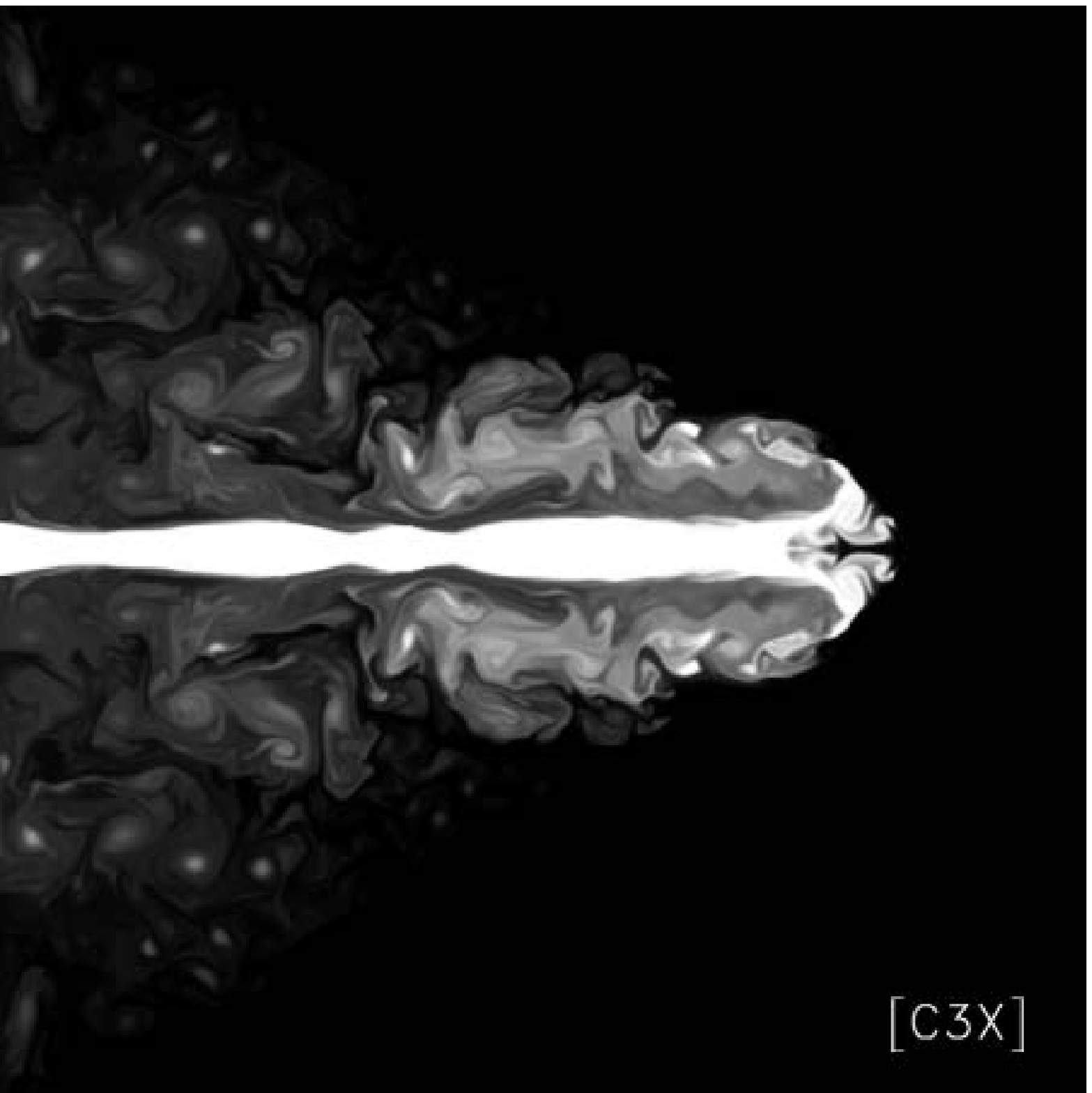}\end{array}
&
\begin{array}{c}\includegraphics[width=6cm]{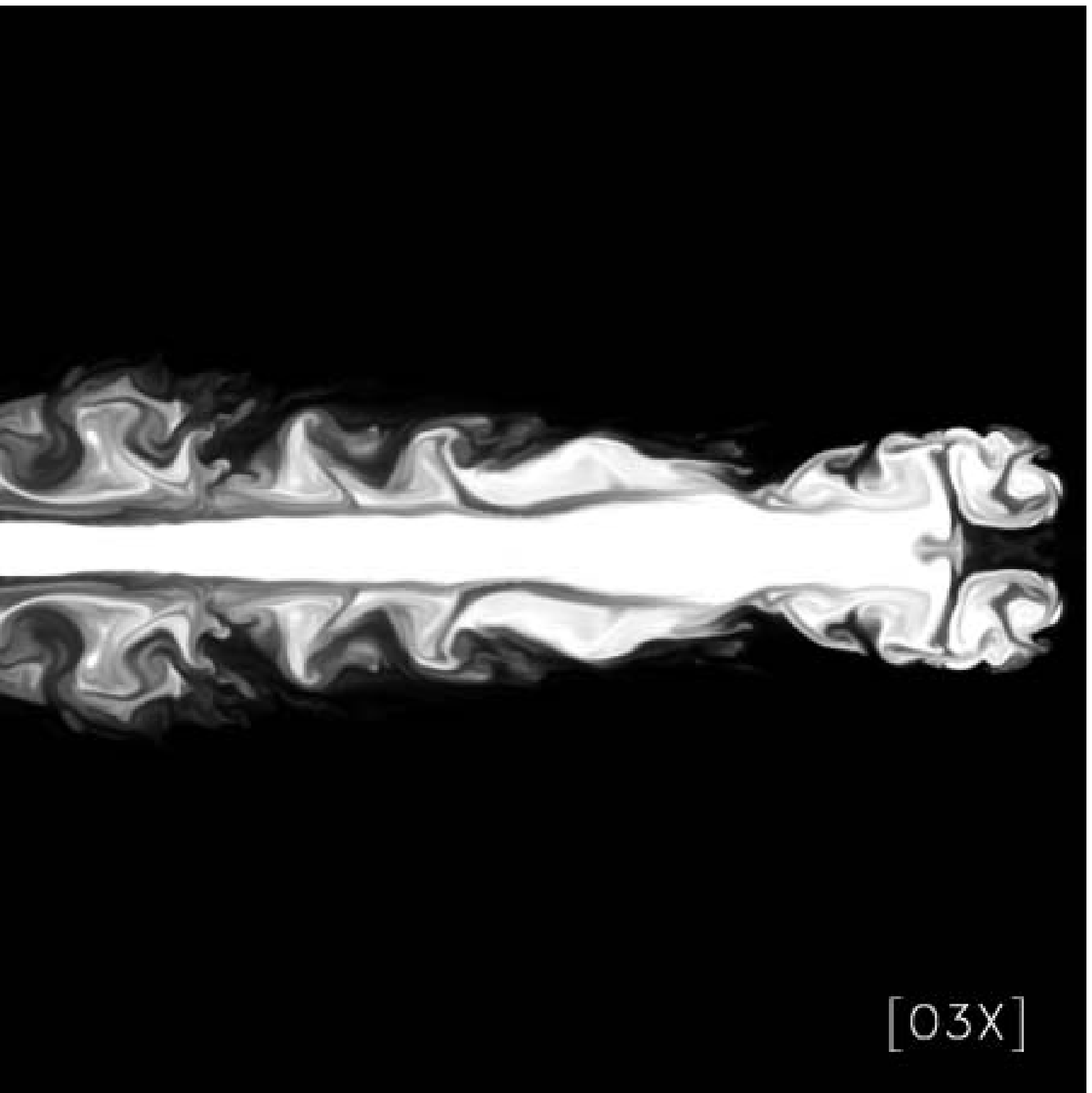}\end{array}
\\
\begin{array}{c}\includegraphics[width=6cm]{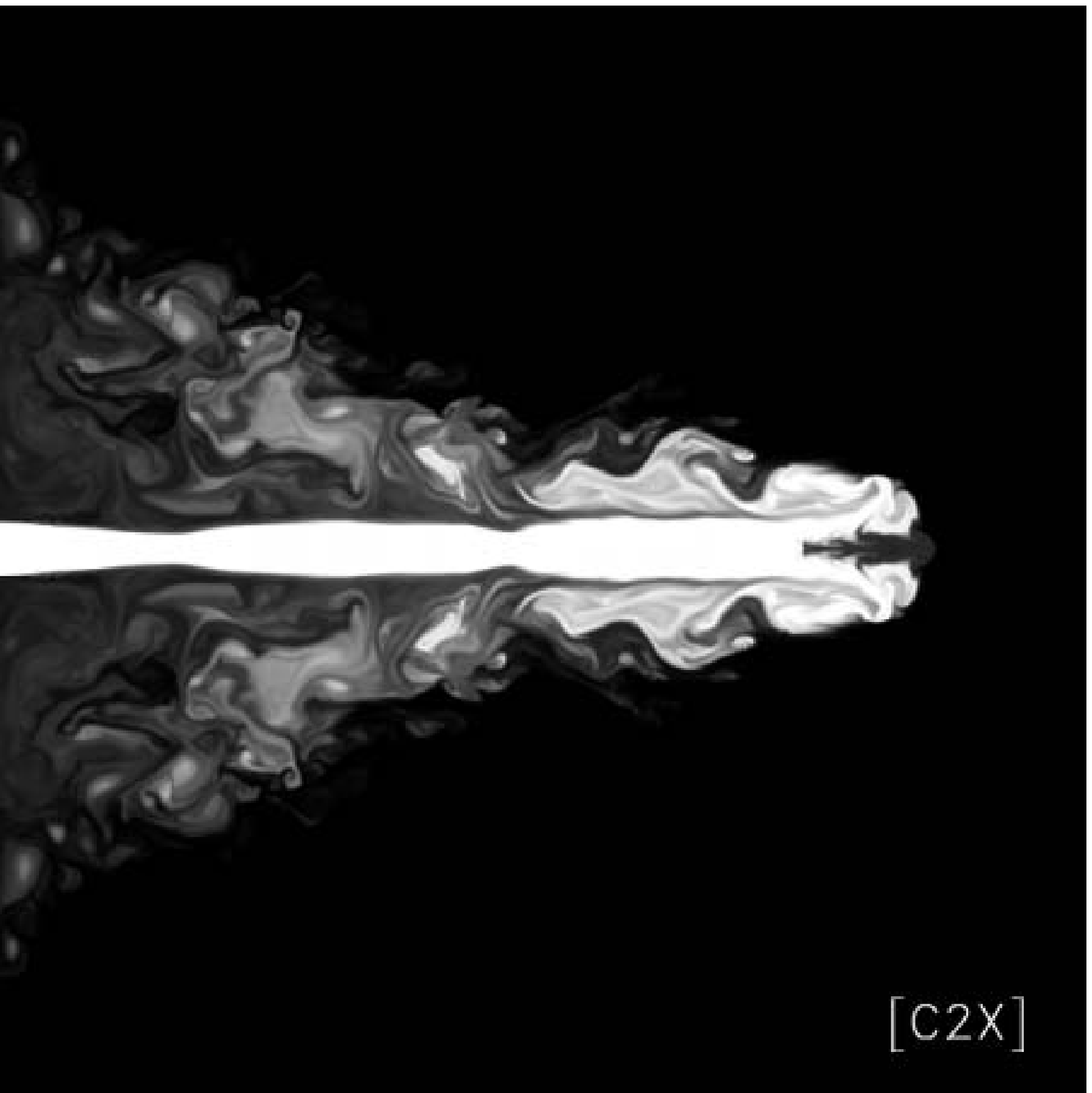}\end{array}
&
\begin{array}{c}\includegraphics[width=6cm]{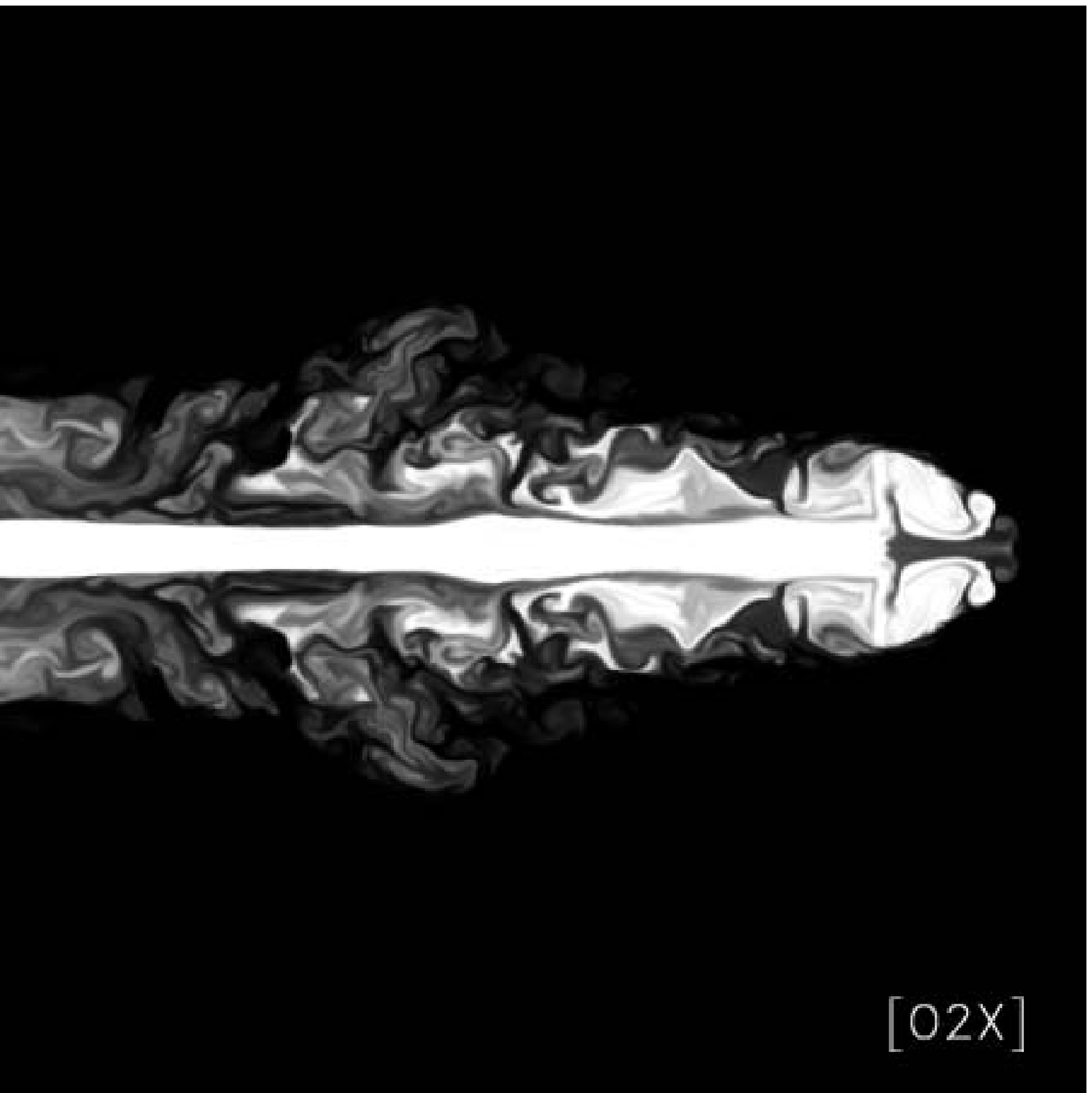}\end{array}
\\
\end{array}$
\caption{
Scalar tracer $\varphi$
showing the distribution of jet material
for the same region, time and choice of parameters
as in Figure~\ref{f:backflow.flow}.
White ($\varphi=1$) represents pure jet plasma;
black ($\varphi=0$) represents gas originating in the ambient medium.
}
\label{f:backflow.tracer}
\end{figure}

The density distribution and velocity fields of representative frames
from six of the simulations are shown in Figure~\ref{f:backflow.flow}.
At the beginning of each simulation,
a bow shock is quickly established around the jet.
The regions just downstream and inside the bow shock
are dominated by the shocked dense interstellar medium (ISM).
The innermost regions immediately surrounding the jet
are occupied by a cocoon of low-density gas
moving in a turbulent backflow opposite the jet direction.
Flow velocities in some regions are comparable to
the initial jet velocity $v_{\rm j}$,
and are much larger than the bow shock velocity
(see Figure~\ref{f:backflow.velocity}).
Turbulent velocities are typically lower in the outer parts of the cocoon,
further from the jet.
The scalar tracer $\varphi$
(see Figure~\ref{f:backflow.tracer})
reveals that jet-derived matter dominates
the inner regions of the backflow,
whereas the outer regions just downstream of the bow shock
are entirely composed of background gas.

The evolution of the backflow depends upon the left boundary condition.
When the left boundary of the grid is open to outflow,
the backflow occupies a cylindrical ``sheath''
surrounding and flowing opposite the jet,
ultimately flowing off the grid.
This ``sheath'' has a radius a few times that of the jet.
It increases gradually as the jet progresses across the grid,
but is insensitive to $\eta$.
Very little of the surrounding gas mixes with
backflowing jet plasma:
most of the fast-moving backflow
still exhibits high concentrations of jet plasma ($\varphi\ga 0.5$)
when it exits the left boundary
(see right column of Figure~\ref{f:backflow.tracer}).

In simulations with a closed, reflecting boundary,
cocoon eddies accumulate inside a roughly conical volume,
which is widest near the left boundary,
and tapers towards the head of the jet.
This ``wake'' cocoon
expands laterally as the jet progresses forwards,
and jet material accumulates and mixes with
the shocked ISM.
The eddies are typically smaller
and have lesser concentrations of jet matter (lower $\varphi$)
with increasing distance from the head of the jet
(see upper panels of Figure~\ref{f:backflow.tracer}).

The density contrast of the jet, $\eta$,
affects the gross structure of
the surrounding radio emitting regions.
If the left boundary is closed,
jets with lower $\eta$ produce wider bow shocks and wider ``wake'' cocoons
(see left panels of Figure~\ref{f:backflow.flow}).
However in cases with an open boundary
the  high-$\varphi$, fast backflowing ``sheath'' cocoon
has approximately the same width,
$\sim 4r_{\rm j} - 5 r_{\rm j}$, for all the values of $\eta$
we have studied: $\eta=10^{-2}, 10^{-3}$ and $10^{-4}$
(see right panels of Figure~\ref{f:backflow.tracer}).
In such systems the bow shock widens marginally with decreasing $\eta$.

The amount of mixing between jet and ISM also shows a dependence upon $\eta$.
For the open boundary mixing per unit length is more pronounced in the backflow
for the larger values of $\eta$ (see the right panels of
Figure~\ref{f:backflow.tracer}).

There is a similar dependence for the case of the closed boundary with some
additional features.
For $\eta=10^{-2}$,
the cocoon widens almost linearly from the head of the jet
(radius $\la3r_{\rm j}$)
and the degree of mixing in the backflow
increases more or less gradually with distance from this point.
However in cases with $\eta=10^{-4}$,
the head of the jet is persistently surrounded by
a broad (radius $\sim6r_{\rm j}$), low-density, high-$\varphi$ ellipsoidal
region  that is essentially the first turbulent cell in the backflow.
(see upper-left panels of
Figure~\ref{f:backflow.flow},~\ref{f:backflow.tracer}).
There is a lot of mixing at the trailing edge of this cavity
and average $\varphi$ values decline rapidly
to the left of that region.
Cases with $\eta=10^{-3}$ show an intermediate behaviour:
the cavity is more extended in the $z$ direction
and includes more fingers of entrained dense gas.

In the ultra-light ($\eta = 10^{-4}$) jets with an enduring cavity,
the head of the jet surges back and forth along the axis
between the front and back ends of the cavity.
Similar surging behaviour occurs in most of the simulations,
to different extents and at different frequencies.
This cyclic process begins when
turbulence in the backflow briefly compresses and constricts the jet
at one point
(see Figure~\ref{f:surging}, panels~2 \& 3).
A portion of the front of jet is cut off,
and its remnant mixes into the front of the cocoon.
Jet plasma accumulates upstream of the obstruction
(Figure~\ref{f:surging}, panel~4),
until it is able to surge through to a point much closer to the bow shock
(Figure~\ref{f:surging}, panels~5 \& 6).
Thus the terminal shock of the jet is not a constant feature:
it may occur close to the bow shock
(when the jet has just undergone a surge),
or it may occur many jet diameters further upstream
(when the jet is being disturbed by turbulence).
The extent and frequency of surging depends on the jet parameters
and the condition of the left boundary.
In the closed-boundary, $\eta=10^{-4}$ cases,
the strong, persistent, frontal eddy
frequently pinches off the jet at the back end of the cavity,
where the backflow converges back towards the surface of the jet.
Cases with $(\eta,M)=(10^{-4},50)$
exhibit unusually extreme surging behaviour,
because the relatively violent turbulence throughout the entire cocoon
can pinch this ultra-light jet at almost any point along its length.

\begin{figure}[h]
\centering \leavevmode
$\begin{array}{cc}
\begin{array}{c}\includegraphics[width=6cm]{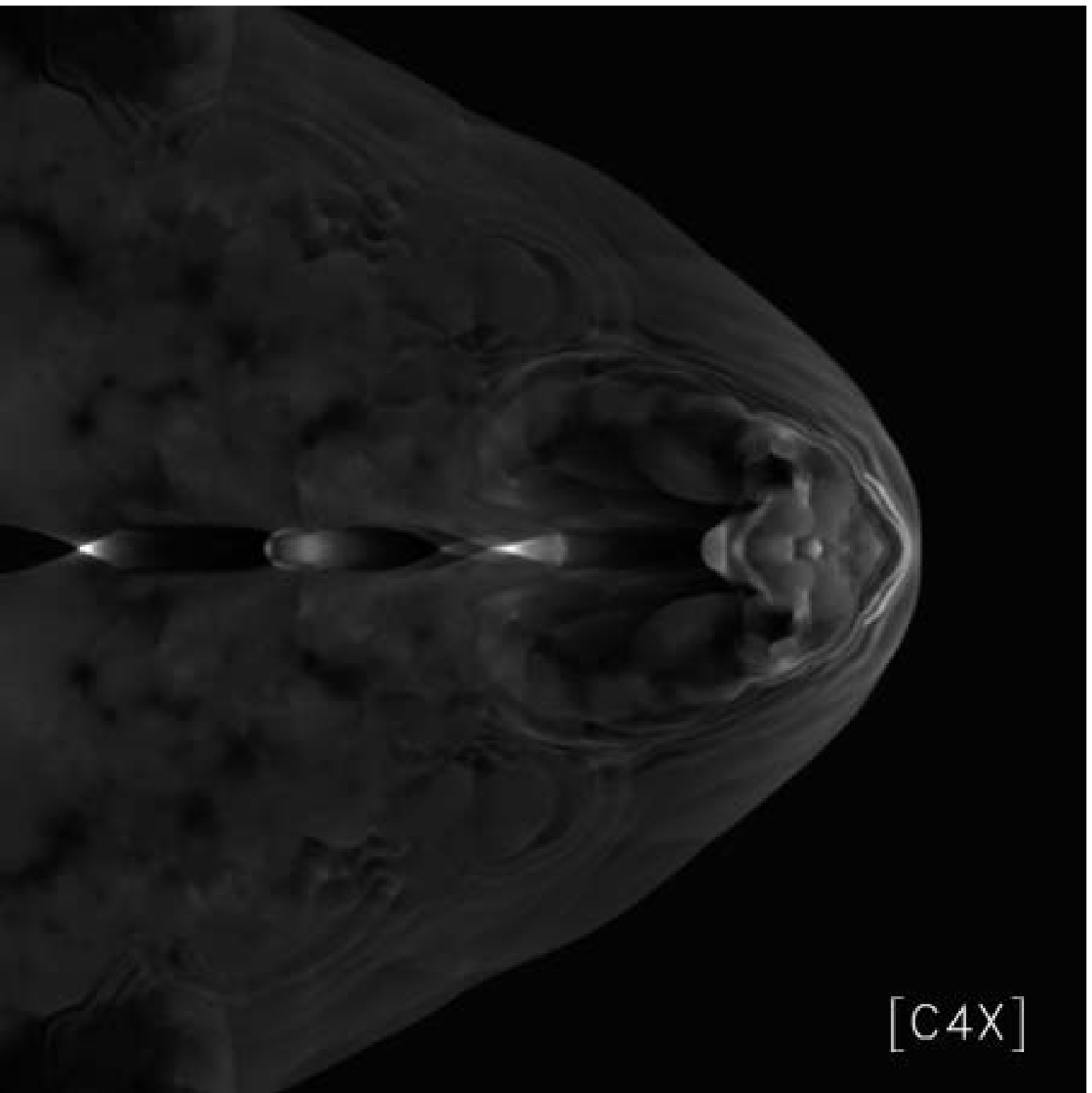}\end{array}
&
\begin{array}{c}\includegraphics[width=6cm]{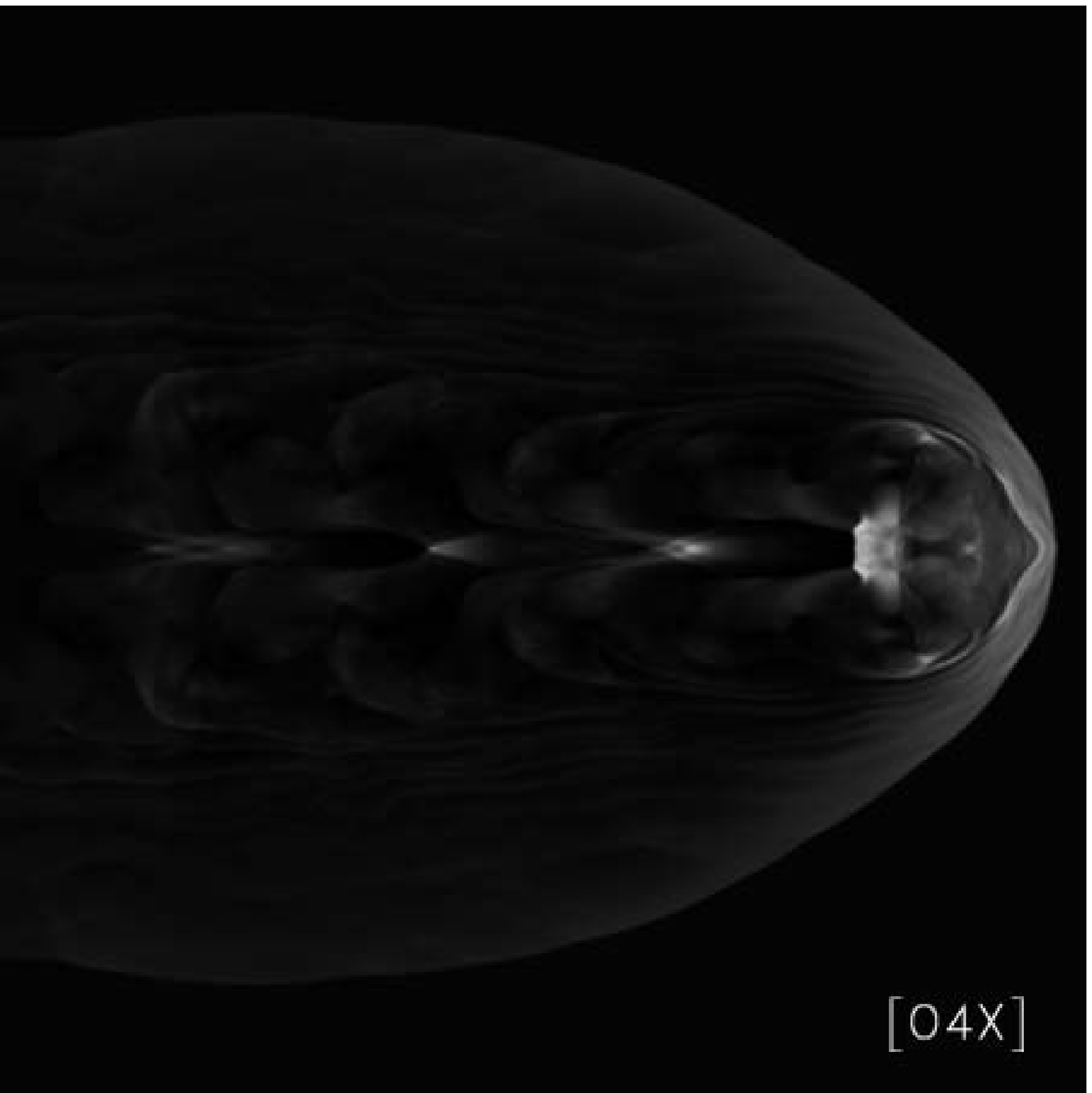}\end{array}
\\
\begin{array}{c}\includegraphics[width=6cm]{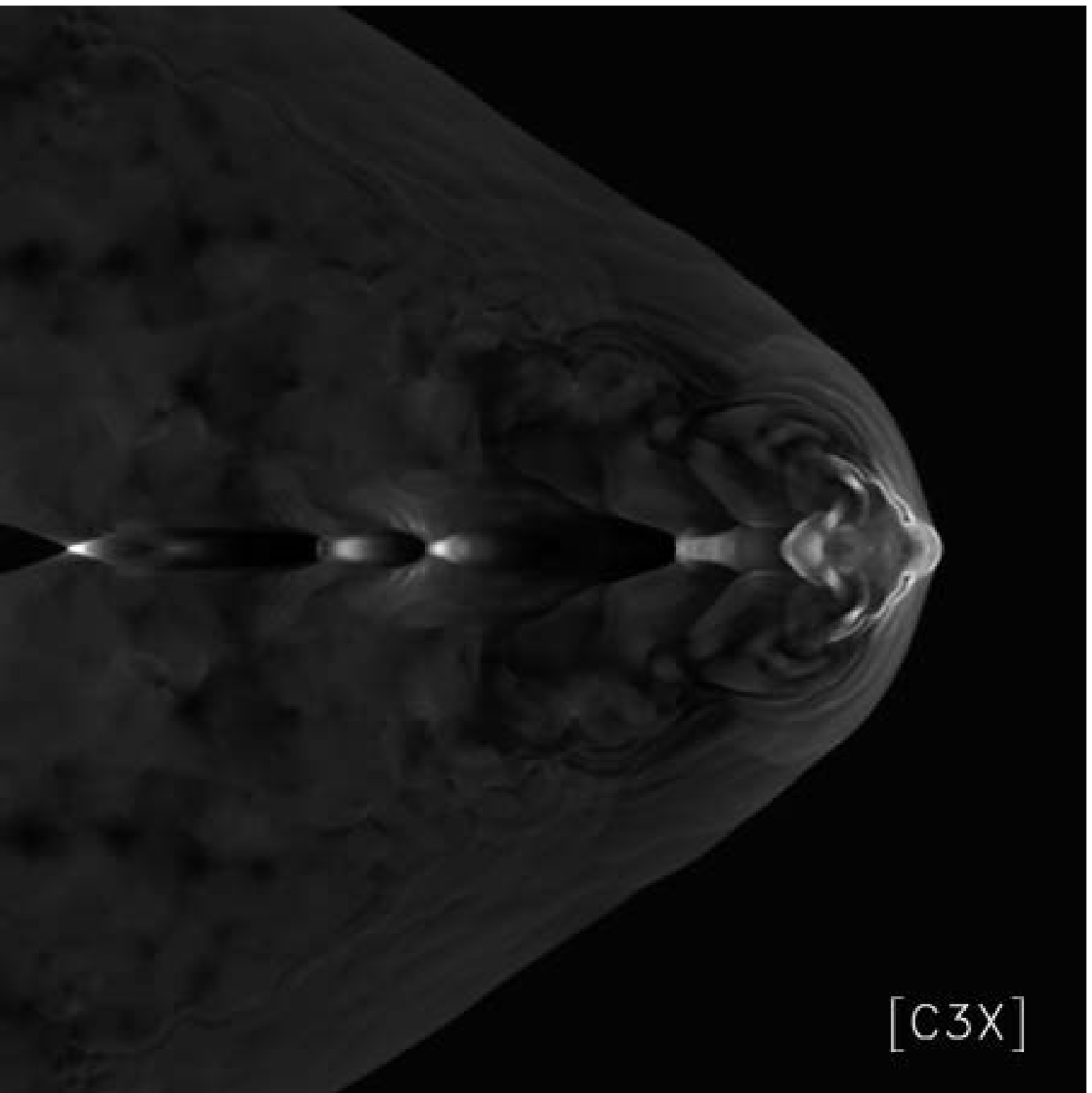}\end{array}
&
\begin{array}{c}\includegraphics[width=6cm]{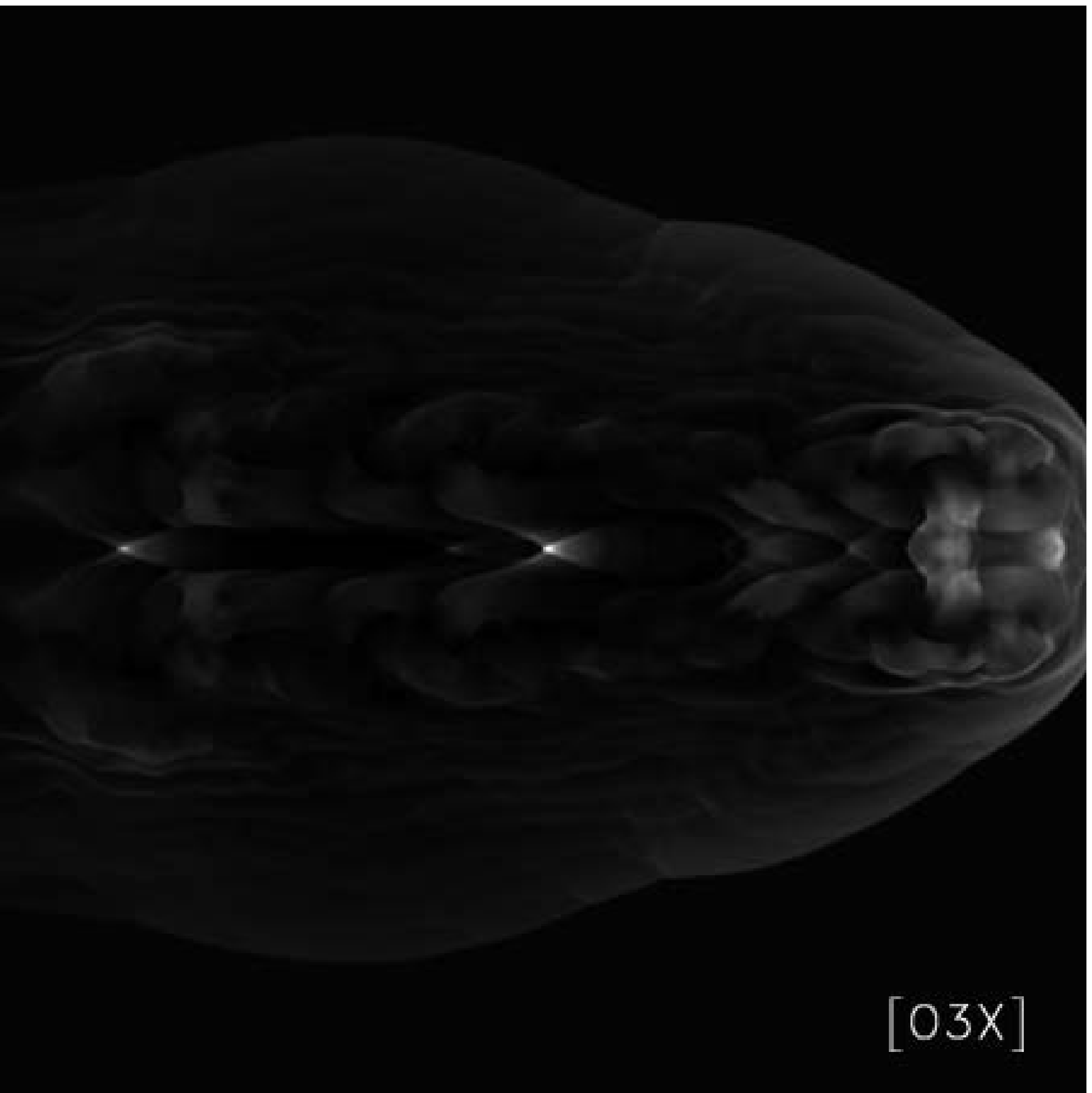}\end{array}
\\
\begin{array}{c}\includegraphics[width=6cm]{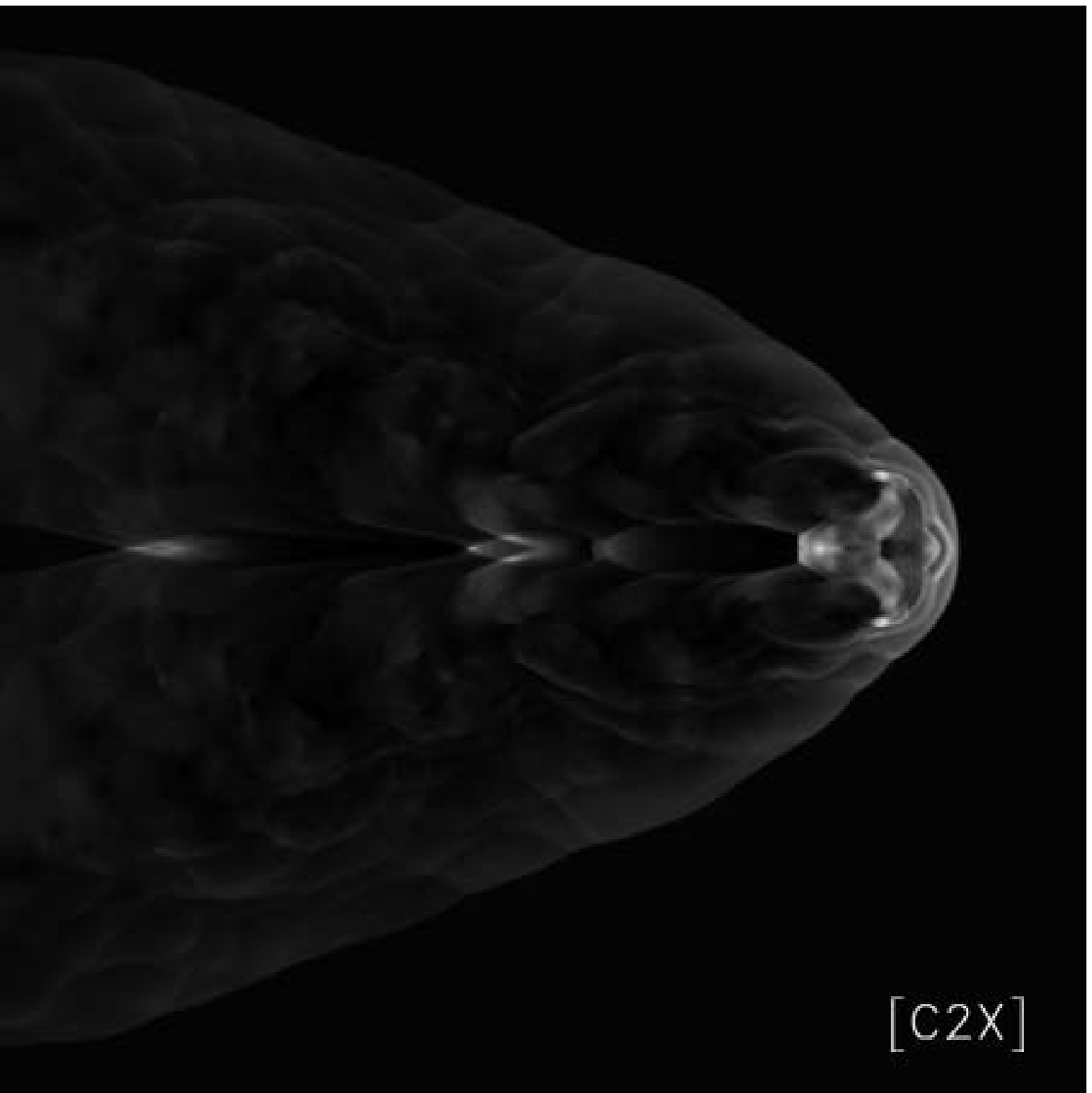}\end{array}
&
\begin{array}{c}\includegraphics[width=6cm]{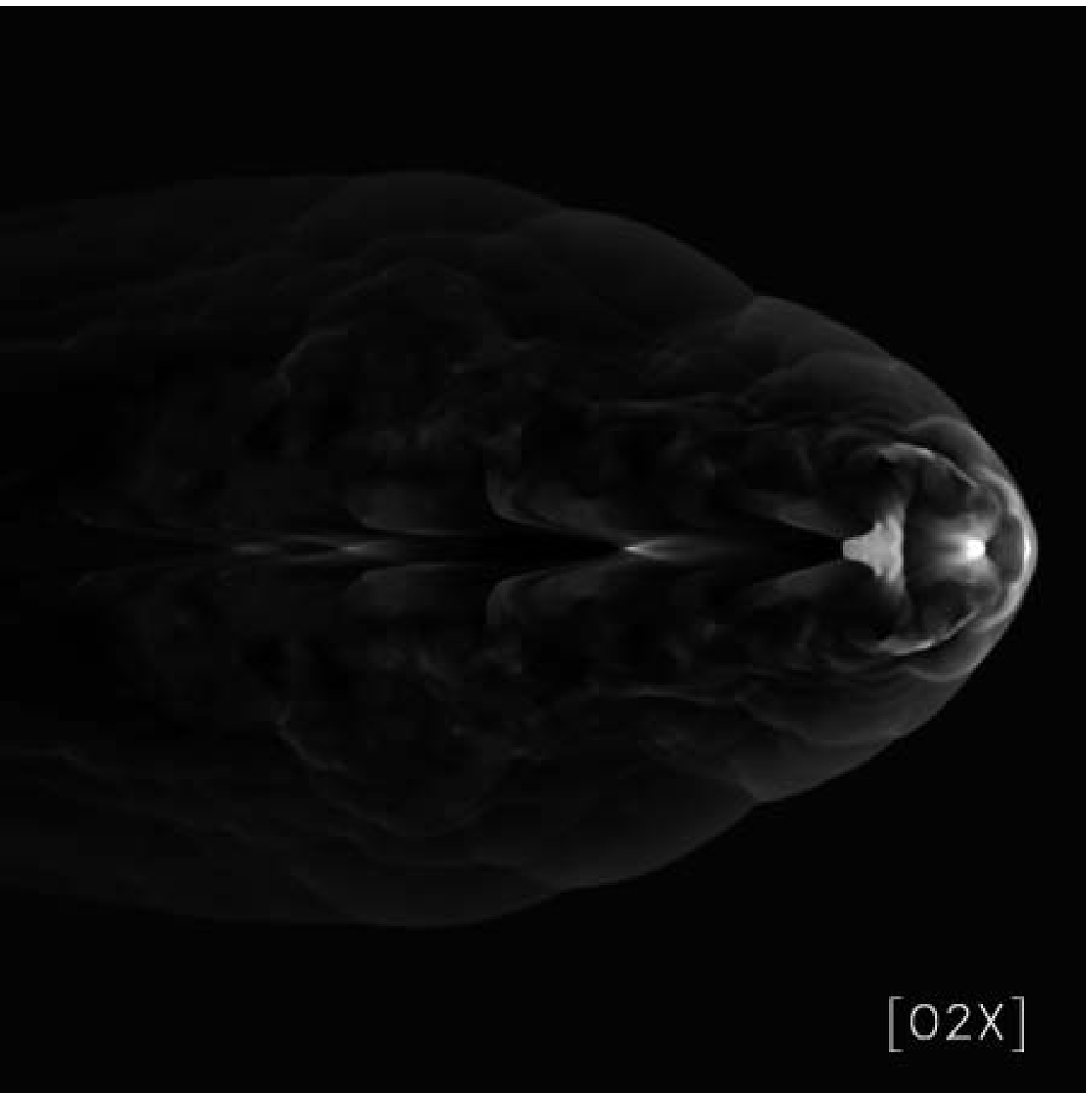}\end{array}
\\
\end{array}$
\caption{
Pressure maps corresponding to the frames in Figure~\ref{f:backflow.flow}.
}
\label{f:backflow.pressure}
\end{figure}

\begin{figure}[h]
\centering \leavevmode
$\begin{array}{ccc}
\begin{array}{c}\includegraphics[width=4cm]{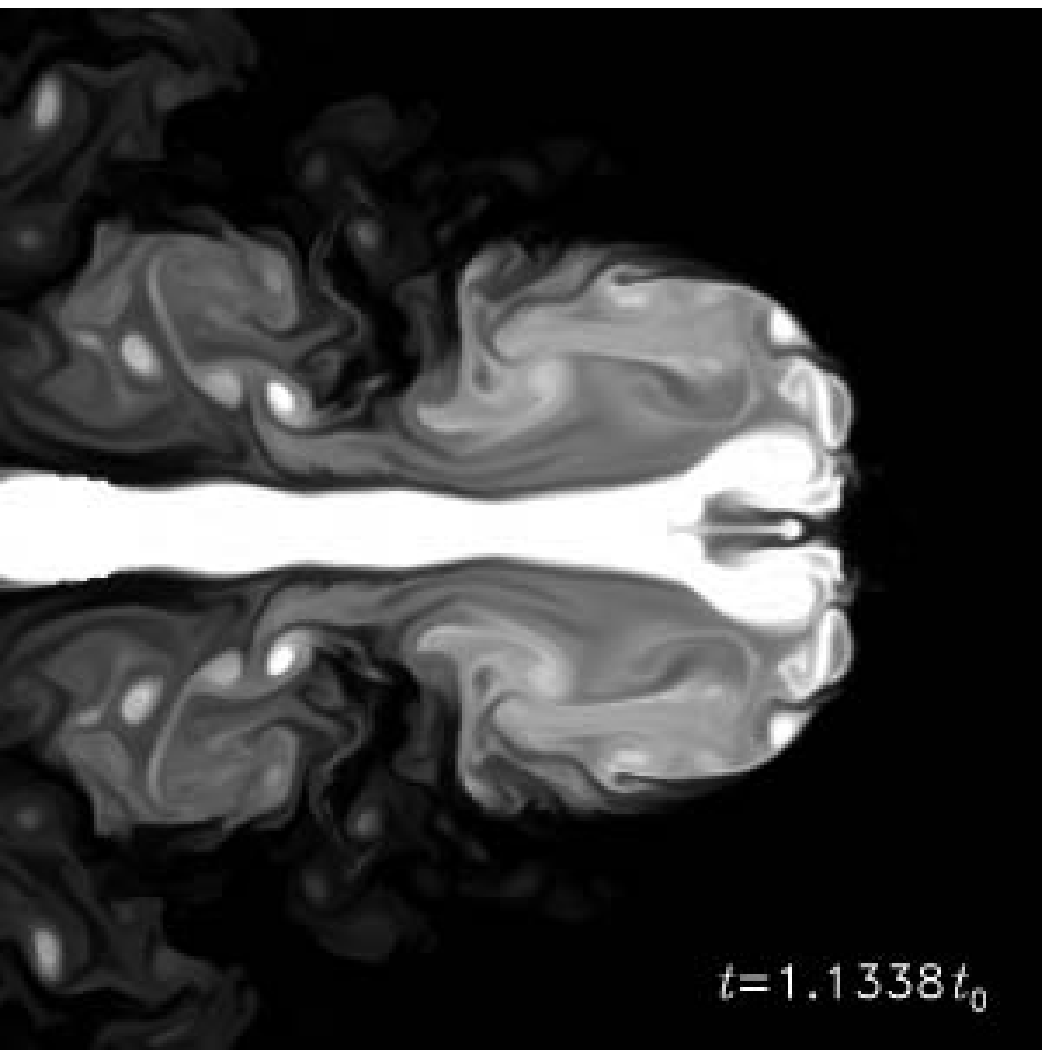}\end{array}
&
\begin{array}{c}\includegraphics[width=4cm]{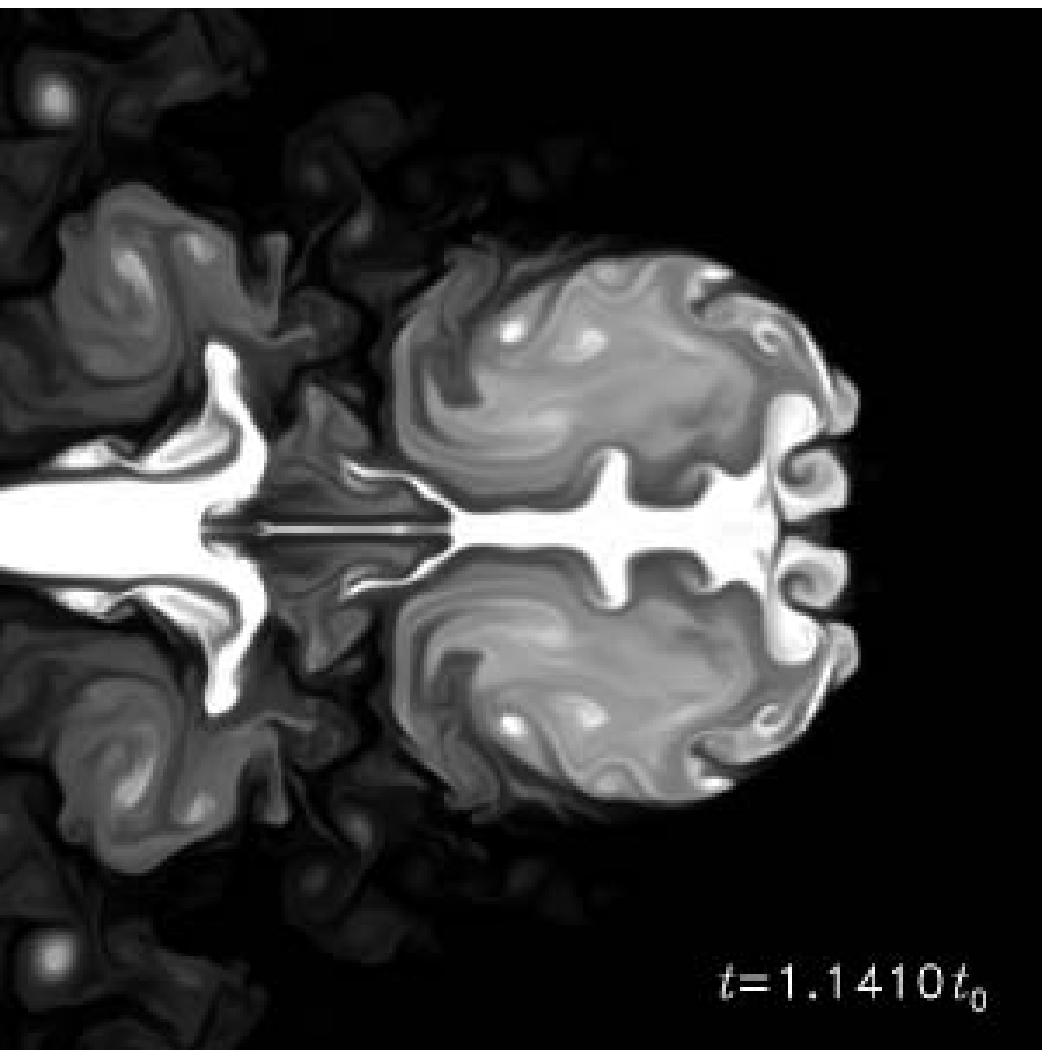}\end{array}
&
\begin{array}{c}\includegraphics[width=4cm]{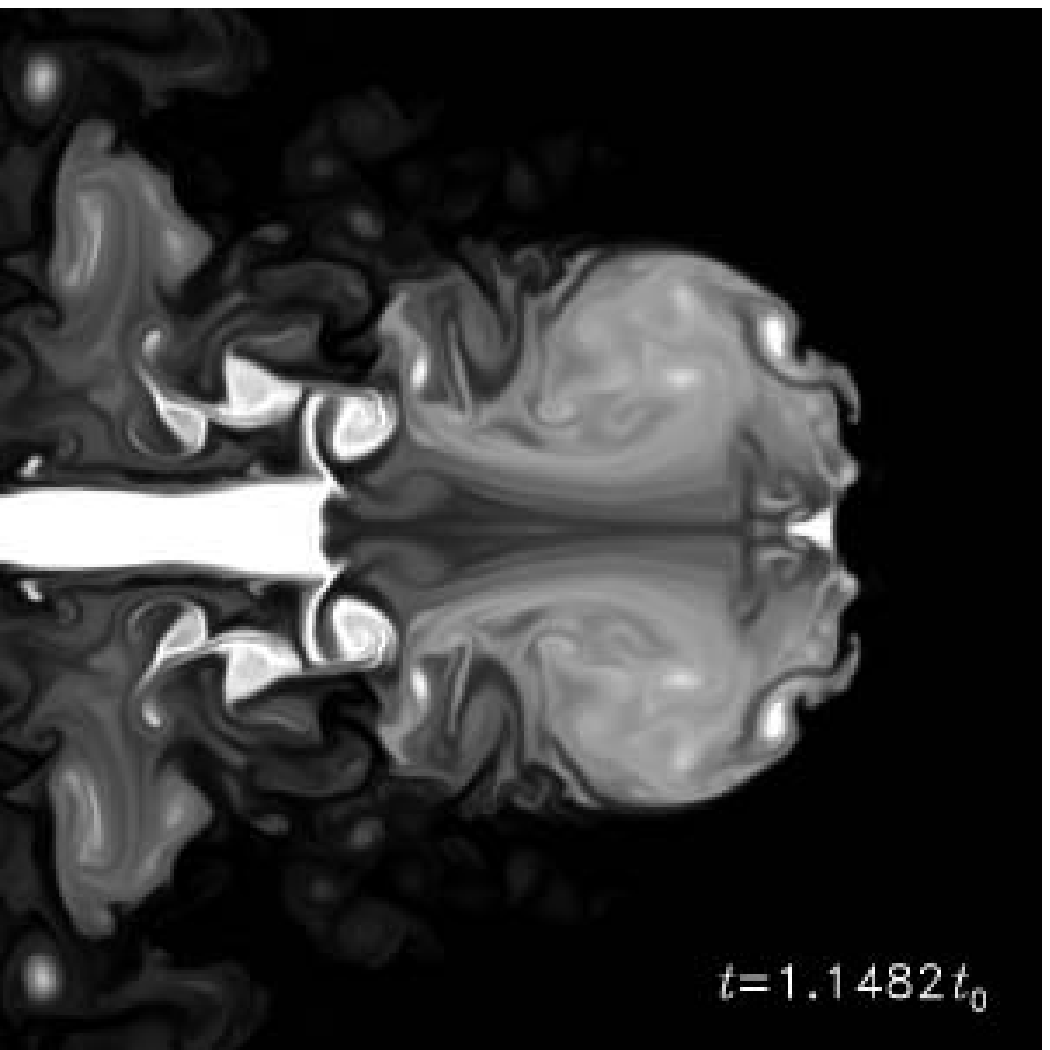}\end{array}
\\
\begin{array}{c}\includegraphics[width=4cm]{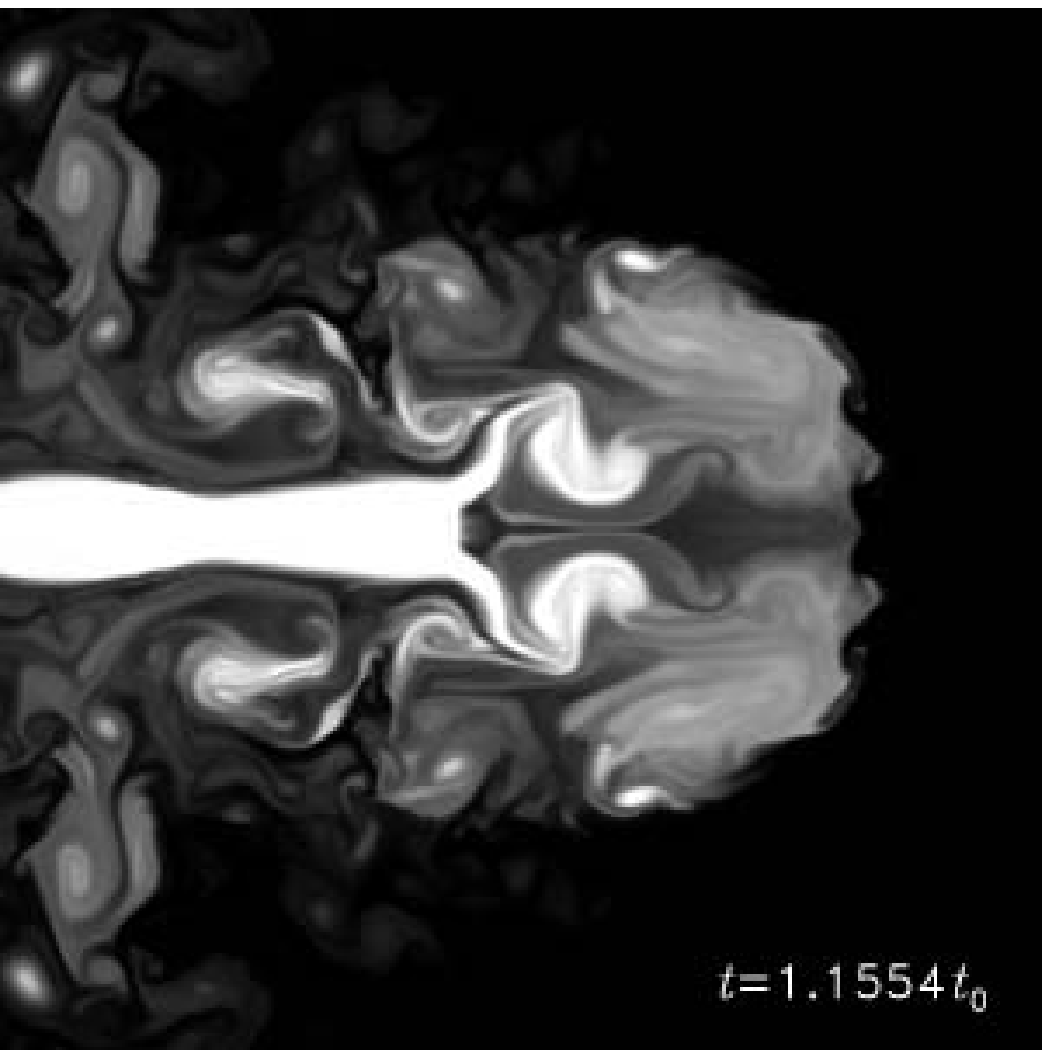}\end{array}
&
\begin{array}{c}\includegraphics[width=4cm]{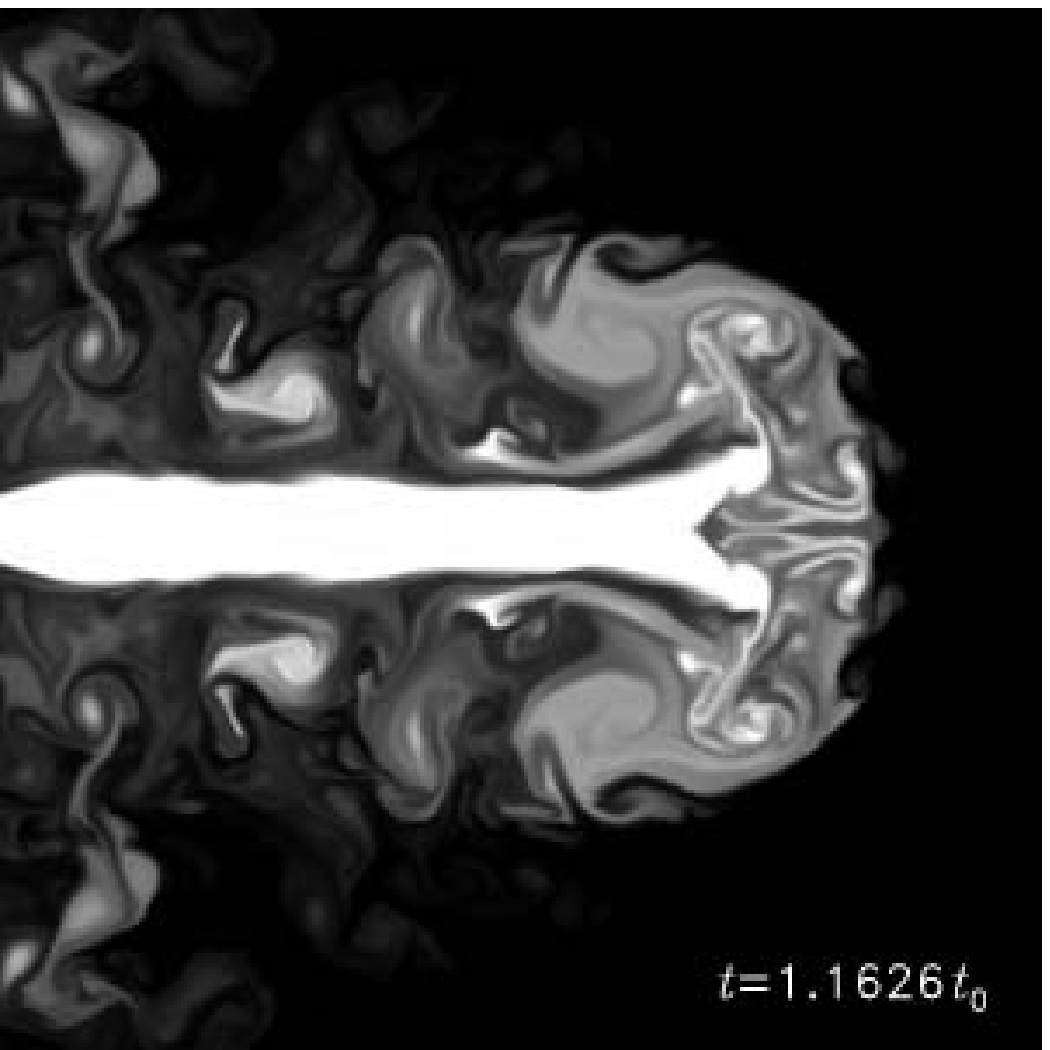}\end{array}
&
\begin{array}{c}\includegraphics[width=4cm]{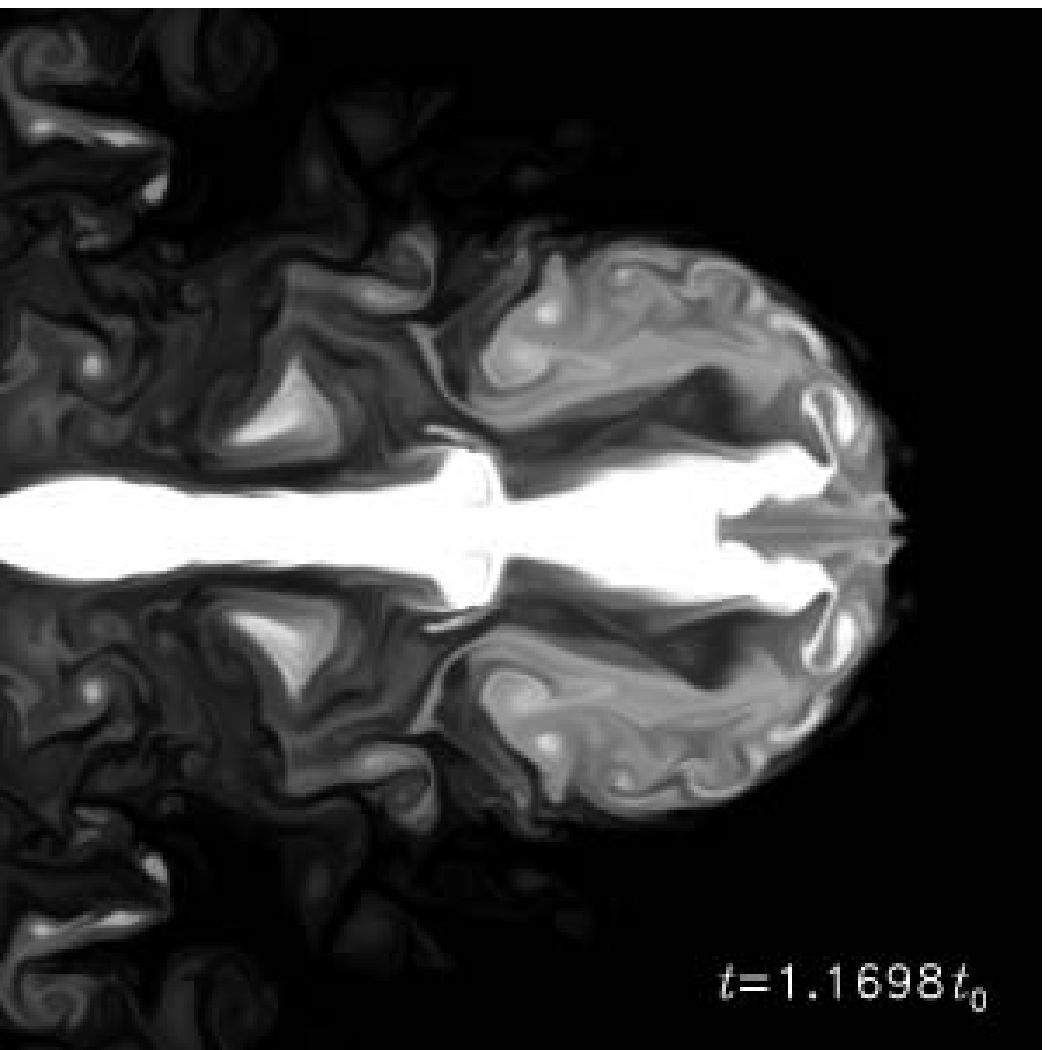}\end{array}
\\
\end{array}$
\caption{
Snapshots of the scalar tracer, $\varphi$,
showing one episode of surging behaviour
in simulation {\tt c4V},
in which $(\eta,M)=(10^{-4},5)$.
The spatial subregion is
${\frac56} x_0 < z < {\frac{35}{6}} x_0$ and $r<2.5x_0$.
White indicates pure jet plasma;
black indicates a complete absence of jet plasma.
}
\label{f:surging}
\end{figure}

\subsection{Shocks \& emission regions}

As discussed above (\S\ref{s.raytracing}),
the regions of strongest emission are
high pressure shock features with
a significant concentration of jet plasma
(\ie large $\varphi$).
In agreement with past studies
\citep{norman82a},
we find several types of shocks
characterising particular regions of the jet and its surroundings.
The plots of $|\nabla p|$ in Figure~\ref{f:grad.p} show the shocks
appearing as intense lines, in several regions of the flow.

We classify the different shocked regions as follows:
\begin{enumerate}
\item
Firstly there is the bow shock driven by and encompassing
the jet and its cocoon.
The bow shock is persistent but is clearly not involved in the radio emission.
\item
The strongest but most ephemeral shocks occur
near the head of the jet where the jet deposits
a large part of its energy and momentum.
This feature pulsates back and forth along the axis
over distances several times the jet diameter,
sometimes disappearing and reforming.
\item
Shocks also propagate in the cocoon and
either originate from near the head of the jet
or are the result of local interactions between
the jet, regions of the turbulent backflow
and the surrounding layer of dense gas.
Annular shocks are seen propagating both forwards and backwards
through the backflow.
\item
The jet itself is criss-crossed by diamond shocks
caused by momentary and local compression by the turbulent backflow.
As implied by the classical perturbation analysis of \citet{birkhoff57a},
the diamond shocks are more widely separated in jets
with greater $M$.
\end{enumerate}

The distribution of $\varphi$ informs us which of the shocks
produce significant synchrotron emission.
Diamond shocks are always bright
because they occur within pure jet plasma.
For the same reason, the jet's terminal shock
yields a prominent hot-spot in most frames of the simulations.
However wider luminous structures are restricted to the
extent of the high-$\varphi$ ``sheath'' cocoon
in the systems with an open boundary,
or the relatively unmixed frontal regions of the ``wake'' cocoons
arising in the systems with a closed boundary.
Thus, depending on the jet parameters and the boundary condition,
there are no bright features exceeding
$\sim 2r_{\rm j} - 7r_{\rm j}$ in radius
(see Figure~\ref{f:backflow.tracer}).

In summary,
allowing for the axial symmetry of our simulations,
the kinds of bright three-dimensional structures
that are possible include:
bright point-like, disk-like and cone-like shocks within the jet;
in the backflow there may occur a diversity of rings and circular ribbons,
with potentially flat, curved or solid cross-sections.
In the following section,
we consider how these structures may appear
in two-dimensional projections onto the sky,
and how these images may compare with observations.

\section{Relationship to Pictor~A}
\label{s.results.raytracing}

In this section we compare rendered images
with the morphology of bright structures
evident in the observations of Pictor~A.
The optical observations (Figure~\ref{f:optical_image})
are a guide to the current positions of the shocks.
Because of the longer lifetime of radio emitting electrons the radio
image (Figure~\ref{f:radio_image})
is indicative of the overall distribution of jet plasma within the cocoon.

In the rendering of our simulations we seek physical configurations
that show a bright hot-spot at the front of the jet
and a transverse bar upstream of the hot-spot.
As refinements, we prefer images in which
the bar appears brightened at its edges.
The optical image (Figure~\ref{f:optical_image})
also suggests faint arcs connecting
the ends of the bar to the hot-spot.
We also have in mind (in the optical data) the bright knot of emission
at the intersection of the jet axis and the bar,
and the faint tail of emission leading from that knot to the east.

We generated ray-traced images of each frame of each simulation,
with the jet oriented at a variety of inclinations,
$\theta=45^\circ, 60^\circ, 70^\circ, 80^\circ, 90^\circ$
($\theta$ is the angle between the jet and the line of sight).
We inspected and compared all the resulting
images  and sorted them somewhat subjectively
according to their consistency with the topology, proportions
and brightness ratios seen in the Pictor~A observations.
In the following we consider the effects of orientation
upon the quality of the matching rendered images,
and the effects of the jet parameters.
We also inspect the underlying physical nature of the bright features
and their temporal evolution.

\subsection{Orientation}

The orientation of the jet has a substantial effect
on the projected appearance of the axially symmetric
luminous features surrounding the jet.

The hot-spot and knot are not decisive characteristics,
since diamond and terminal shocks appear at numerous instances and their
appearance is virtually independent of orientation.
However the transverse bar in the Pictor~A observations
provides a tight constraint on the selection of appropriate rendered images.
This elongated feature must be the projection of
at least one luminous structure
that is three-dimensionally extended.
Therefore it is necessary but not sufficient
to select snapshots that contain extended high-pressure features.
When projected onto the sky at some orientation,
only a subset of these configurations
produce images similar to the observations.

\begin{figure}[h]
\centering \leavevmode
$\begin{array}{ccc}
\includegraphics[width=4.5cm]{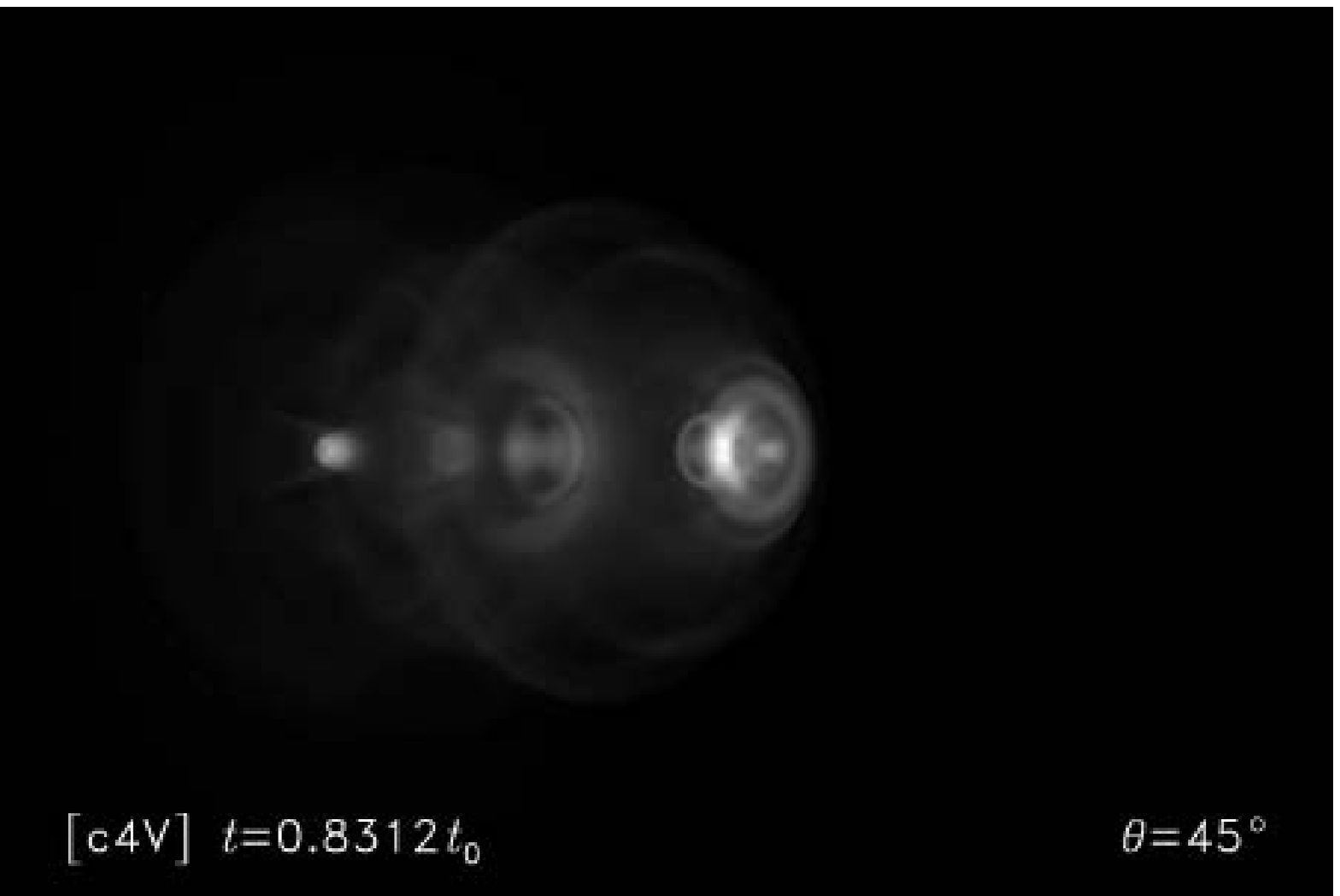}
&
\includegraphics[width=4.5cm]{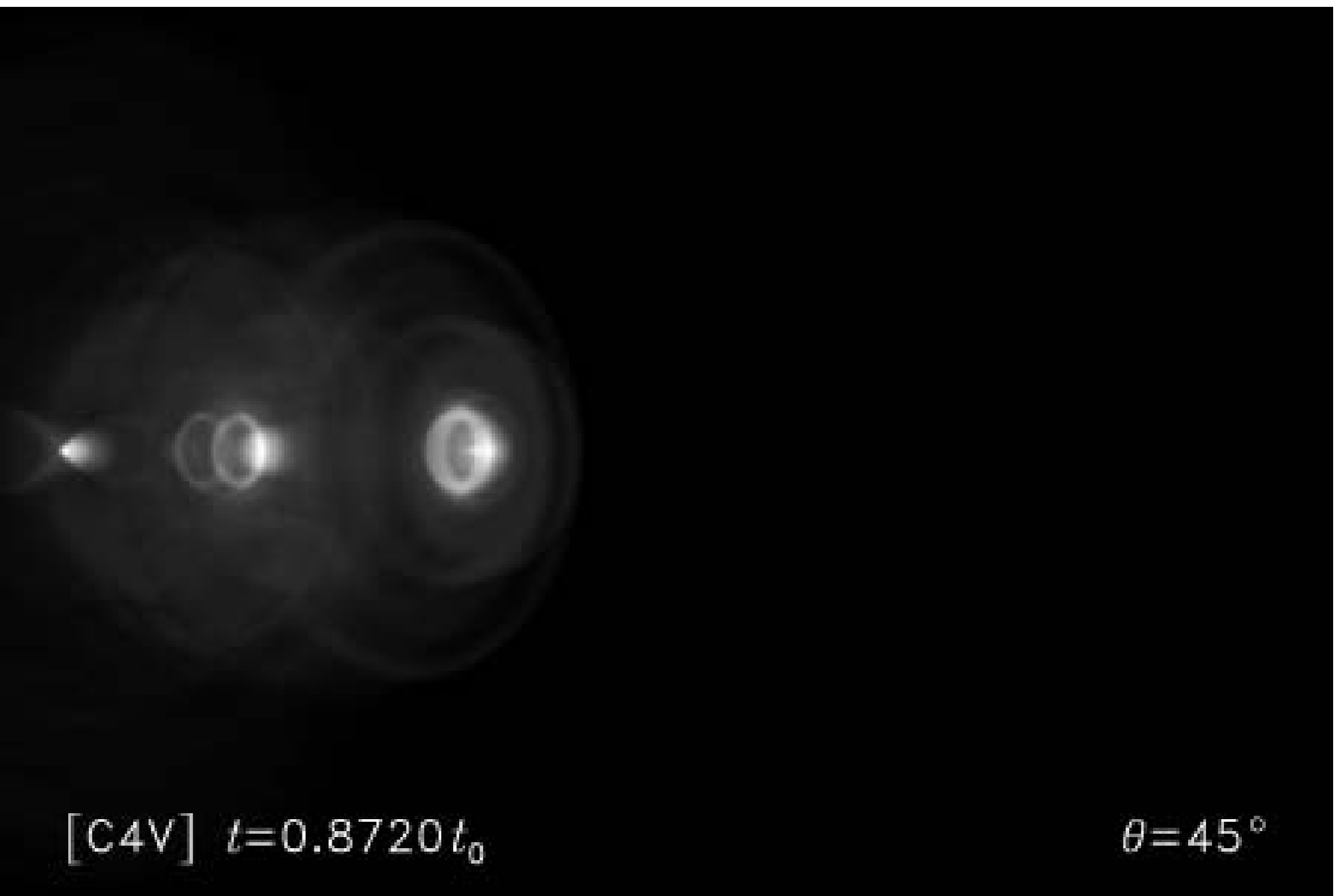}
&
\includegraphics[width=4.5cm]{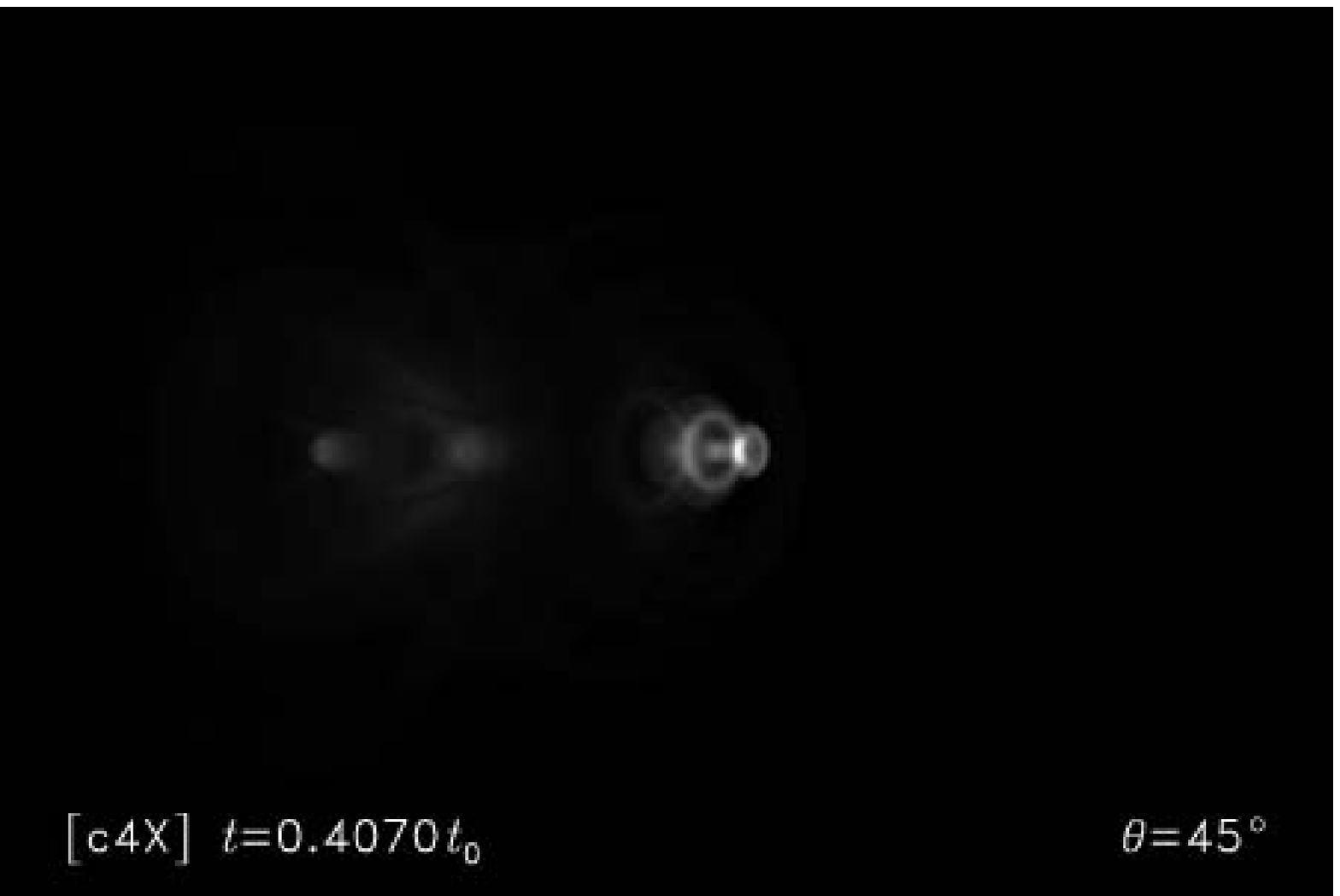}
\\
\includegraphics[width=4.5cm]{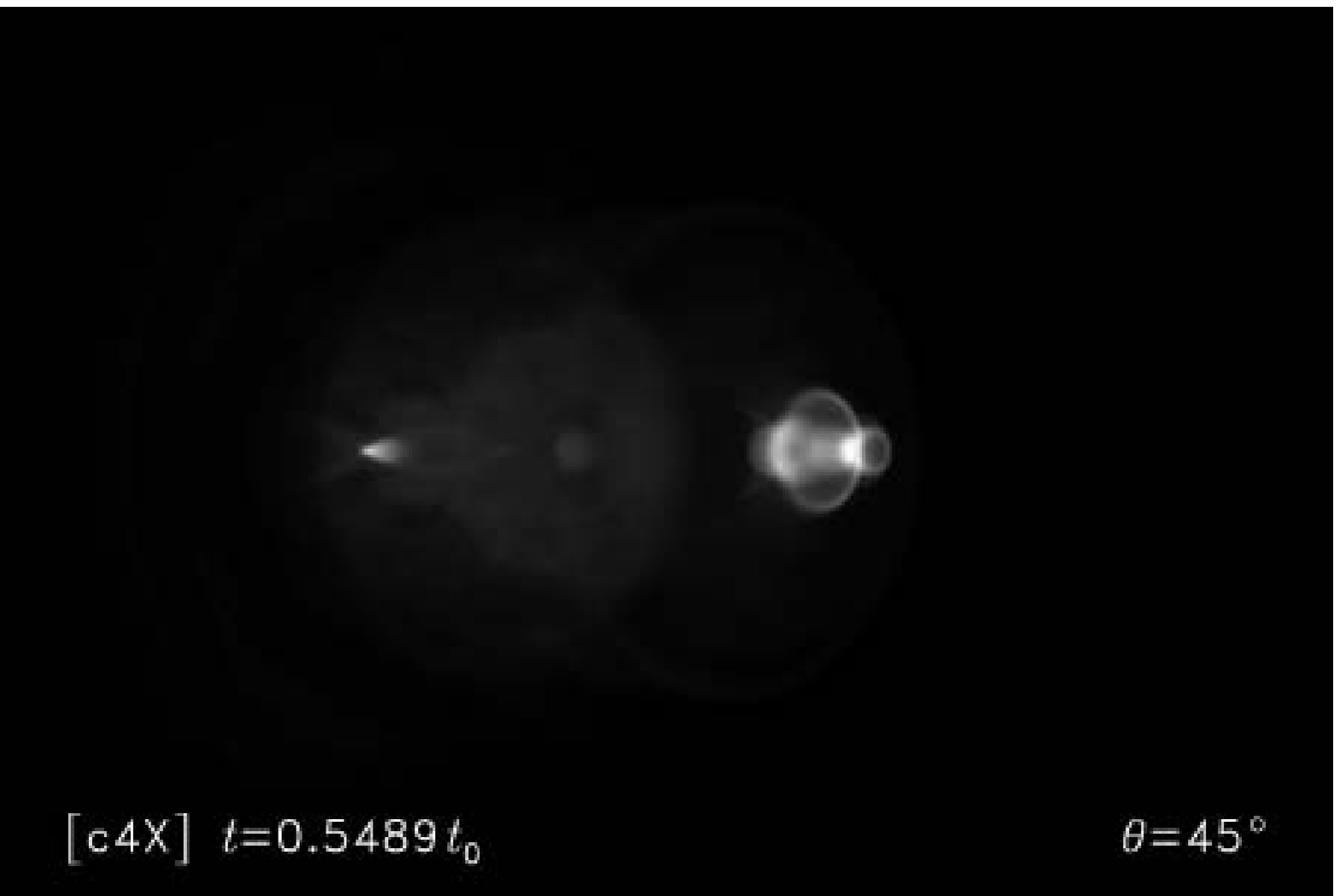}
&
\includegraphics[width=4.5cm]{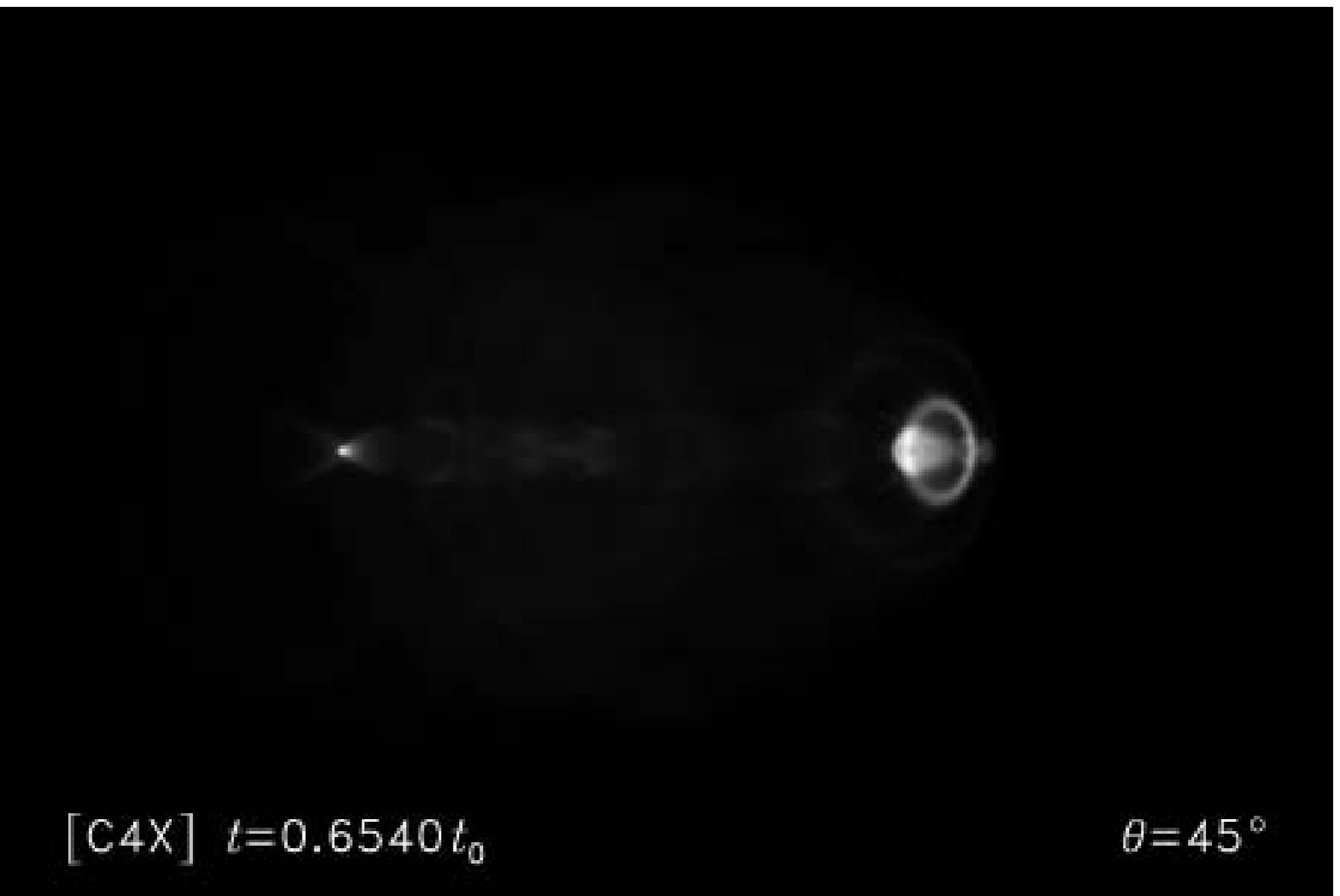}
&
\includegraphics[width=4.5cm]{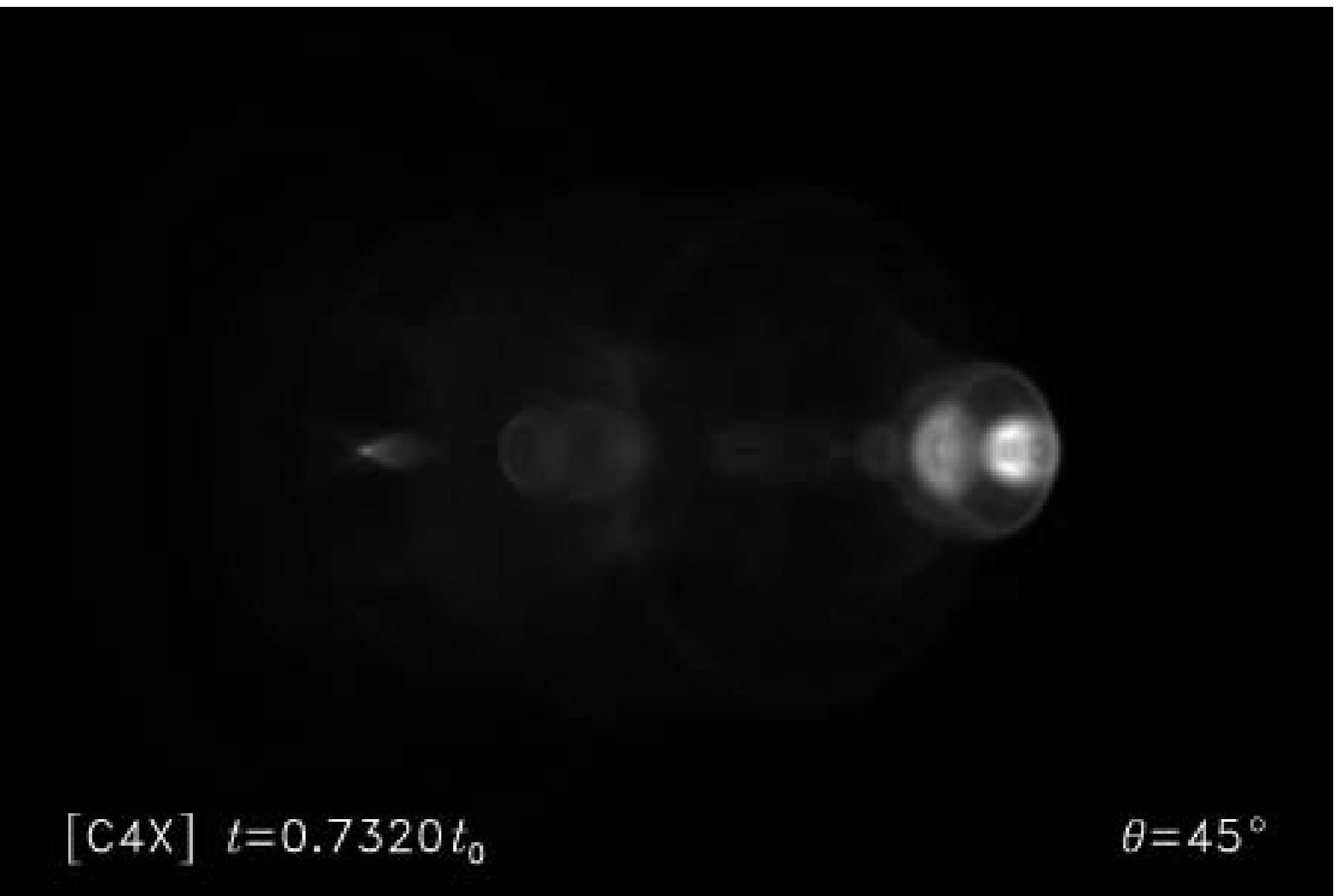}
\\
\includegraphics[width=4.5cm]{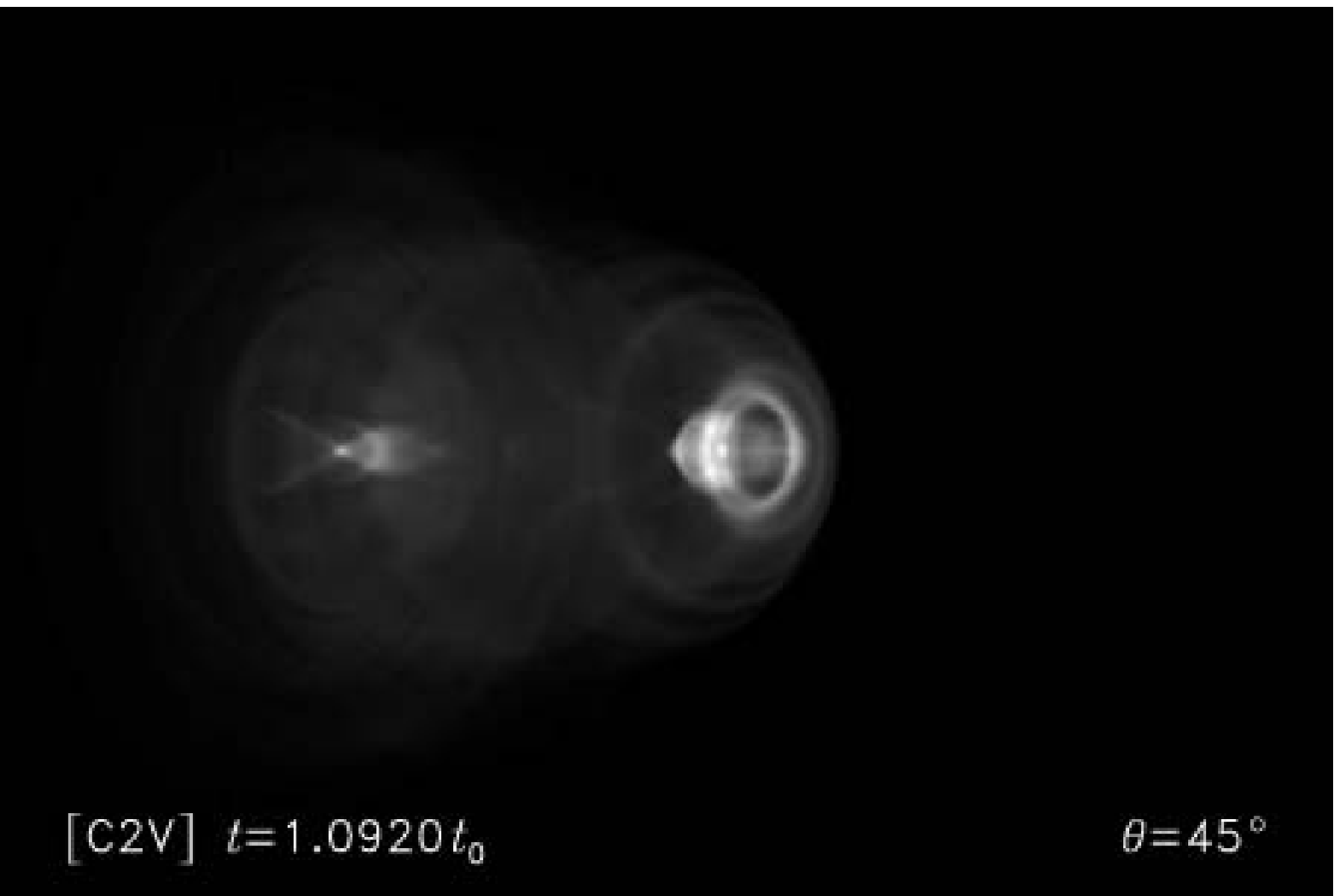}
&
\includegraphics[width=4.5cm]{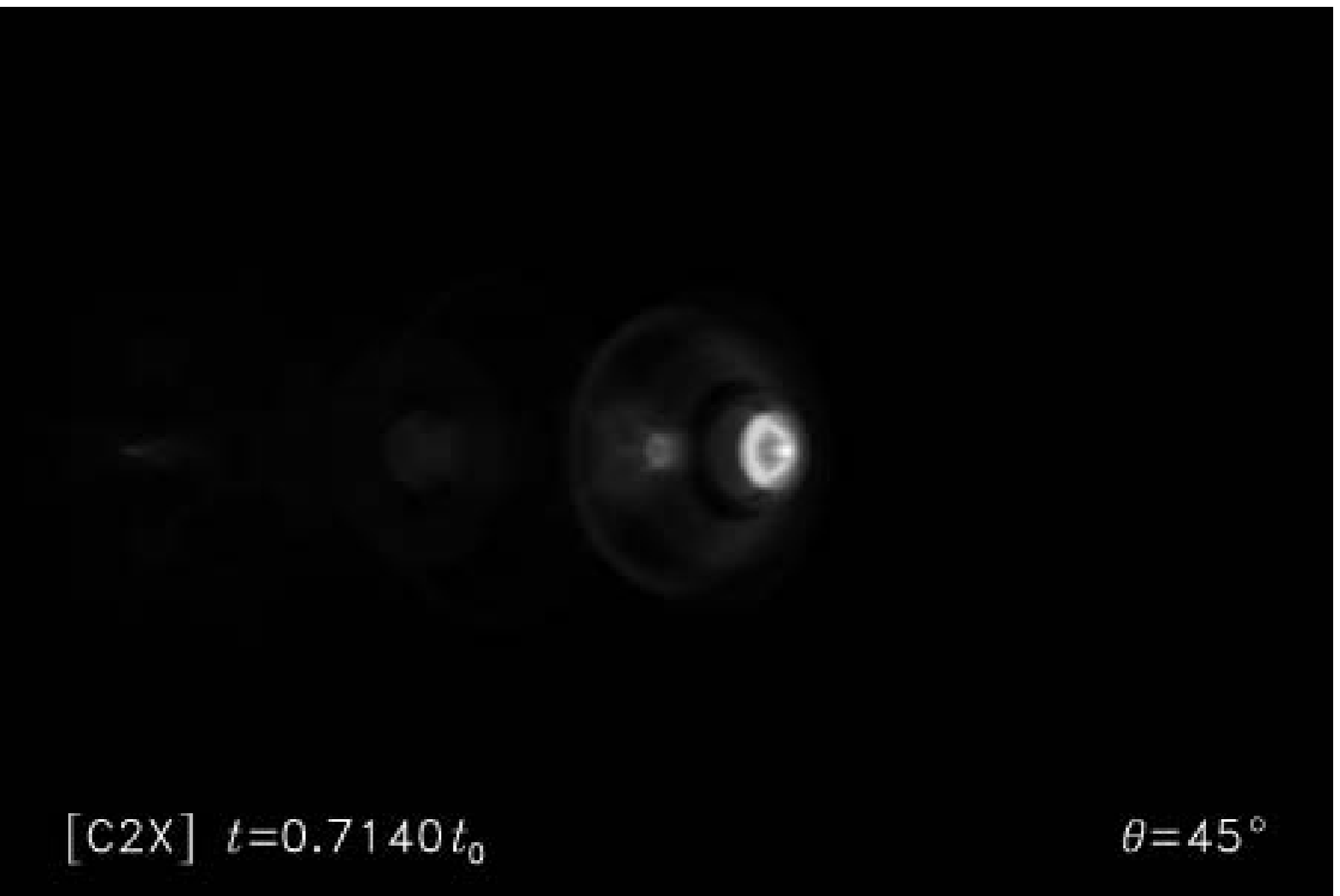}
\\
\end{array}$
\caption{
$450 \times 300$ pixel raytraced images
of a jet inclined at $\theta=45^\circ$ to the line of sight.
None of them show an apparent transverse bar
with a diameter much greater than that of the jet.
The frontal ``hot-spot'' is a shock feature that is smaller
than the initial diameter of the jet.
}
\label{fig.45.best}
\end{figure}

When the jet is at a large angle to the plane of the sky,
\eg $\theta=45^\circ$,
very few frames show the desired topology and brightness:
typically a few times $0.1\%$ of frames
have a subjective resemblance to the images.
The best images of jets at $45^\circ$ inclinations
are shown in Figure~\ref{fig.45.best}.
In these cases,
the appearance of a straight ``bar,'' if any,
is  a coincidence due to the visual superposition of the near and far edges
of unconnected rings in the backflow;
it is not a single physical structure.
One of the positive features of the better renderings at $\theta=45^\circ$
is the abundance of snapshots in which
the hot-spot is projected in front of
an edge of a bright ring in the backflow,
thereby resembling the faint arcs
connecting the hot-spot to the bar in the optical observations.
However the occurence of conspicuously elliptical features
throughout the image tends to ruin the overall resemblence,
and the accidental bars appearing by superposition
are never much wider than the hot-spot.
Indeed the least-bad matches have small features
compared to the initial jet,
$r\la2r_{\rm j}$,
and we speculate that the resemblences might be even less convincing
if these structures could have been resolved more finely
by the simulations.

\begin{figure}[h]
\centering \leavevmode
$\begin{array}{ccc}
\includegraphics[width=4.5cm]{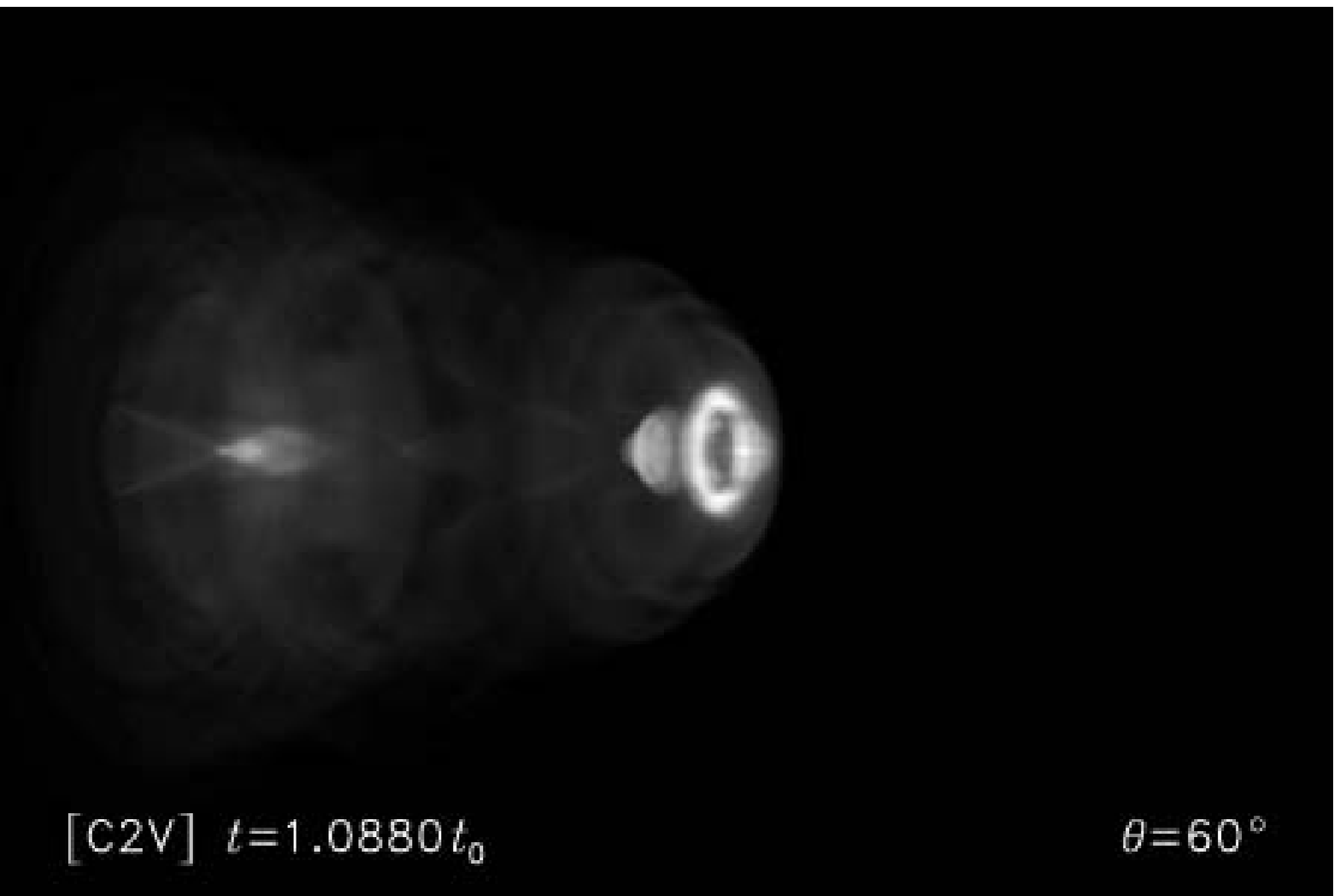}
&
\includegraphics[width=4.5cm]{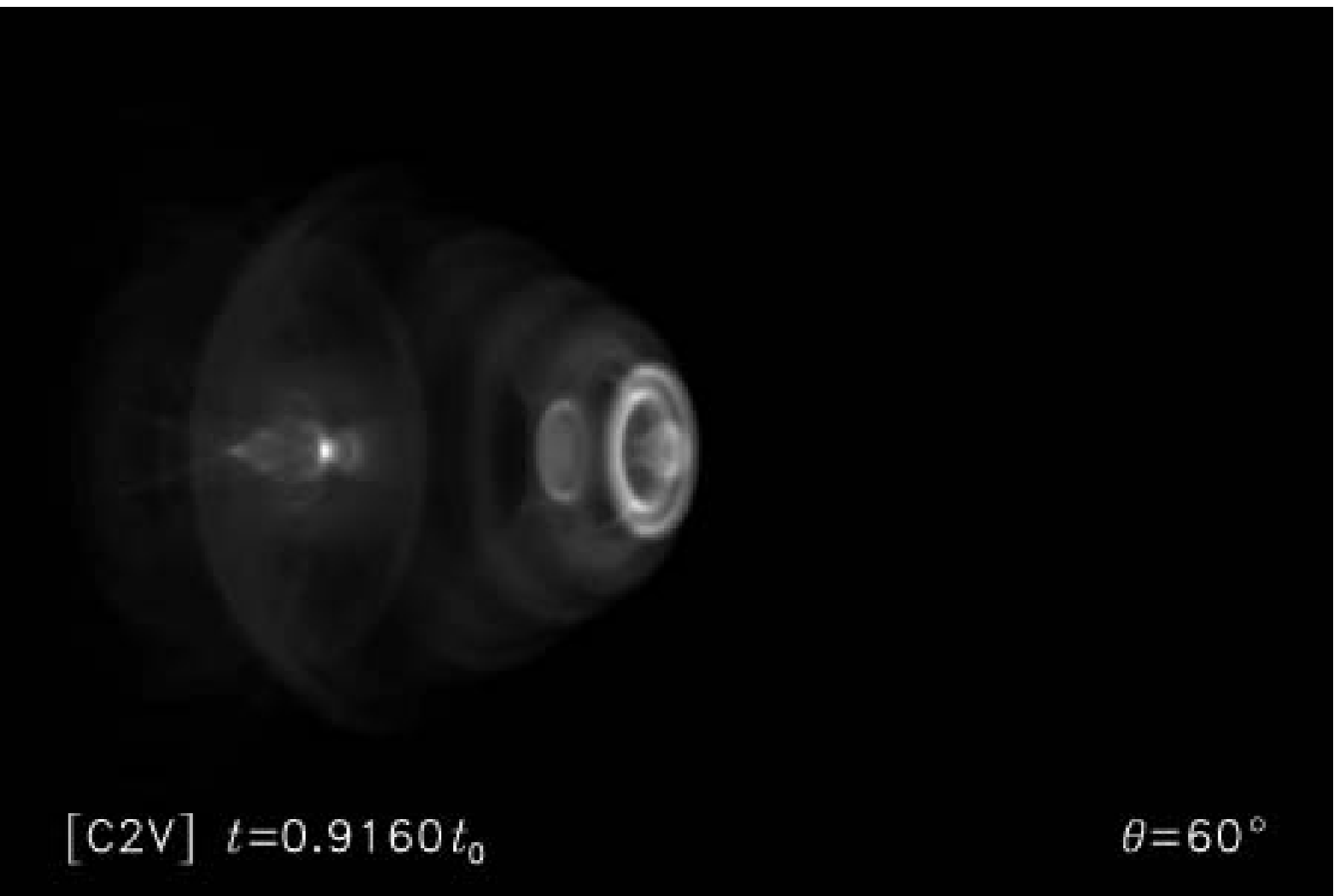}
&
\includegraphics[width=4.5cm]{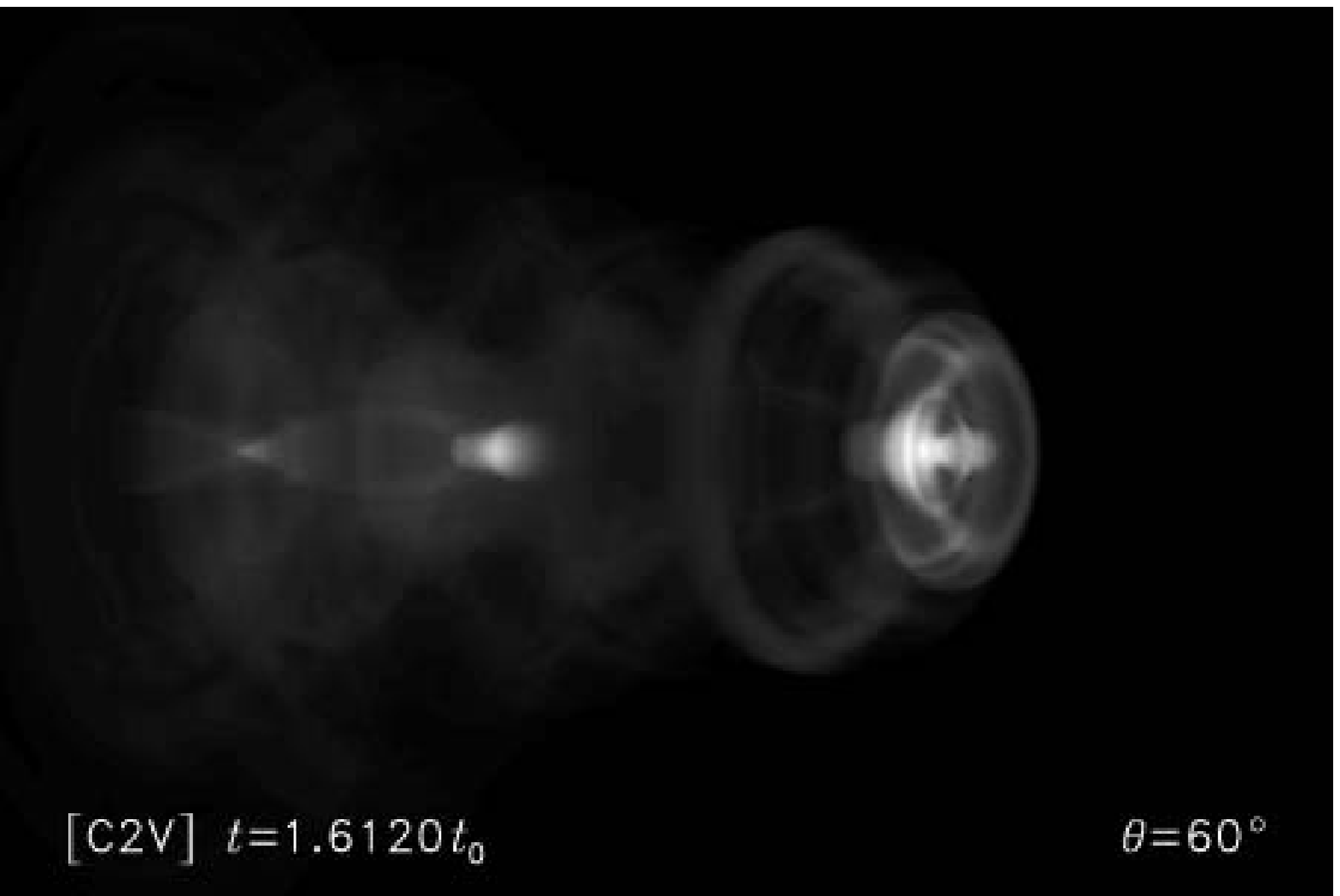}
\\
\includegraphics[width=4.5cm]{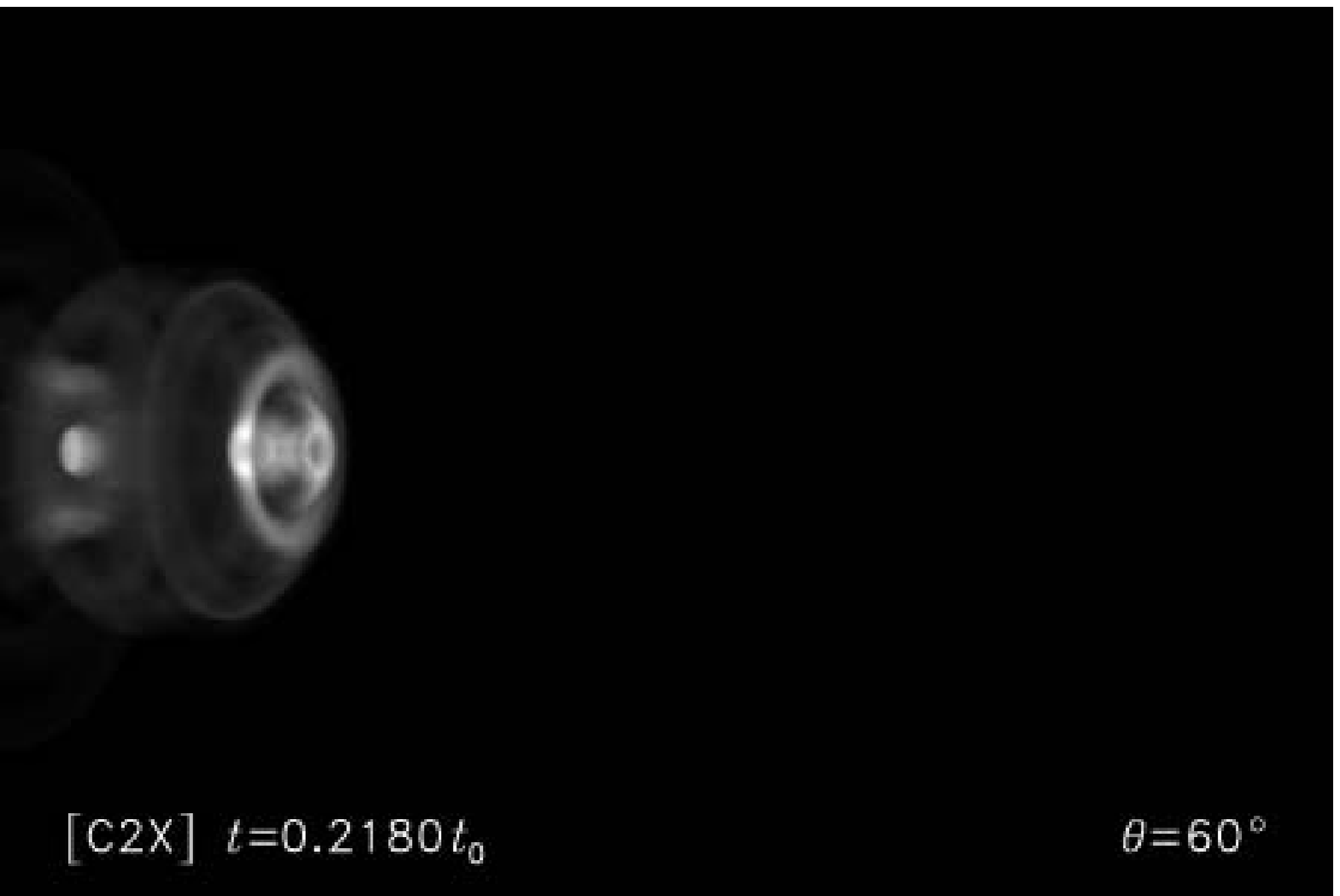}
&
\includegraphics[width=4.5cm]{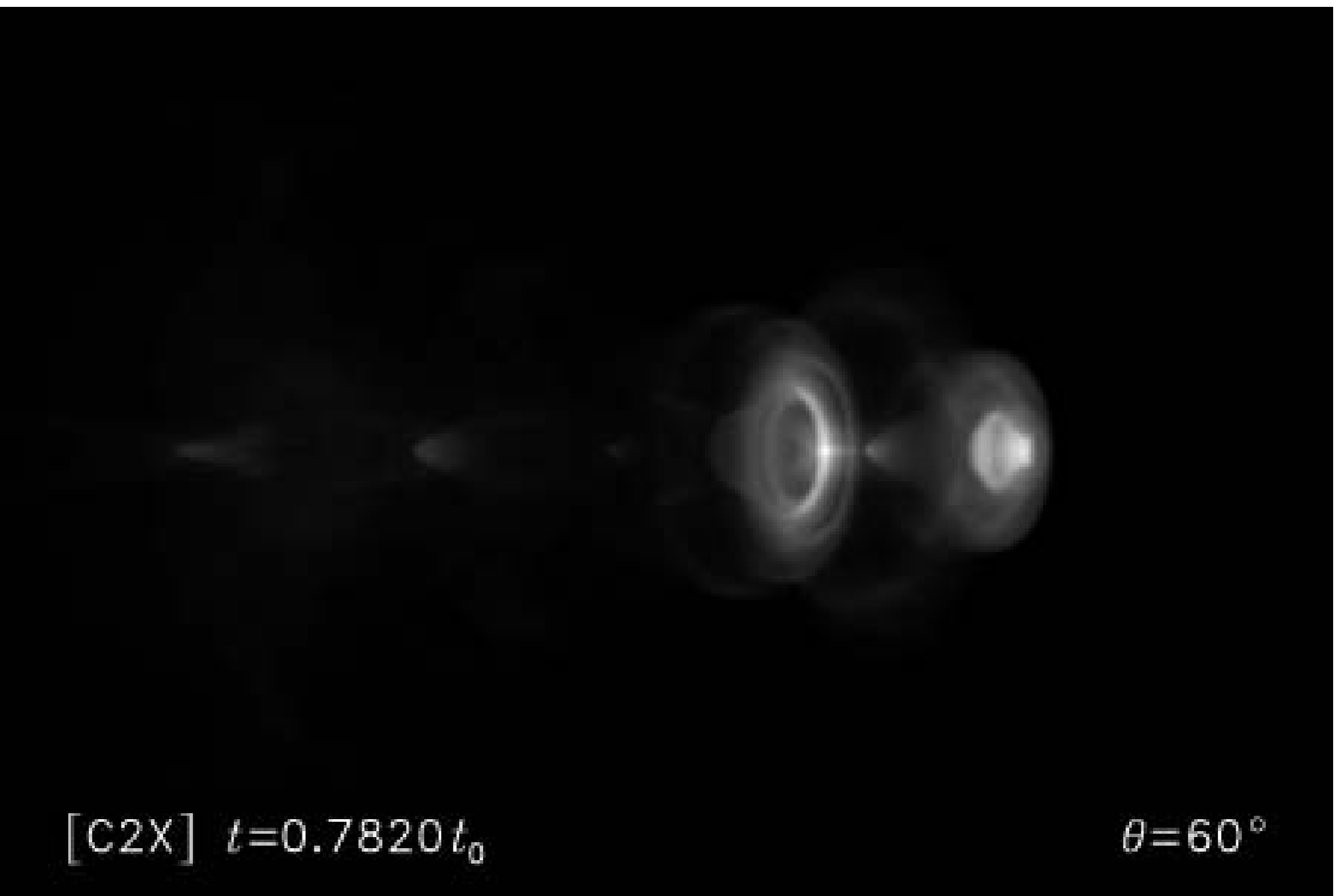}
\\
\includegraphics[width=4.5cm]{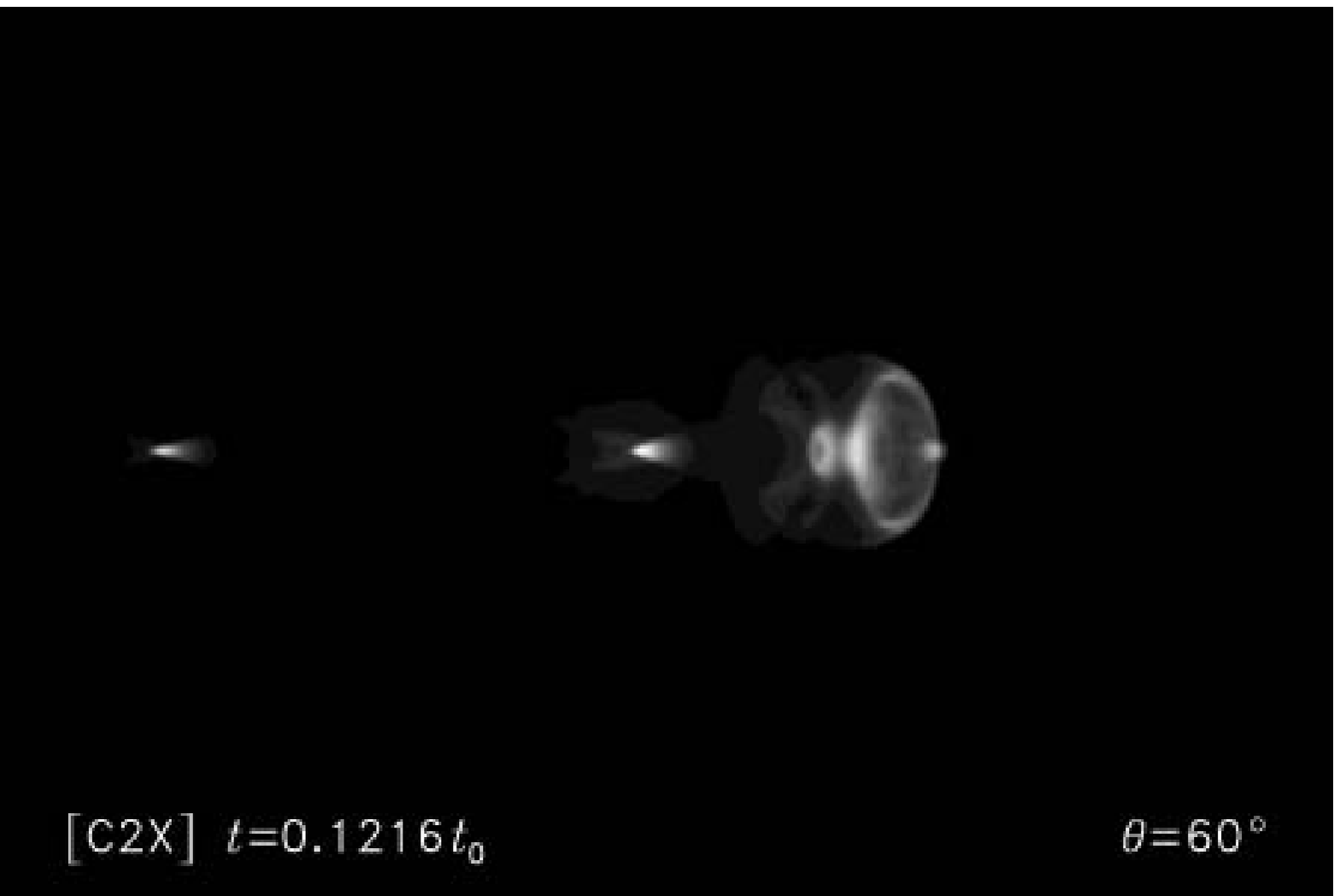}
&
\includegraphics[width=4.5cm]{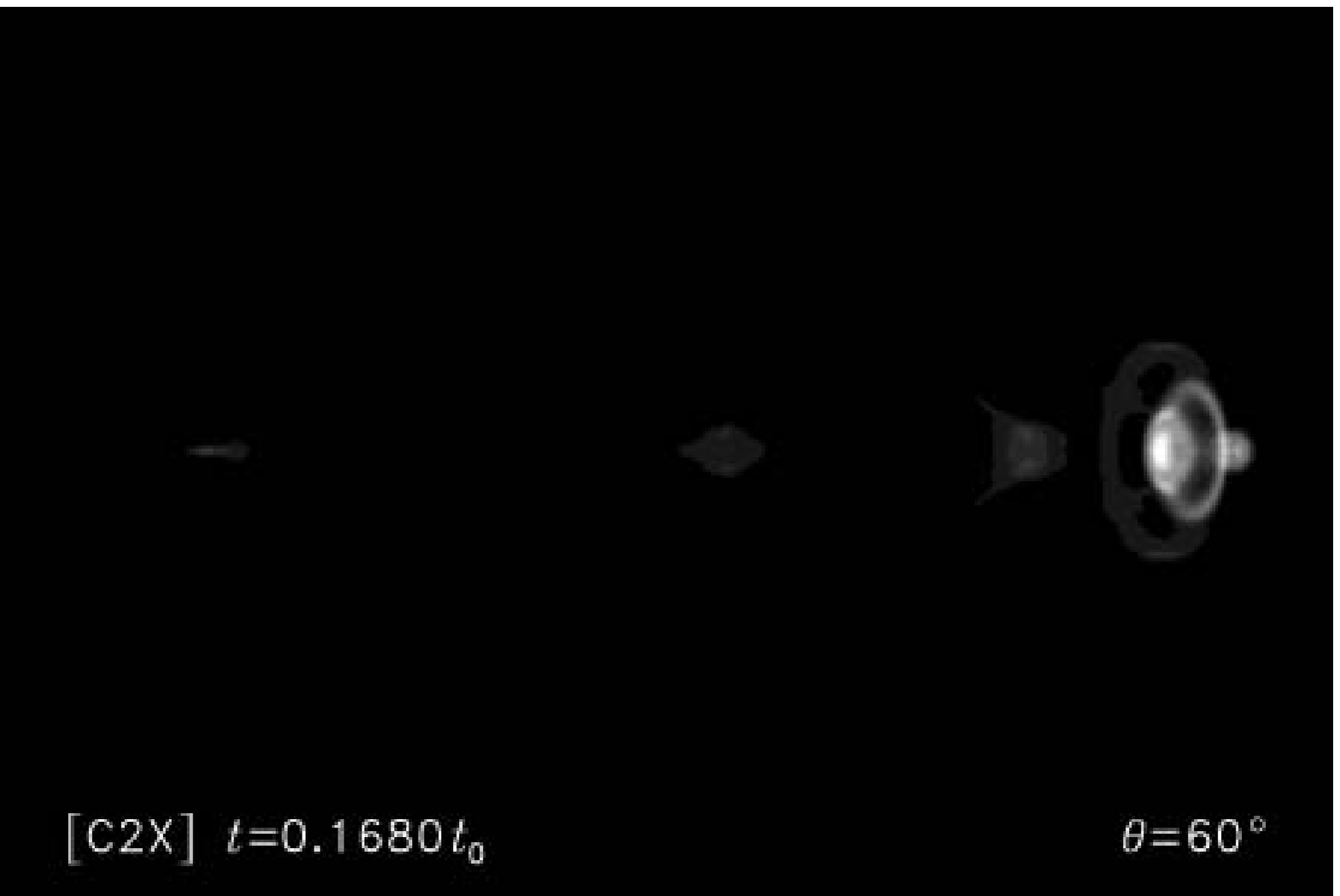}
&
\includegraphics[width=4.5cm]{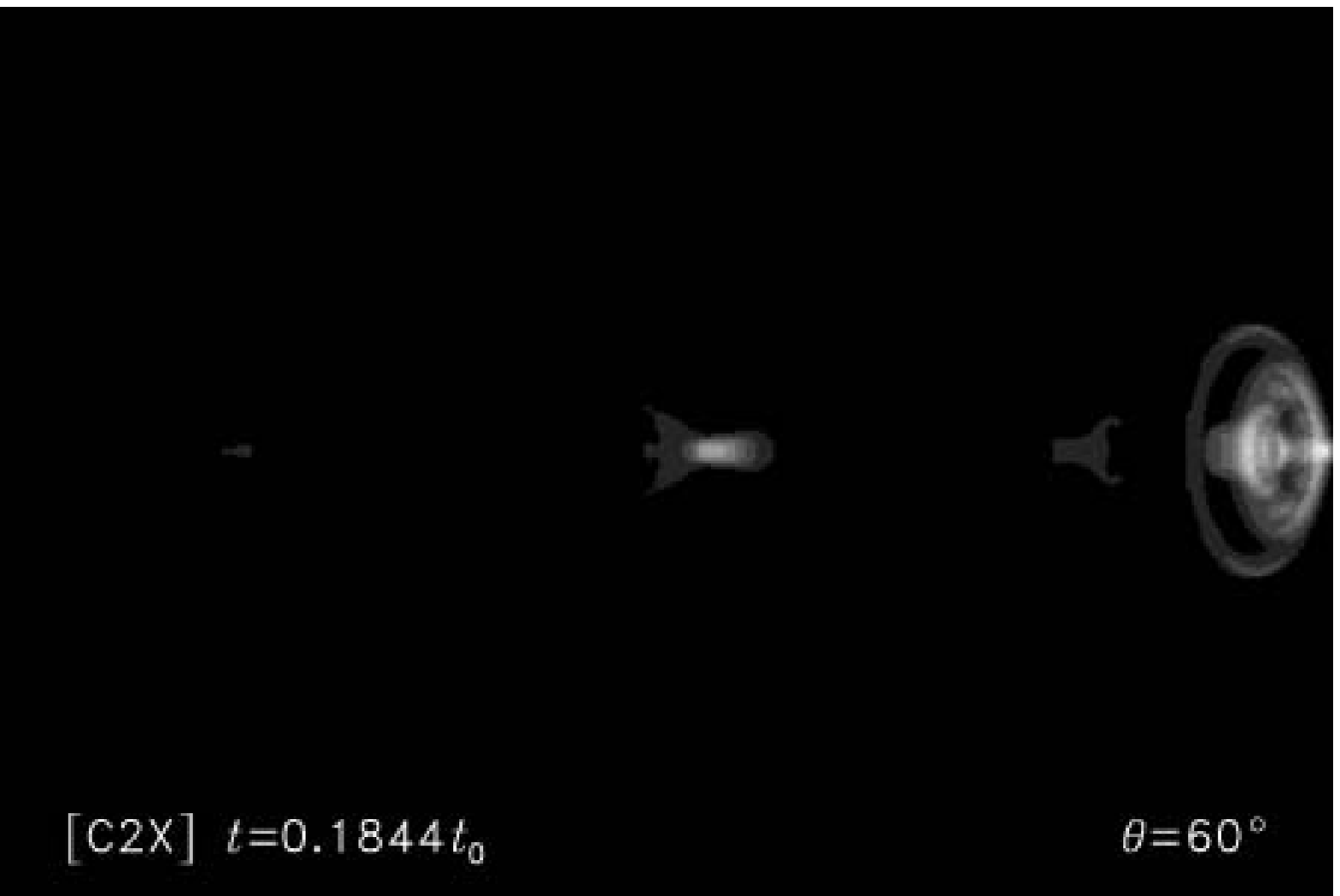}
\\
\end{array}$
\caption{
Some of the best examples of raytraced images
of a jet inclined at $\theta=60^\circ$ to the line of sight.
The displayed subregions are the same size as those
in Figure~\ref{fig.45.best}.
}
\label{fig.60.best}
\end{figure}

When the system is viewed at inclinations
$70^\circ\la \theta \le 90^\circ$,
the matches are more abundant and better in quality.
Approximately $4\%$ of all frames rendered at $\theta=90^\circ$
show features qualitatively similar to
the Pictor~A observations.
Approximately $3\%$ of the frames at $\theta=70^\circ$
show the same quality of resemblance.
Inclinations of $\theta=90^\circ$
produce images with a straight, bright transverse bar
behind the hot-spot.
Lower inclinations, progressively down to $\theta \approx70^\circ$,
yield more images in which arcs appear to connect the ends of the bar
with the hot-spot.
However the trade-off is that the apparent bar
may be less straight and distinct.

At inclinations as low as $\theta=60^\circ$,
frames that discernibly resemble observation
are as rare as $\approx0.6\%$.
Some of the best examples are presented in Figure~\ref{fig.60.best},
but they are not compelling matches.
Like the best morphological matches at $\theta=45^\circ$,
their bar radii are $\la 2r_{\rm j}$,
and they tend to involve coincidental superposition of
the far side of an annulus behind the hot-spot.
There are instances showing a favourable ``pedestal'' or ``arcs''
near the hot-spot,
but the best simulated ``bar'' features
are plainly elliptical rather than straight.

\begin{figure}[h]
\centering \leavevmode
$\begin{array}{ccc}
\includegraphics[width=4.5cm]{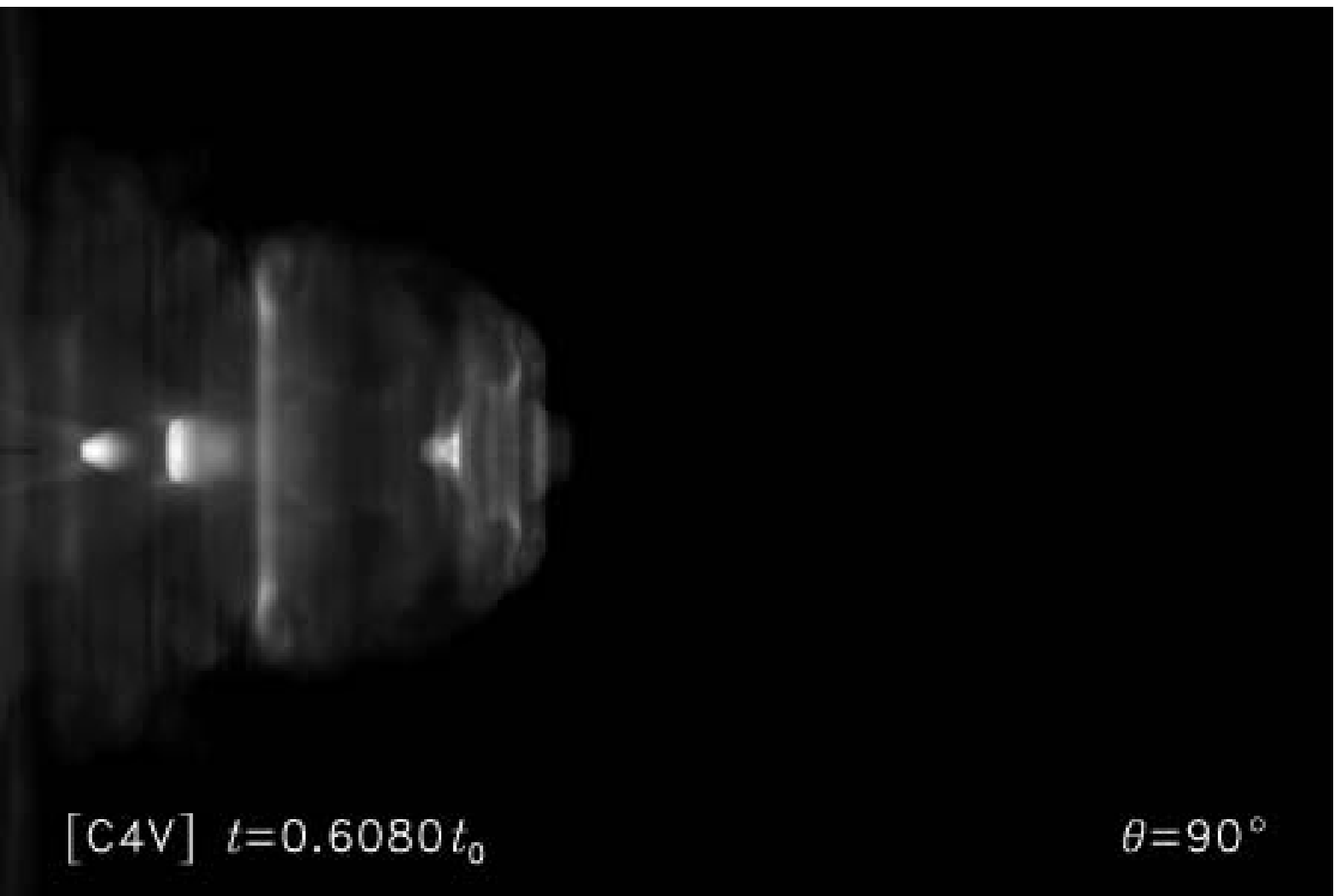}
&
\includegraphics[width=4.5cm]{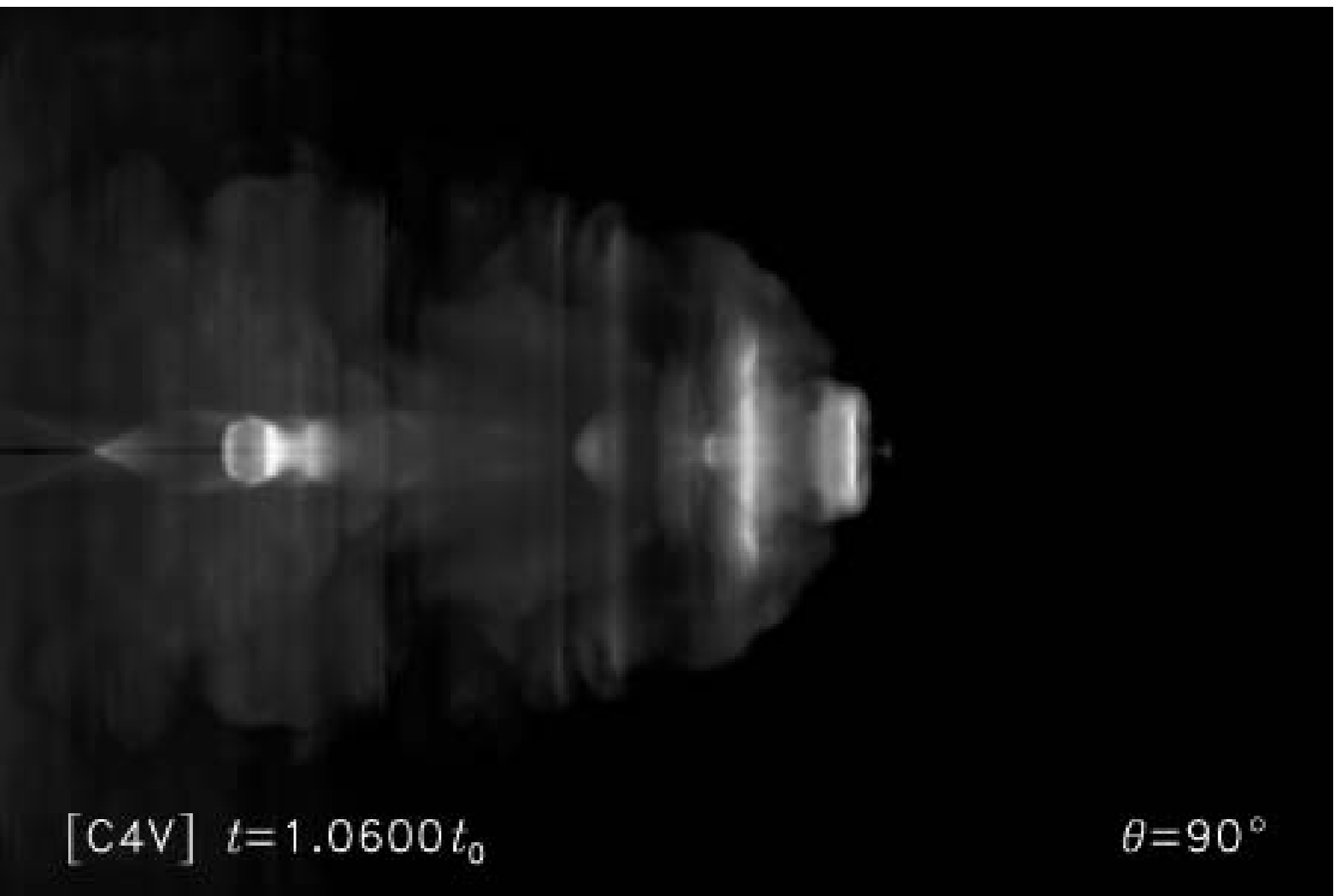}
&
\includegraphics[width=4.5cm]{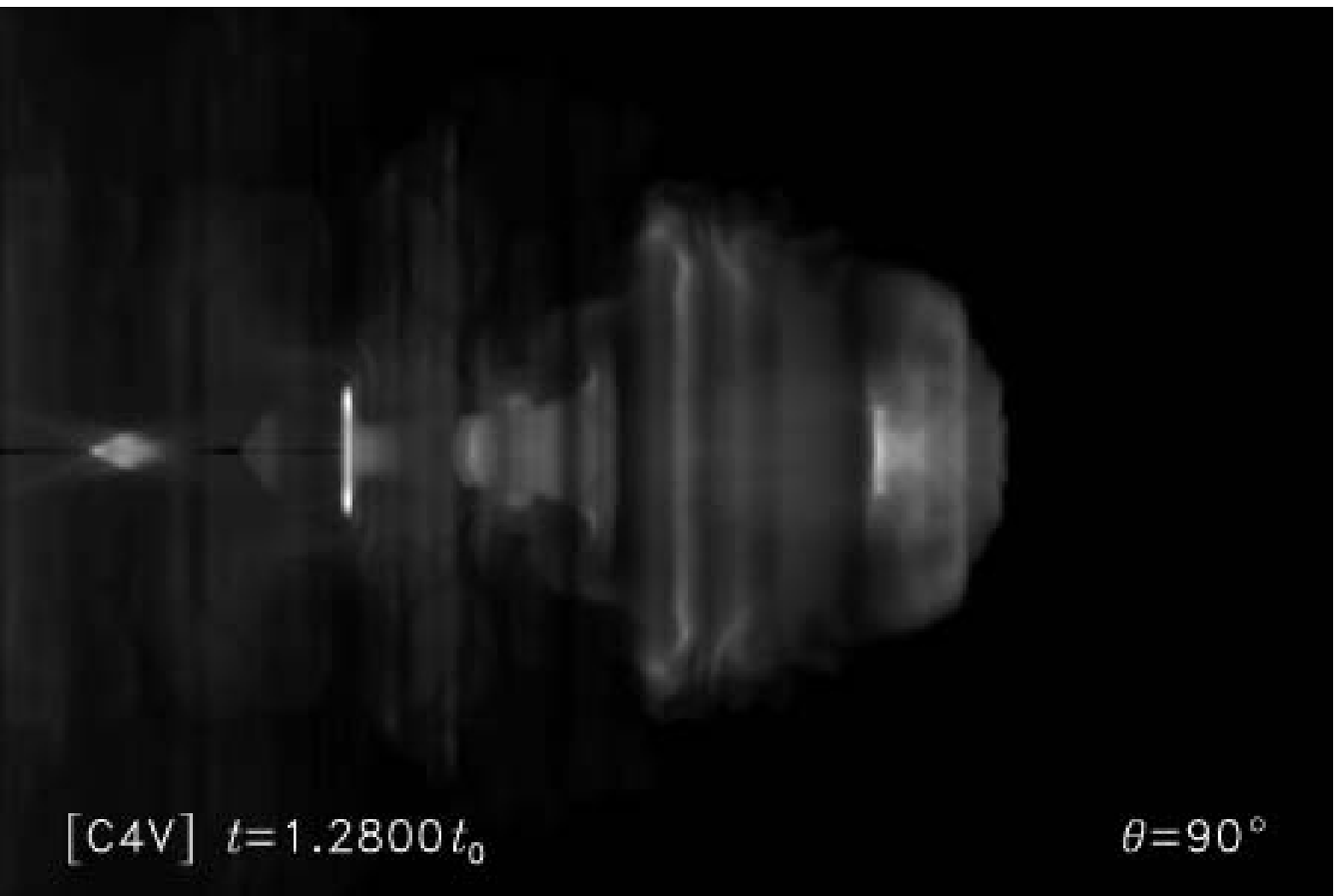}
\\
\includegraphics[width=4.5cm]{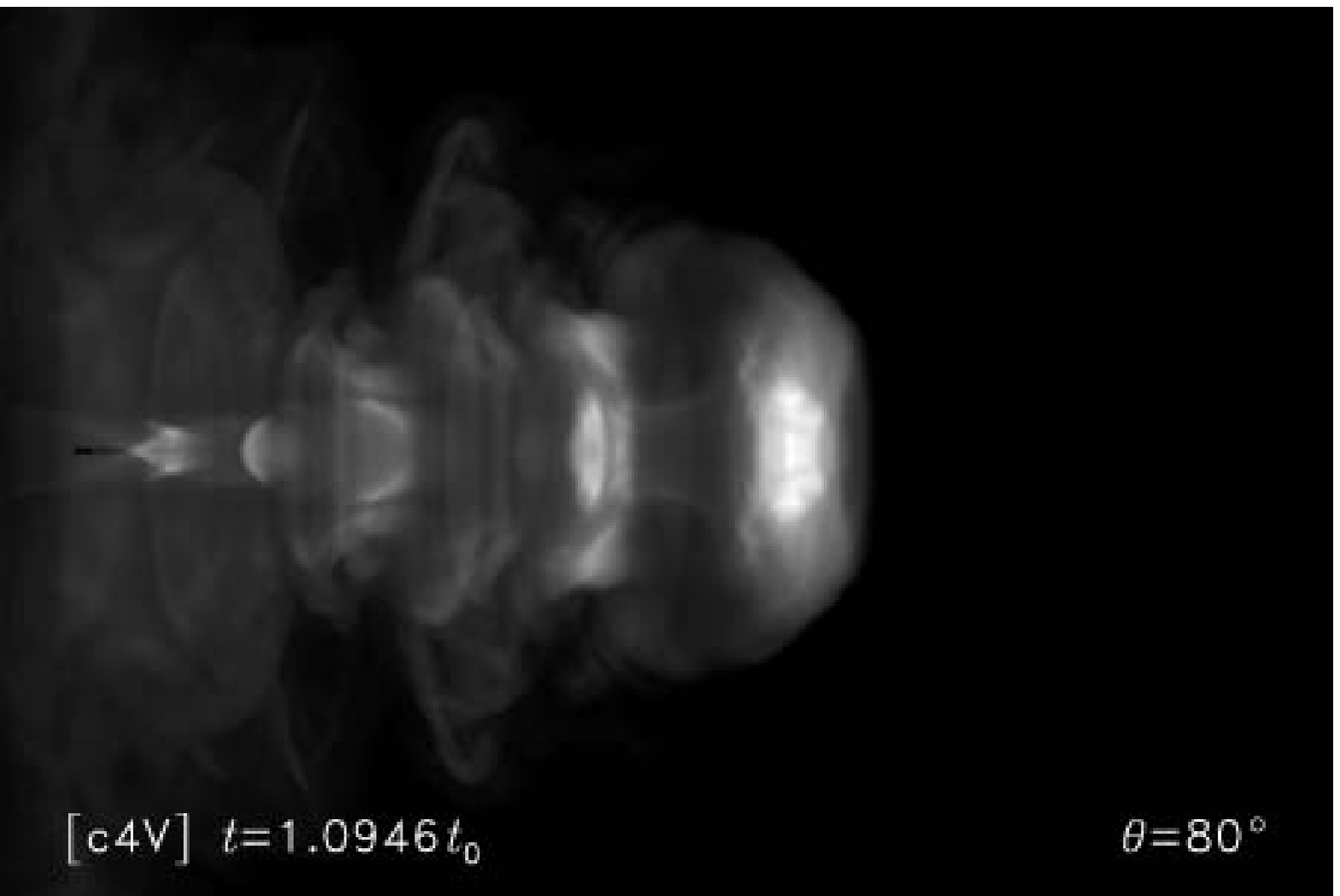}
&
\includegraphics[width=4.5cm]{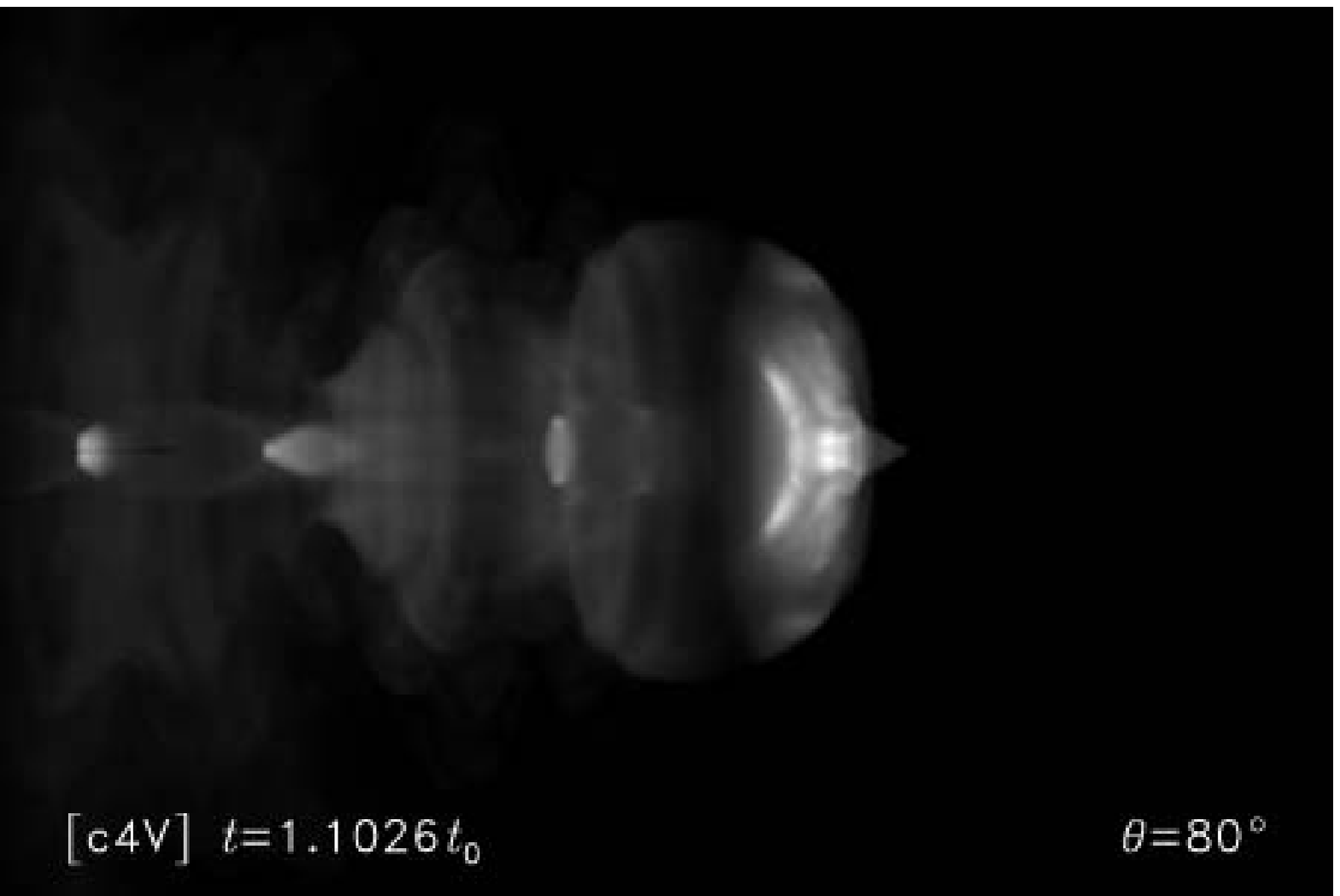}
&
\includegraphics[width=4.5cm]{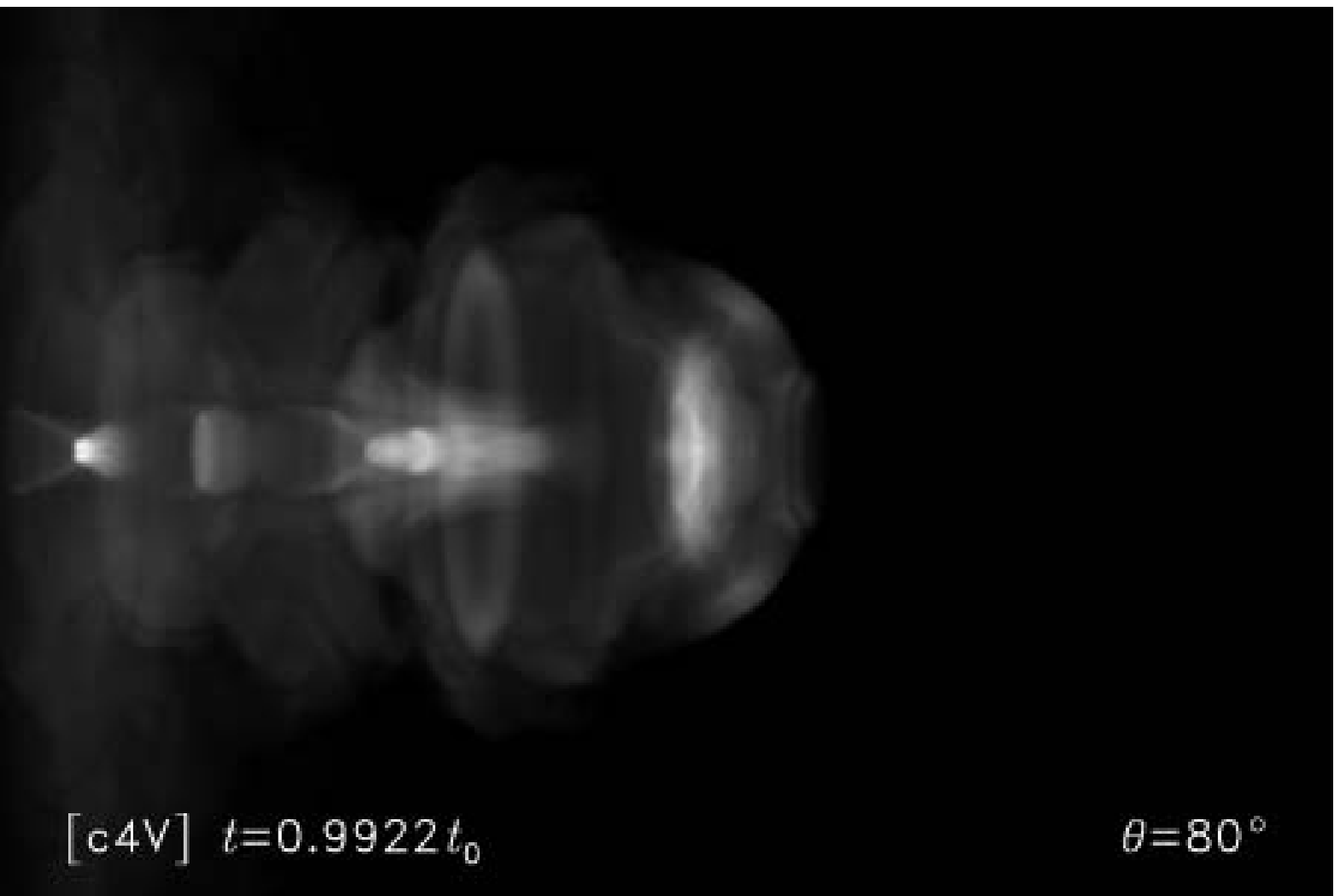}
\\
\includegraphics[width=4.5cm]{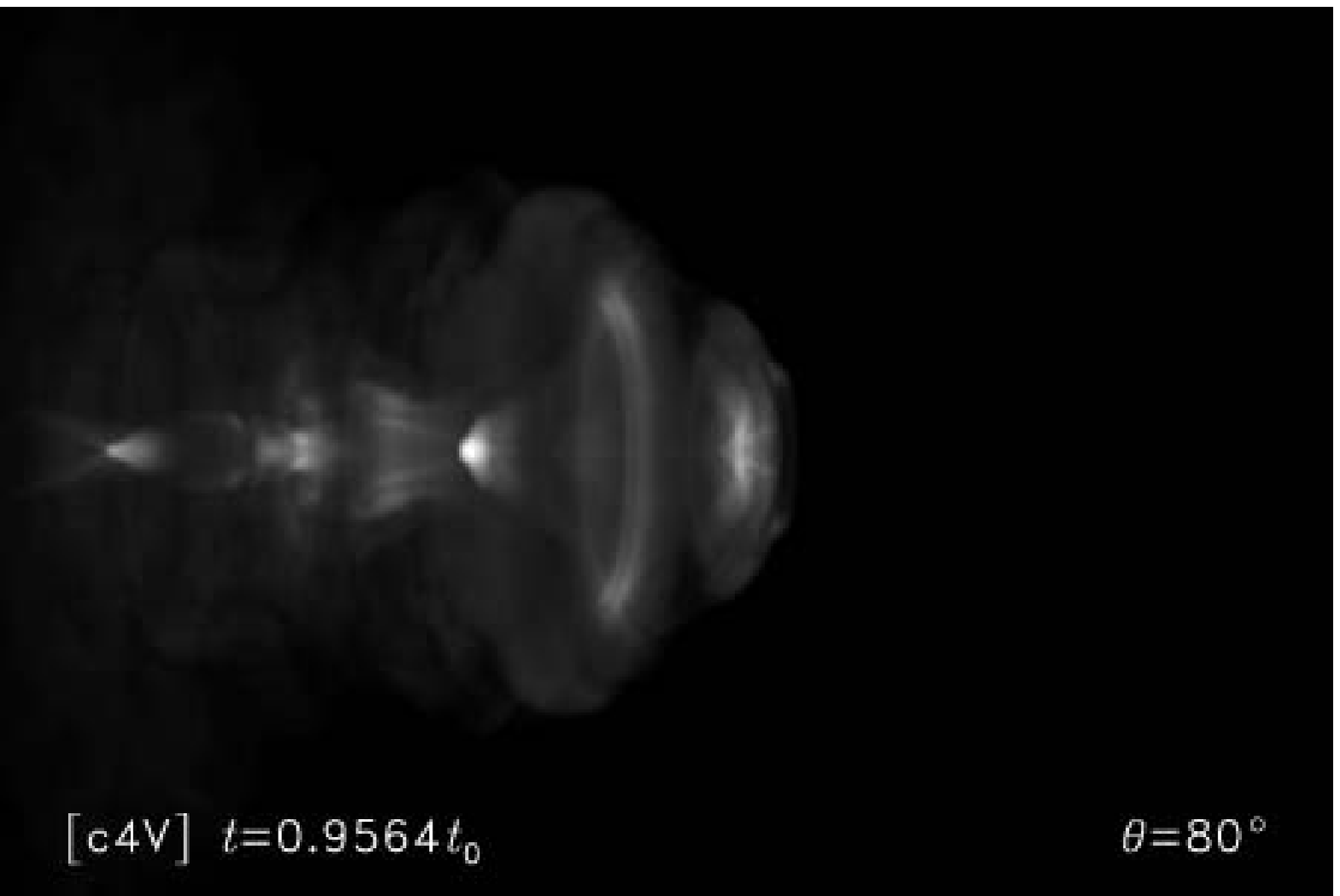}
&
\includegraphics[width=4.5cm]{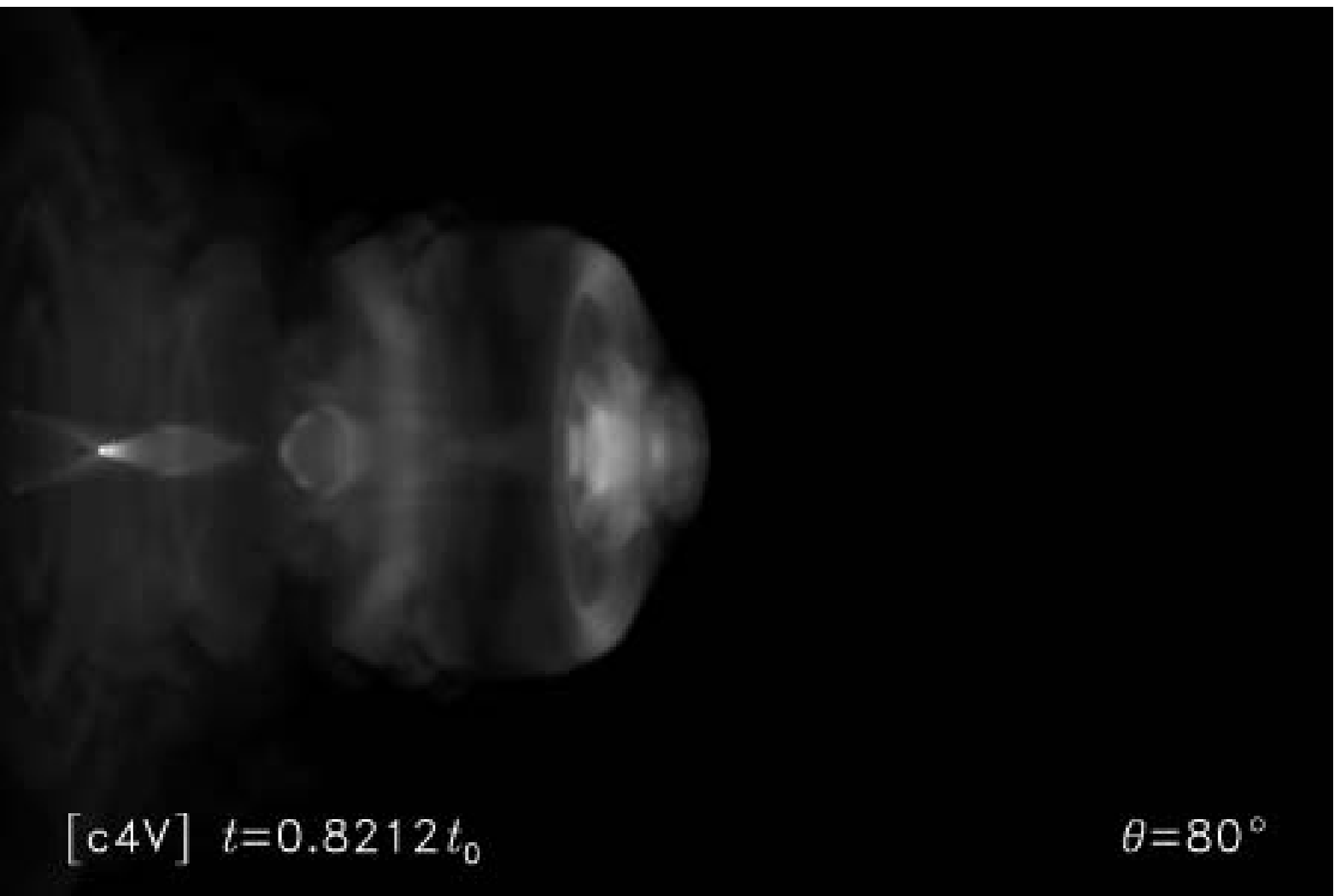}
\\
\end{array}$
\caption{
A selection of morphological matches to Pictor~A,
from ray-traced renderings in a
$450 \times 300$ pixel sub-region
about a jet with parameters
$(\eta,M)=(10^{-4},5)$.
}
\label{f:best_morphologies-4m5}
\end{figure}

\begin{figure}[h]
\centering \leavevmode
$\begin{array}{ccc}
\includegraphics[width=4.5cm]{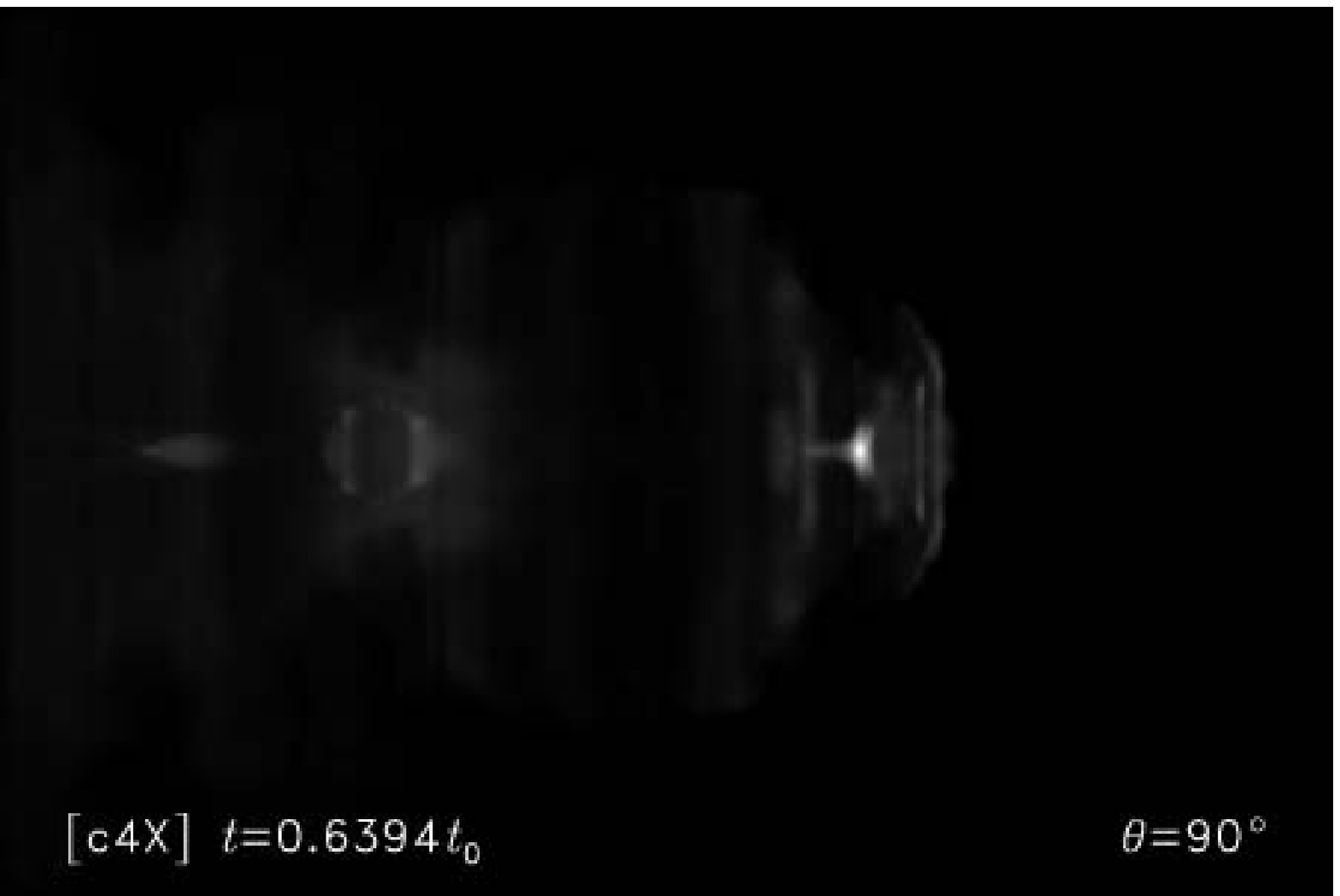}
&
\includegraphics[width=4.5cm]{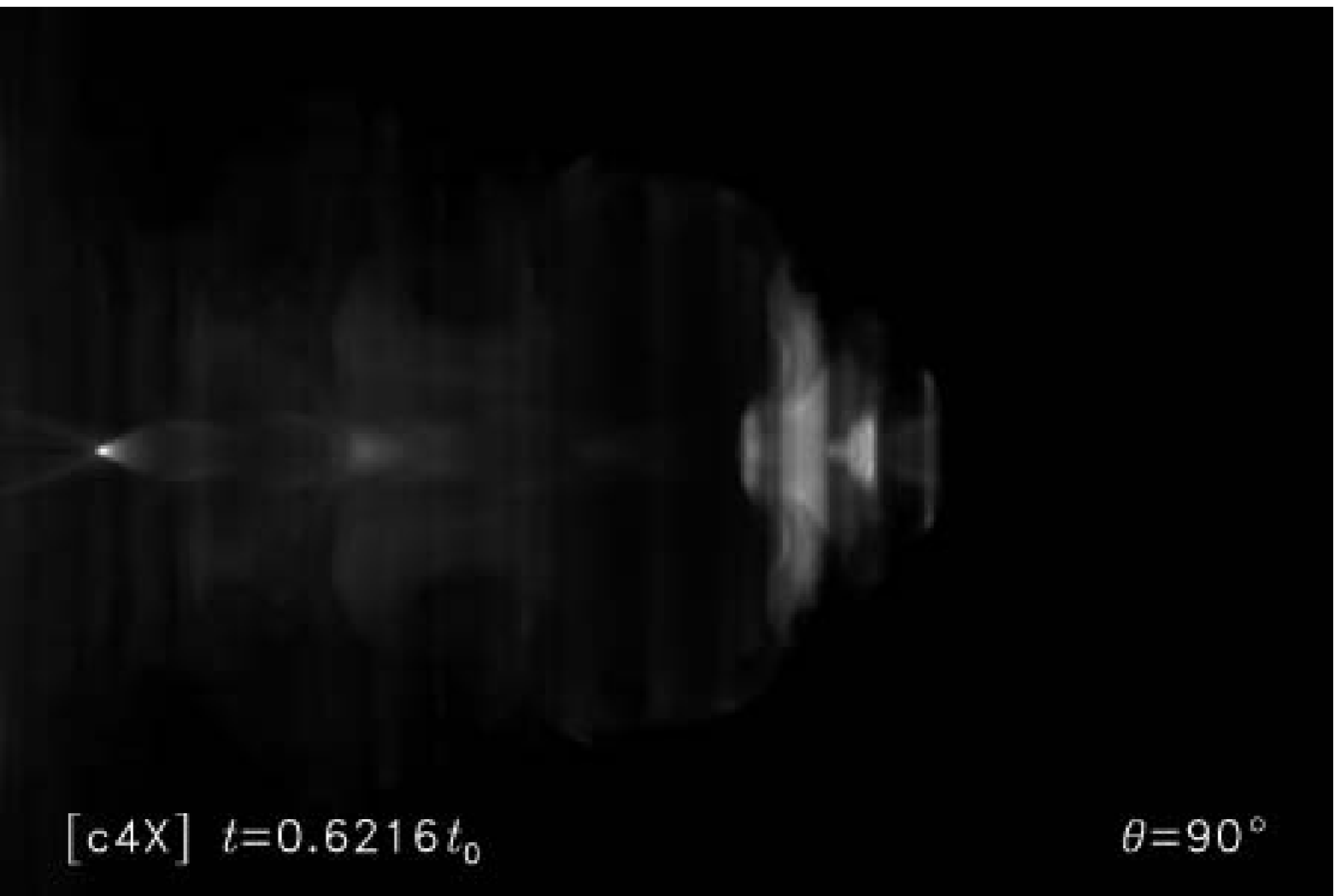}
&
\includegraphics[width=4.5cm]{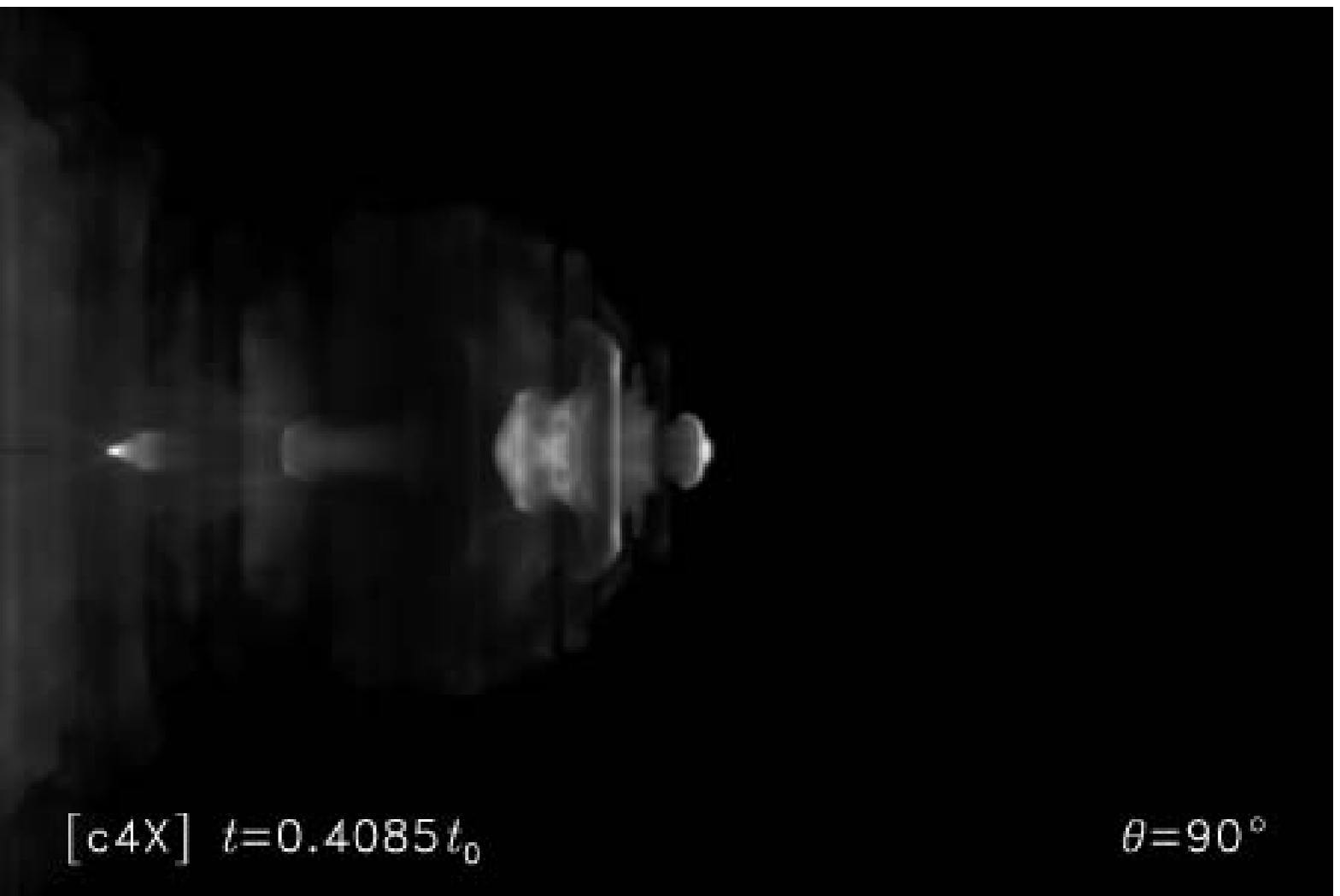}
\\
\includegraphics[width=4.5cm]{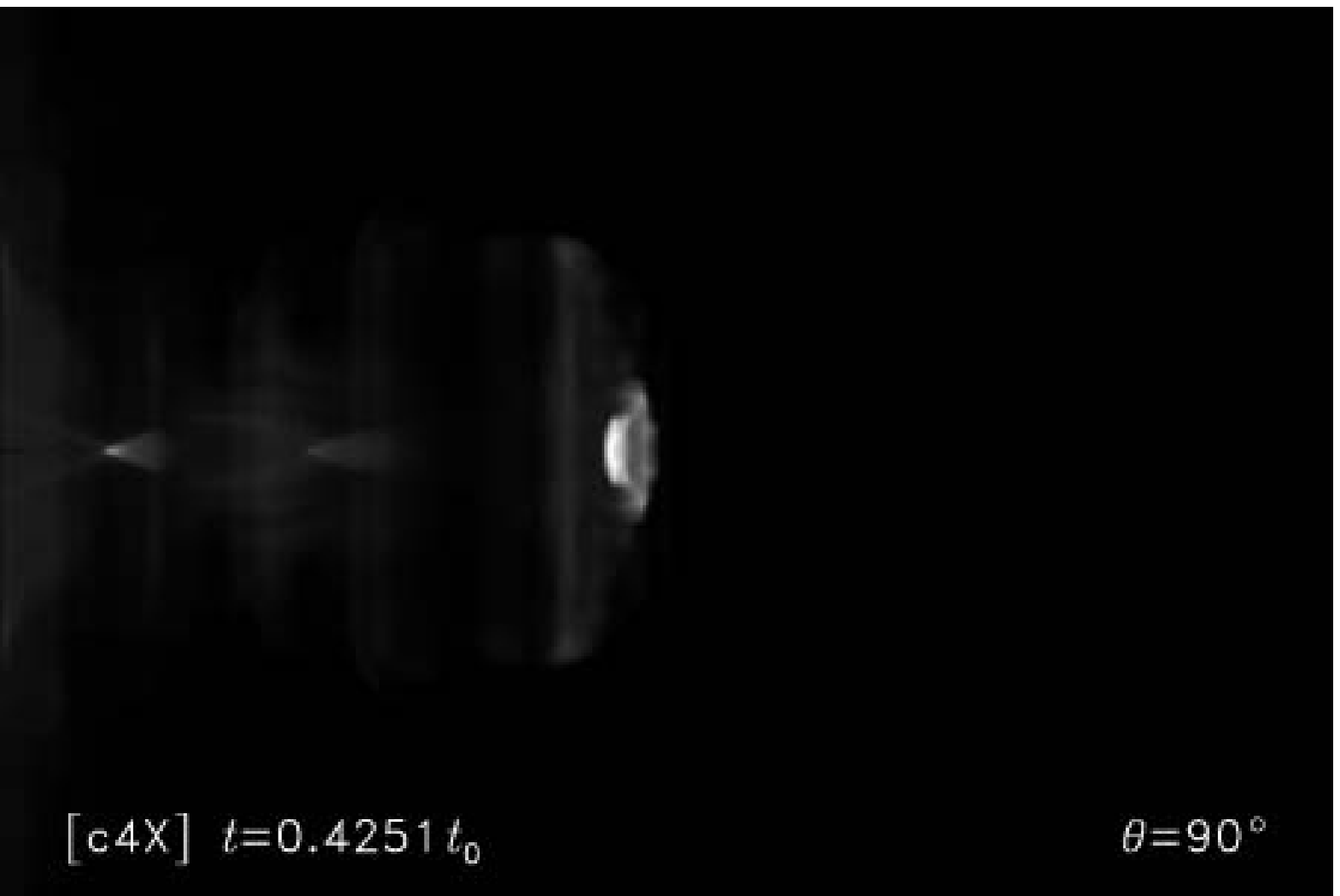}
&
\includegraphics[width=4.5cm]{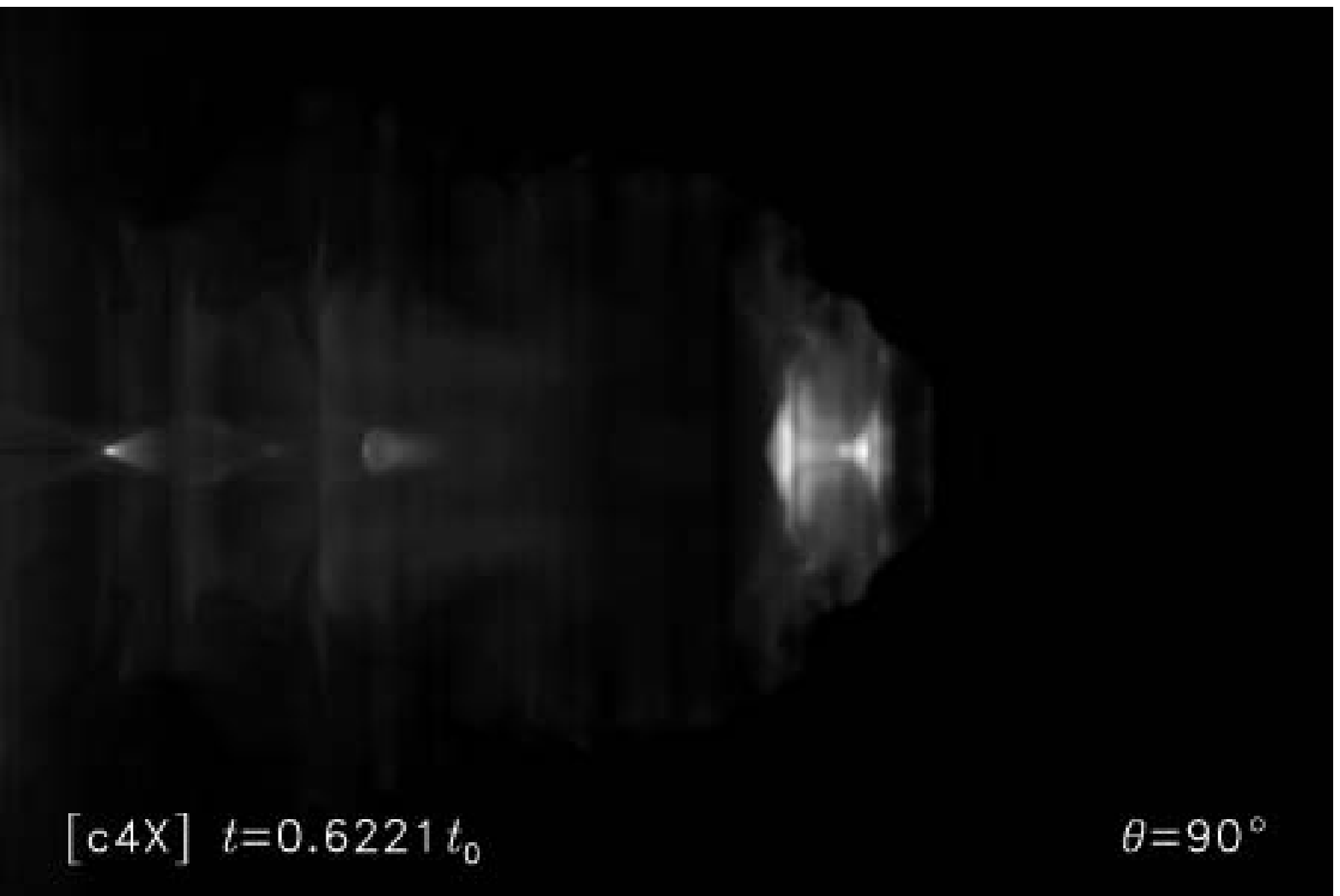}
&
\includegraphics[width=4.5cm]{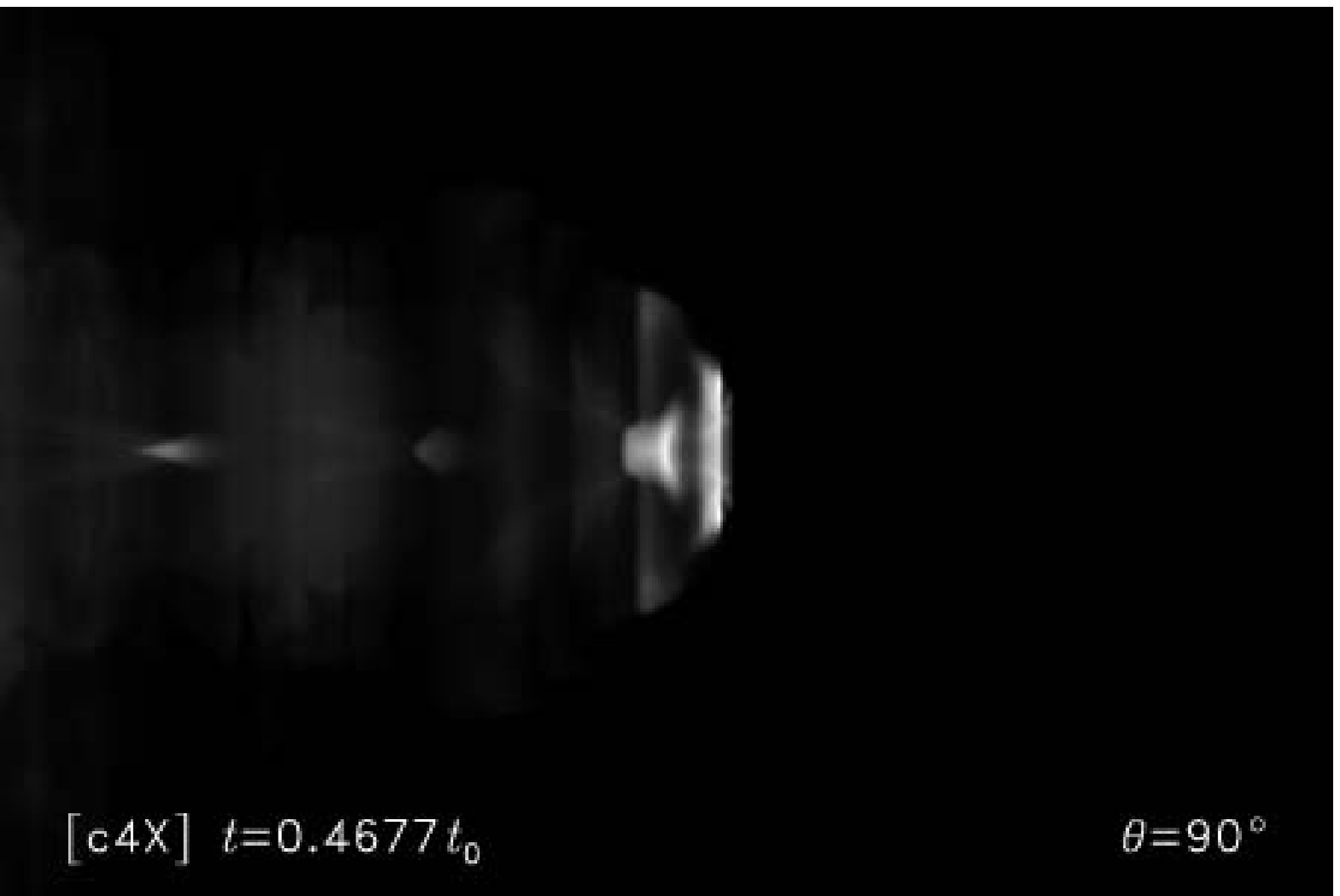}
\\
\end{array}$
\caption{
Morphological matches as in Figure~\ref{f:best_morphologies-4m5}
but with jet parameters
$(\eta,M)=(10^{-4},10)$.
}
\label{f:best_morphologies-4mx}
\end{figure}

\begin{figure}[h]
\centering \leavevmode
$\begin{array}{ccc}
\includegraphics[width=4.5cm]{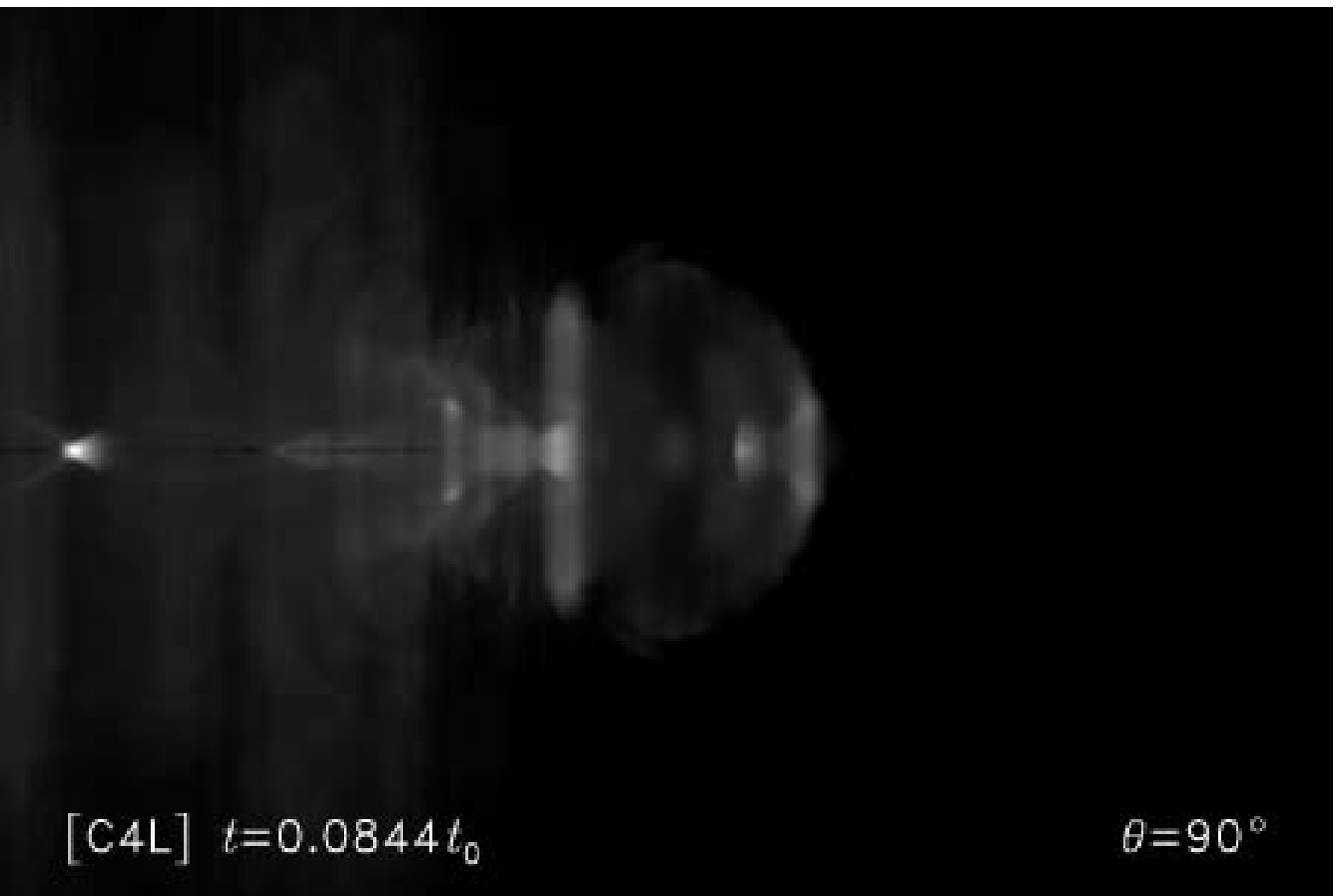}
&
\includegraphics[width=4.5cm]{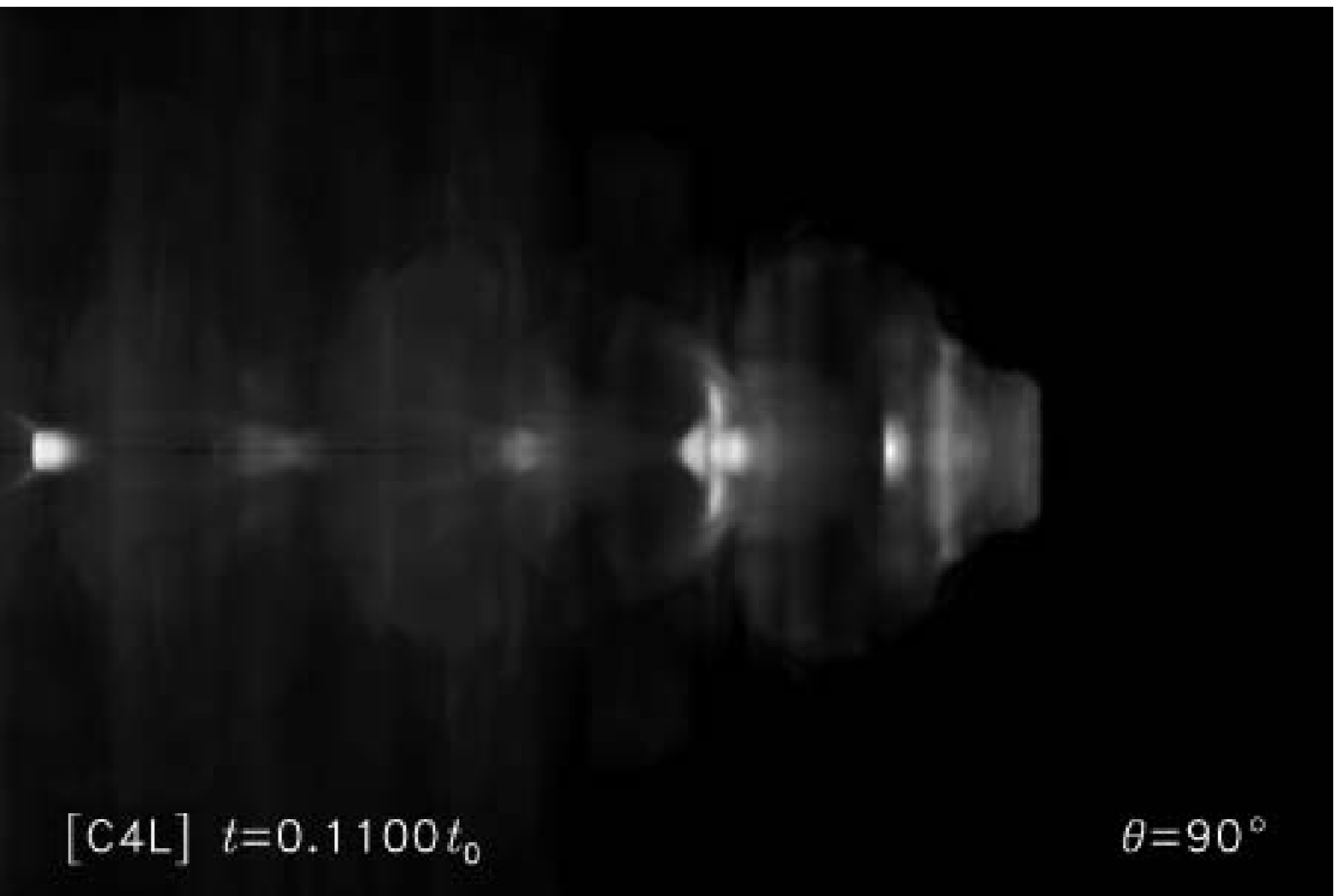}
&
\includegraphics[width=4.5cm]{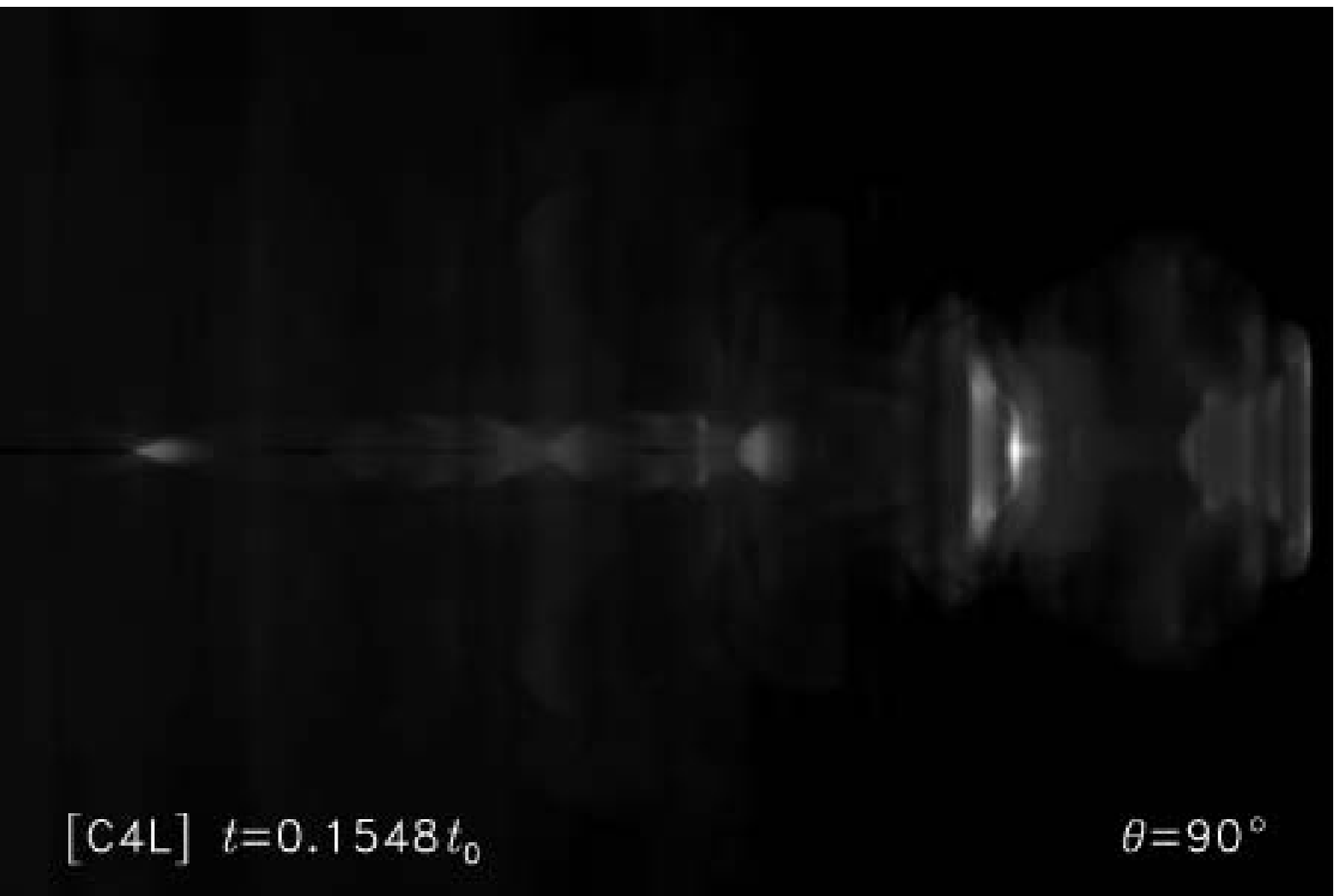}
\\
\end{array}$
\caption{
Morphological matches as in Figure~\ref{f:best_morphologies-4m5}
but with jet parameters
$(\eta,M)=(10^{-4},50)$.
}
\label{f:best_morphologies-4ml}
\end{figure}

\begin{figure}[h]
\centering \leavevmode
$\begin{array}{ccc}
\includegraphics[width=4.5cm]{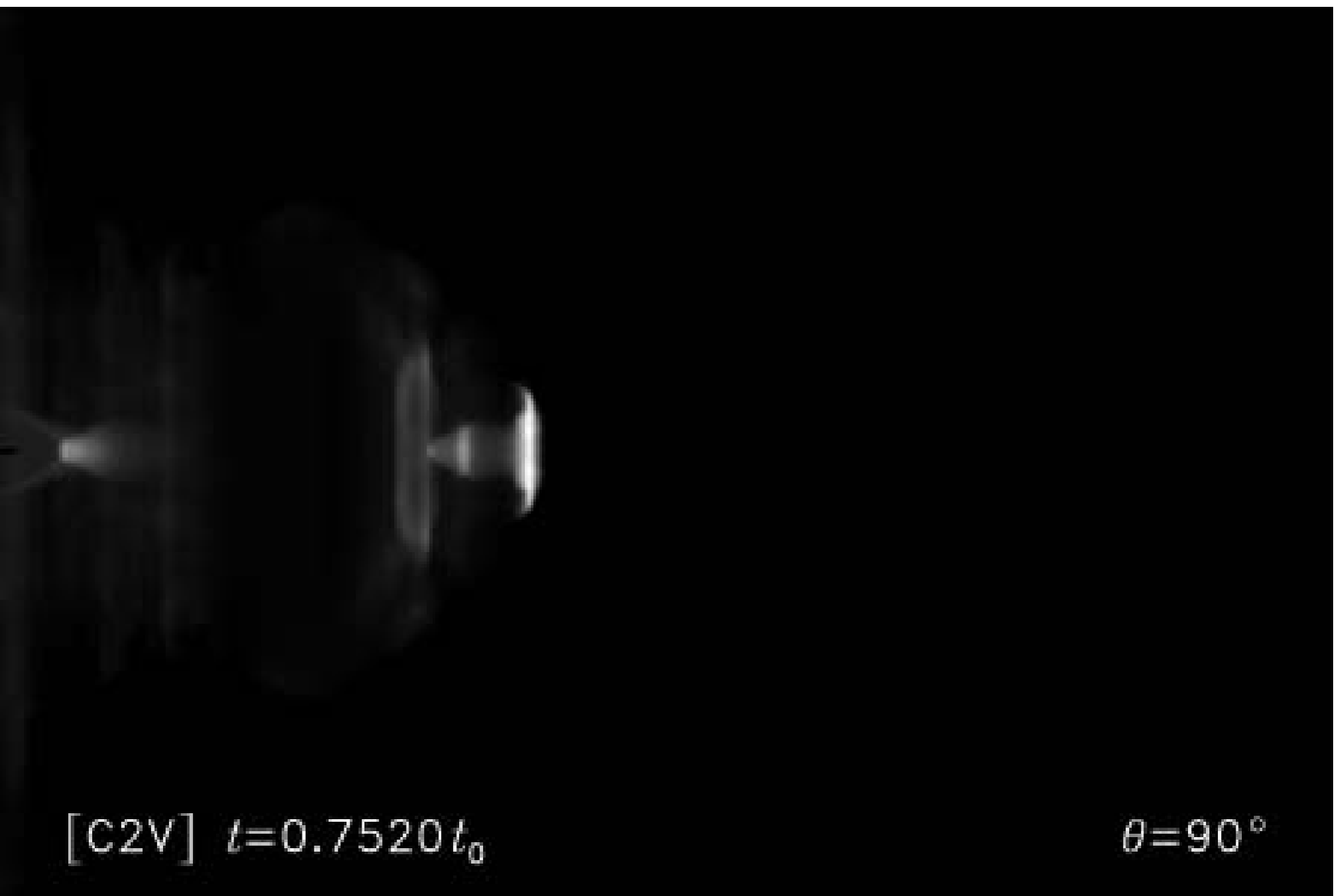}
&
\includegraphics[width=4.5cm]{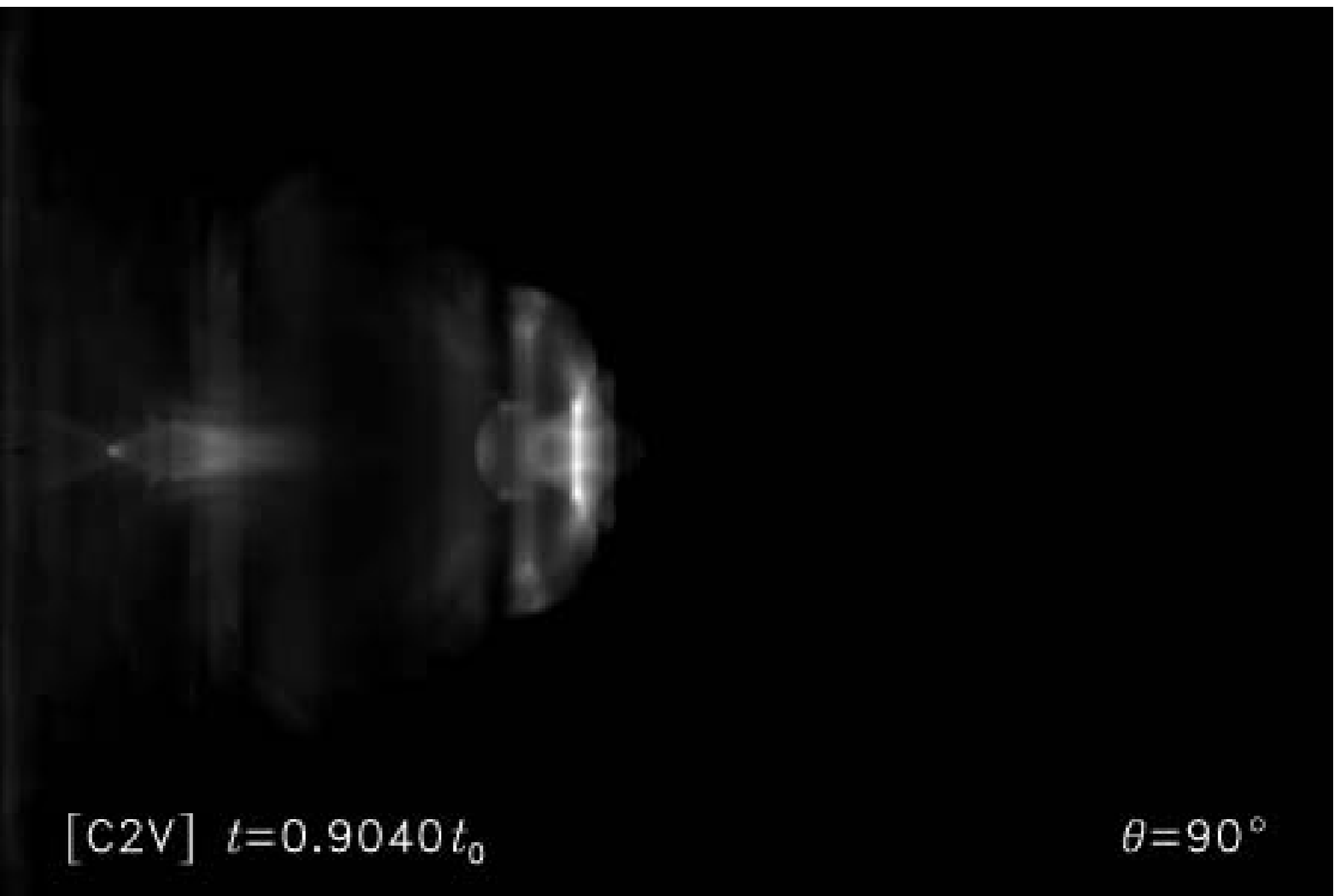}
&
\includegraphics[width=4.5cm]{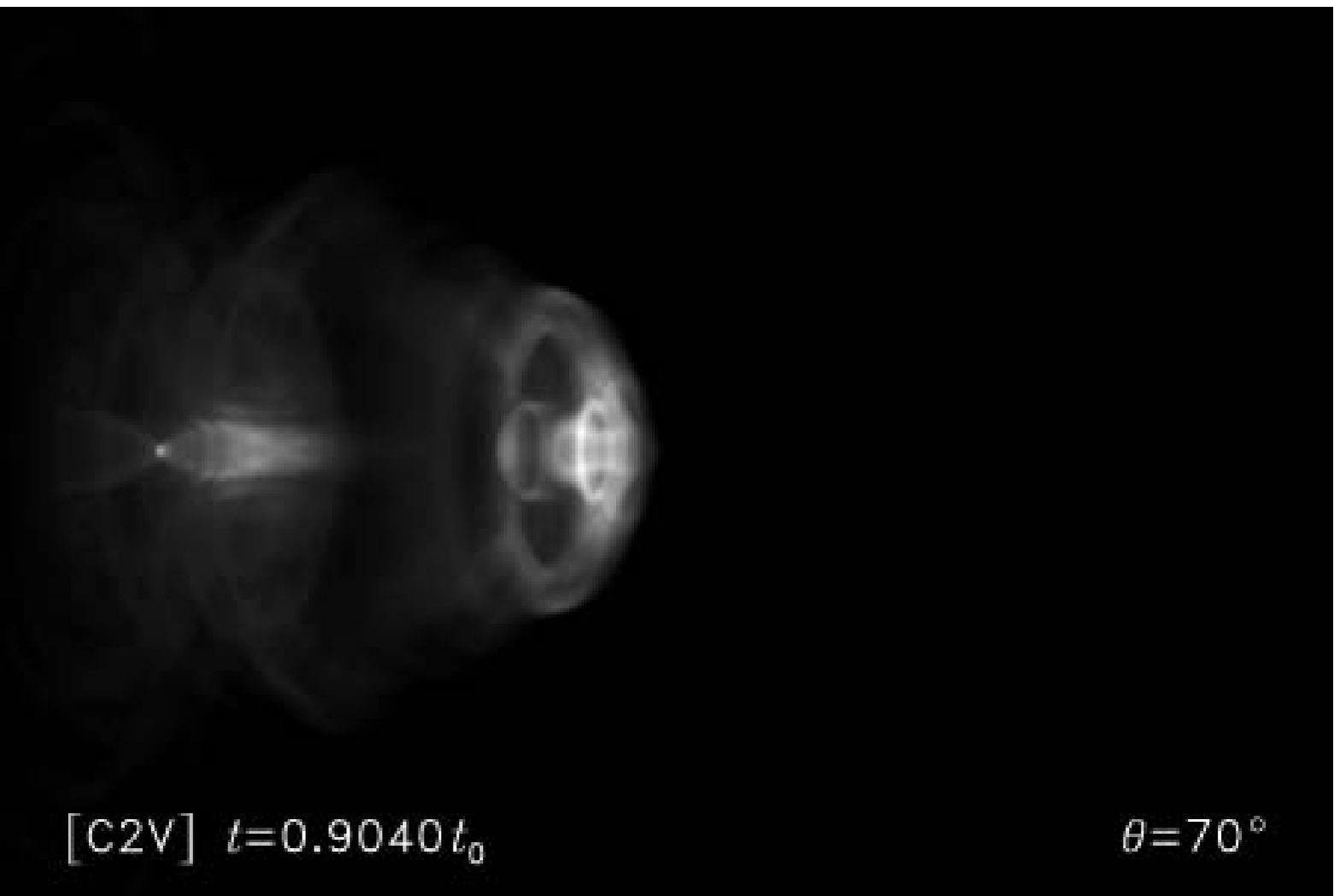}
\\
\end{array}$
\caption{
Morphological matches as in Figure~\ref{f:best_morphologies-4m5}
but with jet parameters
$(\eta,M)=(10^{-2},5)$.
}
\label{f:best_morphologies-2m5}
\end{figure}

\begin{figure}[h]
\centering \leavevmode
$\begin{array}{ccc}
\includegraphics[width=4.5cm]{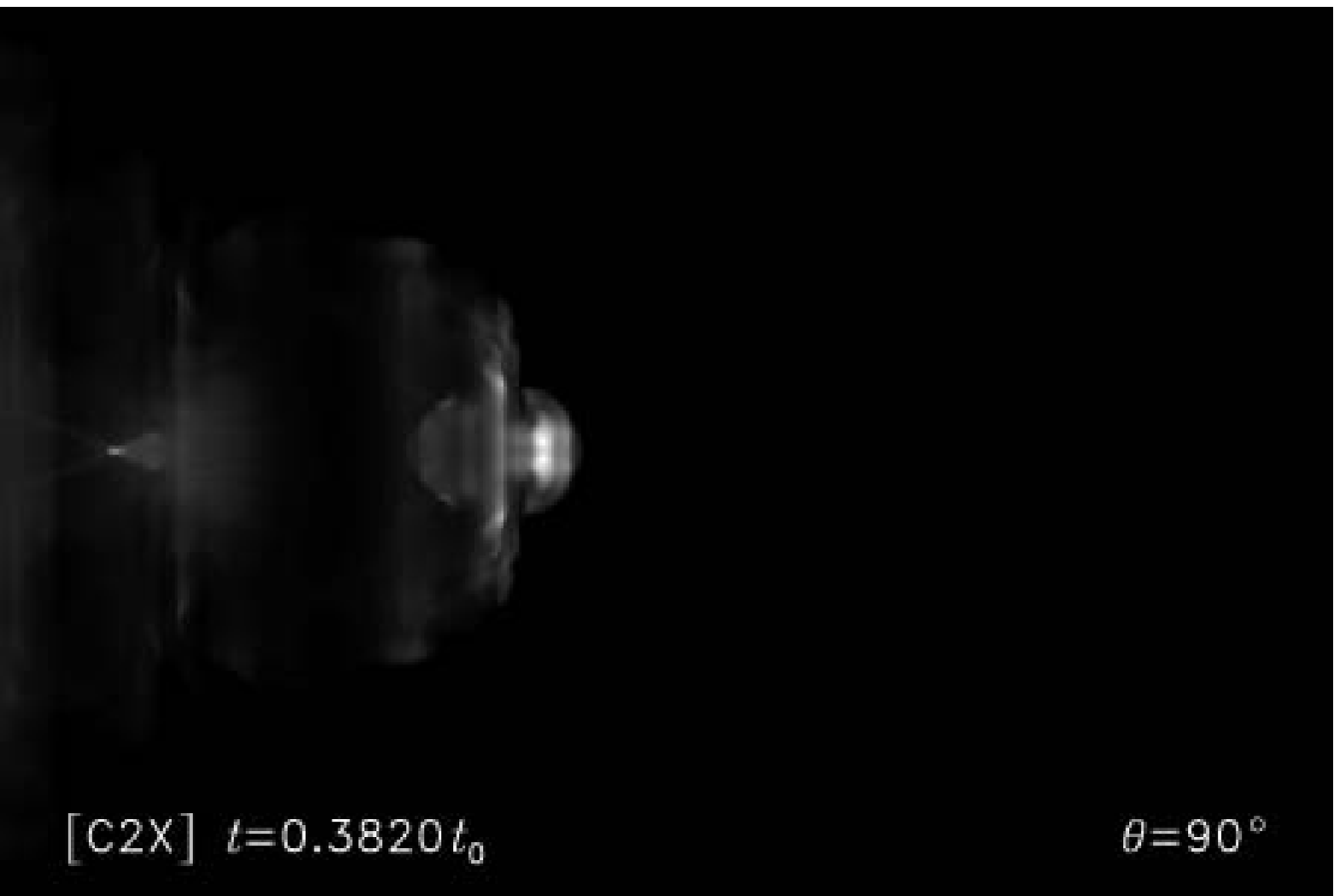}
&
\includegraphics[width=4.5cm]{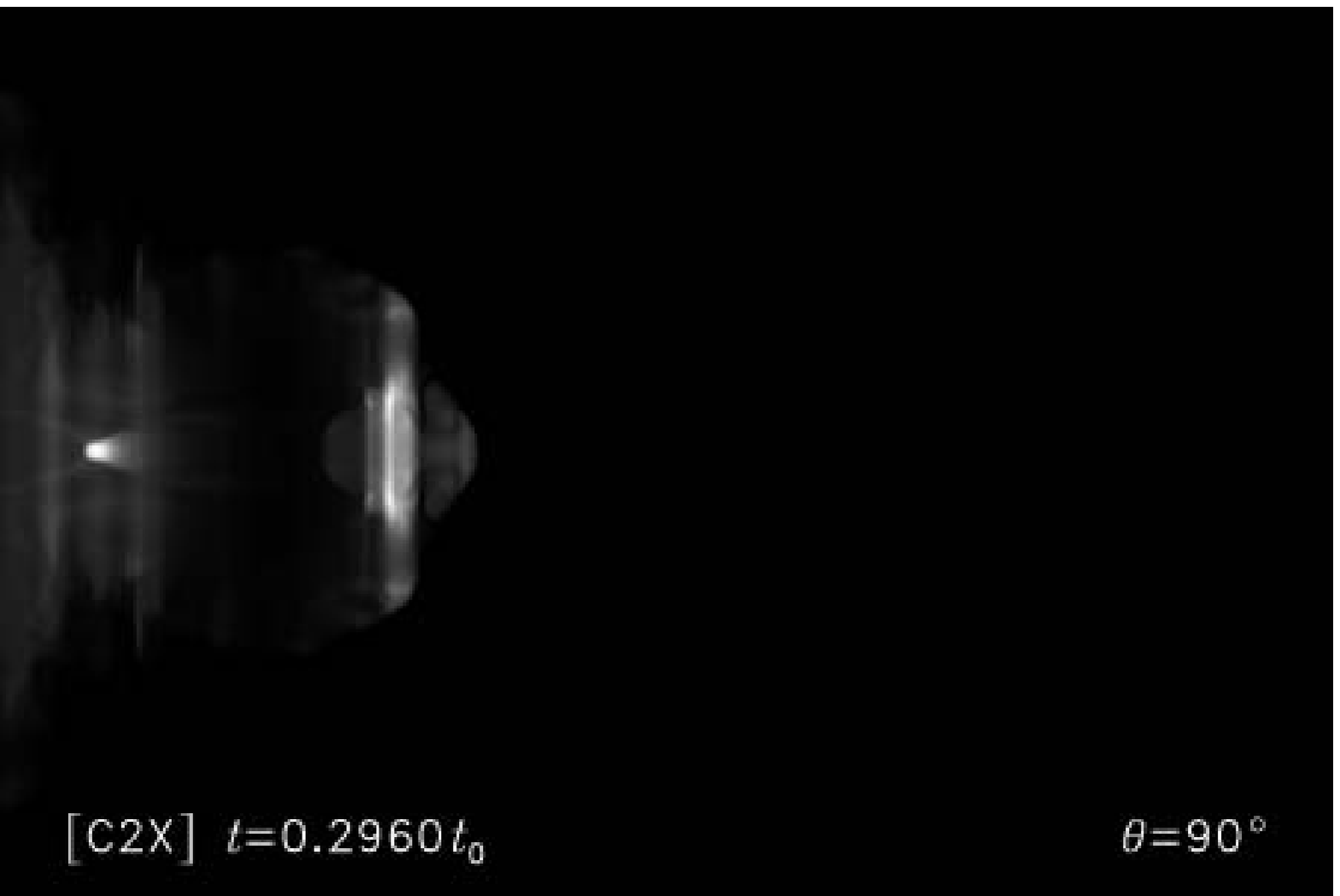}
&
\includegraphics[width=4.5cm]{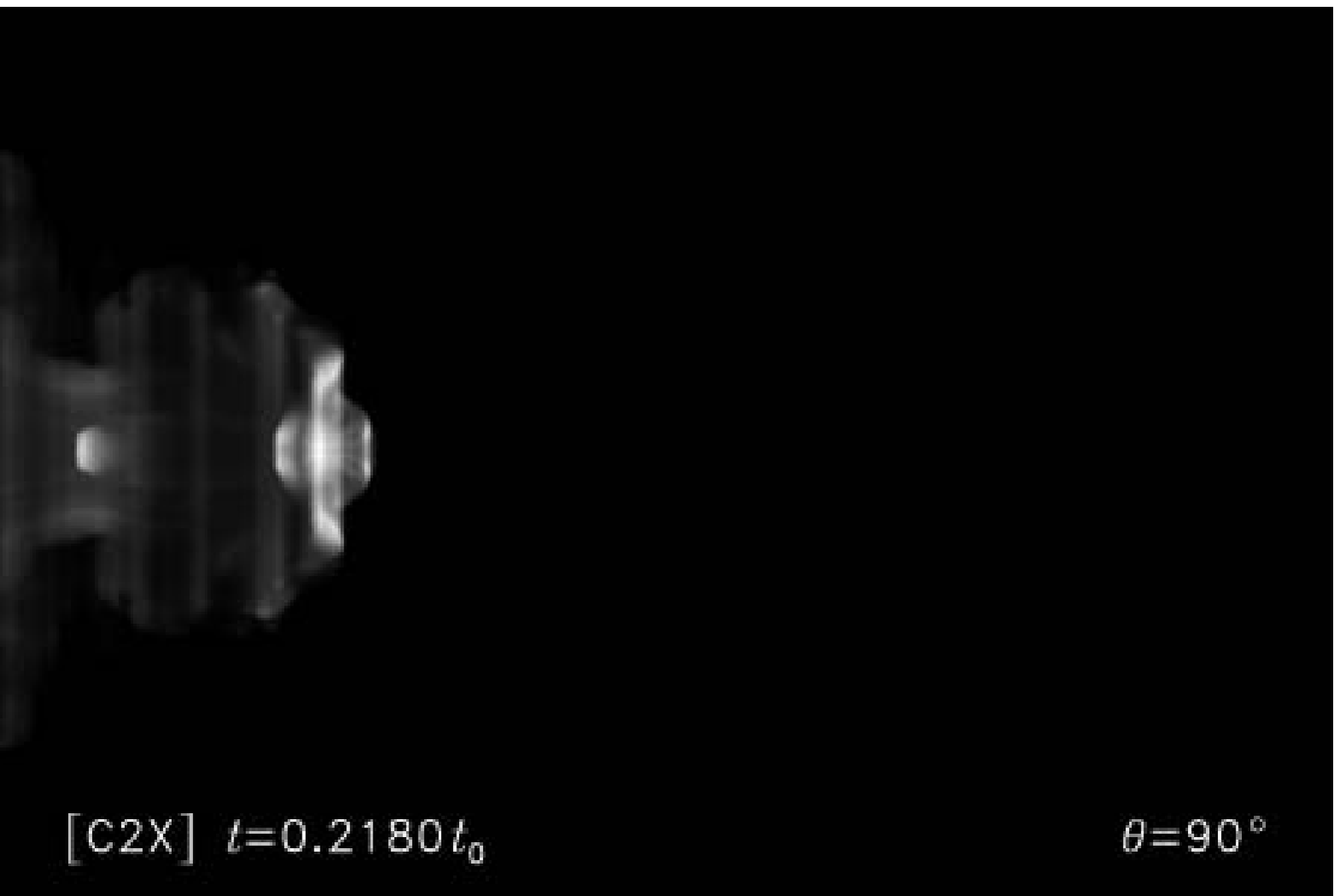}
\\
\includegraphics[width=4.5cm]{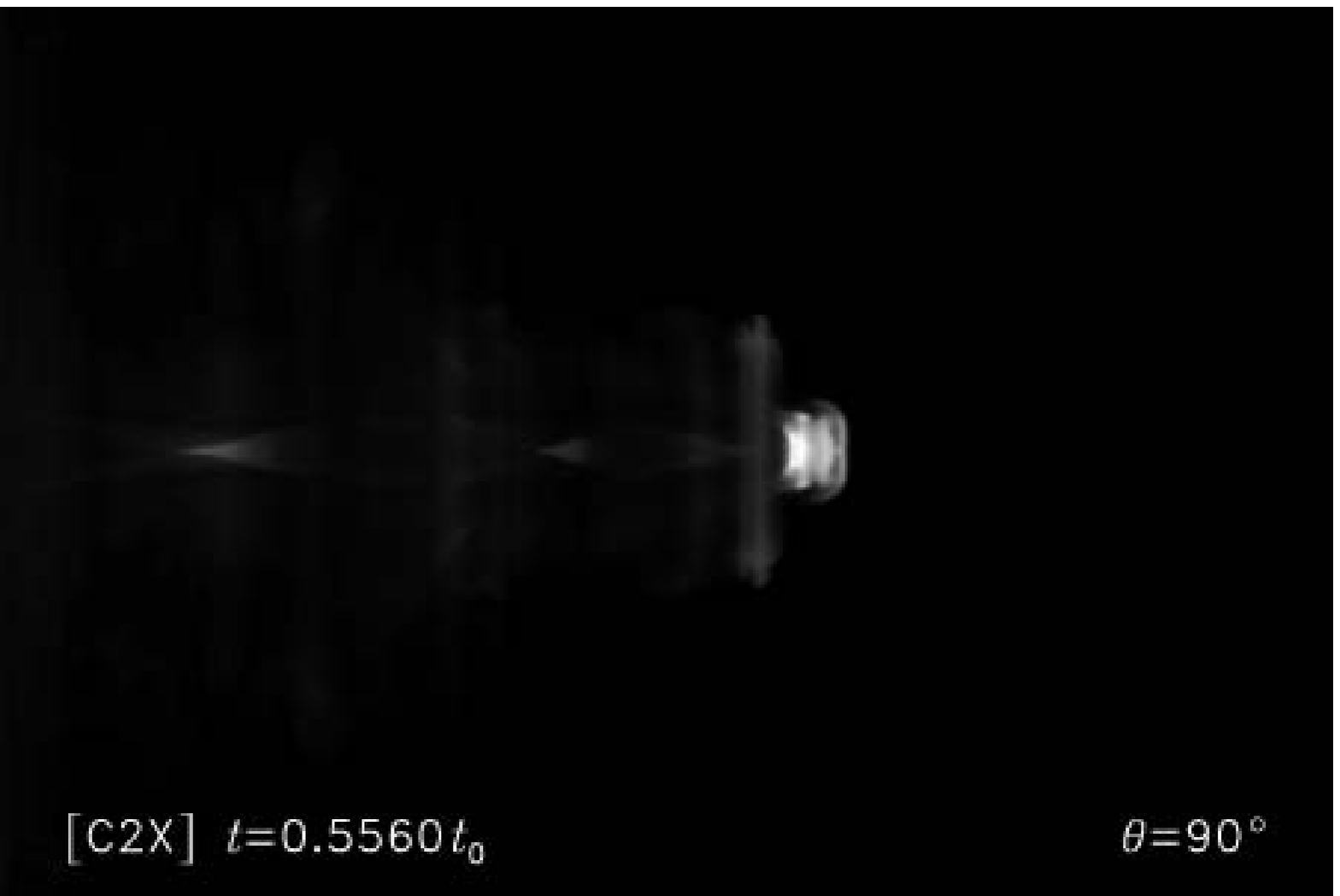}
&
\includegraphics[width=4.5cm]{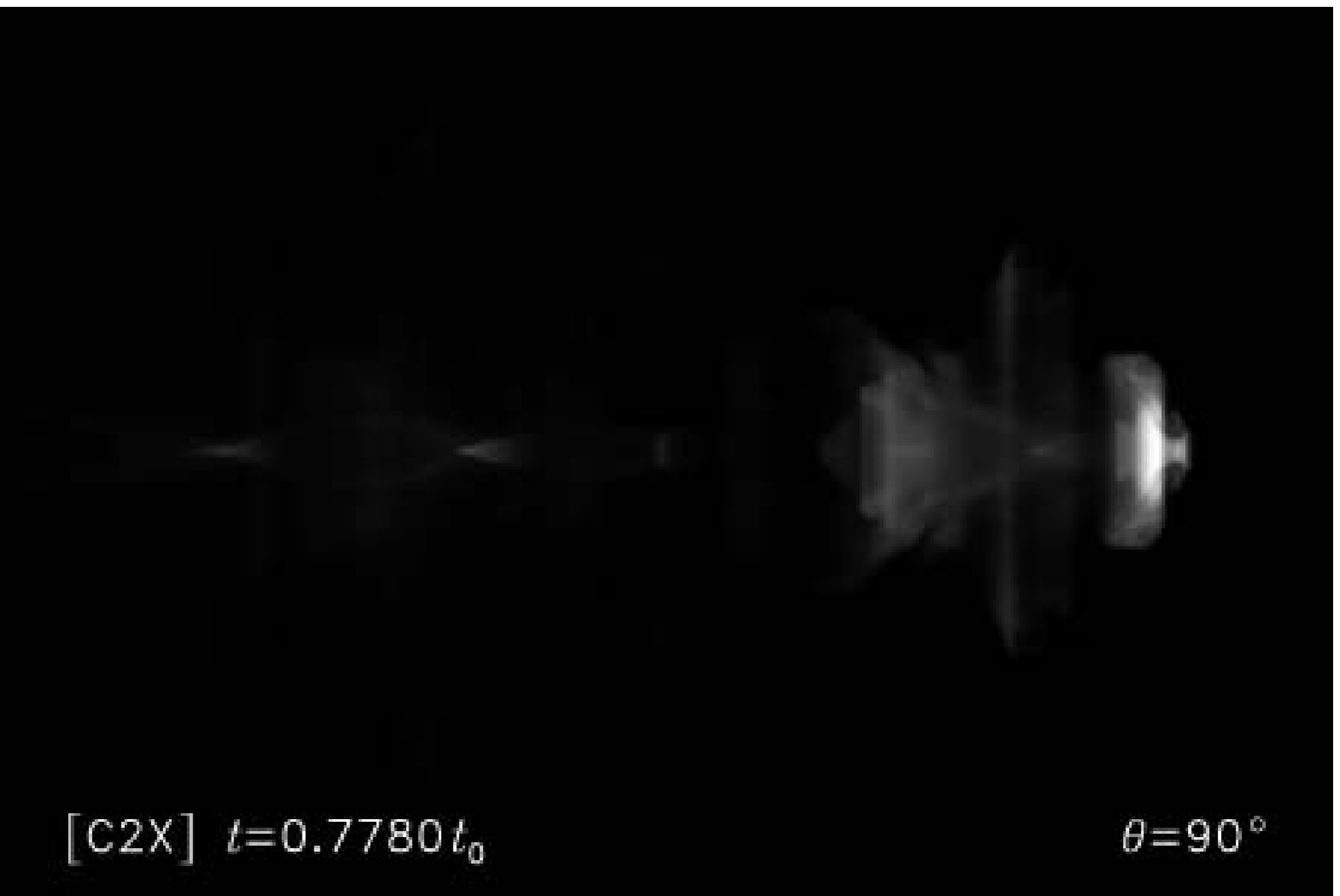}
\end{array}$
\caption{
Morphological matches as in Figure~\ref{f:best_morphologies-4m5}
but with jet parameters
$(\eta,M)=(10^{-2},10)$.
}
\label{f:best_morphologies-2mx}
\end{figure}

\begin{figure}[h]
\centering \leavevmode
$\begin{array}{ccc}
\includegraphics[width=4.5cm]{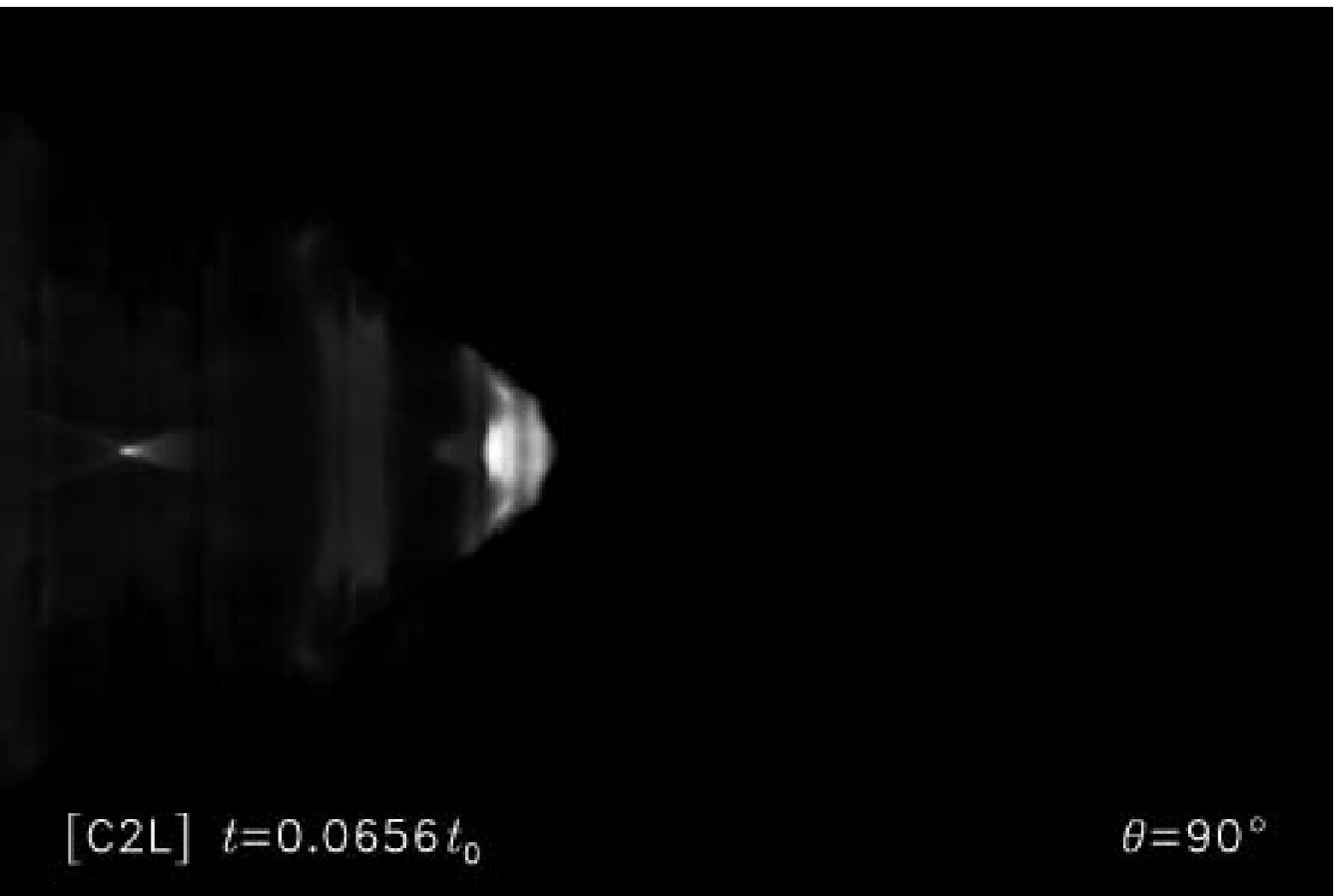}
&
\includegraphics[width=4.5cm]{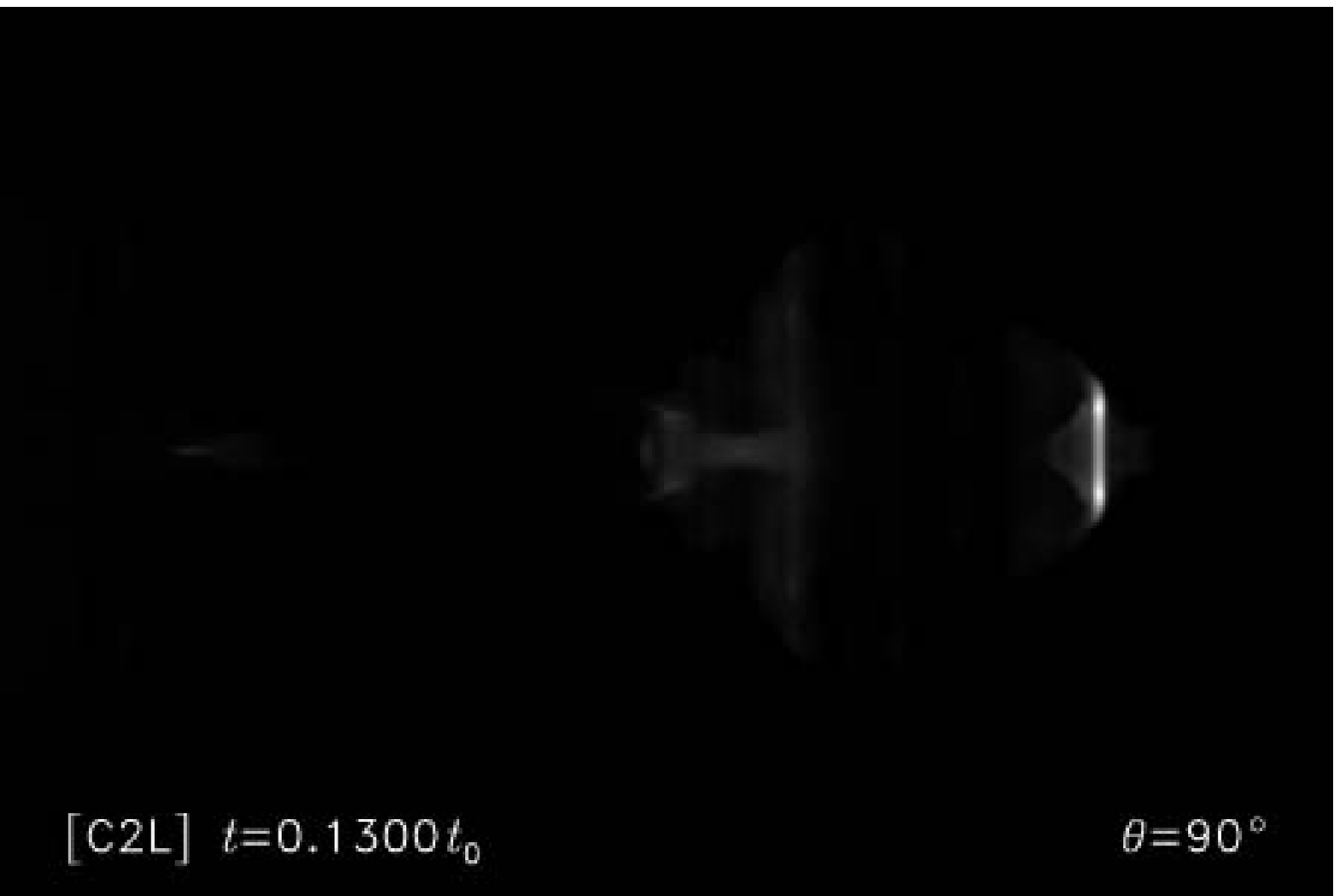}
&
\includegraphics[width=4.5cm]{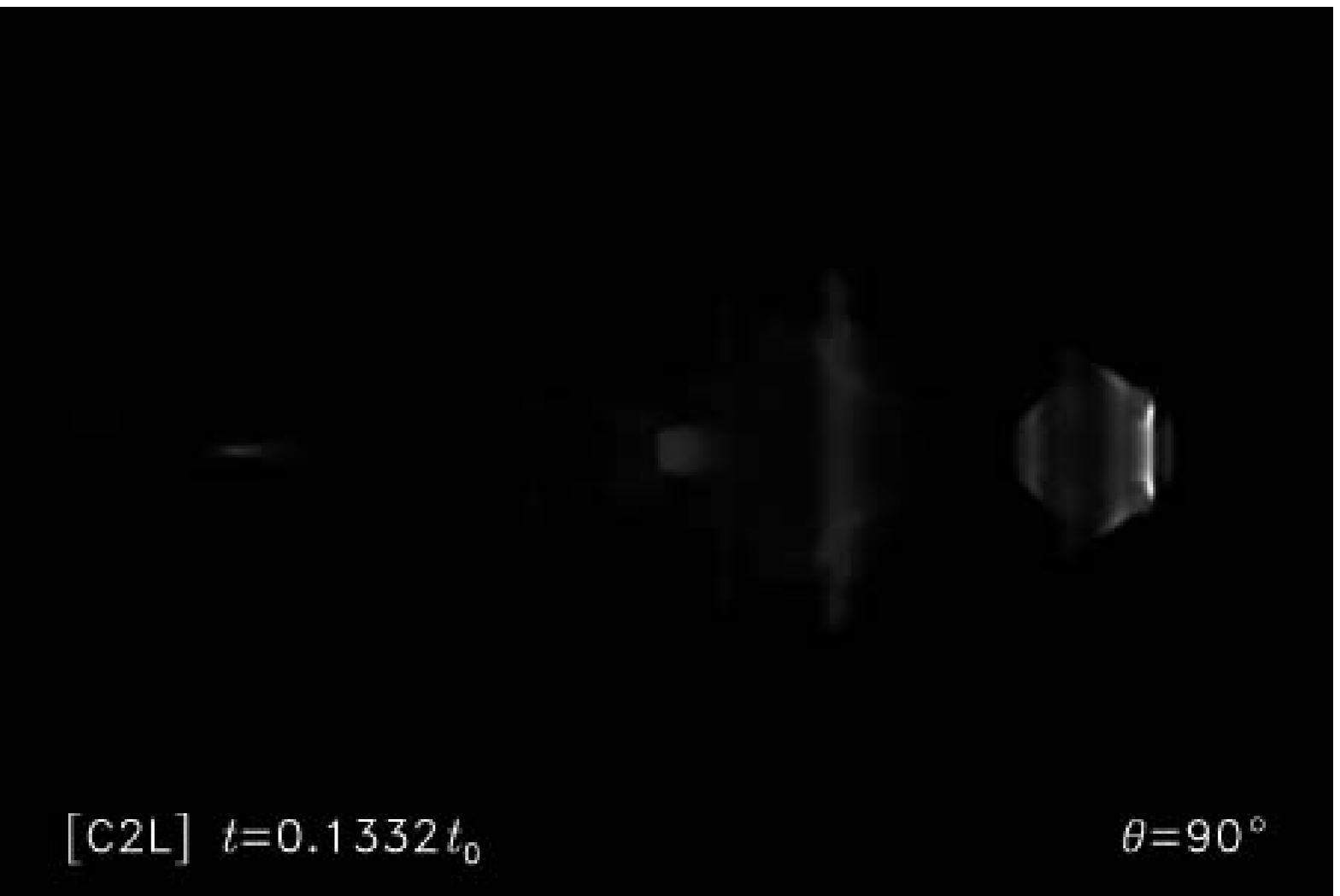}
\\
\includegraphics[width=4.5cm]{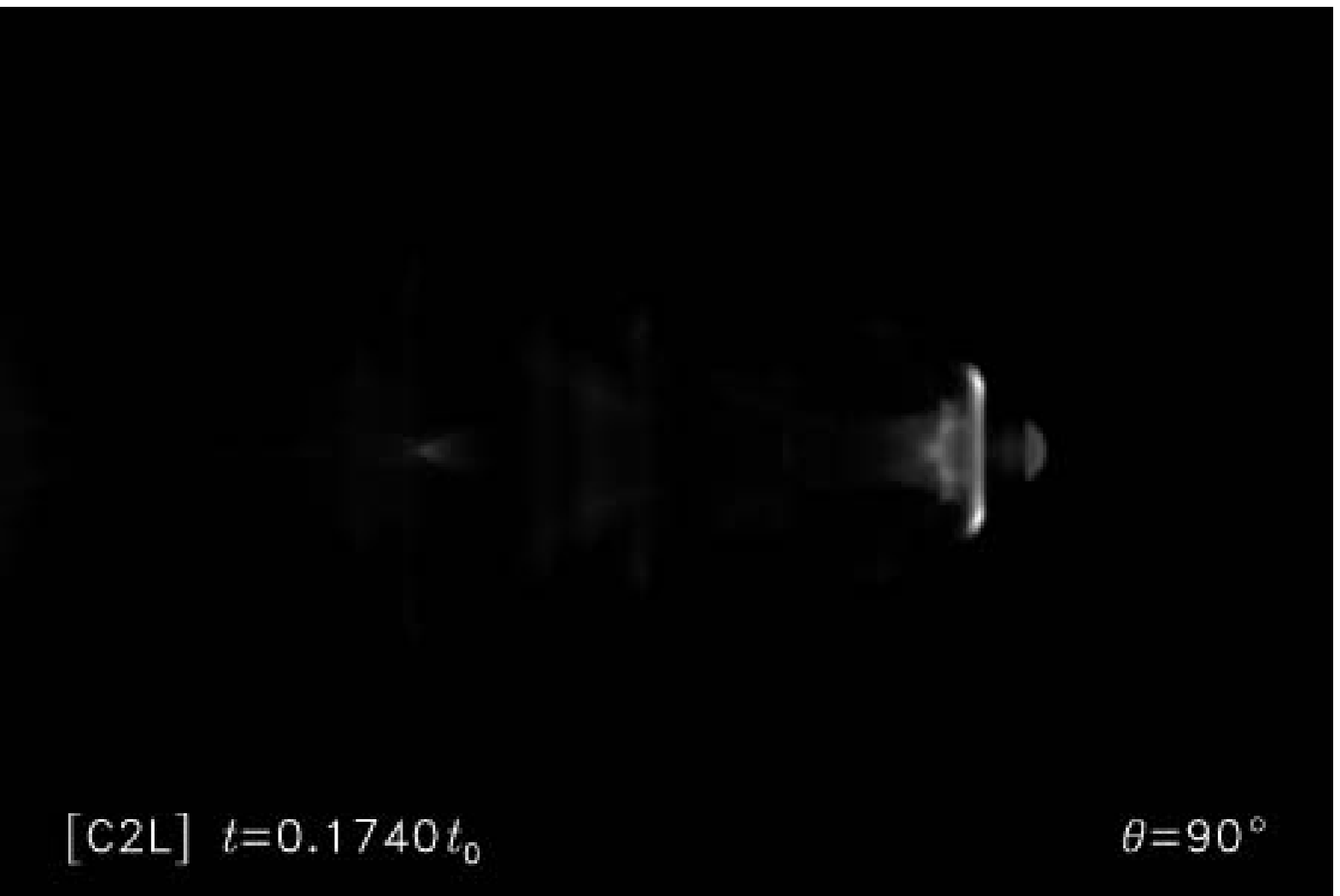}
&
\includegraphics[width=4.5cm]{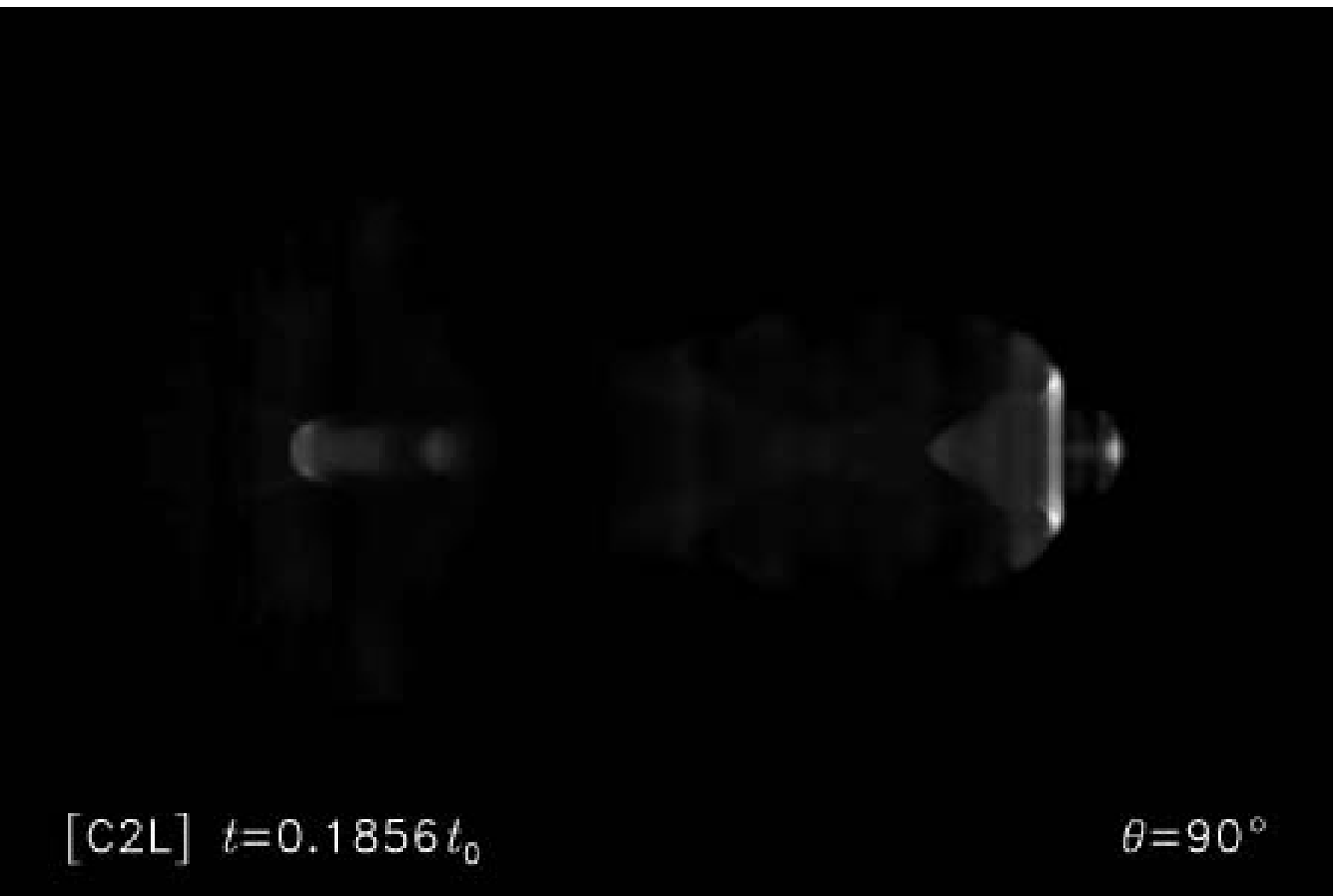}
&
\includegraphics[width=4.5cm]{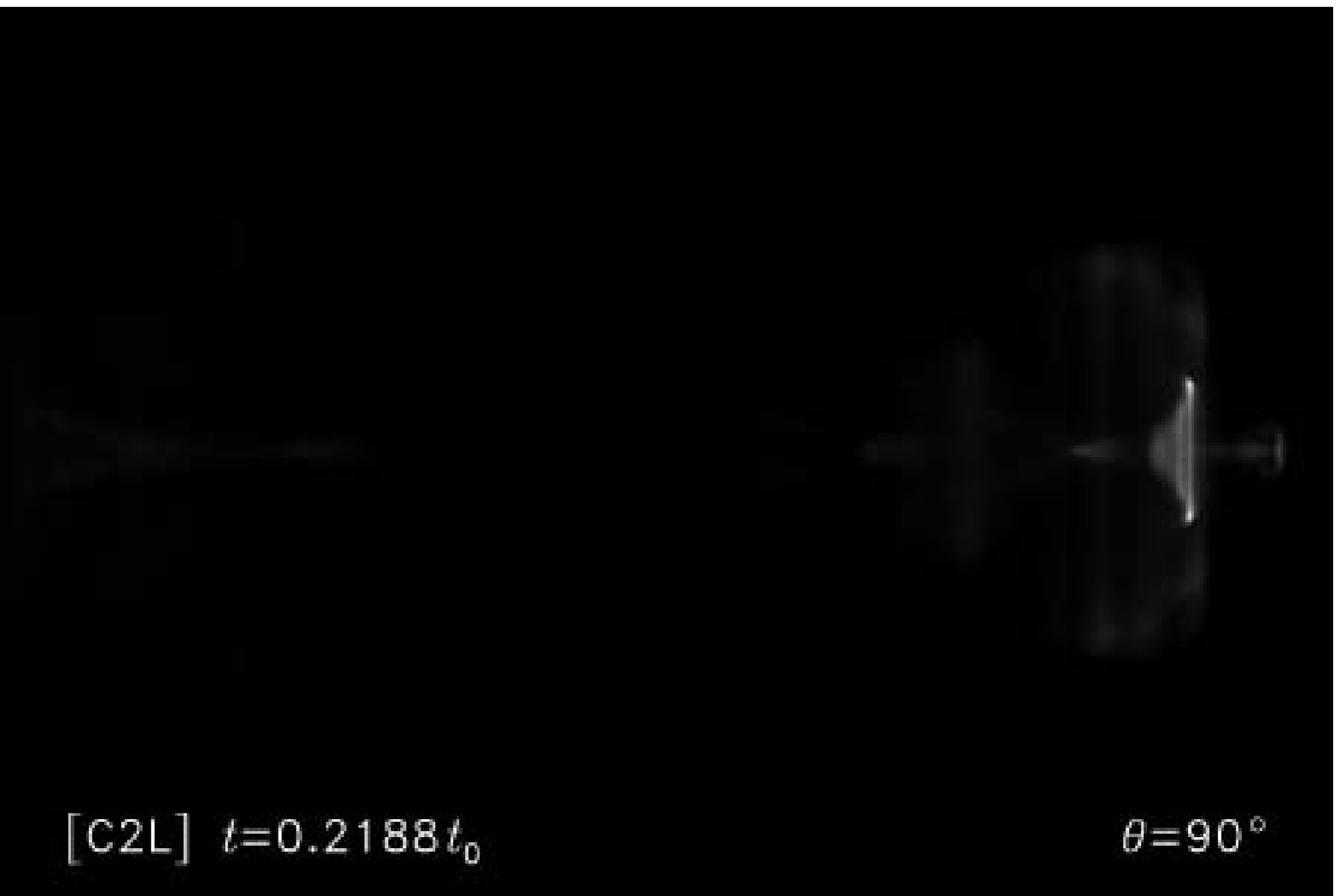}
\\
\end{array}$
\caption{
Morphological matches as in Figure~\ref{f:best_morphologies-4m5}
but with jet parameters
$(\eta,M)=(10^{-2},50)$.
}
\label{f:best_morphologies-2ml}
\end{figure}

\begin{figure}
\centering \leavevmode
$\begin{array}{ccc}
\includegraphics[width=4.5cm]{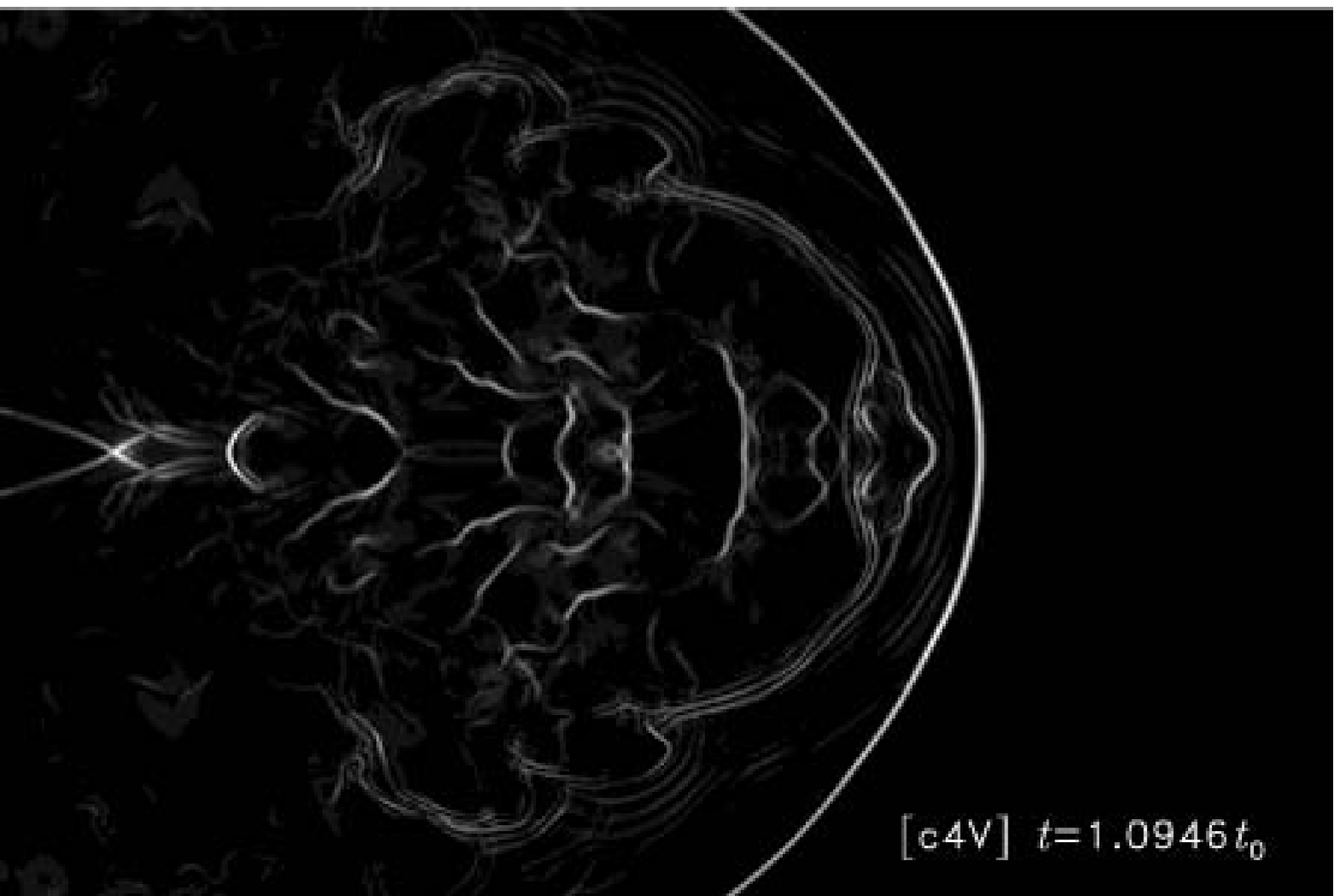}
&
\includegraphics[width=4.5cm]{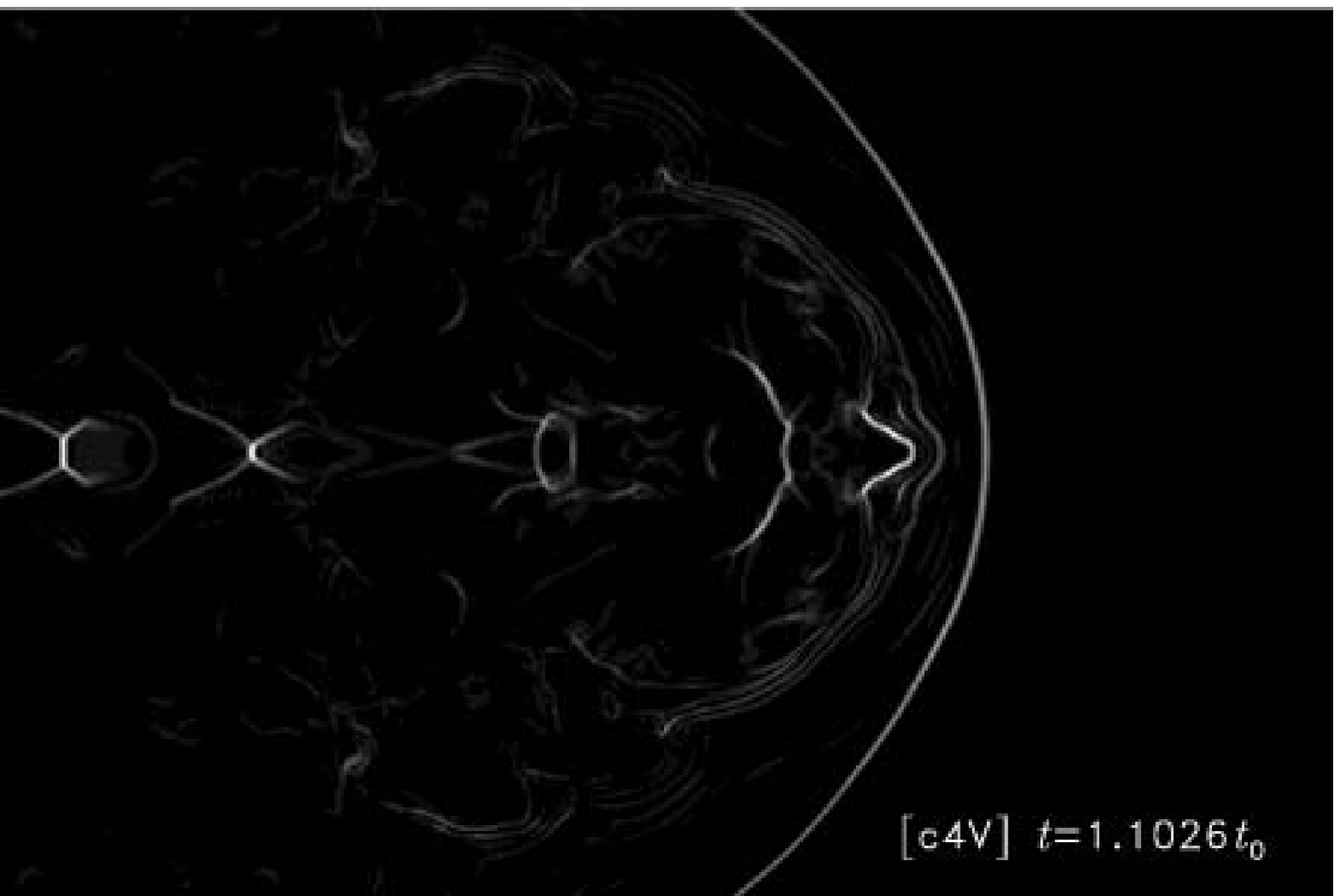}
&
\includegraphics[width=4.5cm]{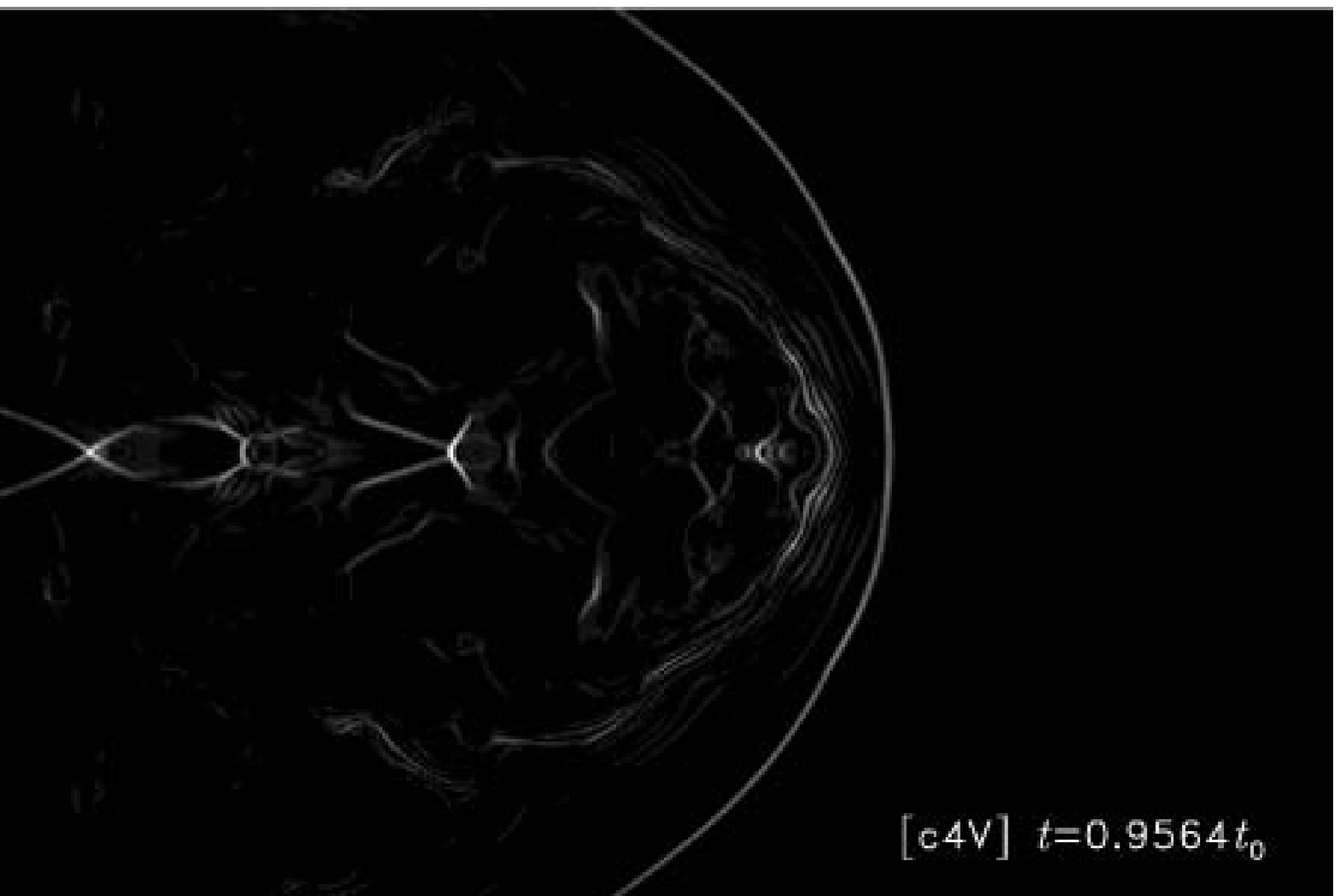}
\\
\includegraphics[width=4.5cm]{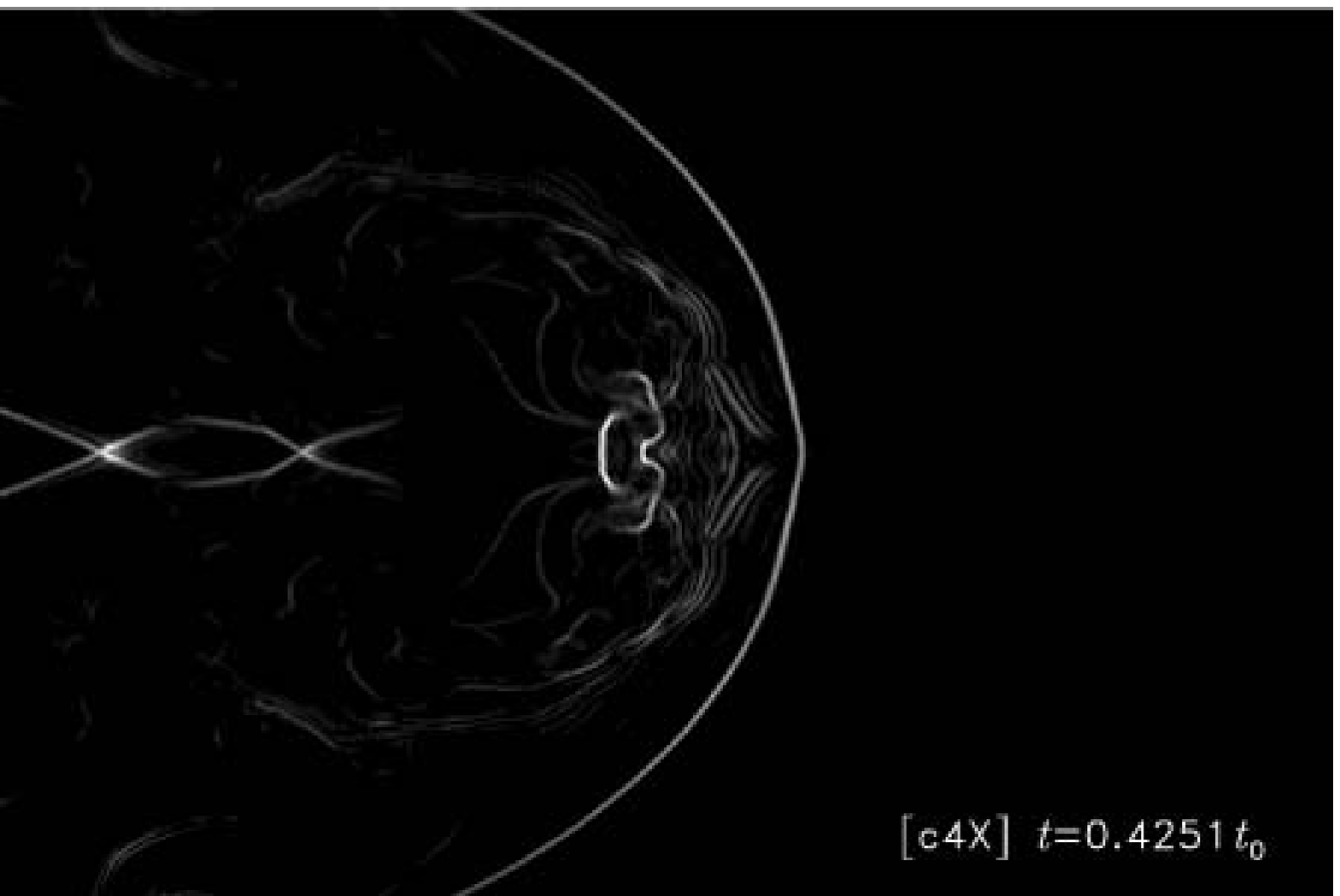}
&
\includegraphics[width=4.5cm]{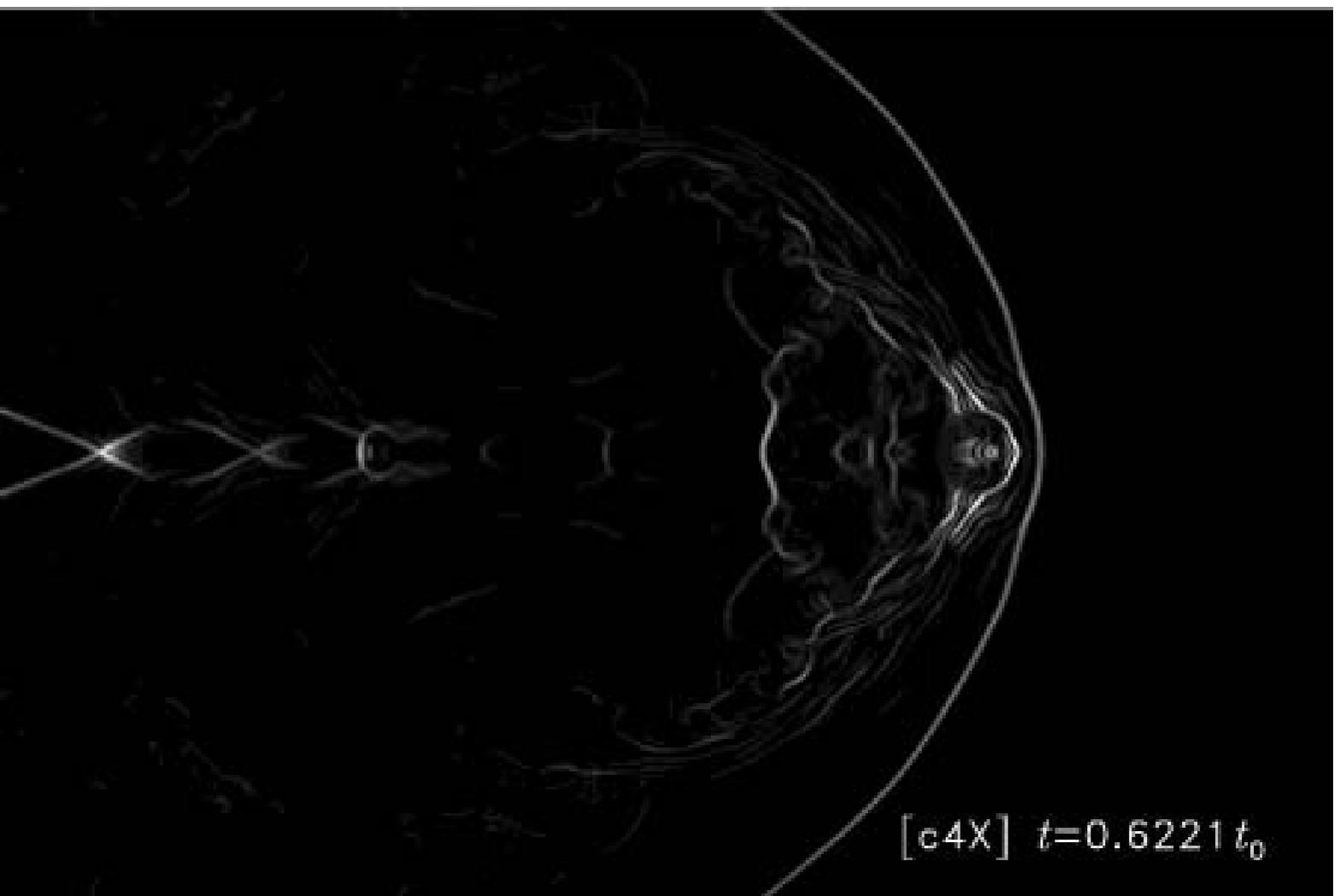}
&
\includegraphics[width=4.5cm]{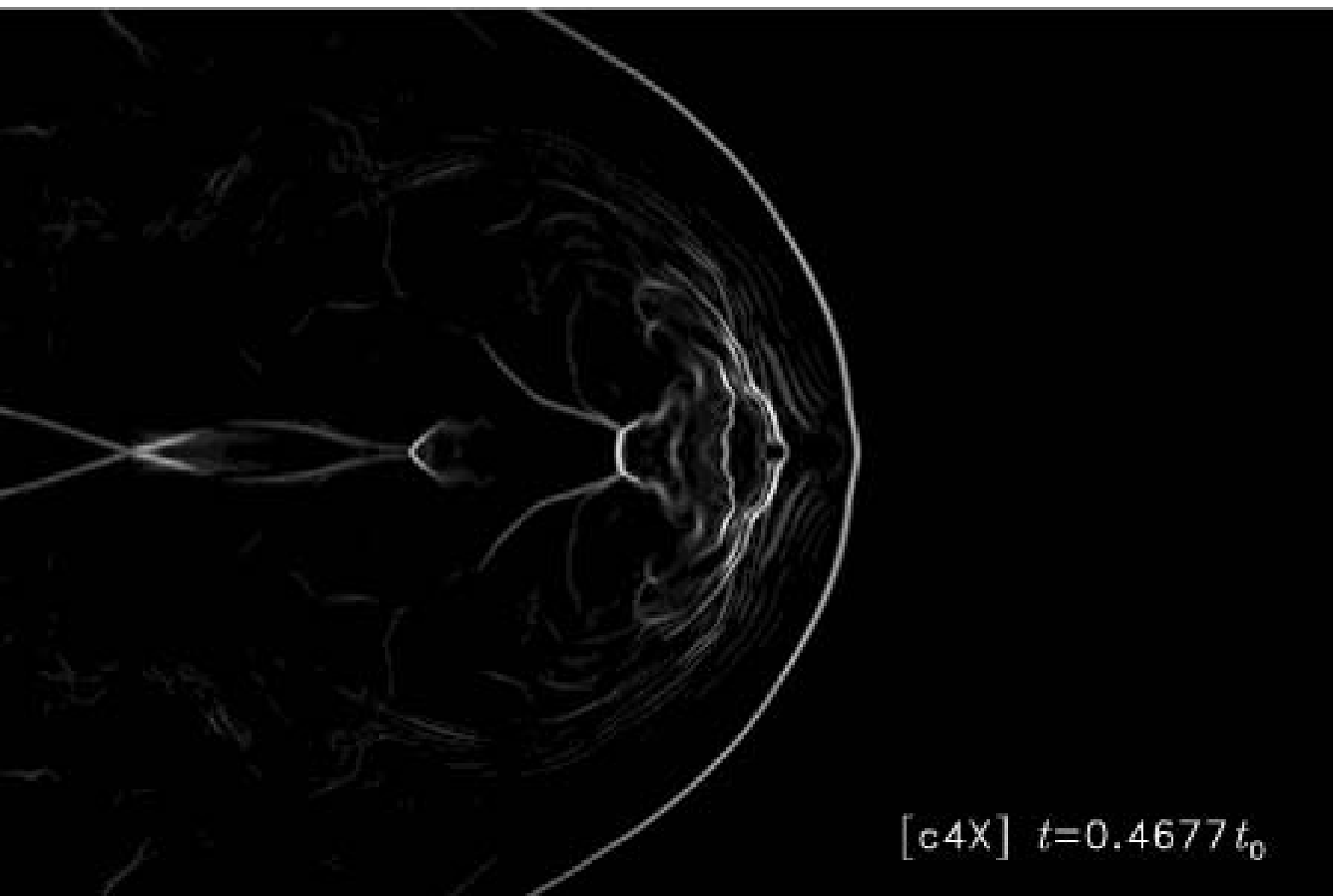}
\\
\includegraphics[width=4.5cm]{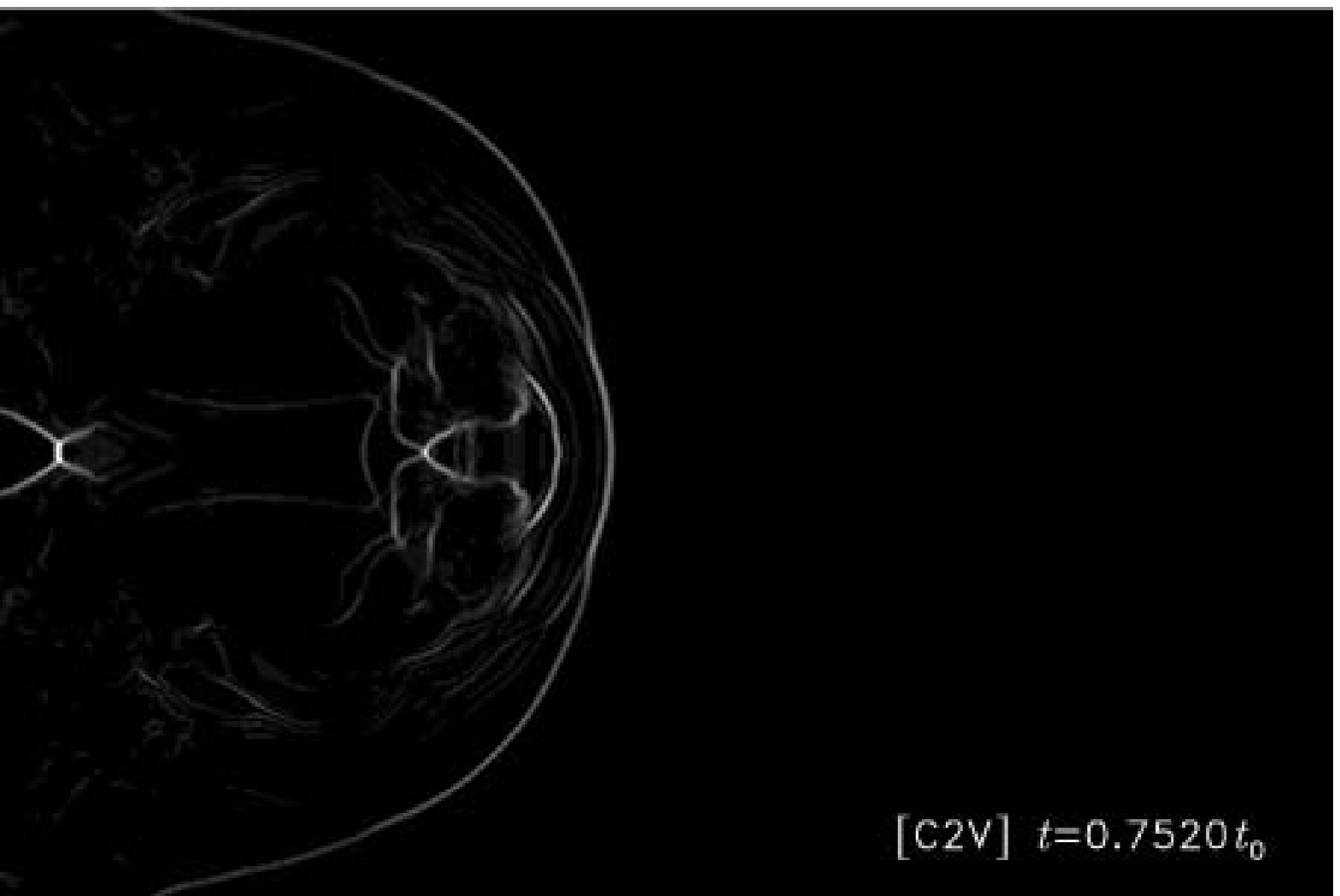}
&
\includegraphics[width=4.5cm]{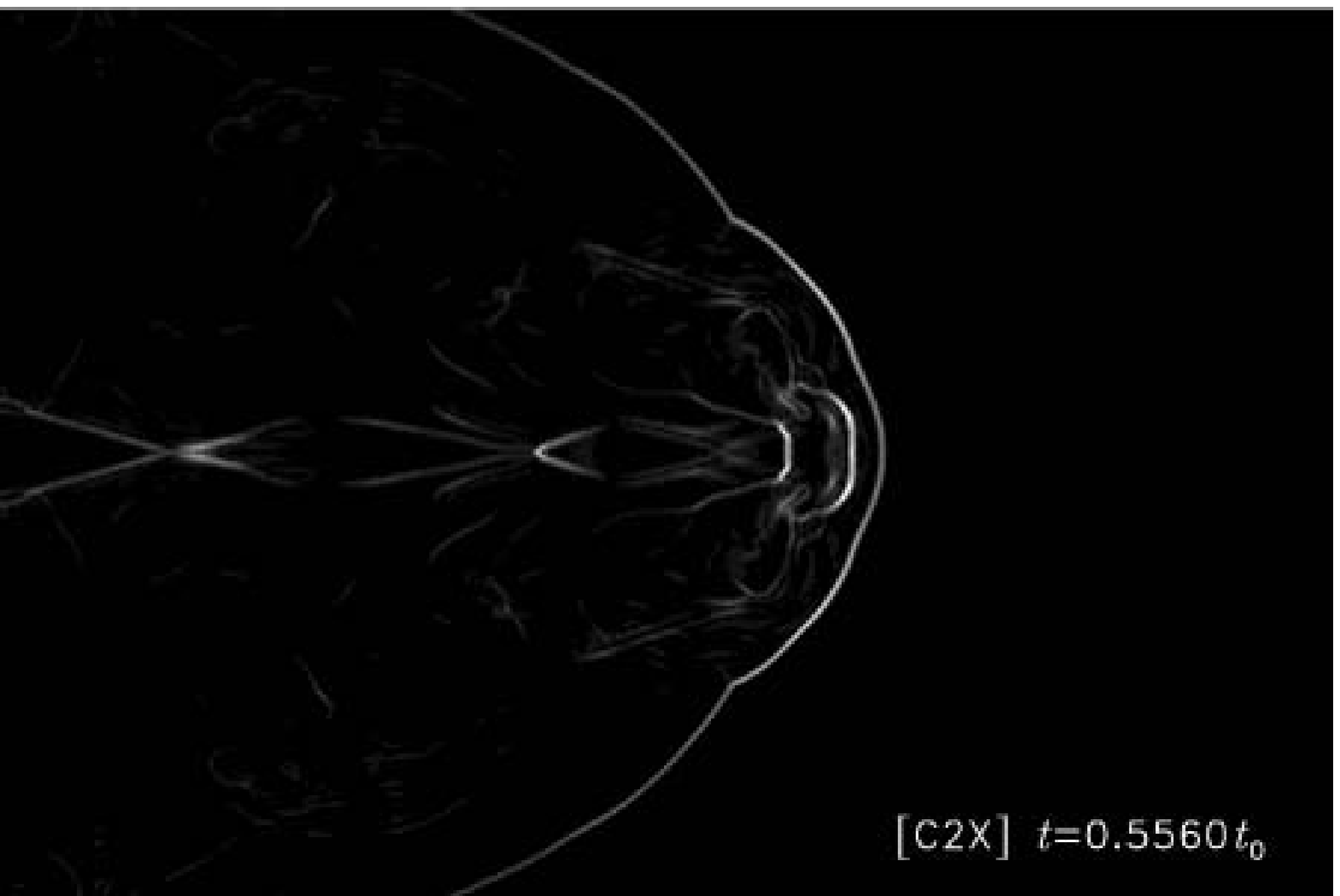}
&
\includegraphics[width=4.5cm]{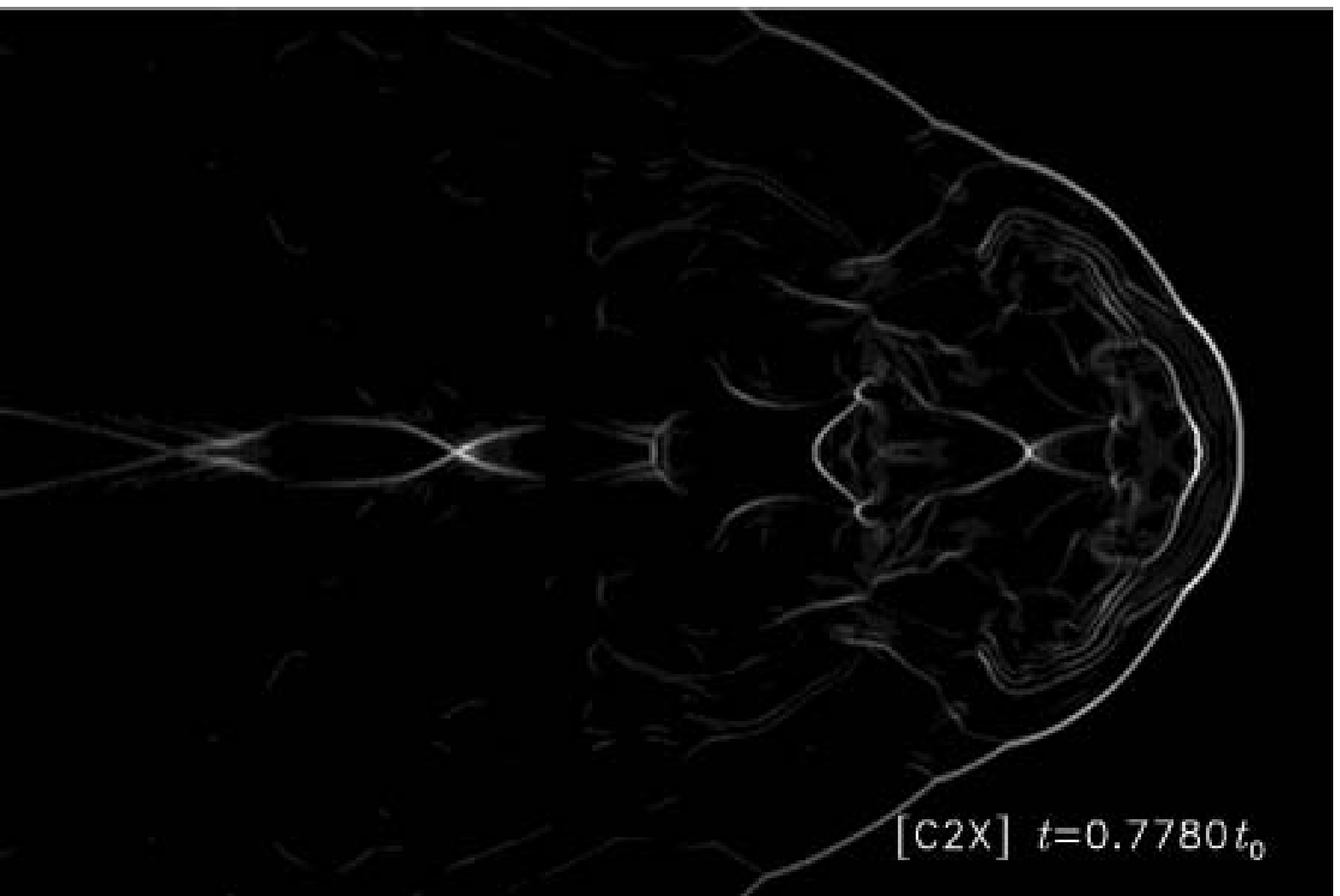}
\end{array}$
\caption{
Plots of $|\nabla p|$,
delineating the presence of shocks.
Panel dimensions are the same as in
Figures~\ref{f:best_morphologies-4m5}-\ref{f:best_morphologies-2ml}.
}
\label{f:grad.p}
\end{figure}

At a given inclination,
good morphologies appear with roughly equal abundance
for all choices of the jet parameters,
with the exception of the extreme case $(\eta,M)=(10^{-4},50)$
where the turbulent backflow and the jet behave abnormally
(see \S\ref{s.spluttering}).
For example renderings,
see Figures~\ref{f:best_morphologies-4m5}-\ref{f:best_morphologies-2ml},
and a selection of corresponding plots of $|\nabla p|$ in
Figure~\ref{f:grad.p},
which indentify the principal intense linear features with shocks.
We display more examples for cases where $\eta=10^{-4}$
but this is only because these simulations produced more frames of data,
and the emission features evolve more rapidly compared to
the jet's advance across the grid
(see \S\ref{s.temporal.variation} below).

As implied by the distribution of $\varphi$
and the commonplace occurrence of annular shocks
throughout the backflow,
the rendered snapshots include
bright bars of diverse widths,
(up to approximately $6r_{\rm j}$ in cases with $\eta = 10^{-4}$).
The observed filament
(Figures~\ref{f:radio_image},~\ref{f:optical_image})
has a radius four times that of the hot-spot.
If the observed hot-spot corresponds to the width of the jet,
then all of our simulations are in principle
able to produce bright shocks in the backflow
that are consistent in size with the observed filament.

\subsection{Brightness distribution}
\label{s:brightness_distribution}

Let us now consider the overall brightness distribution of various features
in the image.
This includes the brightness distribution of the filament,
its brightness relative to the hot-spot and
the overall brightness of the hot-spot and bar relative to the lobe.
Note that in the observations, the brightest part of the filament
is approximately $7\%$ of the peak brightness in the hot-spot,
in the radio and about $2\%$ in the optical. 

Some frames rendered from the simulations have shapes
that are consistent with the observations of Pictor~A,
in terms of qualitative topology and approximate proportions,
but may differ from the observations in terms of brightness.
For instance, in some imperfect examples,
the bar is dim on the edges rather than being edge-bright,
\eg $(\eta,M,t,\theta)=(10^{-4},10,0.6216t_0,90^\circ)$
(in centre panel, upper row of Figure~\ref{f:best_morphologies-4mx},
and
right column, Figure~\ref{f:pageant-4mx.1})
and
$(\eta,M,t,\theta)=(10^{-4},5,0.6221t_0,90^\circ)$
(centre panel, lower row of Figure~\ref{f:best_morphologies-4mx}).
In these cases,
the bar is the image of a disk-like (rather than annular) shock
that cuts the jet and a high-$\varphi$ transverse stream
diverging from the jet.

In some cases in which the morphology is approximately correct,
the ratio of filament to hot-spot brightness is larger than observed:
\eg $(\eta,M,t,\theta)=(10^{-2},10,0.296t_0,90^\circ)$
shown in the central panel of Figure~\ref{f:best_morphologies-2mx};
and
$(\eta,M,t,\theta)=(10^{-2},10,0.8212t_0,90^\circ)$
(right column, Figure~\ref{f:example.08}),
in which the wide bar is a shock cutting through
both the jet and cocoon.
However, we cannot rule out cases with overly bright bars
since relativistic beaming may be important in the hot-spot. There are also examples in
which the ratio of filament to hot-spot brightness is approximately correct, e.g. the
three examples in the uper row of Figure~\ref{f:best_morphologies-2ml}. Here the
brightness ratios are approximately $12\%$, $8\%$ and $12\%$
(ordered from left to right).
The appropriate ratios are preferentially found
in the highest Mach number cases.
A filament at the $7\%$ level
is generally more difficult to find in a low Mach number
lobe because of the lack of contrast with the background cocoon.
The physics of
the contrast between hot-spot and cocoon is discussed further below.

The observed filament is extended more to the south of the inferred jet
than the north.
Because of the axisymmetry of the simulations,
this feature is not replicated.
However, an axisymmetric approximation is justified
in view of the very straight jet evident in the X-ray image
\citep{wilson01a}.
Nevertheless, in most of the matches that we have presented
in Figures~\ref{f:best_morphologies-4m5}-\ref{f:best_morphologies-2ml},
the simulated filament is seen to be brighter towards the ends.
This is the result of the
annular shocks arising in the region of the hot-spot.

A feature of the observations that is replicated
in approximately $40\%$ of the matches in
Figures~\ref{f:best_morphologies-4m5}-\ref{f:best_morphologies-2ml}
is a small ``knot'' of emission at the centre of the bar.
Such a feature is often the result of the superposition 
the bright annulus with a section of the jet that is also bright,
such as a shock upstream of and related to the terminal shock
of the hot-spot.

Yet another observational feature that appears in almost every panel in
Figures~\ref{f:best_morphologies-4m5}-\ref{f:best_morphologies-2ml}
is the radio knot south-east of the bar
along the inferred direction of the jet.
These knots correspond to diamond shocks in the jet;
their exact location depends stochastically upon
the details of the turbulence in the cocoon.

In the radio observations (Figure~\ref{f:radio_image}),
diffuse emission from the radio lobe in regions near the bar and hot-spot
is $\la 0.5\%$ of the maximum surface brightness of the hot-spot.
As Perley et~al. noted, such a large contrast is highly unusual
for a radio galaxy.
Ideally, our ray-traced images should produce a similar contrast
between the brightness of bar and hot-spot
and the wispy rings and dim emission pervading the rest of the cocoon.

Ideally, higher Mach number jets produce hot-spots with a higher contrast to
the lobe. In simple terms, if the jet terminates in a single steady shock,
the pressure contrast between the hot-spot
and the cocoon is given by
$p_{\rm hs}/p_{\rm cocoon}\sim M^2$.
This aspect of jet physics lead Perley et~al.
to suggest that the Mach number may be as high as $40$.
In terms of a realtivistic jet,
this is equivalent to a Lorentz factor of 24.
However, the inferred parameters of the hot-spot
(especially the pressure)
could be significantly affected by beaming.
The beaming pattern may be quite complex
in view of the velocity field produced by the terminal oblique shock
(see \S\ref{s.activity}).

\begin{figure}
\centering \leavevmode
$\begin{array}{lcc}
&\eta=10^{-2}&\eta=10^{-4}
\\
M=5&
\begin{array}{c}\includegraphics[width=5cm]{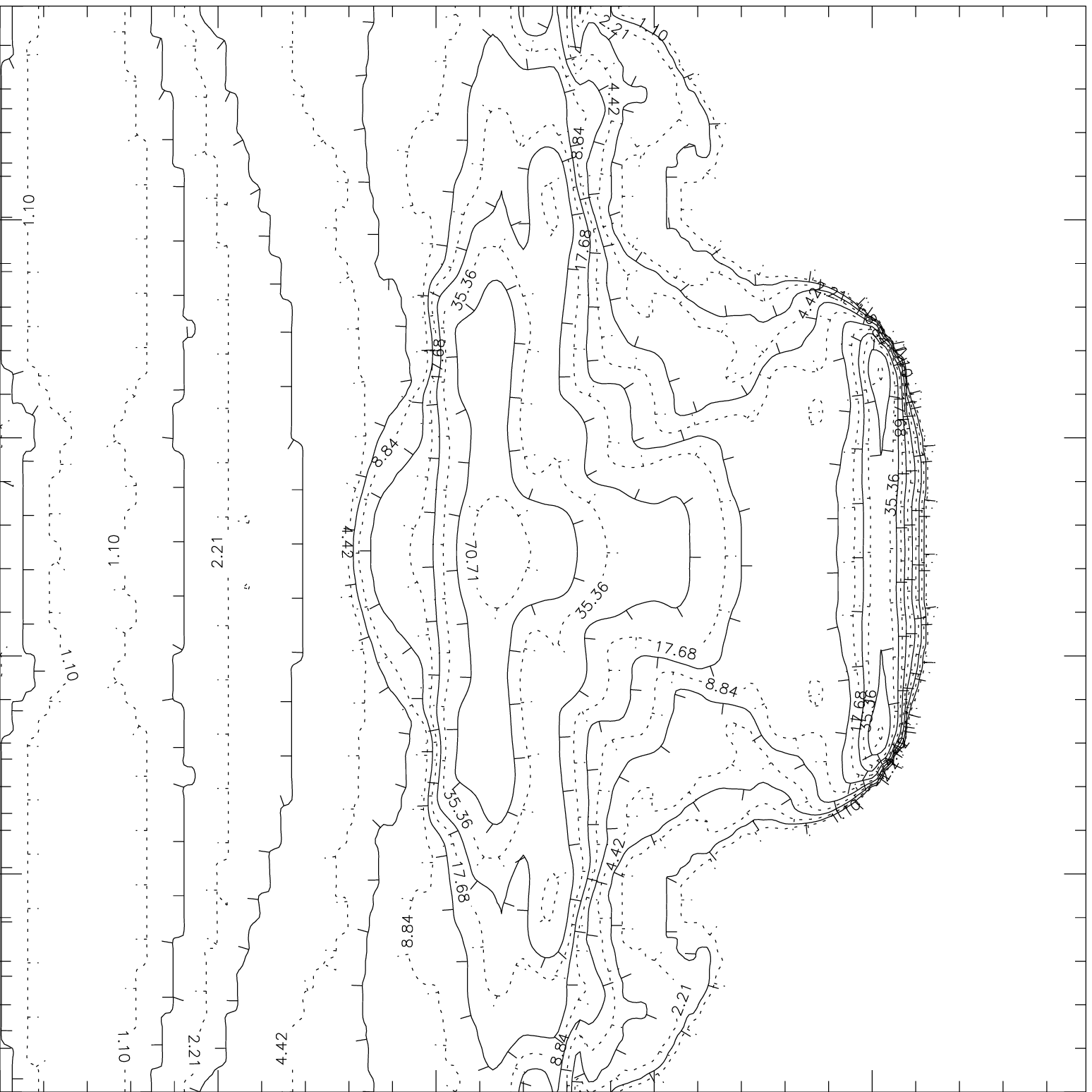}\end{array}
&
\begin{array}{c}\includegraphics[width=5cm]{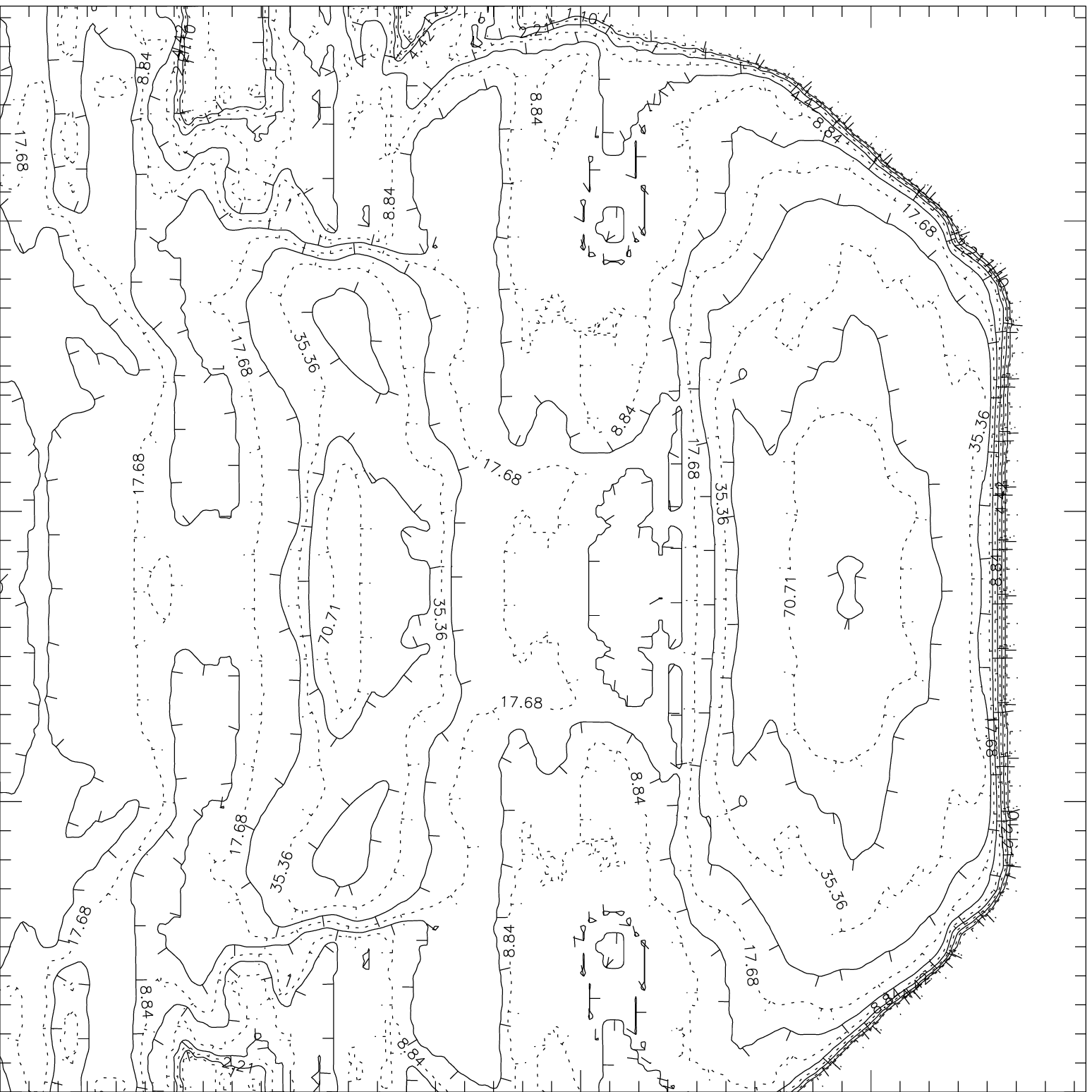}\end{array}
\\
M=10&
\begin{array}{c}\includegraphics[width=5cm]{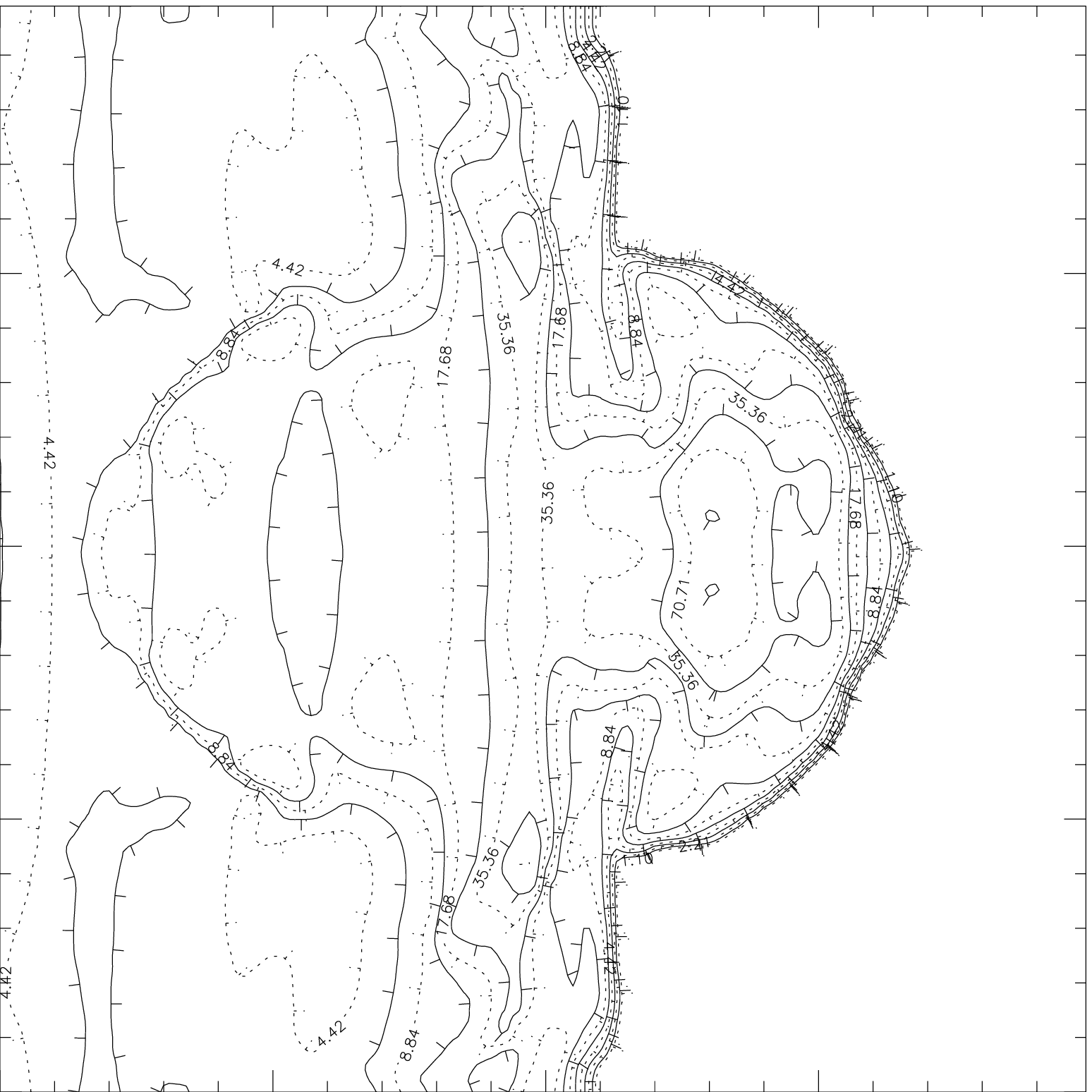}\end{array}
&
\begin{array}{c}\includegraphics[width=5cm]{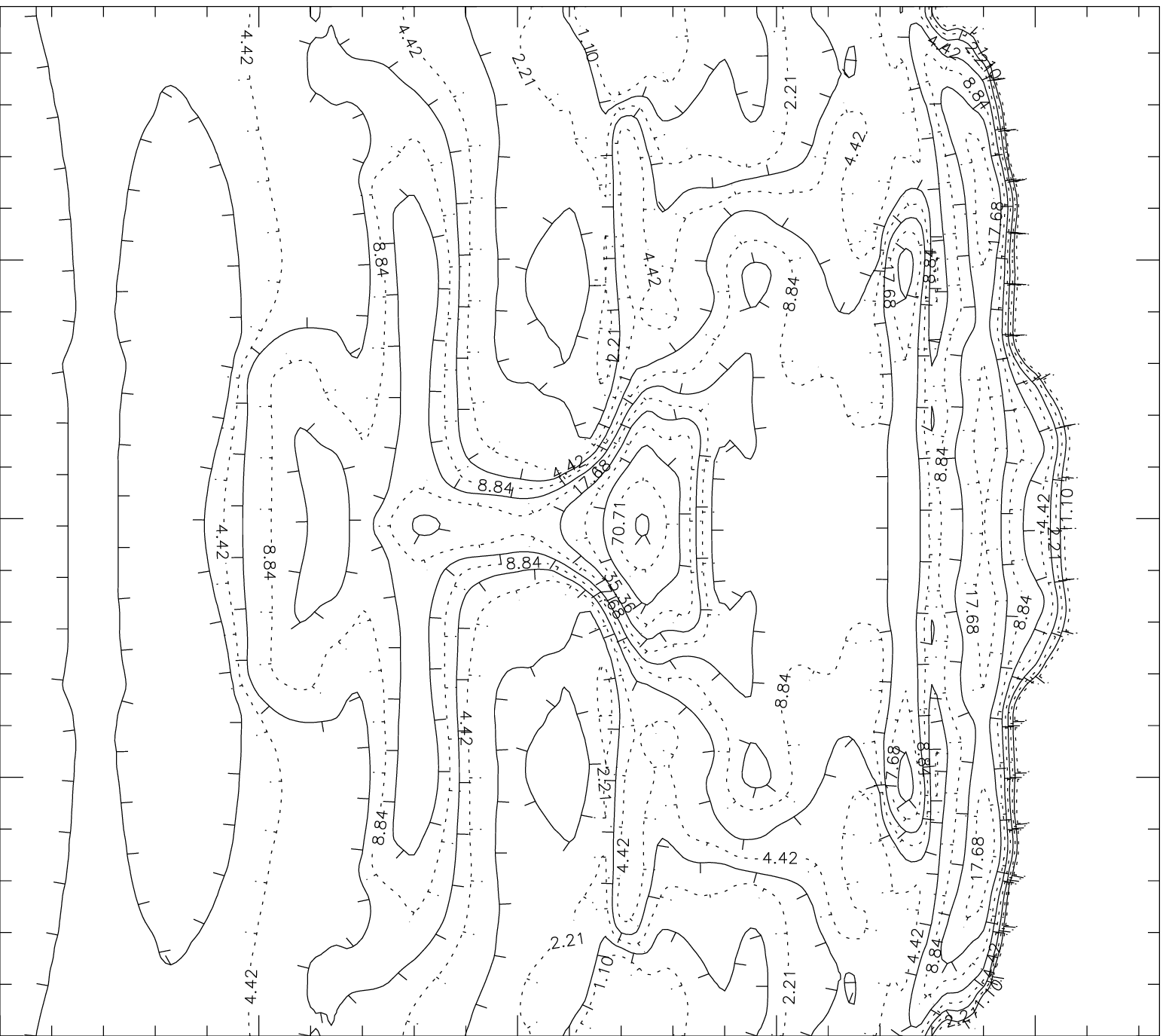}\end{array}
\\
M=50&
\begin{array}{c}\includegraphics[width=5cm]{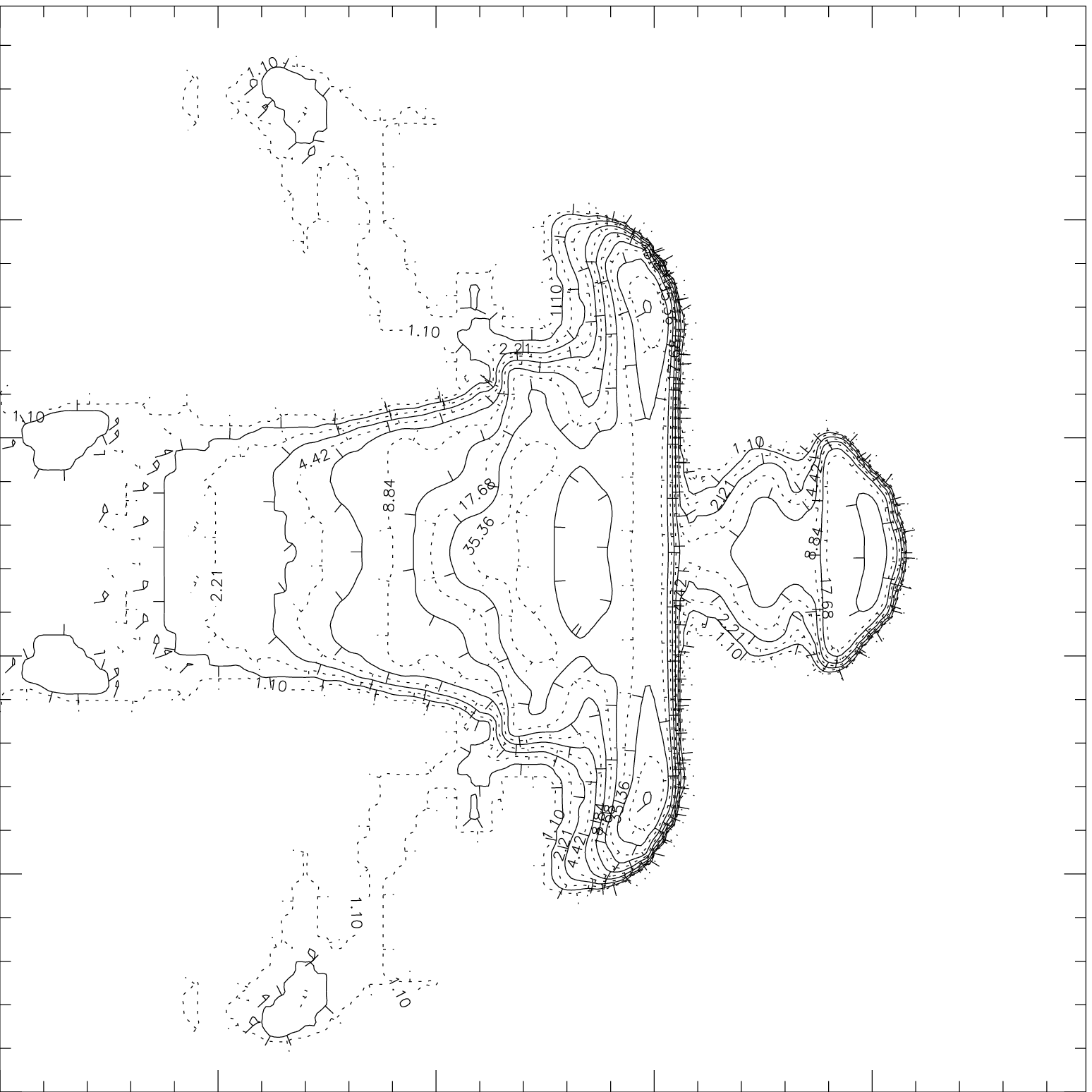}\end{array}
&
\begin{array}{c}\includegraphics[width=5cm]{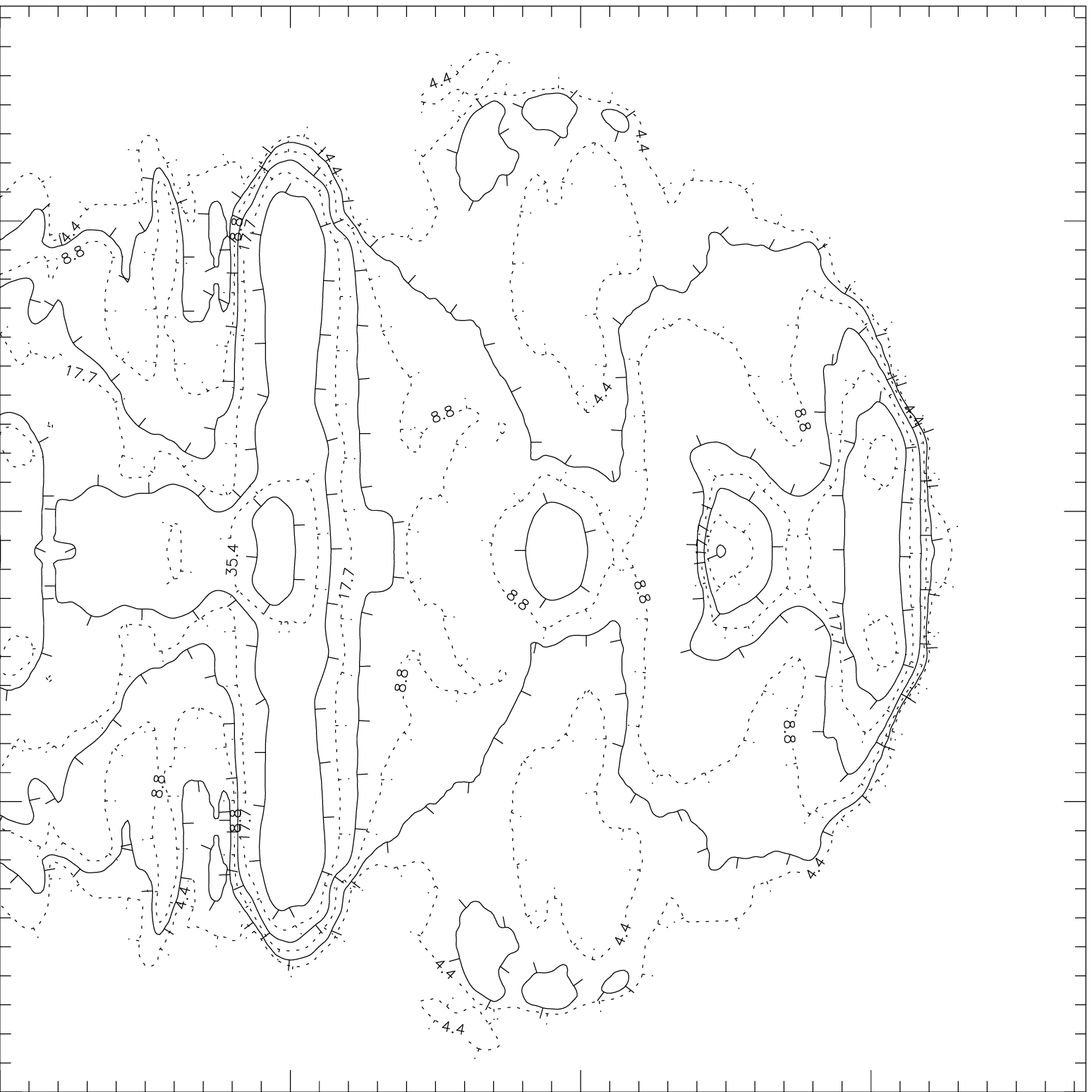}\end{array}
\\
\end{array}$
\caption{
Ray-traced surface brightness contour maps of the hot-spot vicinity,
selected from the simulations with a closed left boundary.
Contour levels are at consecutive multiples of
$1/\sqrt{2}$ times the peak value.
Each column shows a different case of $\eta$;
each row corresponds to a different $M$.
For the sake of clarity, contour lines are alternately dotted and solid,
and show normal marks directed along the ``downhill'' gradient.
}
\label{f:contours}
\end{figure}

Ray-traced images from the simulations with parameters
$(\eta,M)=(10^{-4},5)$
(see Figure~\ref{f:best_morphologies-4m5})
give bar and hot-spot features that are typically
only $\sim 10$ times brighter than the background cocoon.
The brightness contrast is greater
in simulations with $(\eta,M)=(10^{-4},10)$
(see Figure~\ref{f:best_morphologies-4mx}).
In this case the contrast varies
bewteen $10^{1.3}$ and $10^{1.7}$. Simulations of very fast jets,
(\eg case {\tt C2L}, with parameters $(\eta,M)=(10^{-2},50)$)
can yield brightness contrasts in the range, i.e $\sim 10^{2-2.5}$ -- 
consistent
with the radio images of Perley et al.
The tendency towards increasing contrast in the hot-spot and bar
can be seen in the example contour maps of surface brightness
for $\eta=10^{-2}$ and for the lower-$M$ cases of $\eta=10^{-4}$
in Figure~\ref{f:contours}.

\label{s.spluttering}
However this trend breaks
in the case {\tt C4L} where $(\eta,M)=(10^{-4},50)$;
the brightness contrast is exceptionally low
-- typically $\la 30$ --
compared to the more sharply defined features
in the $(\eta,M)=(10^{-2},50)$ and $(\eta,M)=(10^{-3},50)$ simulations.
This is an extreme manifestation of the surging behaviour
that we have already noted.
In this case the backflow turbulence is particularly vigorous
and remains so at locations well behind the front of the bow shock.
Strong eddies pinch off
the extremely light jet at numerous locations along its length,
and the jet's penetration into the cocoon
varies dramatically within a dynamical timescale.
Consequently, jet energy is deposited widely throughout the cocoon,
rather than being concentrated at a quasi-stable head;
moderately bright but transient rings and hot-spots
occur throughout a large volume. This aspect may well be a feature of the
restriction to axisymmetry causing the two-dimensional turbulence within the
cocoon to be focussed on the jet.

\subsection{Temporal variation}
\label{s.temporal.variation}

Physical configurations producing
the desired morphology of surface brightness
are ephemeral products of the
jet's pulsating head and turbulent motions within the cocoon.
It is therefore interesting to consider
how these transient features vary in time. 
These considerations are also
motivated by the potential of using hot-spot brightness ratios as a
measure of relativistic beaming. 
Although these calculations are not
relativistic we {\em can} examine the intrinsic distribution of hot-spot
intensities. 
If it were to be the case, for example, 
that order of magnitude fluctuations in the ratio of hot-spot luminosities
were to be likely, then
the use of luminosity ratios as a measure of relativistic speeds 
within hot-spots would be limited. 
We address this point below following consideration
of the structure function of individual hot-spots.

\begin{figure}[h]
\centering \leavevmode
{
$\begin{array}{cc}
\mbox{closed boundary}&\mbox{open boundary}
\\
\begin{array}{c}\includegraphics[width=6cm]{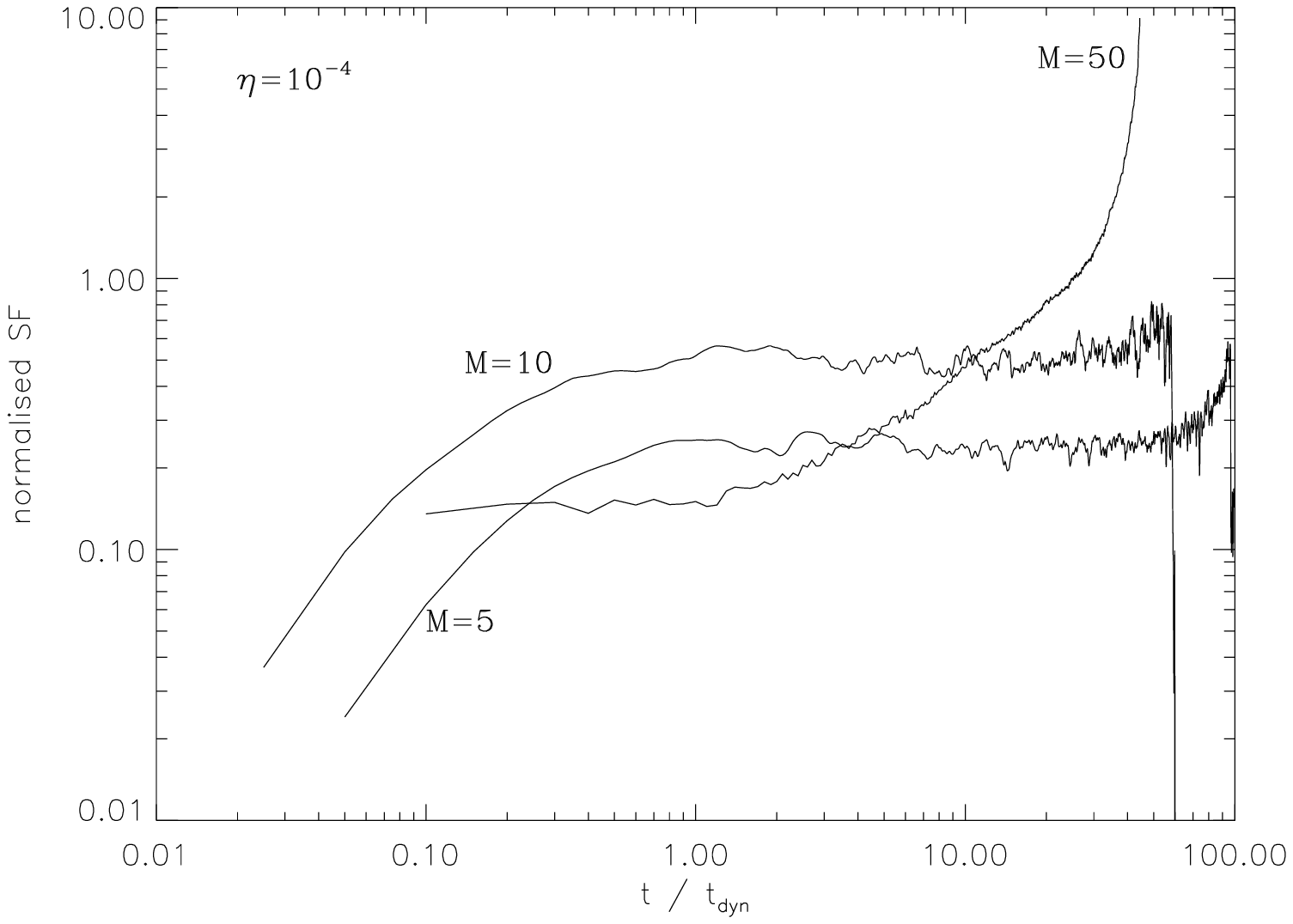}\end{array}
&
\begin{array}{c}\includegraphics[width=6cm]{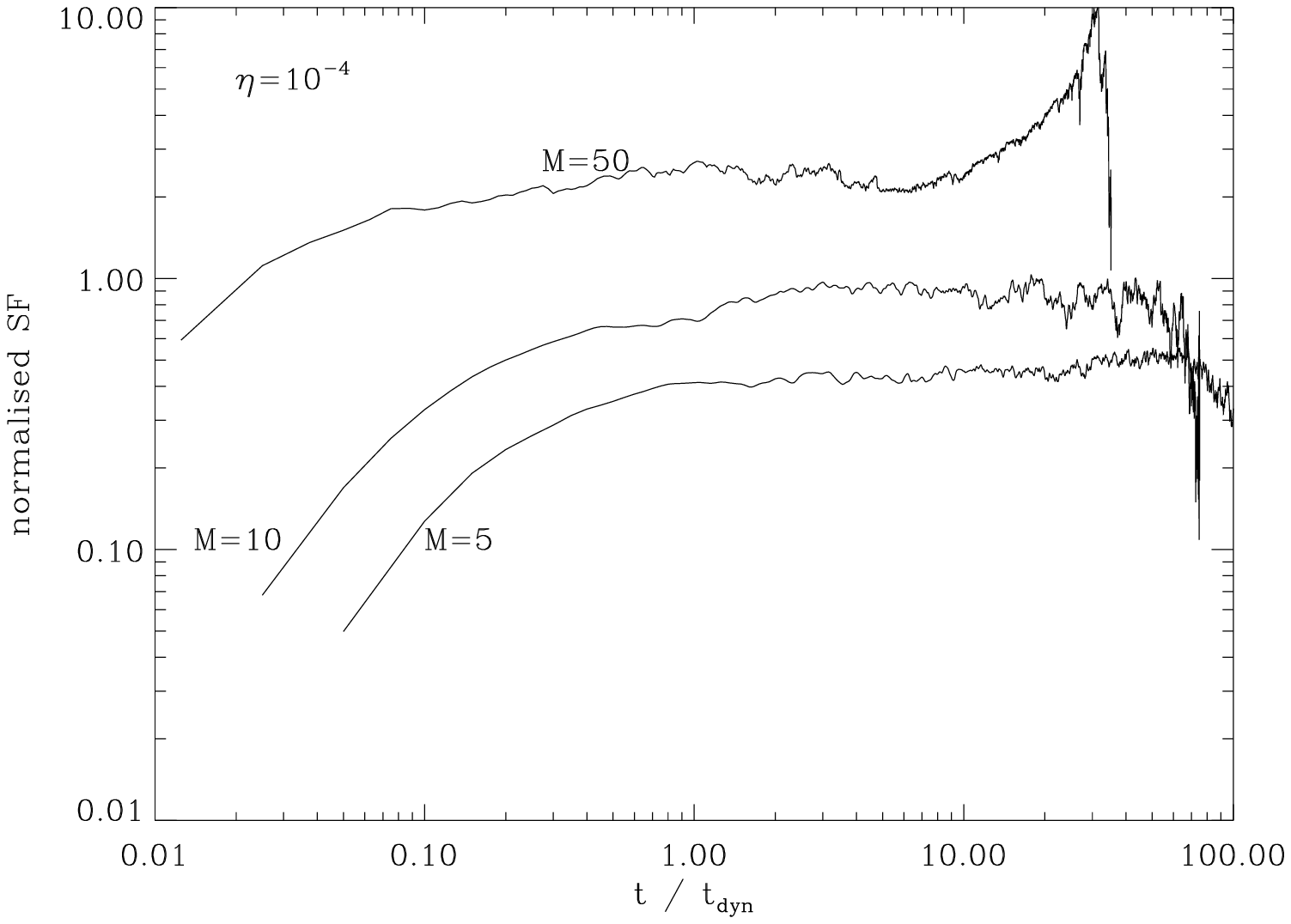}\end{array}
\\
\begin{array}{c}\includegraphics[width=6cm]{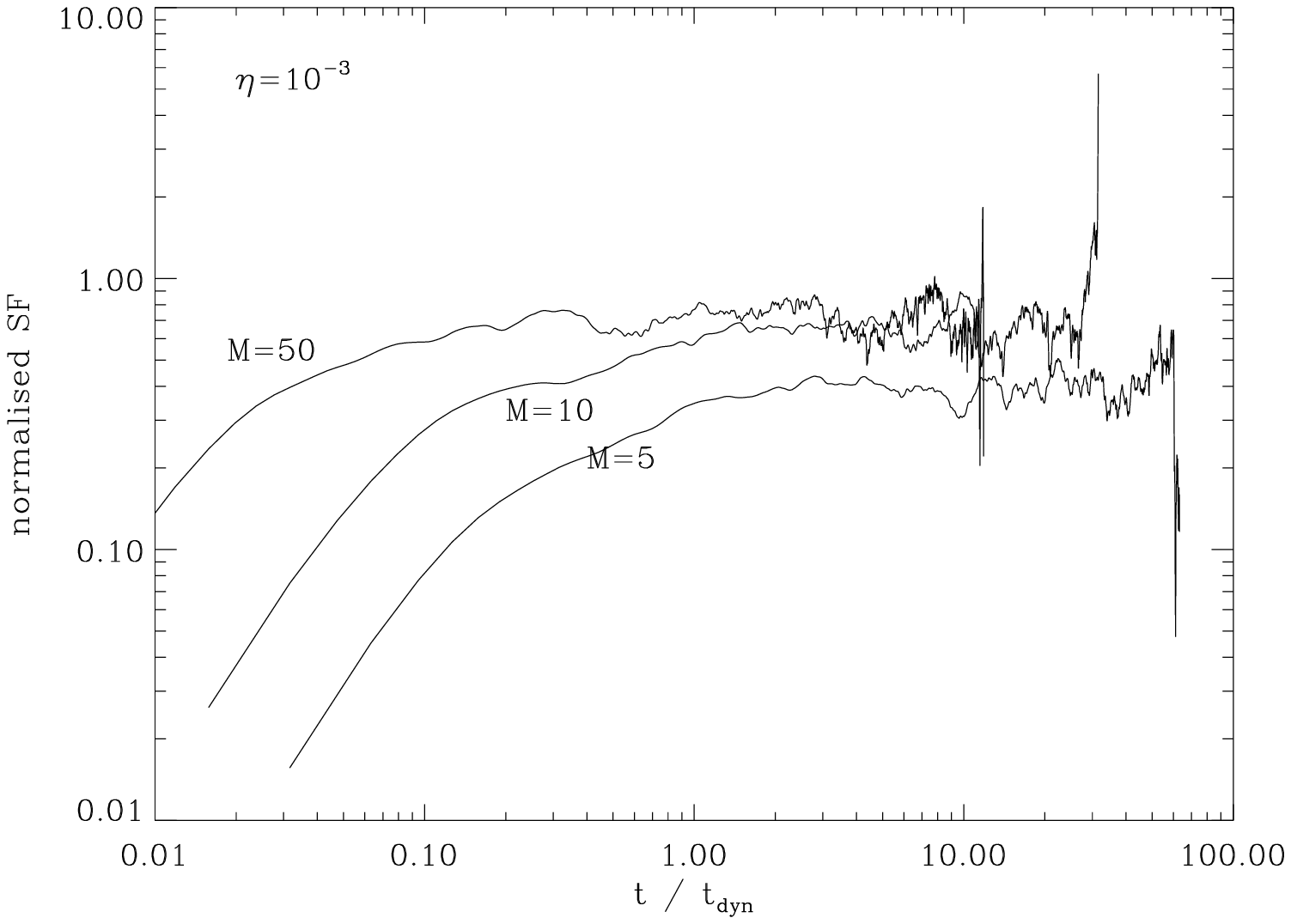}\end{array}
&
\begin{array}{c}\includegraphics[width=6cm]{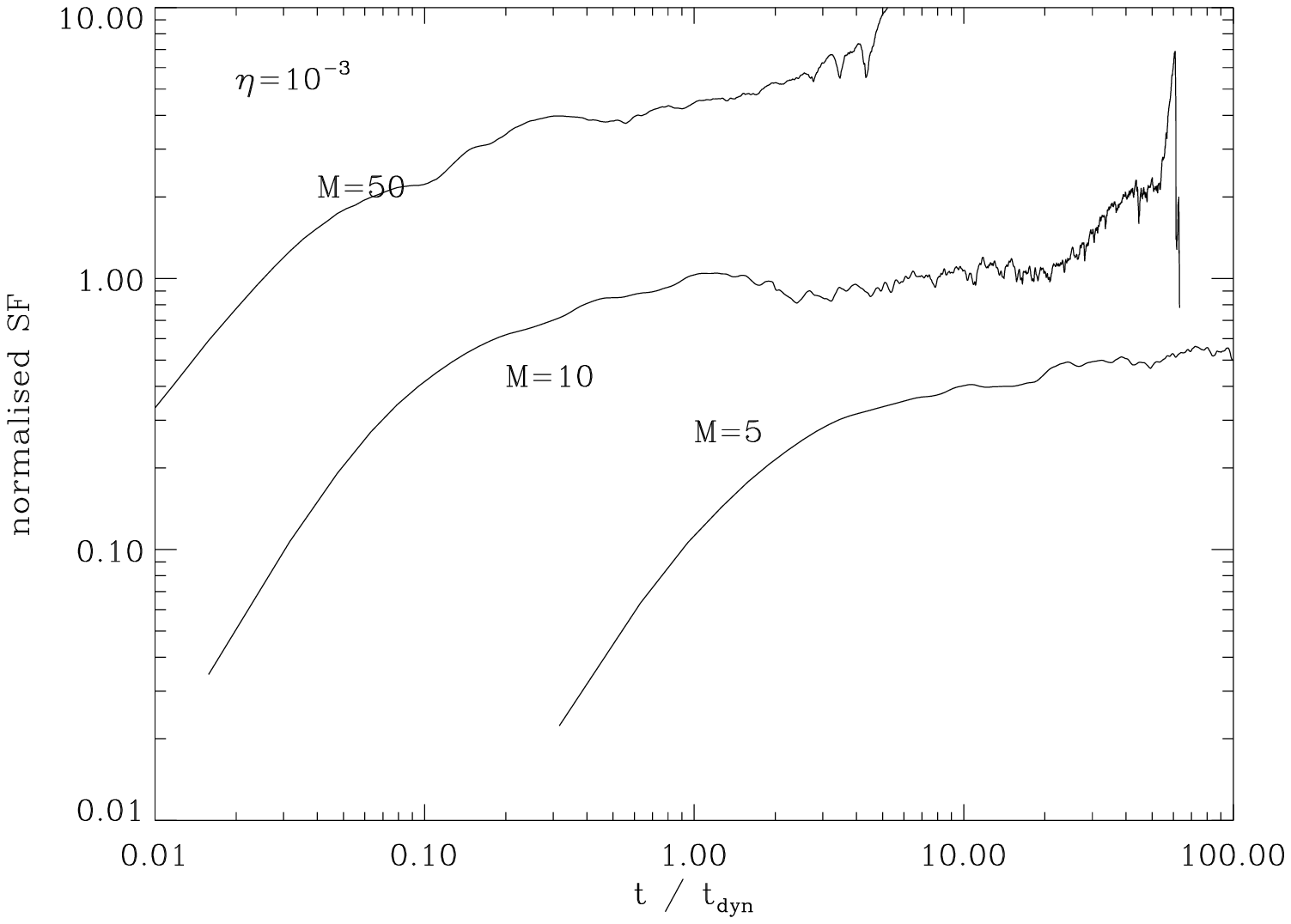}\end{array}
\\
\begin{array}{c}\includegraphics[width=6cm]{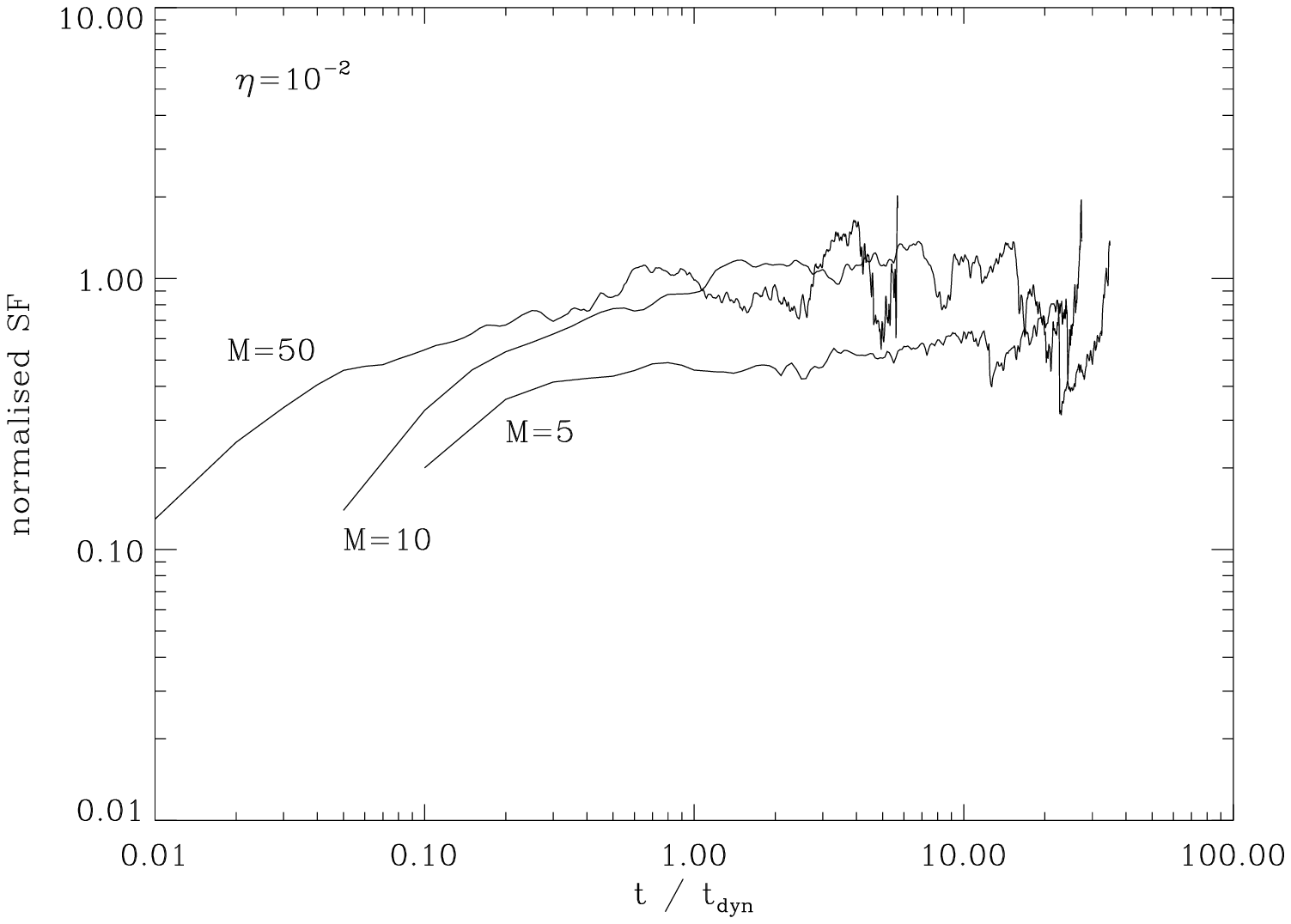}\end{array}
&
\begin{array}{c}\includegraphics[width=6cm]{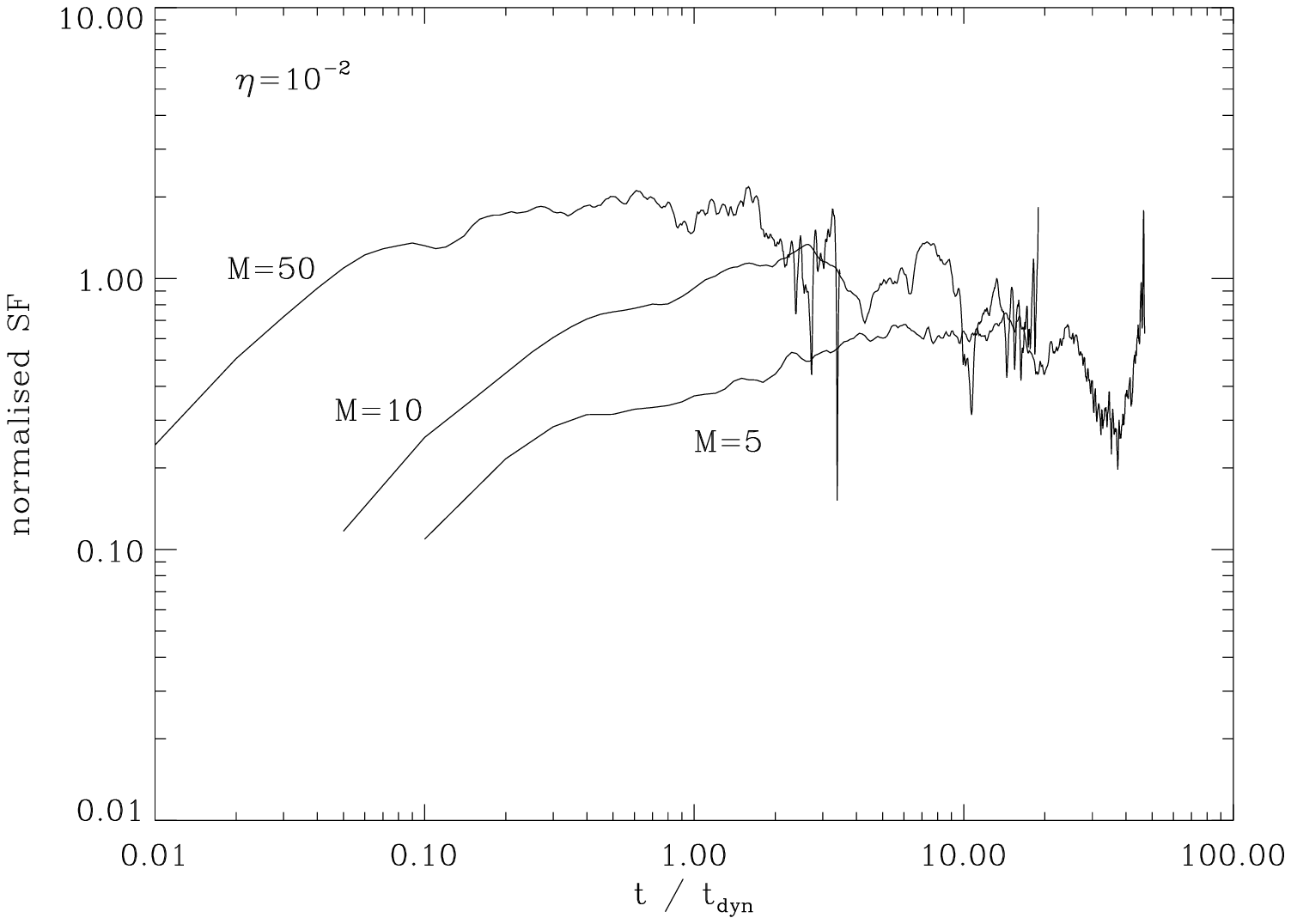}\end{array}
\\
\end{array}$
}
\caption{
Normalised structure functions evaluated for flickering of the peak
intensity of the emission regions.
The left and right columns show
different cases of the left boundary condition.
For comparison of relative variability,
the raw time series were normalised to the respective mean intensities.
We excluded
data points from during
the early establishment of the cocoon
and the eventual exit of the cocoon and jet through the right boundary.
}
\label{f:structure.function}
\end{figure}

The dynamical timescale of structures in the jet material,
such as the hot-spot and shocks in the jet and backflow,
characteristically depends upon the jet's diameter and internal sound speed,
\begin{equation}
t_{\rm dyn} \equiv {{2r_{\rm j}}\over{c_{\rm s,j}}}
\approx0.39\sqrt{\eta} t_0
\ .
\end{equation}
Thus $t_{\rm dyn}\approx0.04t_0$ for jets with $\eta=10^{-2}$,
and $t_{\rm dyn}\approx0.004t_0$ for jets with $\eta=10^{-4}$.
Since the brightest feature is usually a terminal hot-spot
or some other similarly compact structure,
we expect the instantaneous peak surface brightness
to vary in time on a time scale on the order of magnitude of $t_{\rm dyn}$.
Indeed this is the case,
as shown by the structure functions \citep{simonetti85a}
of the time-variation of the peak brightness
of images rendered at $\theta=90^\circ$
(see Figure~\ref{f:structure.function}. The structure functions were
calculated from a time series based upon the peak intensity in each frame
divided by the long term mean. (Frames containing initial transients
were omitted.) Hence the amplitude of the structure function gives one an
estimate of the square of the relative variabiliity on the relevant
timescale.

On time scales longer than $\sim t_{\rm dyn}$
the behaviour of most structure functions is effectively white noise
(power index $\approx 0$);
whereas the behaviour is like ``red noise'' on shorter time-scales.
The transition timescale between these behaviours is of order $0.1-0.5 \>
t_{\rm dyn}$ and depends little on $M$.
However jets with greater $M$ generally show
greater relative variability.

The one exceptional case is the $(\eta,M)=(10^{-4},50)$ simulation with the
closed boundary. 
In this simulation there is long term evolution 
(evident in the upturned curve at long time-scales in
the upper left panel of
Figure~\ref{f:structure.function})
related to the highly turbulent cocoon 
upon which we have already commented in
\S\ref{s.spluttering}.

In all cases bar the exceptional case, the left boundary condition has
negligible effect on  the variation of peak surface brightness
(compare left and right columns of Figure~\ref{f:structure.function}).
This is the result of
the proximity of the brightest structures to the terminal
jet shock, rather than parts of the flow
directly affected by the left boundary.

\begin{figure}[h]
\centering \leavevmode
{
$\begin{array}{cc}
\mbox{closed boundary}&\mbox{open boundary}
\\
\begin{array}{c}\includegraphics[width=6cm]{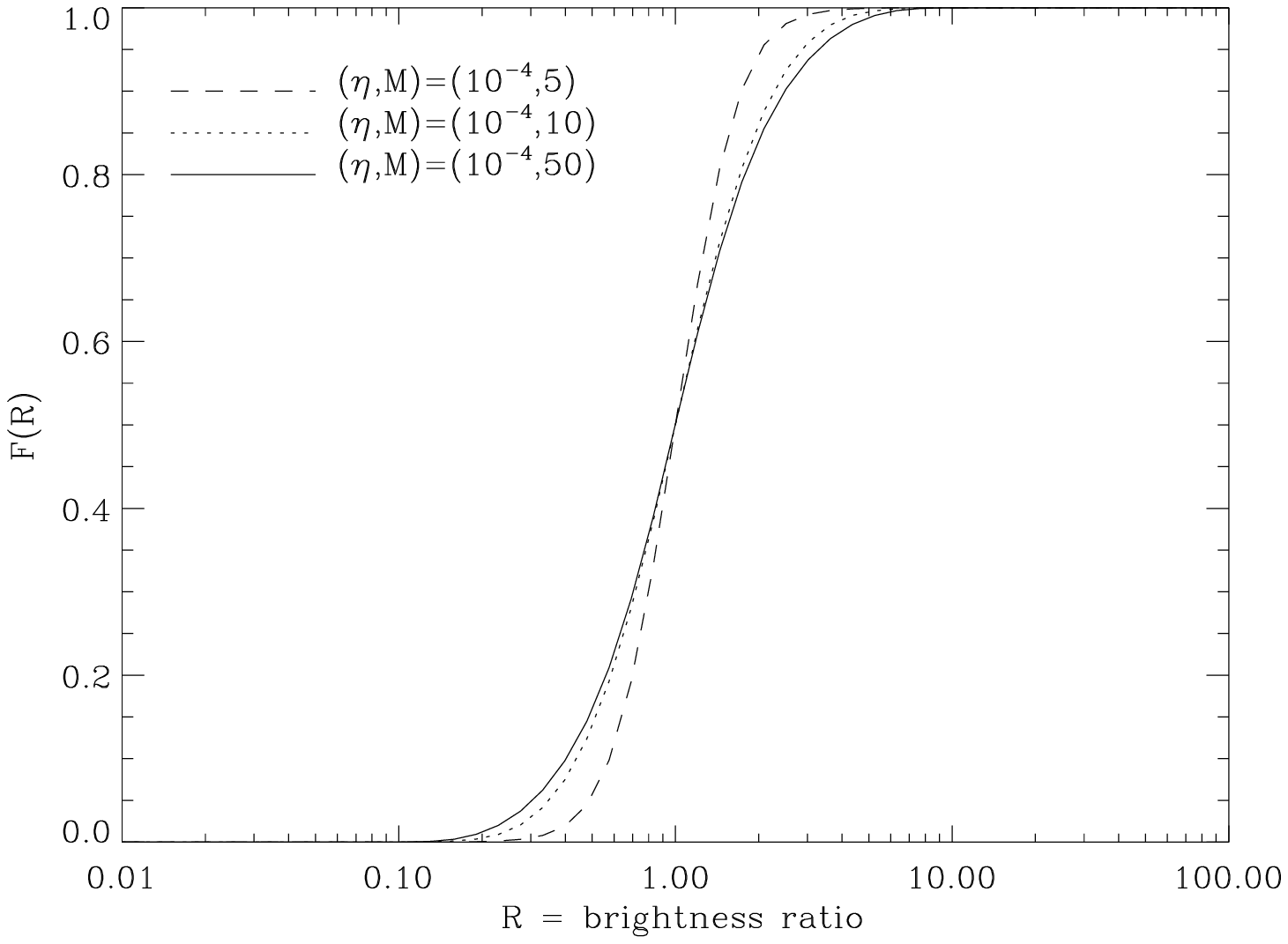}\end{array}
&
\begin{array}{c}\includegraphics[width=6cm]{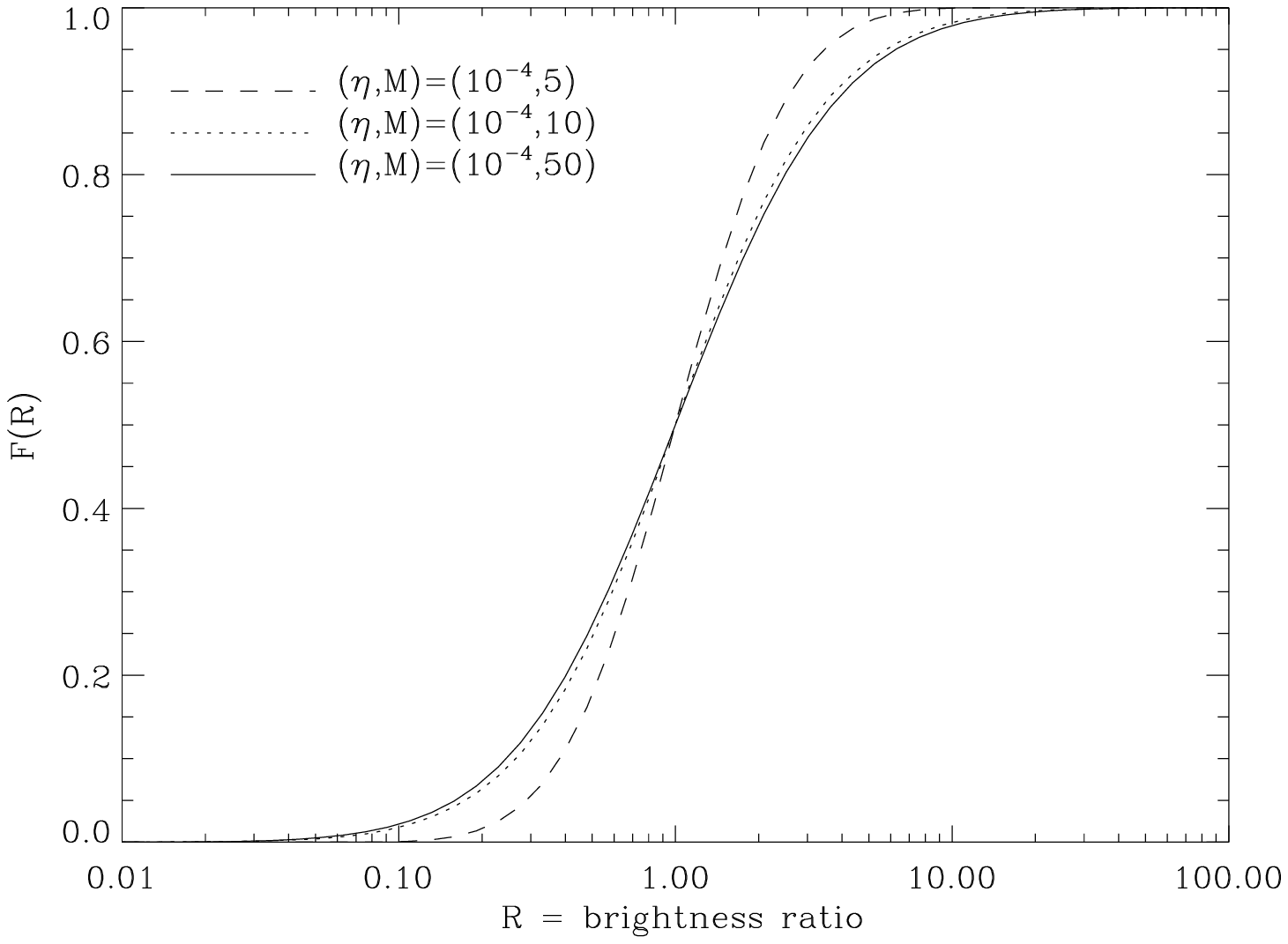}\end{array}
\\
\begin{array}{c}\includegraphics[width=6cm]{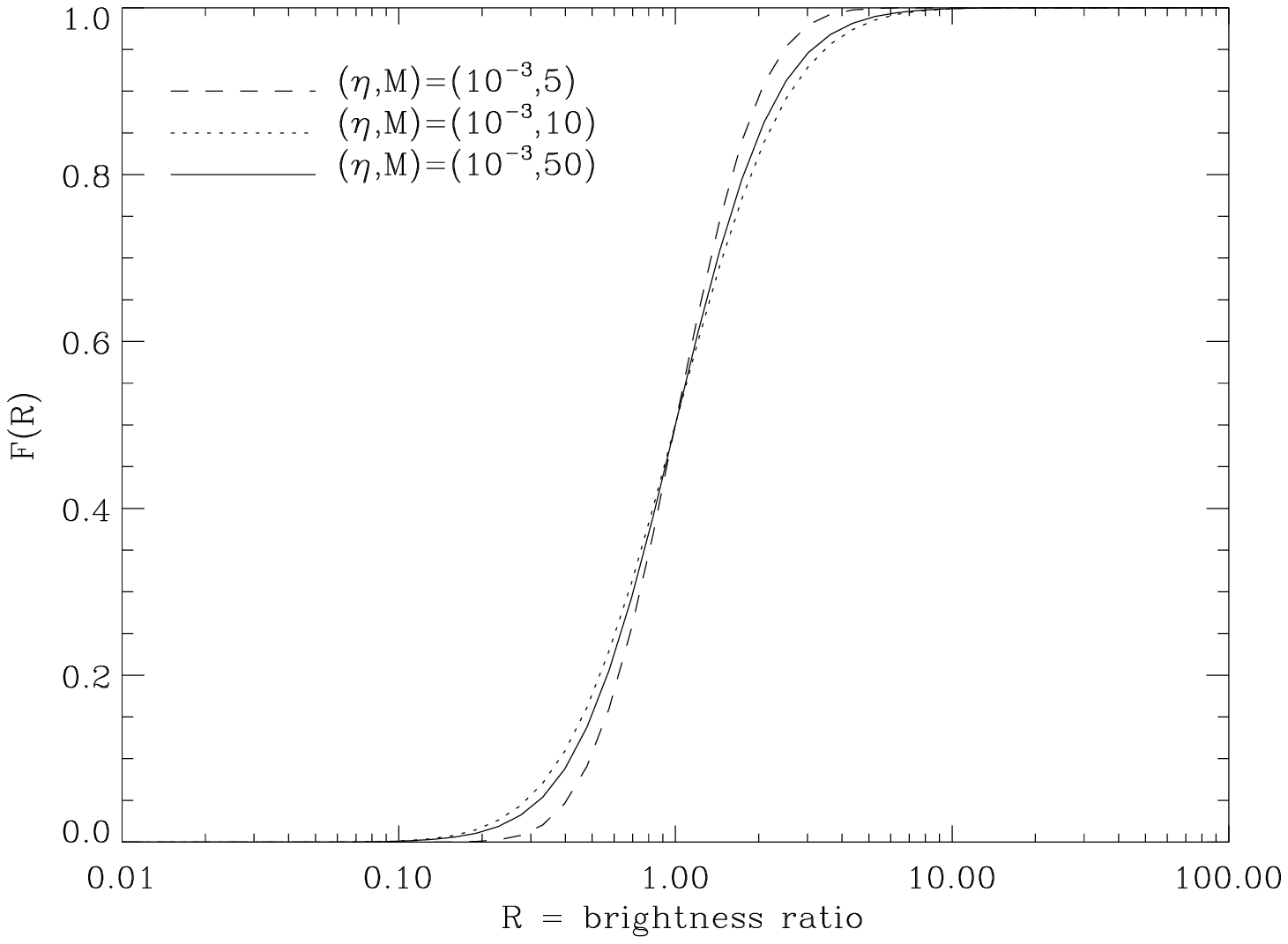}\end{array}
&
\begin{array}{c}\includegraphics[width=6cm]{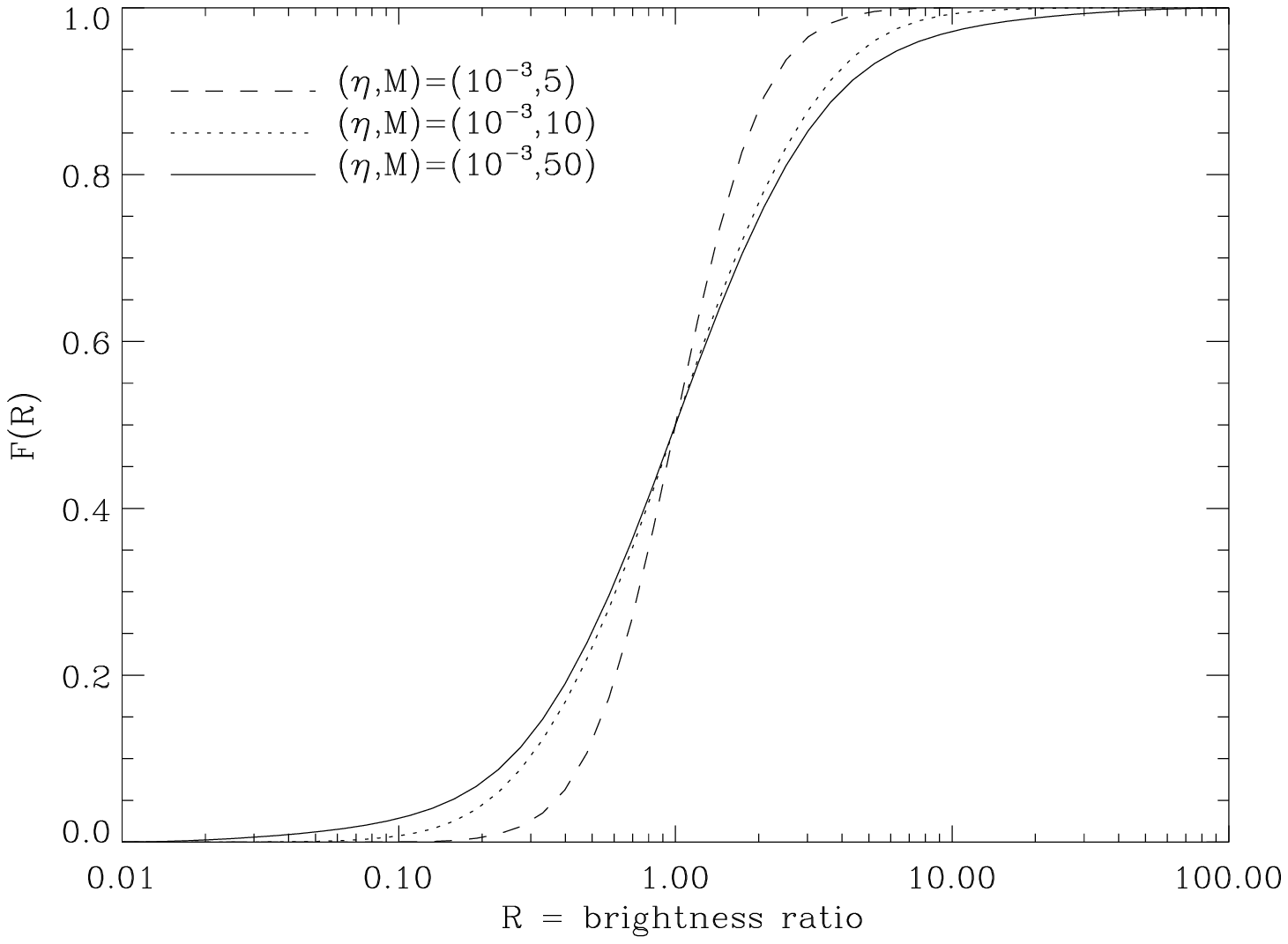}\end{array}
\\
\begin{array}{c}\includegraphics[width=6cm]{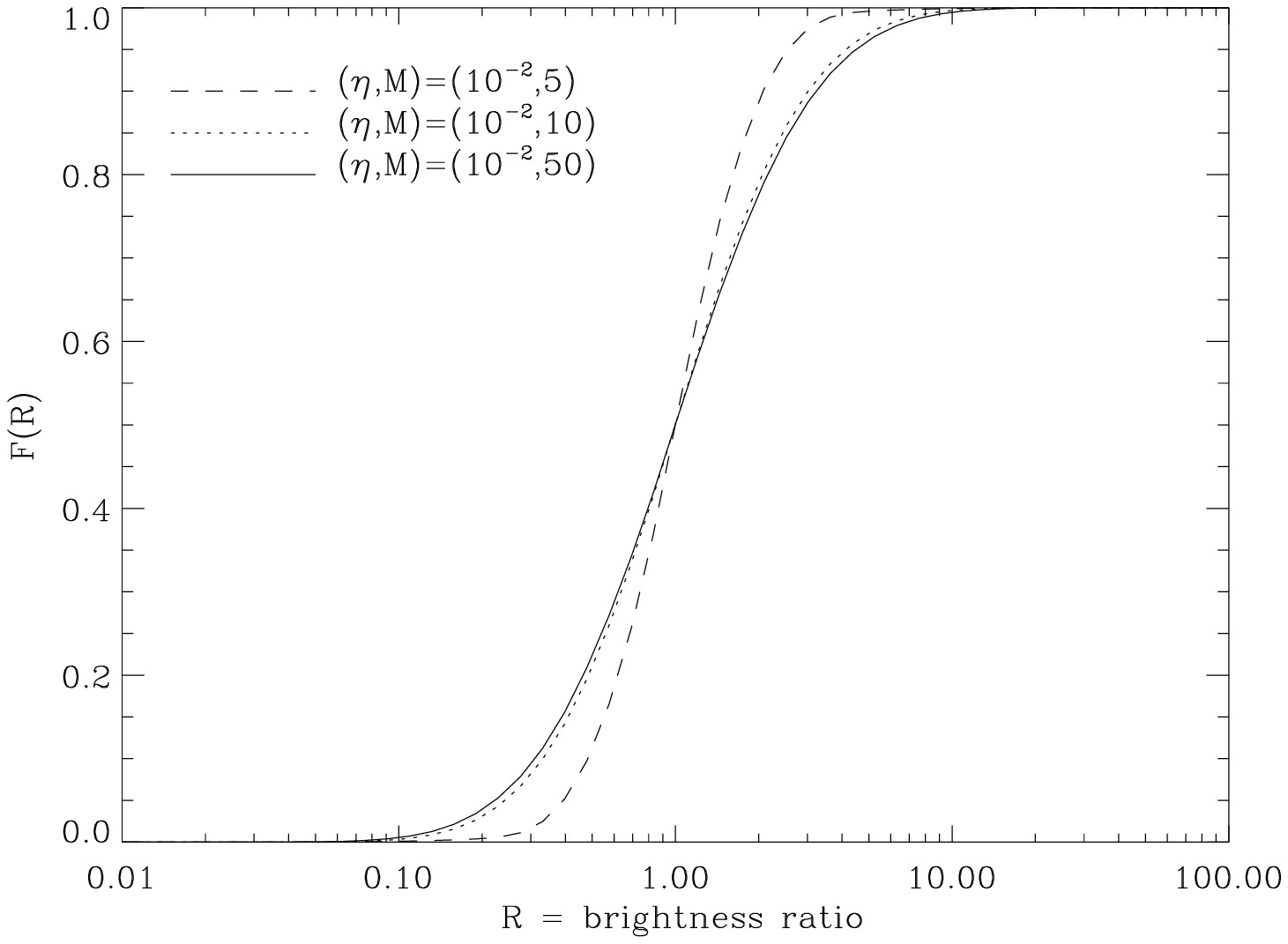}\end{array}
&
\begin{array}{c}\includegraphics[width=6cm]{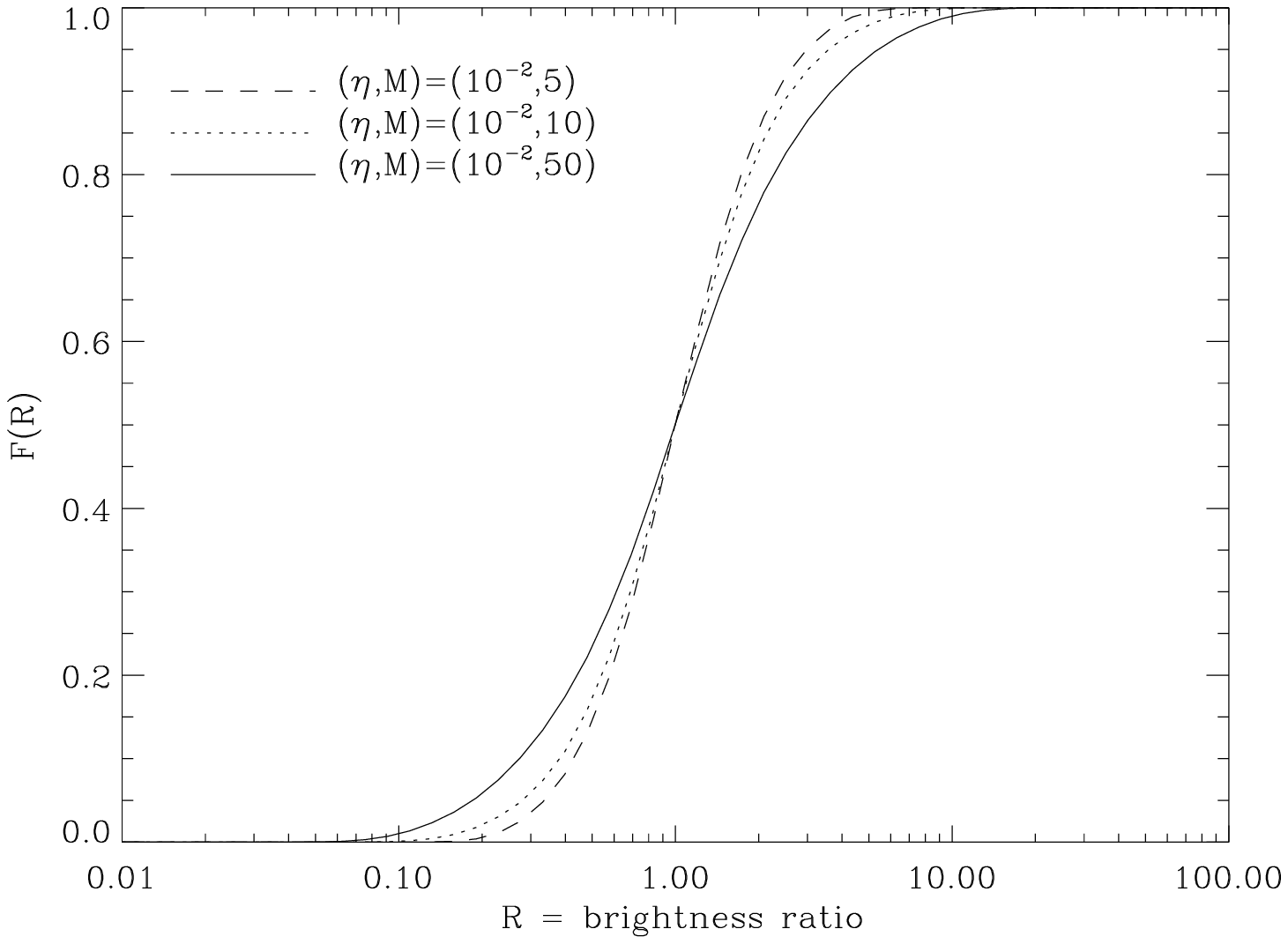}\end{array}
\\
\end{array}$
}
\caption{
Cumulative probability distribution functions, 
$F(R) =\mathrm{Pr}(R'\leq R)$,  
for the ratio of peak surface brightnesses of two
hot-spots which are flickering independently.
Left and right columns respectively show cases
with closed and open left boundaries.
Within each panel, the three curves correspond to cases of
$M=5, 10, 50$ for dashed, dotted and solid lines respectively.
}
\label{f:brightness.ratio}
\end{figure}

In physical units,
with $t_0\sim10^7 T_7^{-1/2}{\rm yr}$,
the dynamical timescale is
$t_{\rm dyn}\sim4\times10^{5(4)}{\rm yr}$
for a jet with $\eta=10^{-2(4)}$.
These time-scales are much shorter than the time taken for
light or jet plasma to traverse
the $2.1\times10^2/h \> {\rm kpc}
= 6.4\times10^5/h \> {\rm ly}$ projected distance
between the nucleus and either hot-spot.
Therefore the intrinsic flickering of two opposite hot-spots
are causally independent.
Figure~\ref{f:brightness.ratio}
shows the cumulative distribution function of the ratio
of the peak brightness of one spot with respect to the other,
$R = I_1 / I_2$,
calculated from simulation output according to the derivation
in Appendix~\ref{appendix.hot-spot.ratio}.
The ratio has a more extended distribution 
in cases with larger Mach number,
consistent with the larger amplitude of the structure function. 
For a given choice of jet parameters $(\eta,M)$,
an open boundary condition gives rise to substantially greater variability.
For instance, the probability that $R>4$ or $R<{\frac14}$
(see Table~\ref{t:hot-spot.ratio})
is approximately $20\%$ for cases with $M=50$ and an open boundary
but only ranges from $\approx 4 - 13\%$ 
in the cases we studied with a closed boundary.
Likewise for small Mach number cases,
for example with $(\eta,M)=(10^{-3},5)$ 
the probability that $R>4$ or $R<{\frac14}$
ranges from $3 - 7\%$ with an open boundary
but only $0.4 - 1.5\%$ with a closed boundary.
Thus in cases where the jet has a low Mach number,
the relative brightnesses of hot-spots on different
sides of a source is a reasonable measure of non-intrinsic effects, 
including relativistic beaming,
however intrinsic variability may be significant
if the Mach number is large ($M\ga10$). 
In gemeral terms these distributions provide a means
of quantifying the uncertainty in using hot-spot ratios in various
applications.

\begin{table}
\caption{
Probability of a hot-spot ratio of at least $4:1$,
$\mathrm{Pr}(R\geq4)+\mathrm{Pr}(R\leq{\frac14}) = 2F({\frac14})$,
occuring at any instant as a result of intrinsic variability,
calculated from representative time sequences
of simulations that are finely resolved in time
(frame interval $\ll t_\mathrm{dyn}$.
}
\begin{center}
$\begin{array}{cc}
\begin{array}{ccrc}
\mbox{left B.C.}&\eta&M&2F({\frac14})\\
\\
\hline
\\
\mbox{closed}&10^{-4}& 5&0.004
\\
\mbox{closed}&10^{-4}&10&0.027
\\
\mbox{closed}&10^{-4}&50&0.055
\\
\mbox{closed}&10^{-3}& 5&0.009
\\
\mbox{closed}&10^{-3}&10&0.068
\\
\mbox{closed}&10^{-3}&50&0.049
\\
\mbox{closed}&10^{-2}& 5&0.015
\\
\mbox{closed}&10^{-2}&10&0.107
\\
\mbox{closed}&10^{-2}&50&0.128
\\
\\\hline
\end{array}
&
\begin{array}{ccrc}
\mbox{left B.C.}&\eta&M&2F({\frac14})\\
\\
\hline
\\
\mbox{open}&10^{-4}& 5&0.065
\\
\mbox{open}&10^{-4}&10&0.183
\\
\mbox{open}&10^{-4}&50&0.206
\\
\mbox{open}&10^{-3}& 5&0.026
\\
\mbox{open}&10^{-3}&10&0.144
\\
\mbox{open}&10^{-3}&50&0.197
\\
\mbox{open}&10^{-2}& 5&0.034
\\
\mbox{open}&10^{-2}&10&0.076
\\
\mbox{open}&10^{-2}&50&0.173
\\
\\\hline
\end{array}
\\
\end{array}$
\end{center}
\label{t:hot-spot.ratio}
\end{table}

\subsection{Underlying structure and activity}
\label{s.activity}

\begin{figure}[h]
\centering \leavevmode
$\begin{array}{cc}
\includegraphics[width=7cm]{fine-4m5_010f1674.eps}
&
\includegraphics[width=7cm]{fine-4m5_010f1714.eps}
\\
\includegraphics[width=7cm]{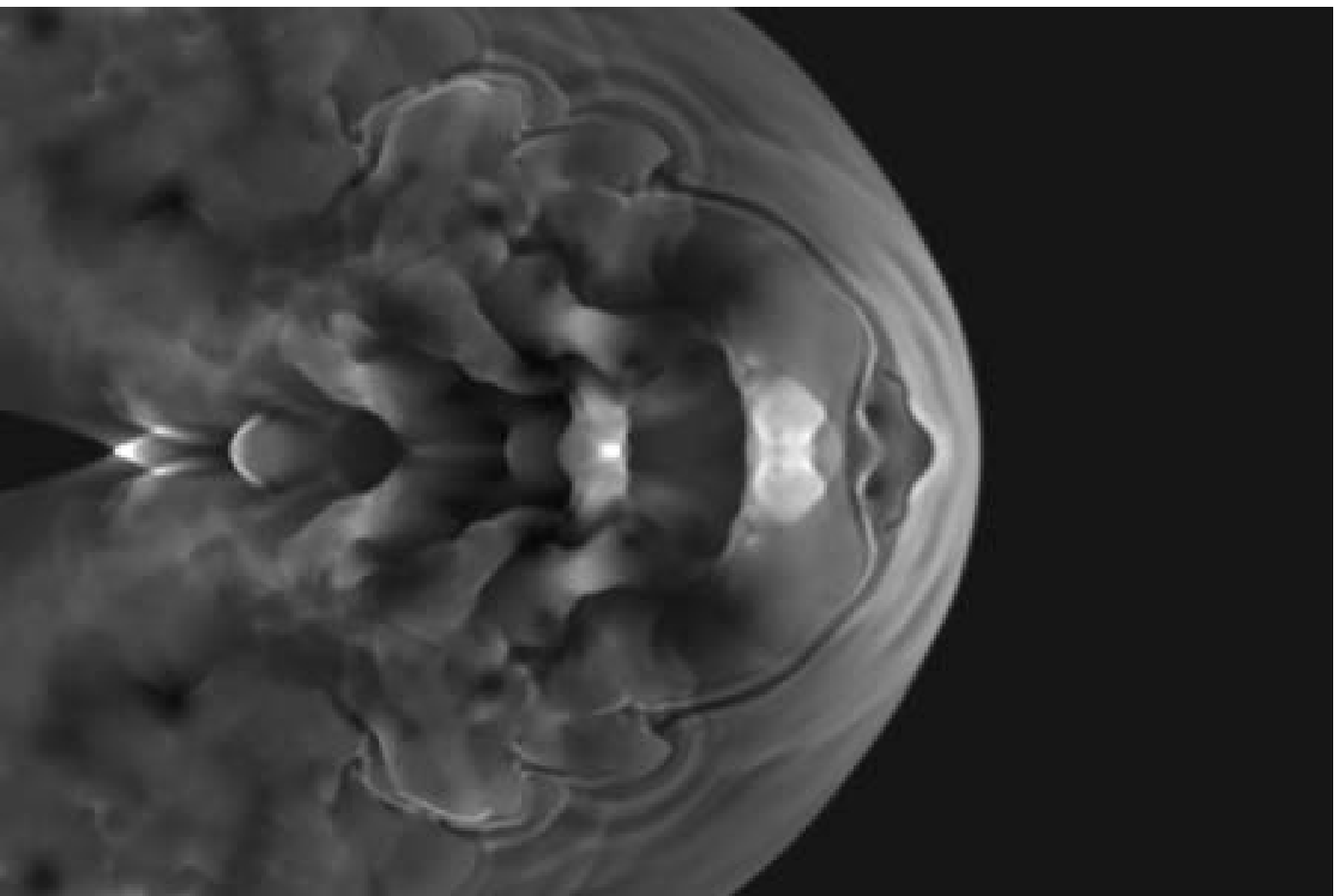}
&
\includegraphics[width=7cm]{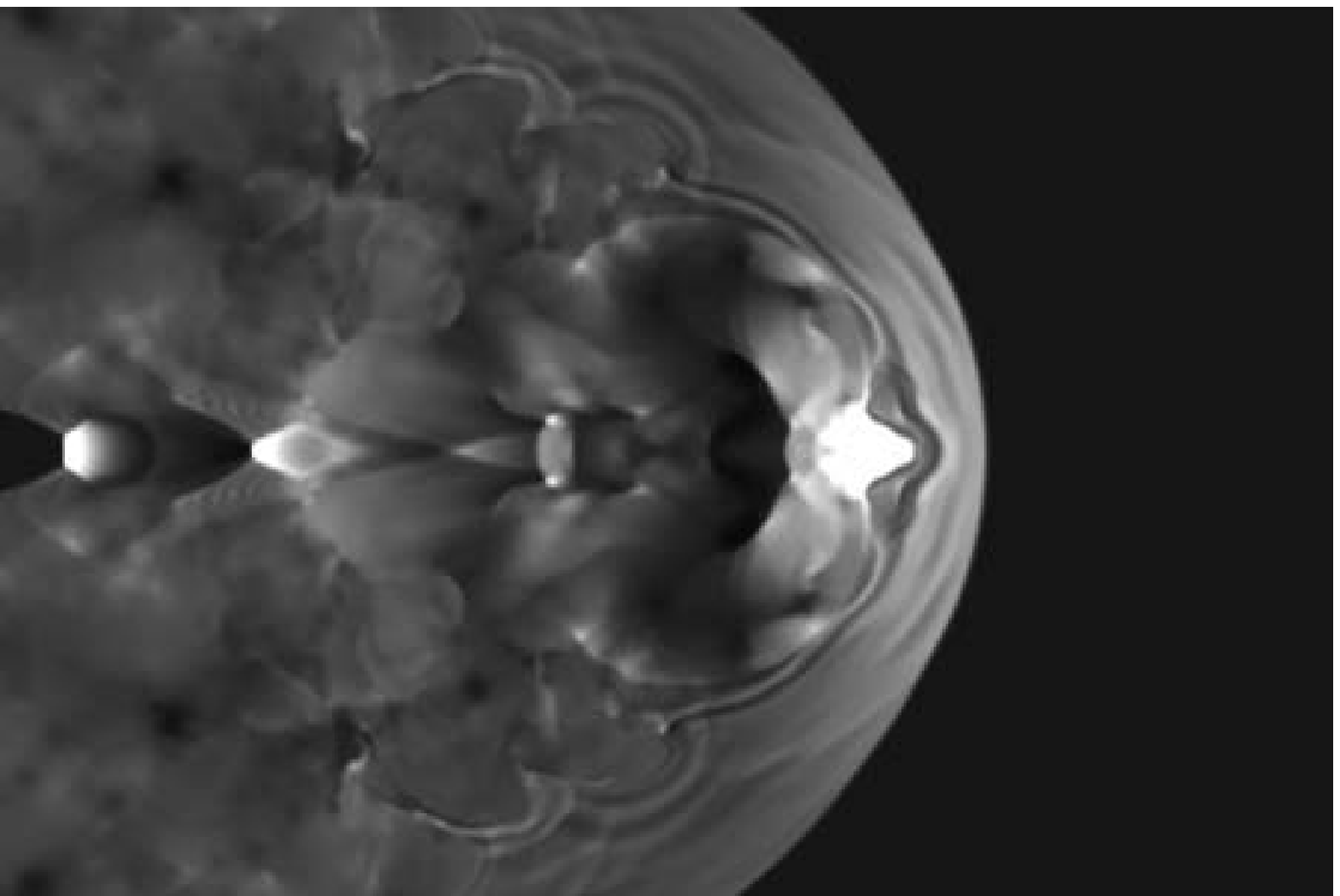}
\\
\includegraphics[width=7cm]{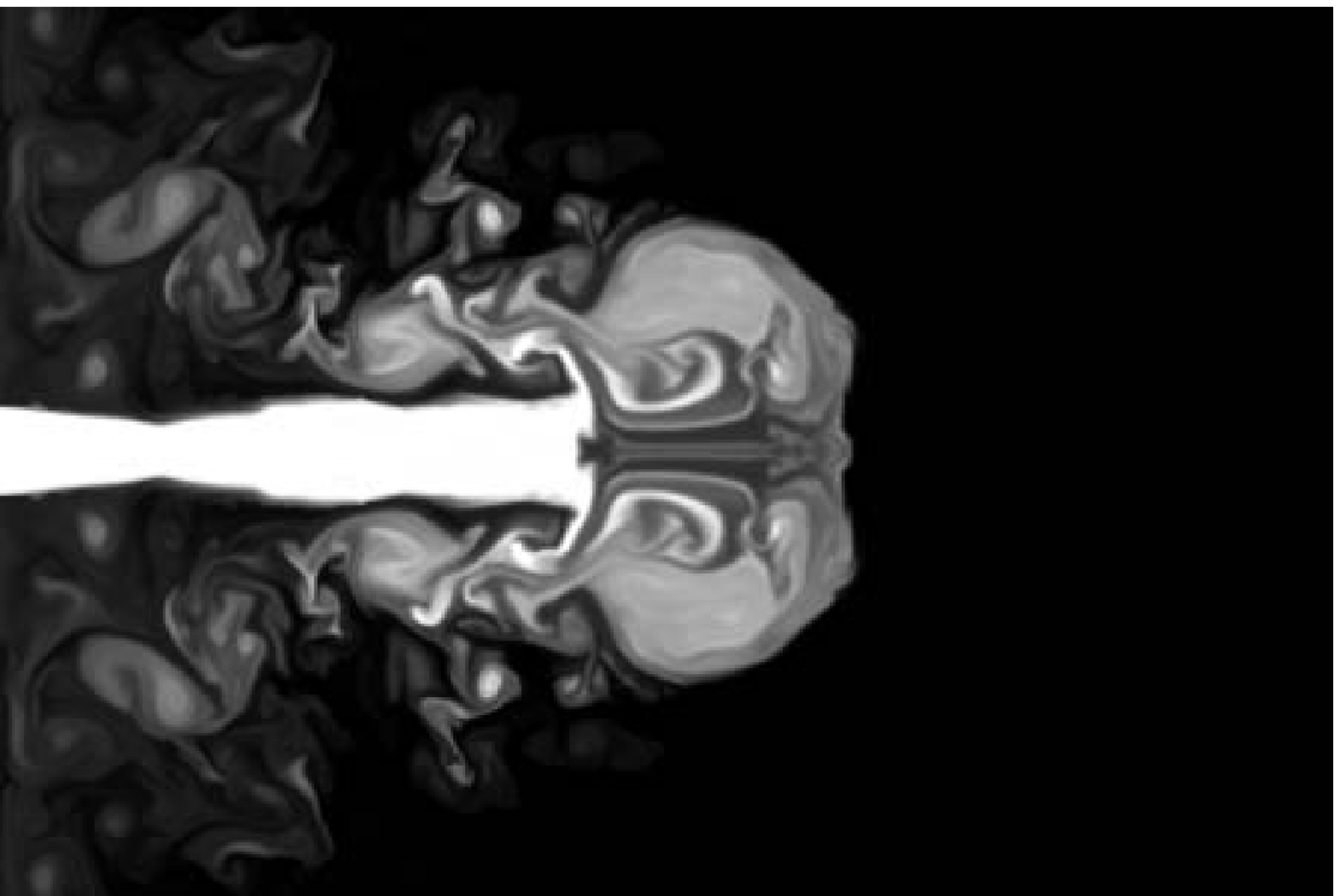}
&
\includegraphics[width=7cm]{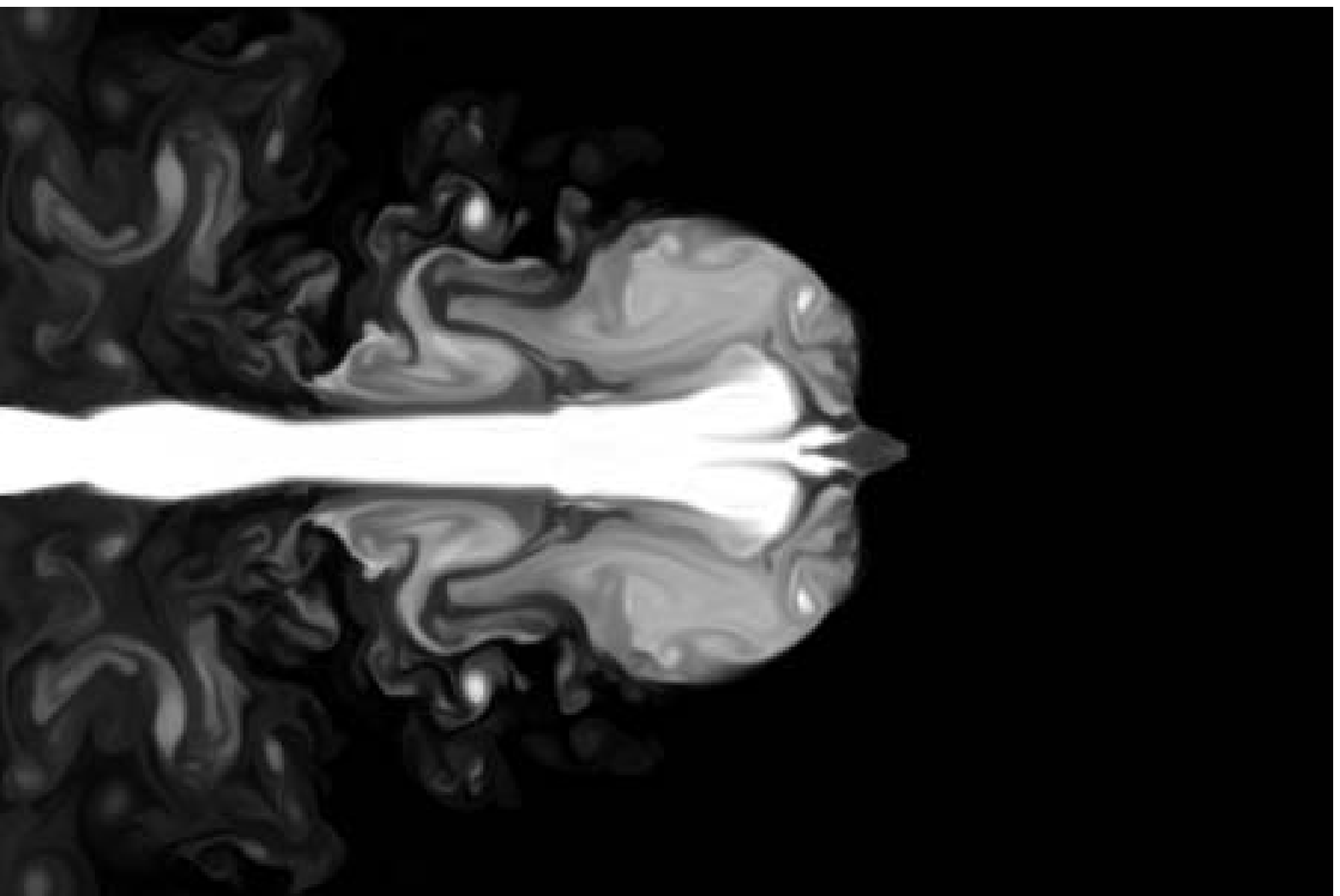}
\\
\includegraphics[width=7cm]{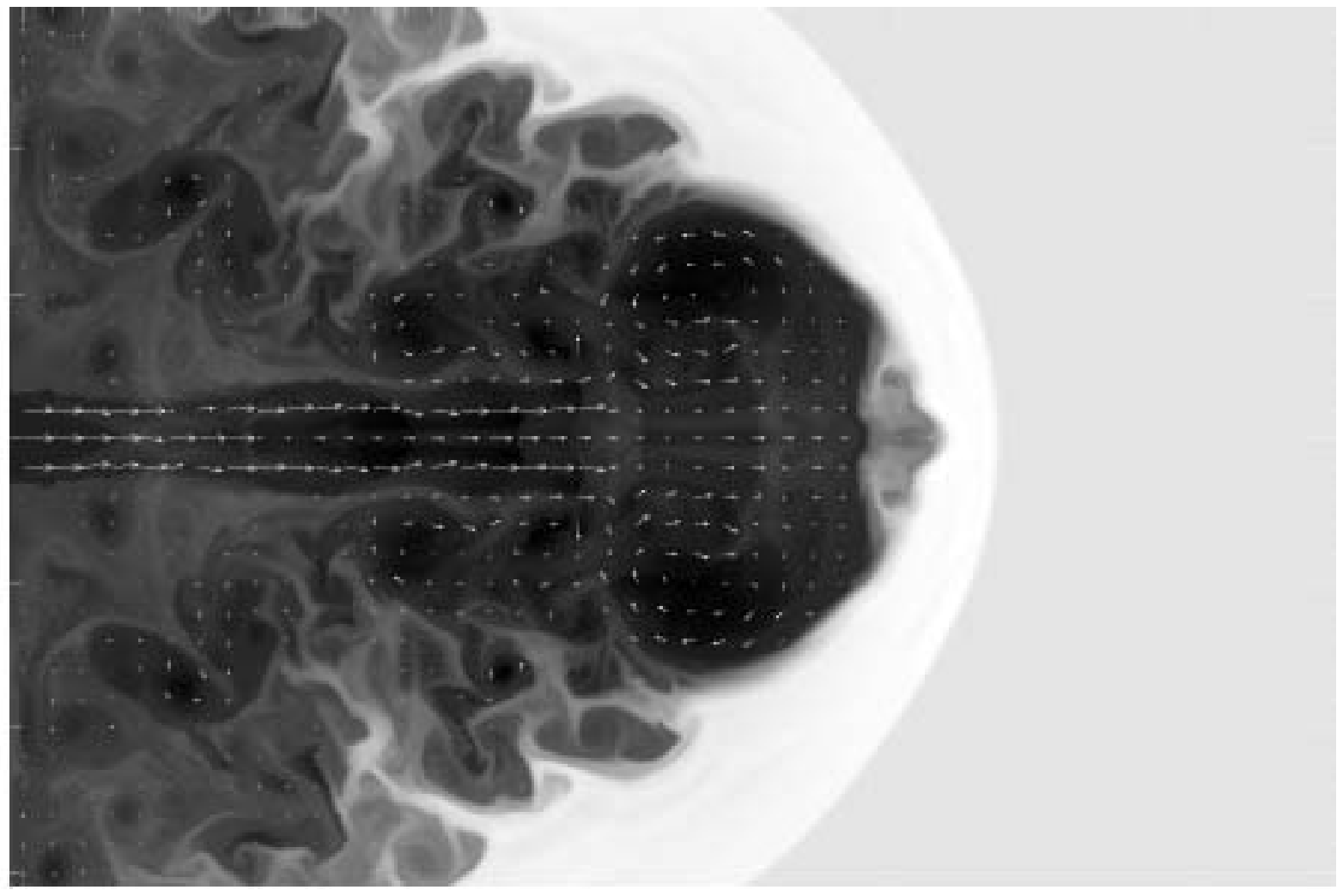}
&
\includegraphics[width=7cm]{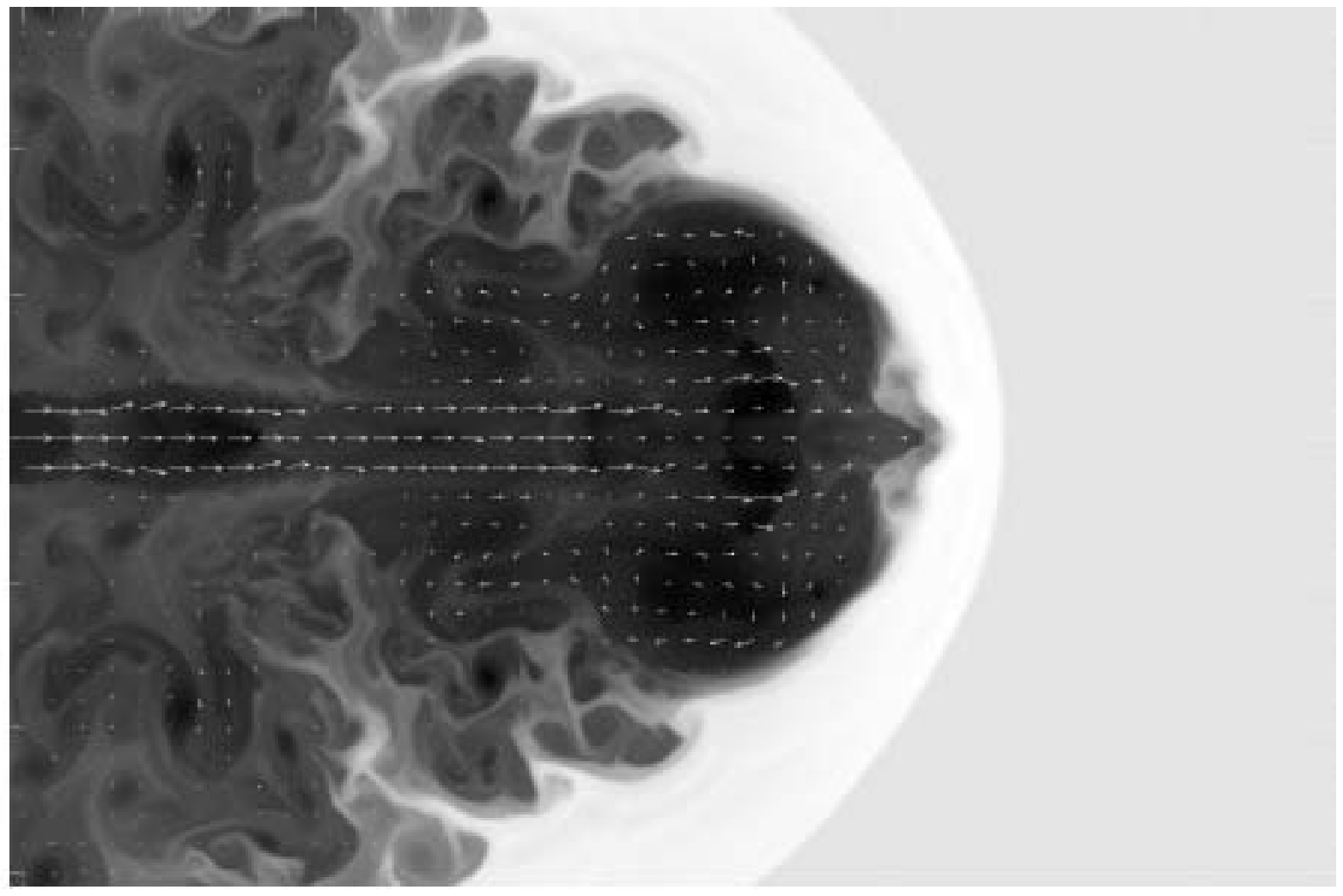}
\\
\end{array}$
\caption{
$450 \times 300$ pixel sub-region
around a jet with parameters 
$(\eta,M)=(10^{-4},5)$.
Rows from top to bottom show:
raytraced images;
pressure;
the tracer of jet material, $\varphi$;
and logarithmic density plots overlain with flow velocity vectors.
}
\label{f:pageant-4m5.1}
\label{f:example.01}
\label{f:example.02}
\end{figure}

\begin{figure}[h]
\centering \leavevmode
$\begin{array}{cc}
\includegraphics[width=7cm]{jet-4m5_000f0152.eps}
&
\includegraphics[width=7cm]{jet-4m5_000f0265.eps}
\\
\includegraphics[width=7cm]{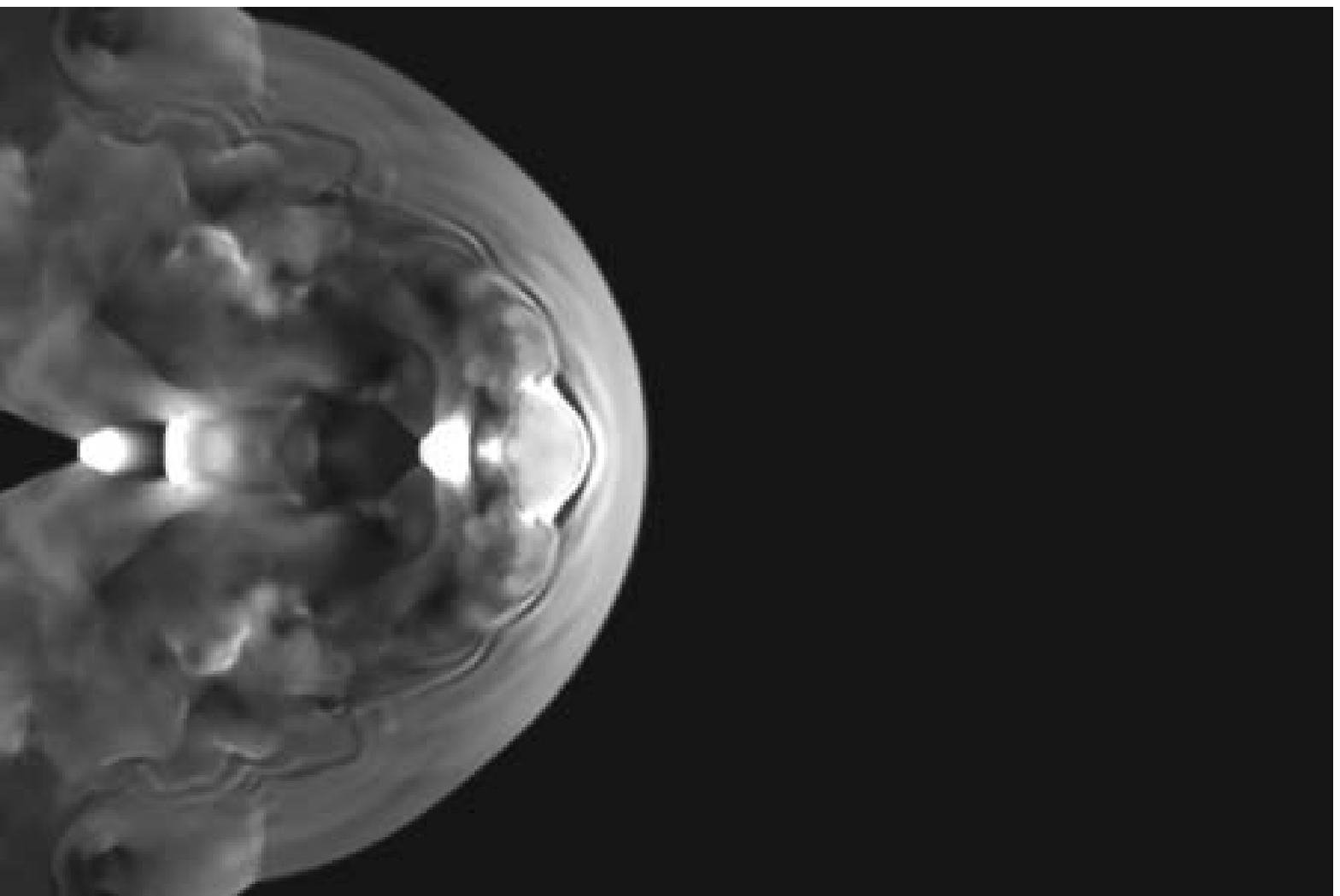}
&
\includegraphics[width=7cm]{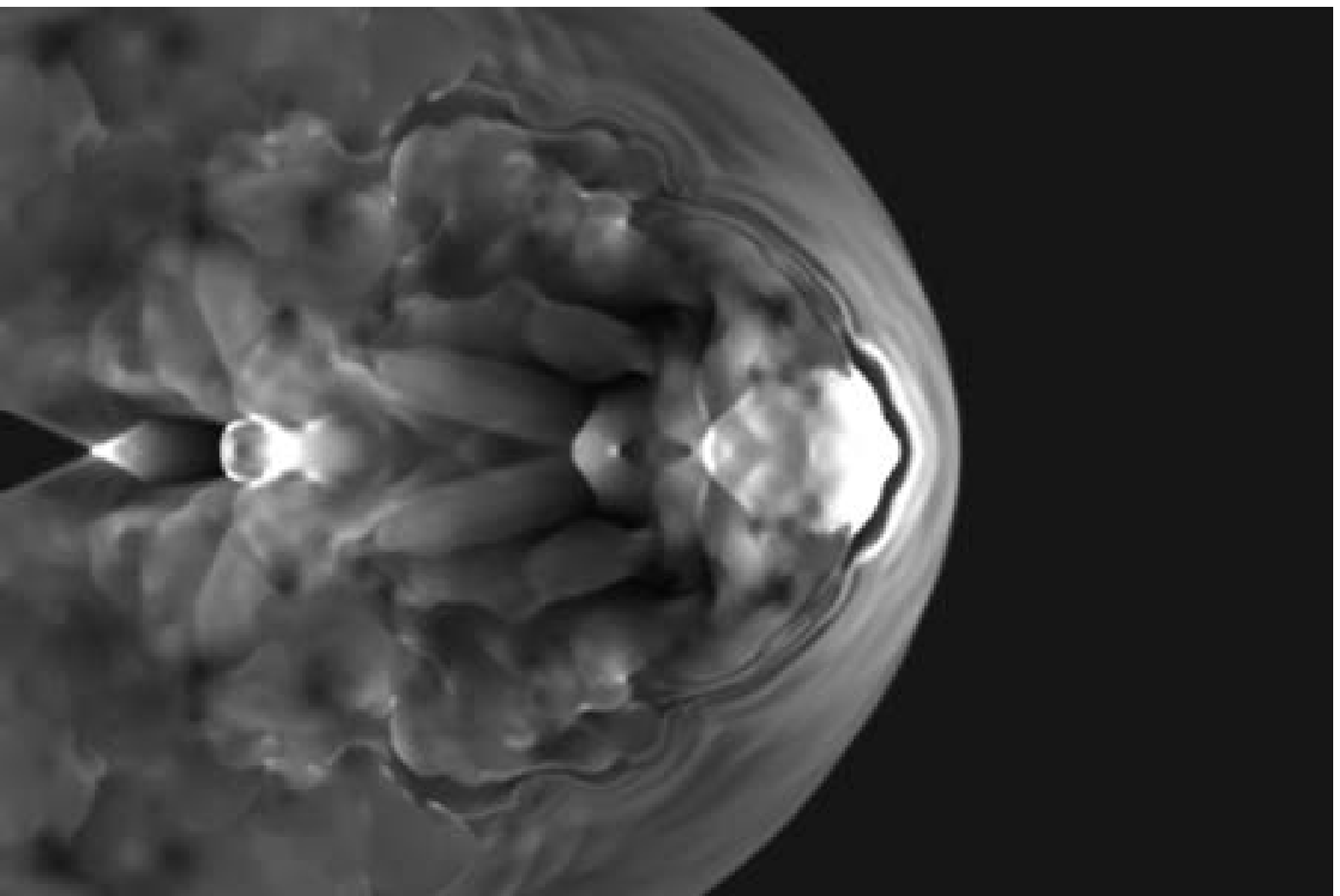}
\\
\includegraphics[width=7cm]{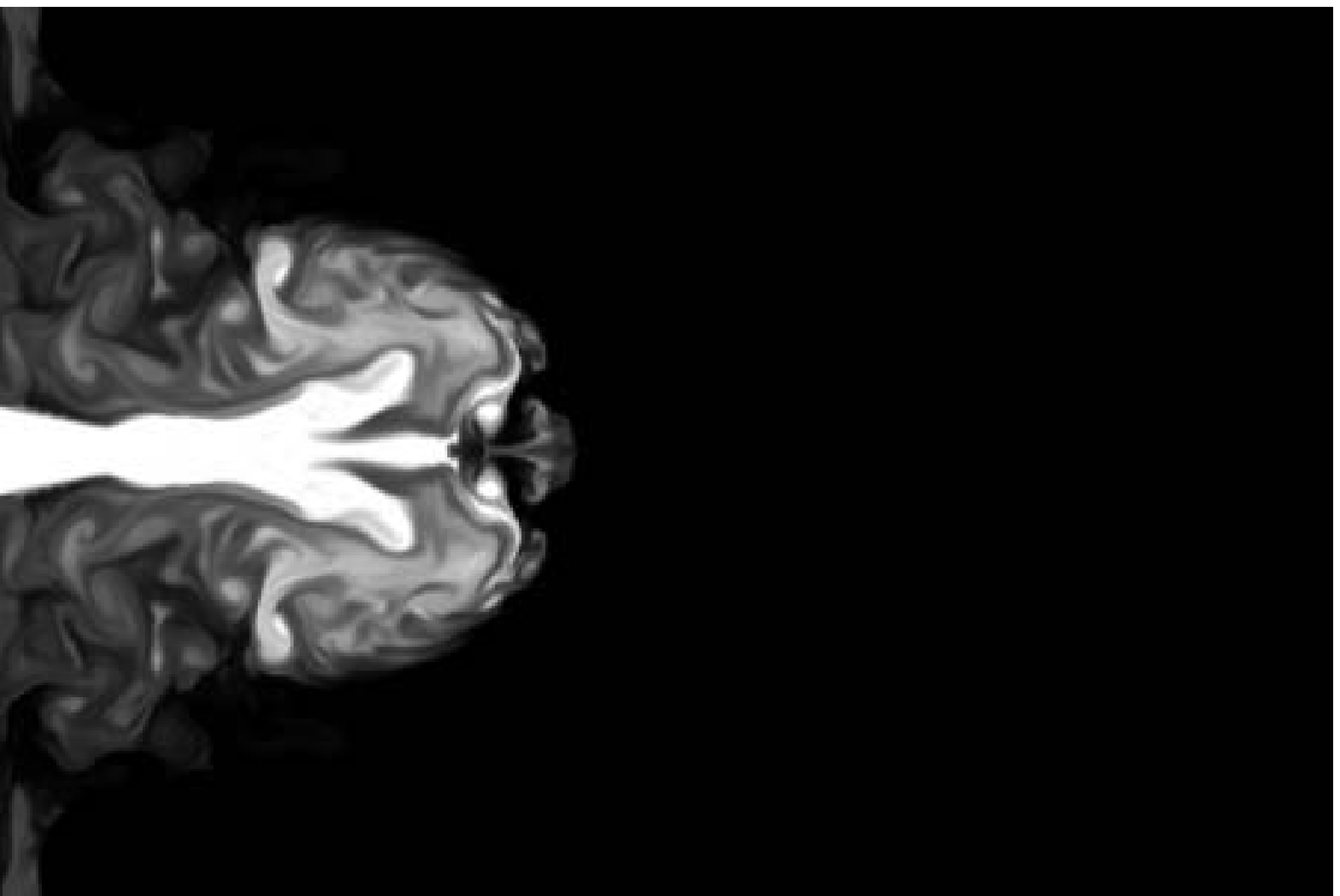}
&
\includegraphics[width=7cm]{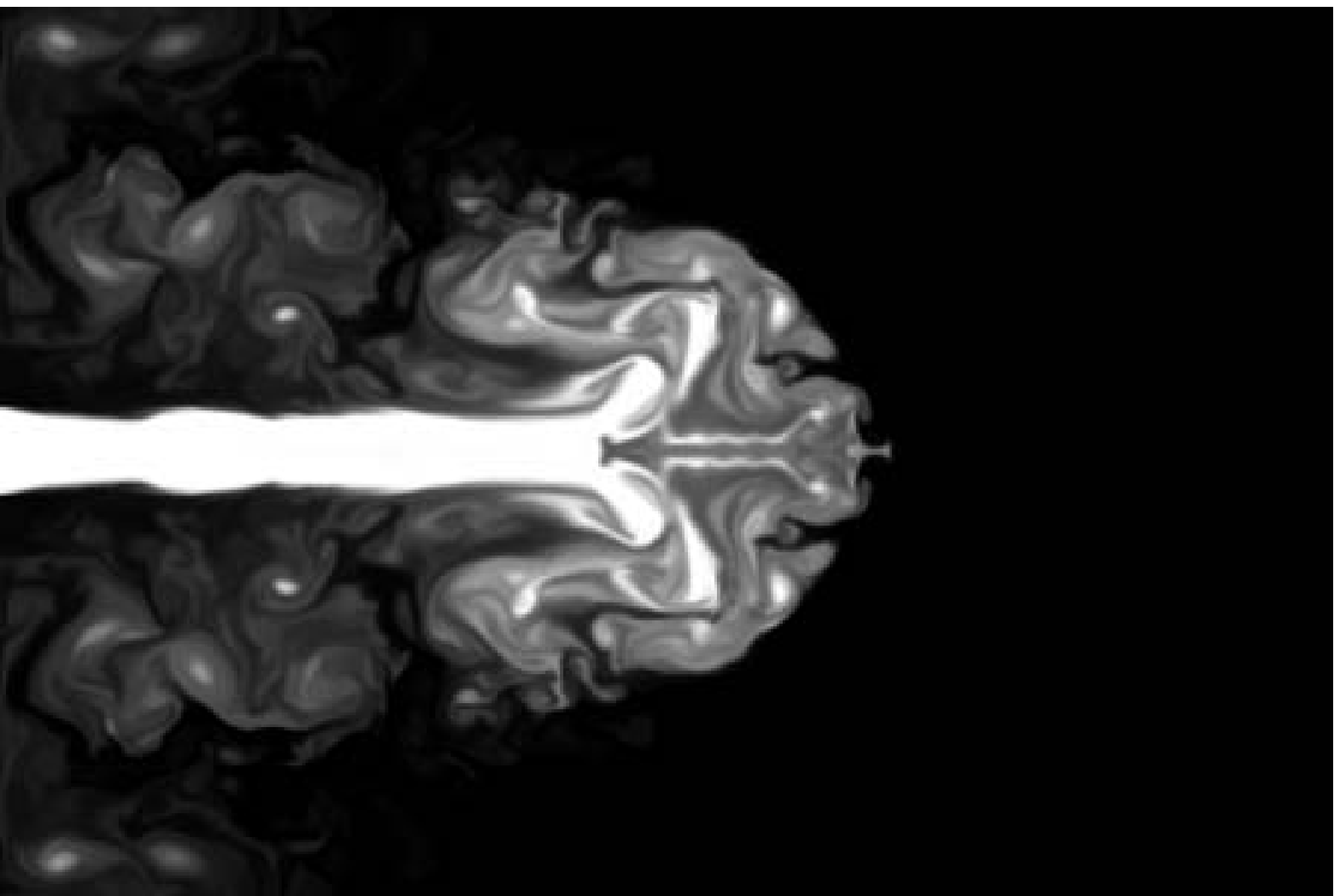}
\\
\includegraphics[width=7cm]{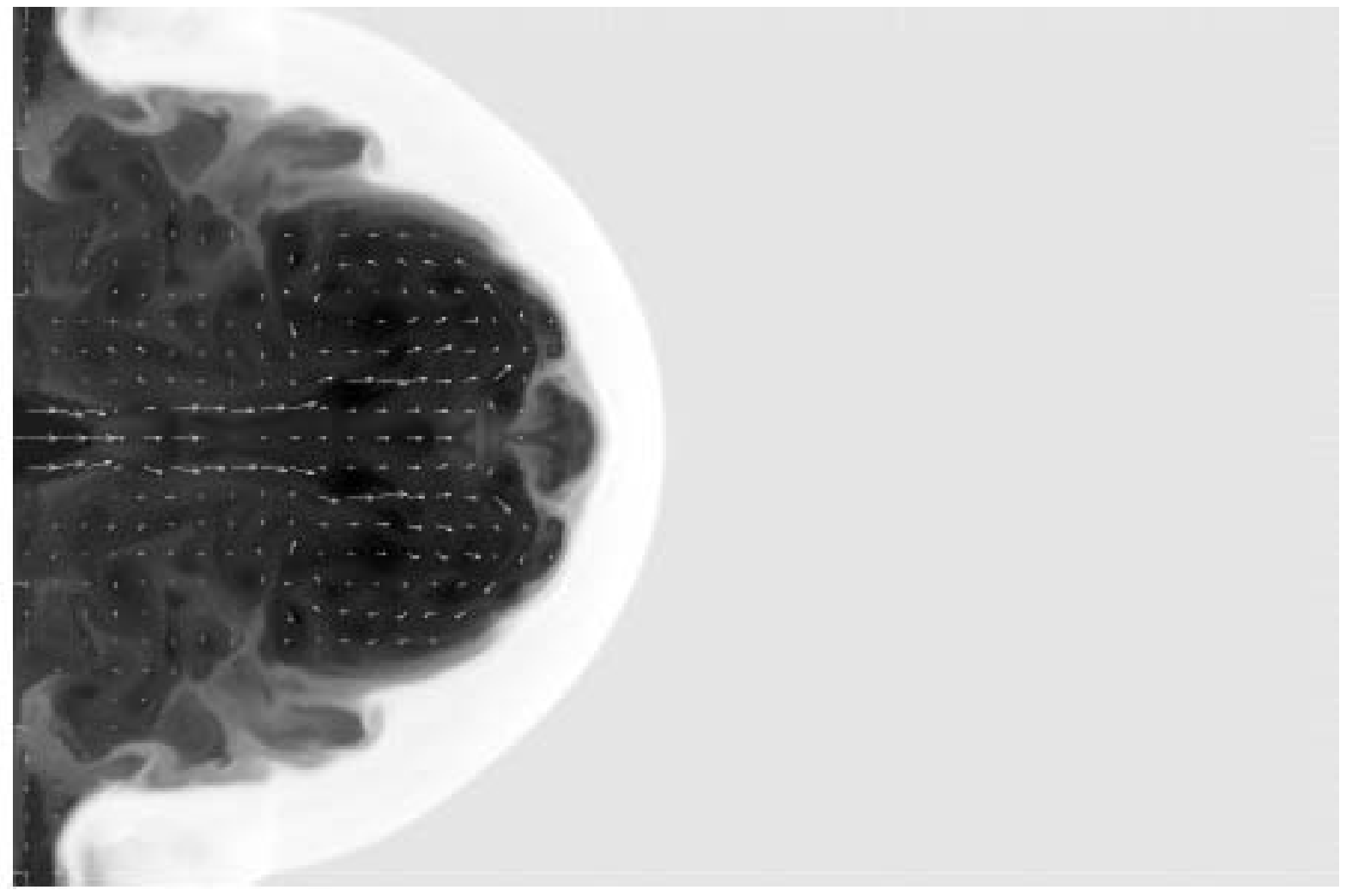}
&
\includegraphics[width=7cm]{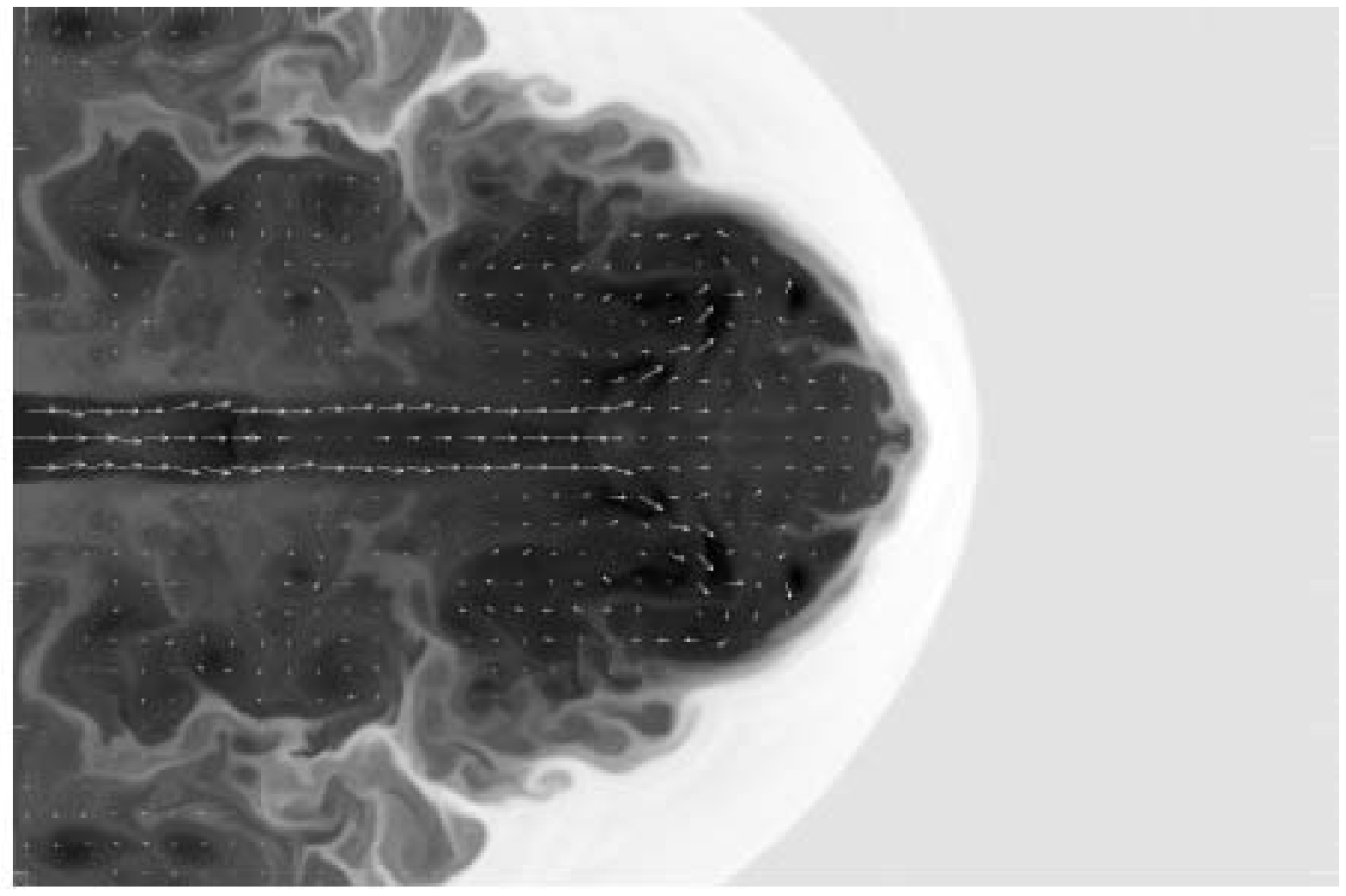}
\\
\end{array}$
\caption{
Further examples of ray-traced frames resembling Pictor~A,
with jet parameters 
$(\eta,M)=(10^{-4},5)$.
As in Figure~\ref{f:pageant-4m5.1},
the rows from top to bottom show
renderings, pressure, the tracer $\varphi$ and flow pattern.
}
\label{f:pageant-4m5.3}
\label{f:example.05}
\label{f:example.06}
\end{figure}

\begin{figure}[h]
\centering \leavevmode
$\begin{array}{cc}
\includegraphics[width=7cm]{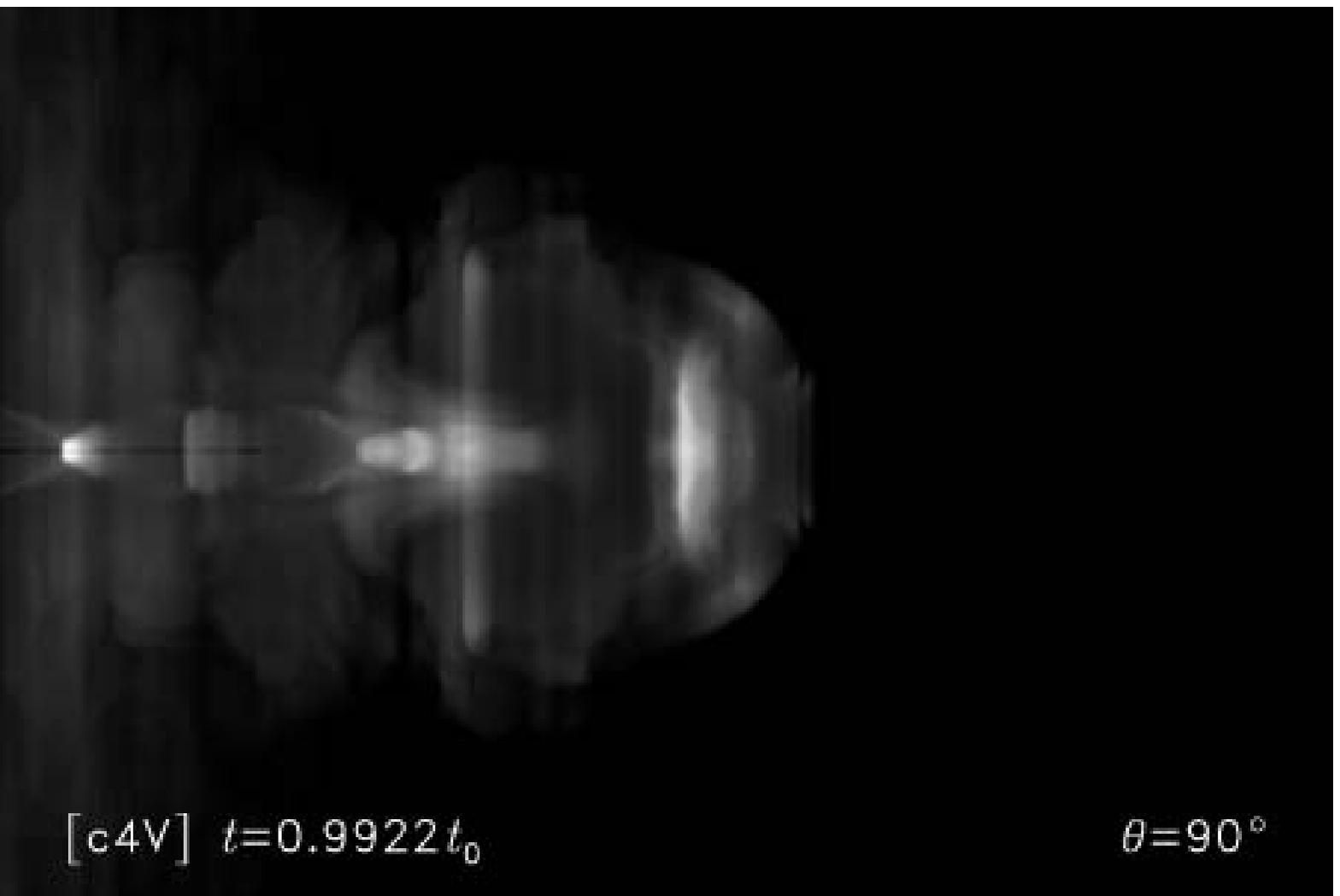}
&
\includegraphics[width=7cm]{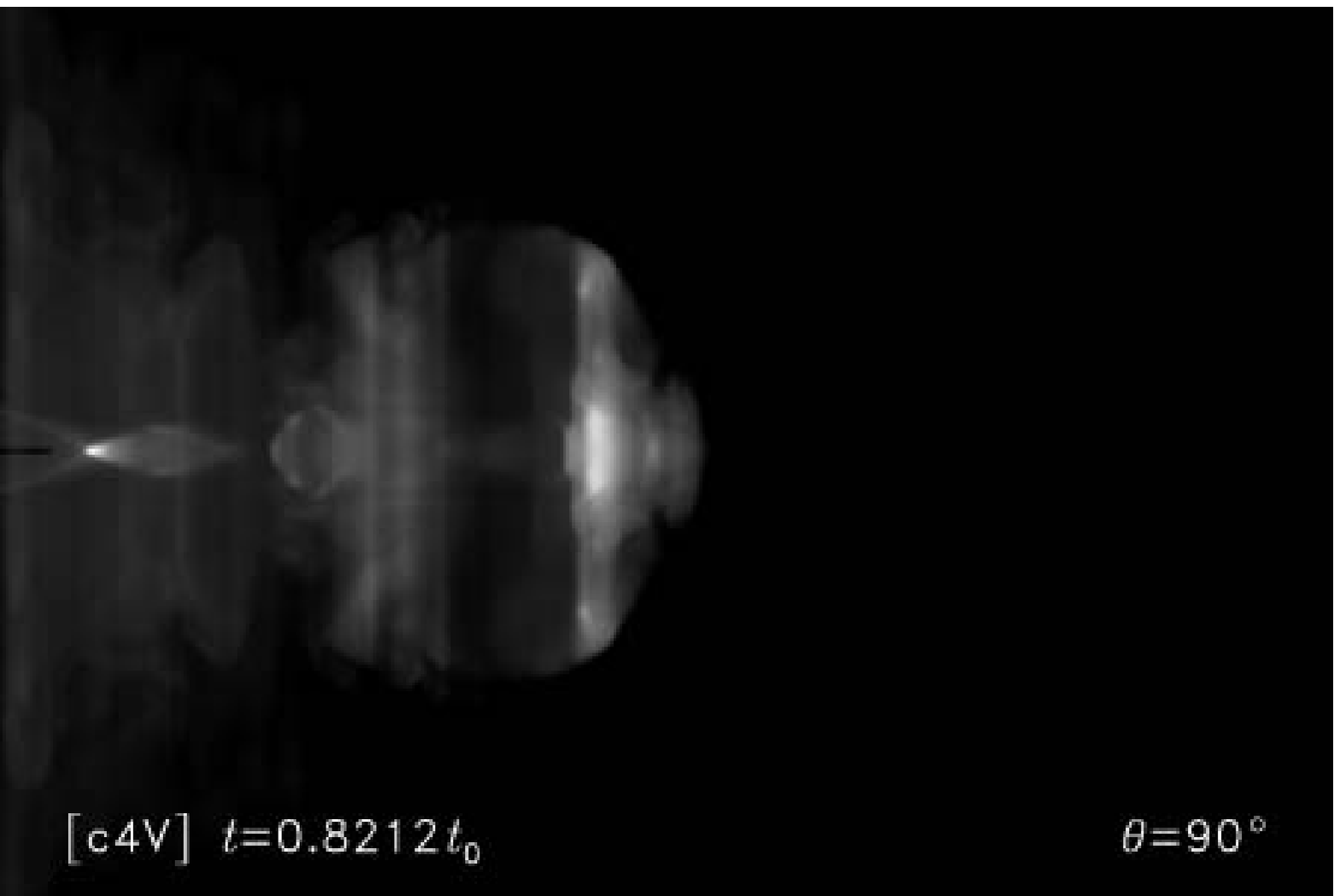}
\\
\includegraphics[width=7cm]{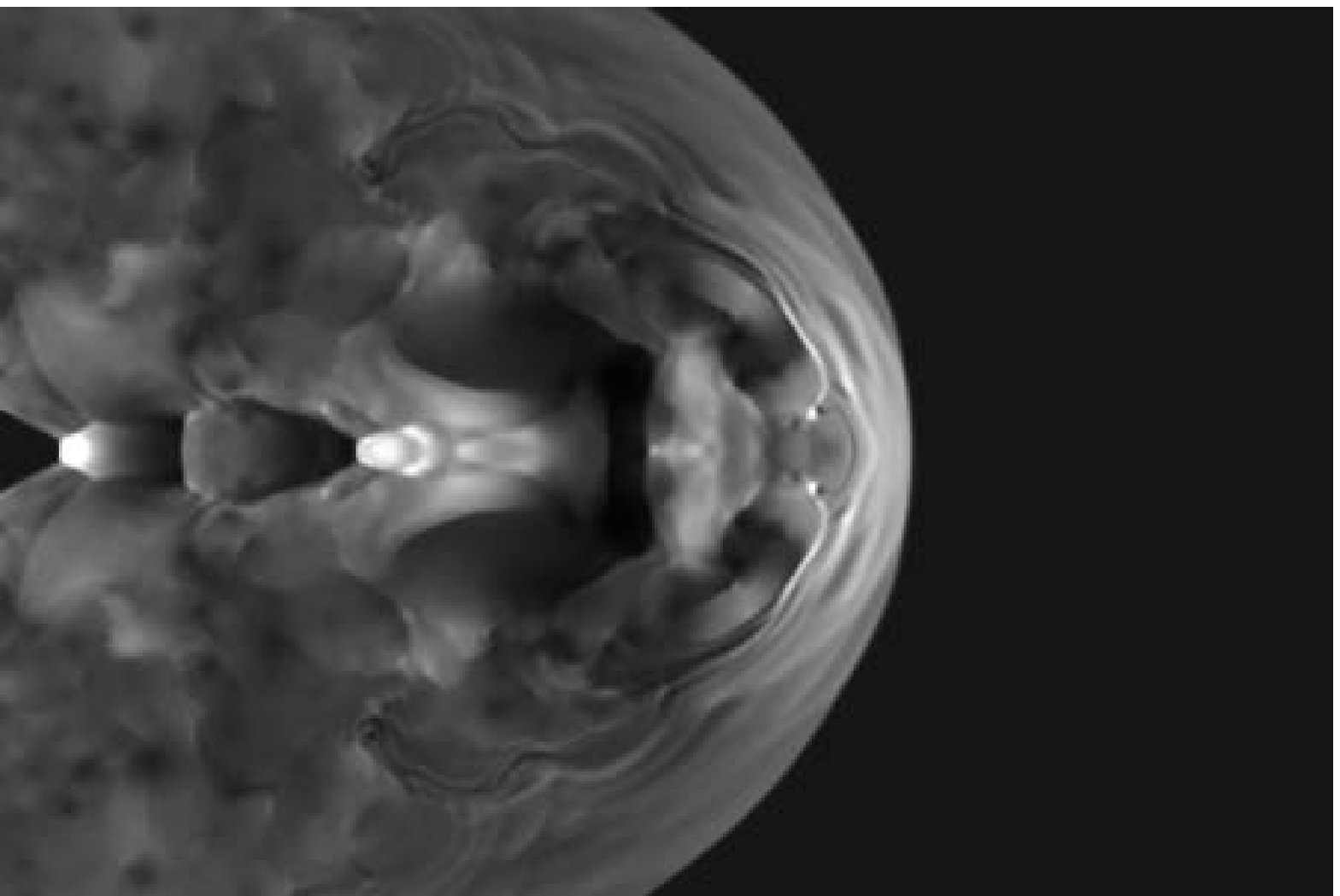}
&
\includegraphics[width=7cm]{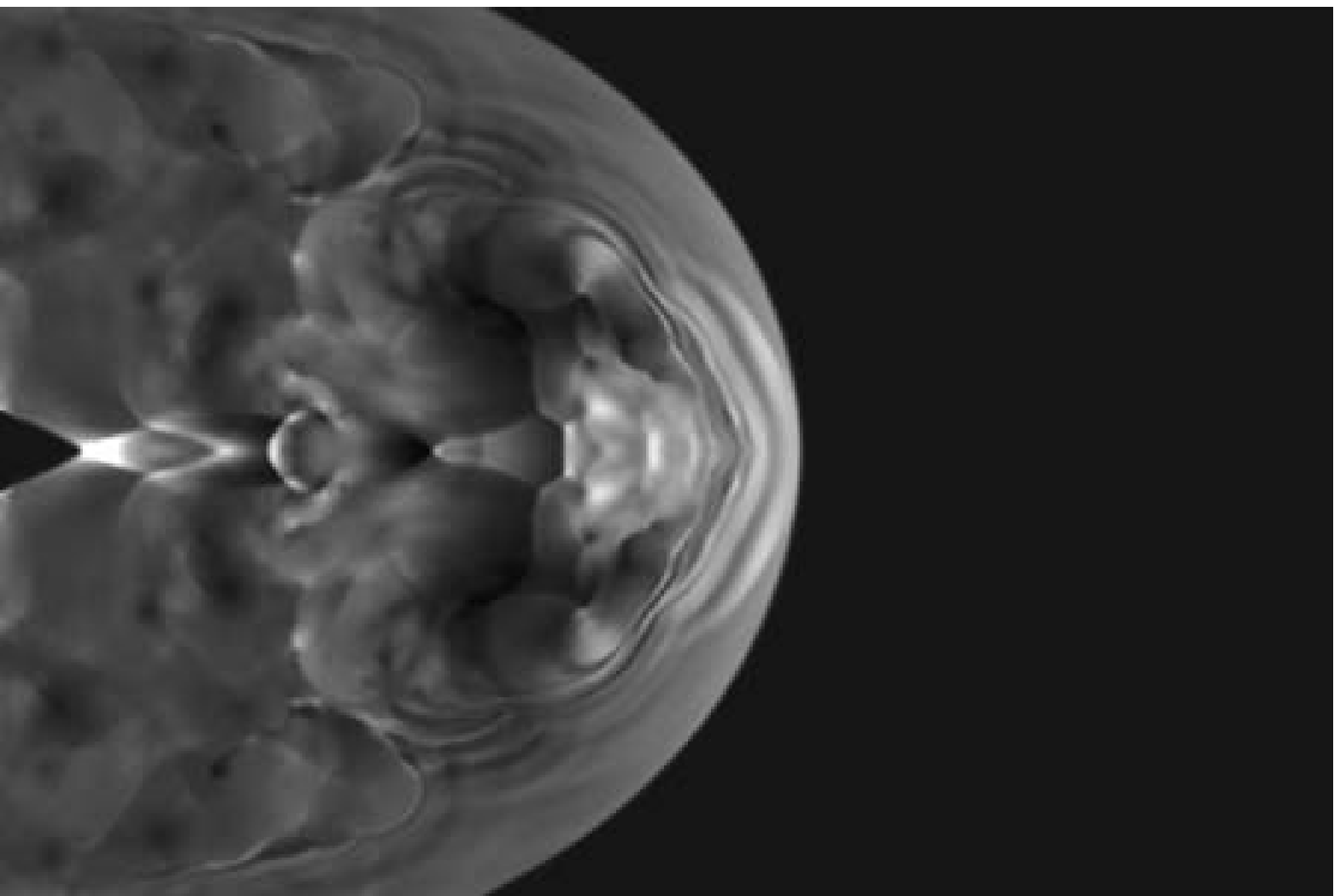}
\\
\includegraphics[width=7cm]{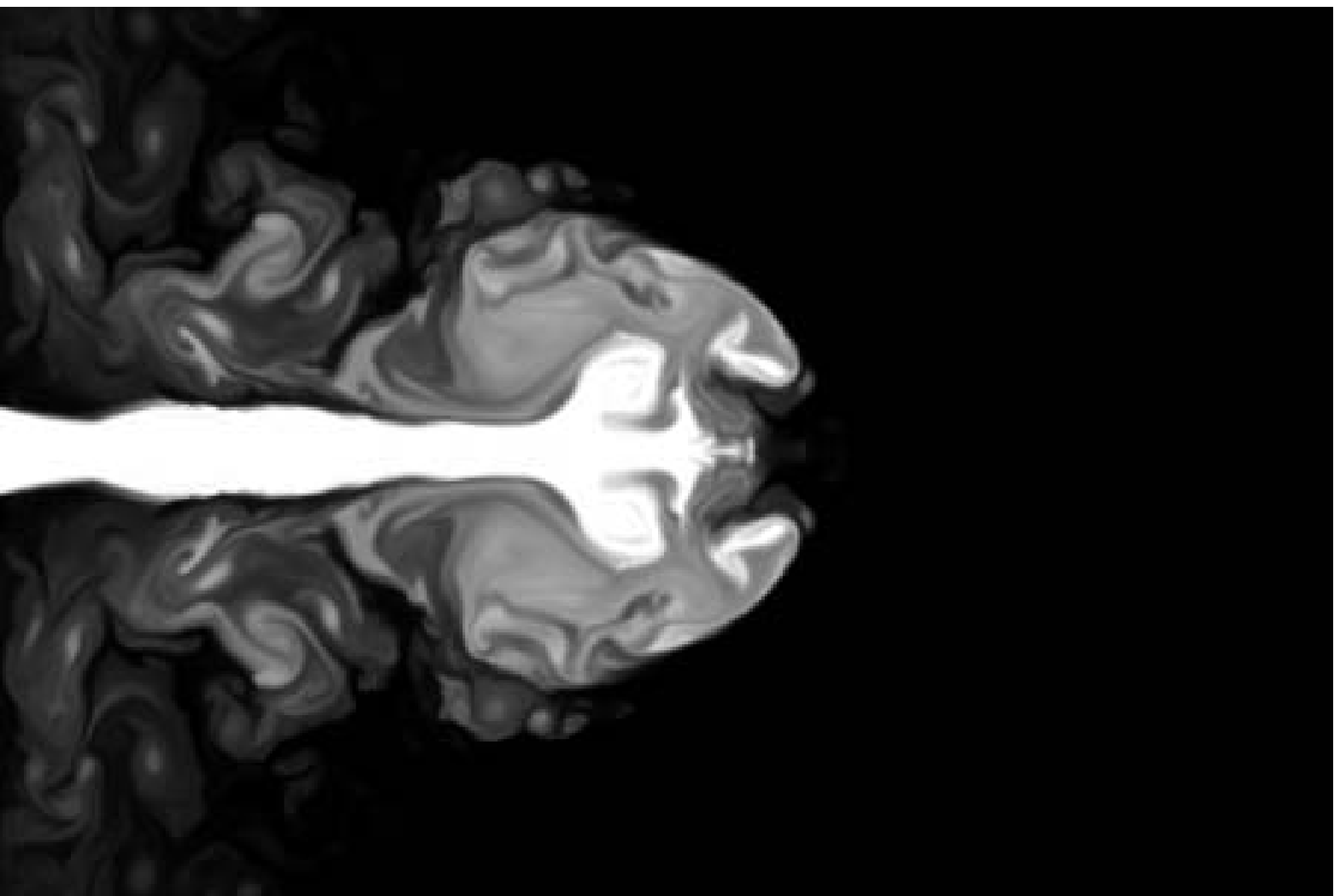}
&
\includegraphics[width=7cm]{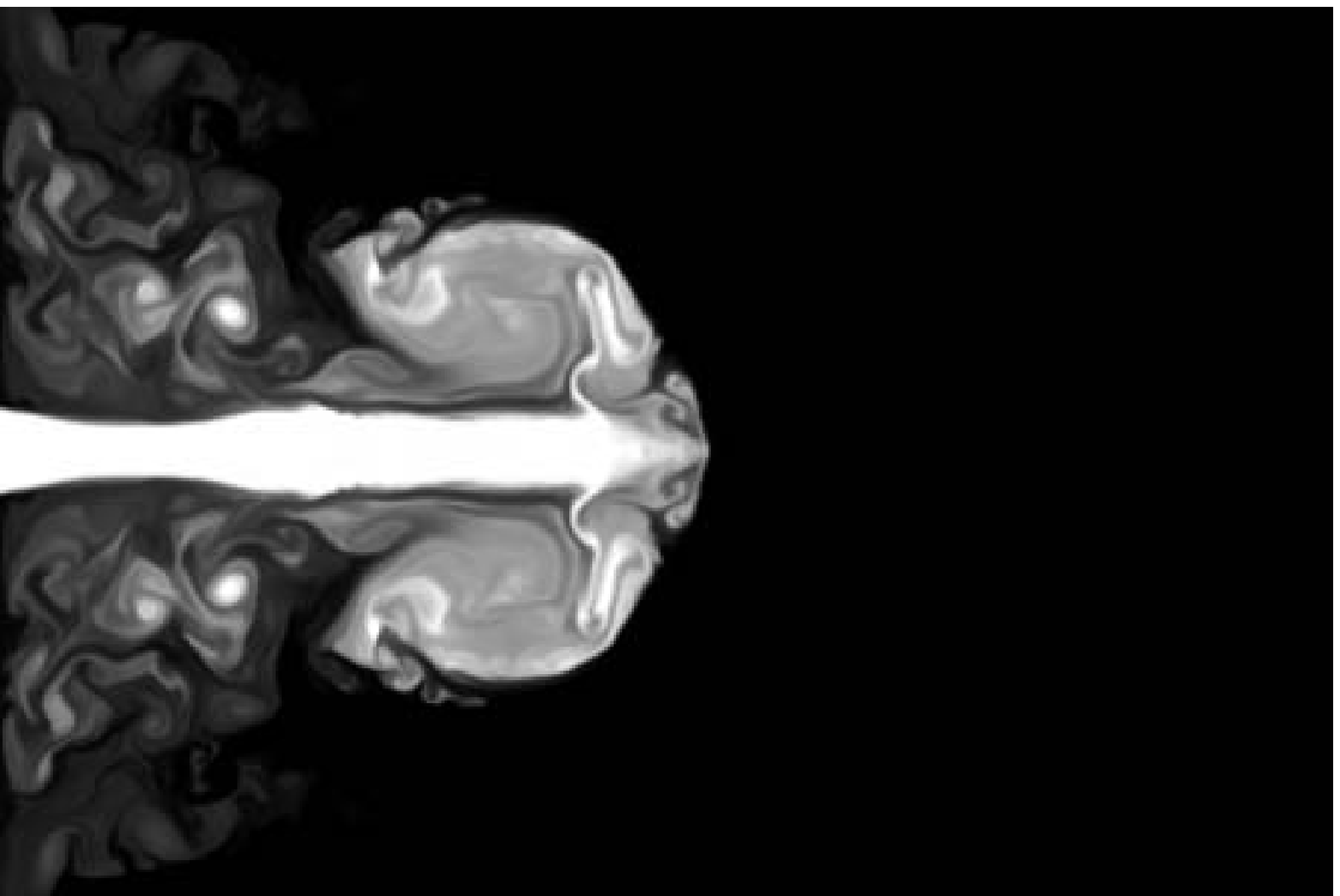}
\\
\includegraphics[width=7cm]{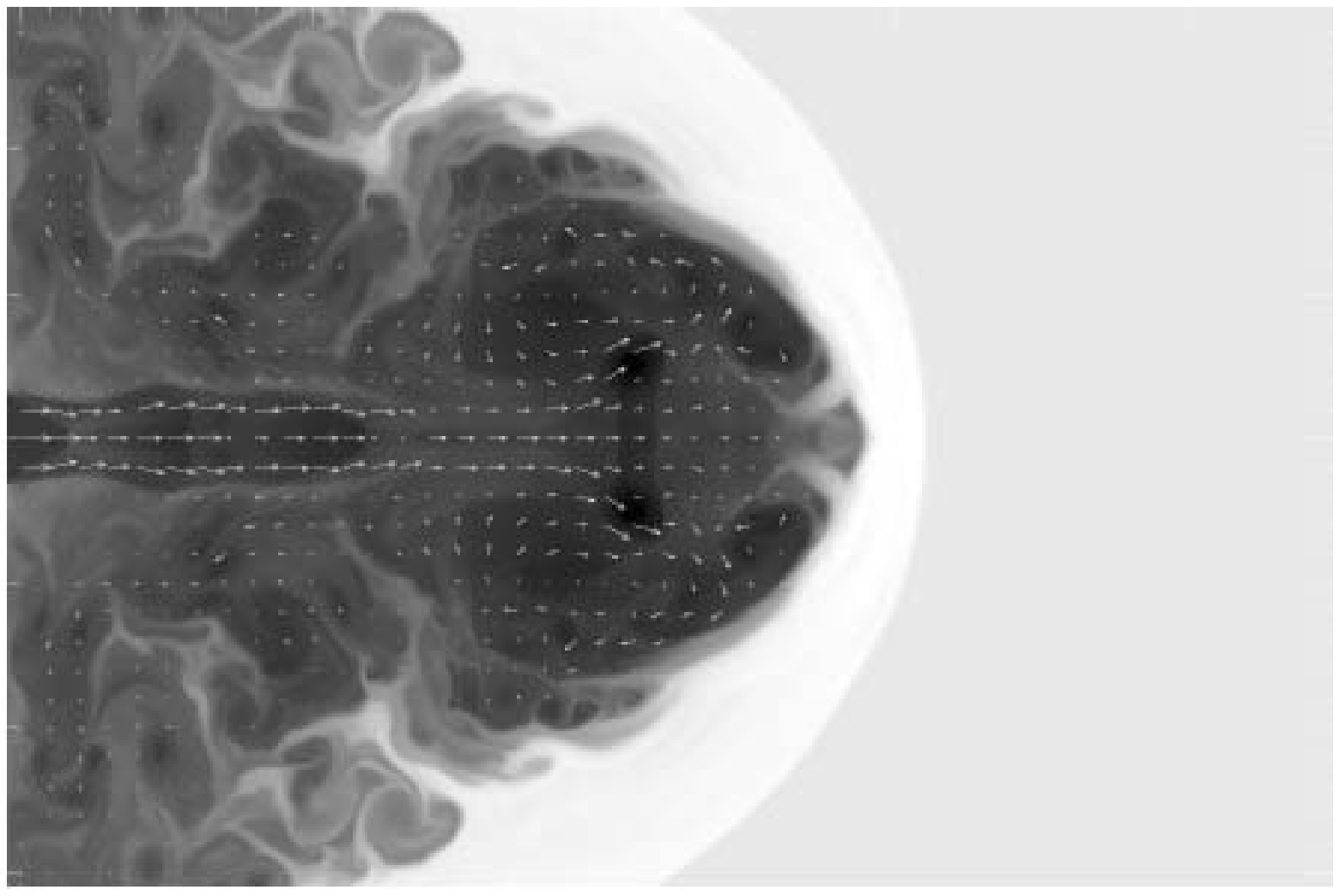}
&
\includegraphics[width=7cm]{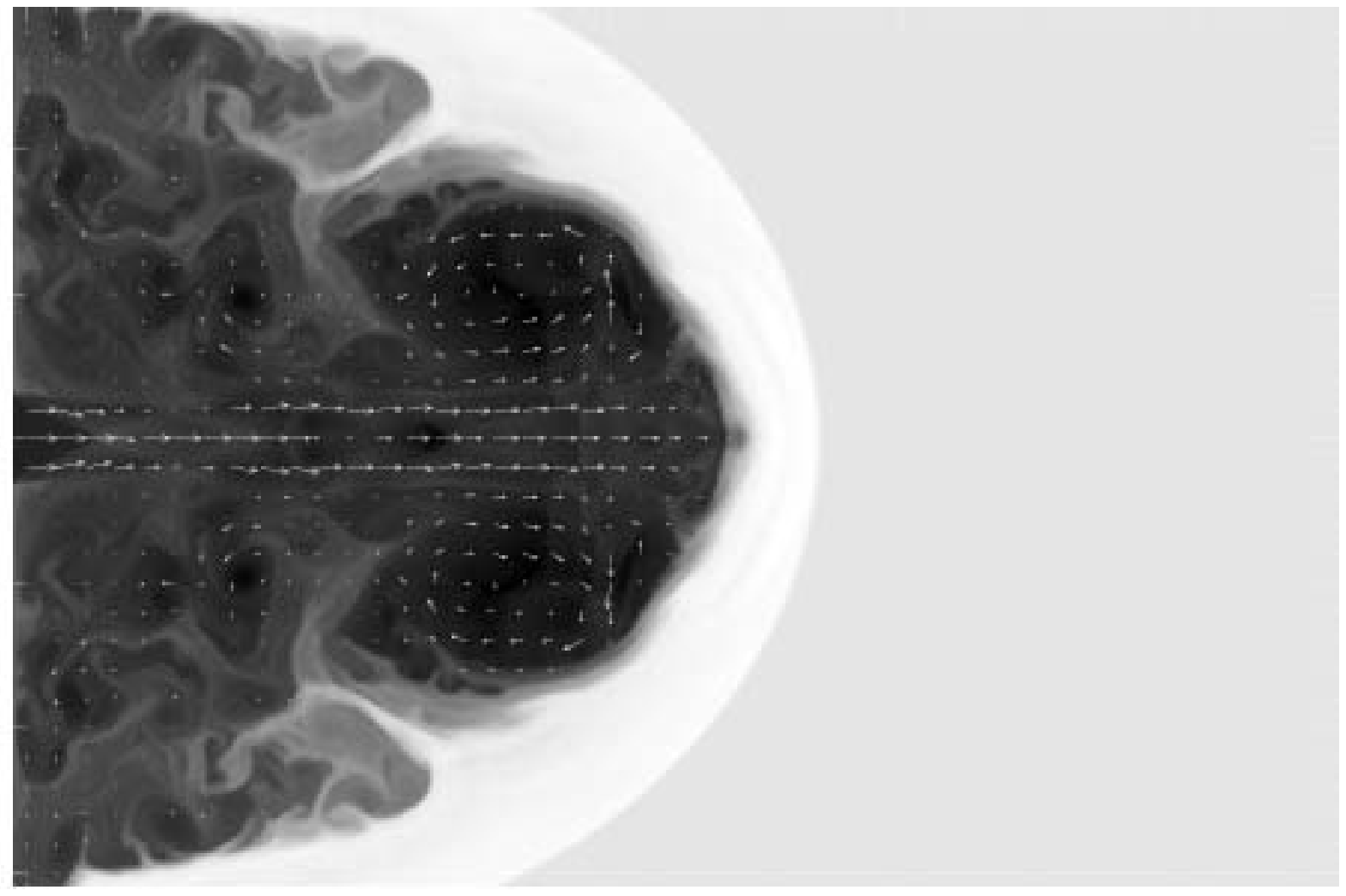}
\\
\end{array}$
\caption{
Further examples of ray-traced frames resembling Pictor~A,
with jet parameters 
$(\eta,M)=(10^{-4},5)$.
As in Figure~\ref{f:pageant-4m5.1},
the rows from top to bottom show
renderings, pressure, the tracer $\varphi$ and flow pattern.
}
\label{f:pageant-4m5.4}
\label{f:example.04}
\label{f:example.08}
\end{figure}

\begin{figure}[h]
\centering \leavevmode
$\begin{array}{cc}
\includegraphics[width=7cm]{fine-4mx_000f0286.eps}
&
\includegraphics[width=7cm]{fine-4mx_000f2417.eps}
\\
\includegraphics[width=7cm]{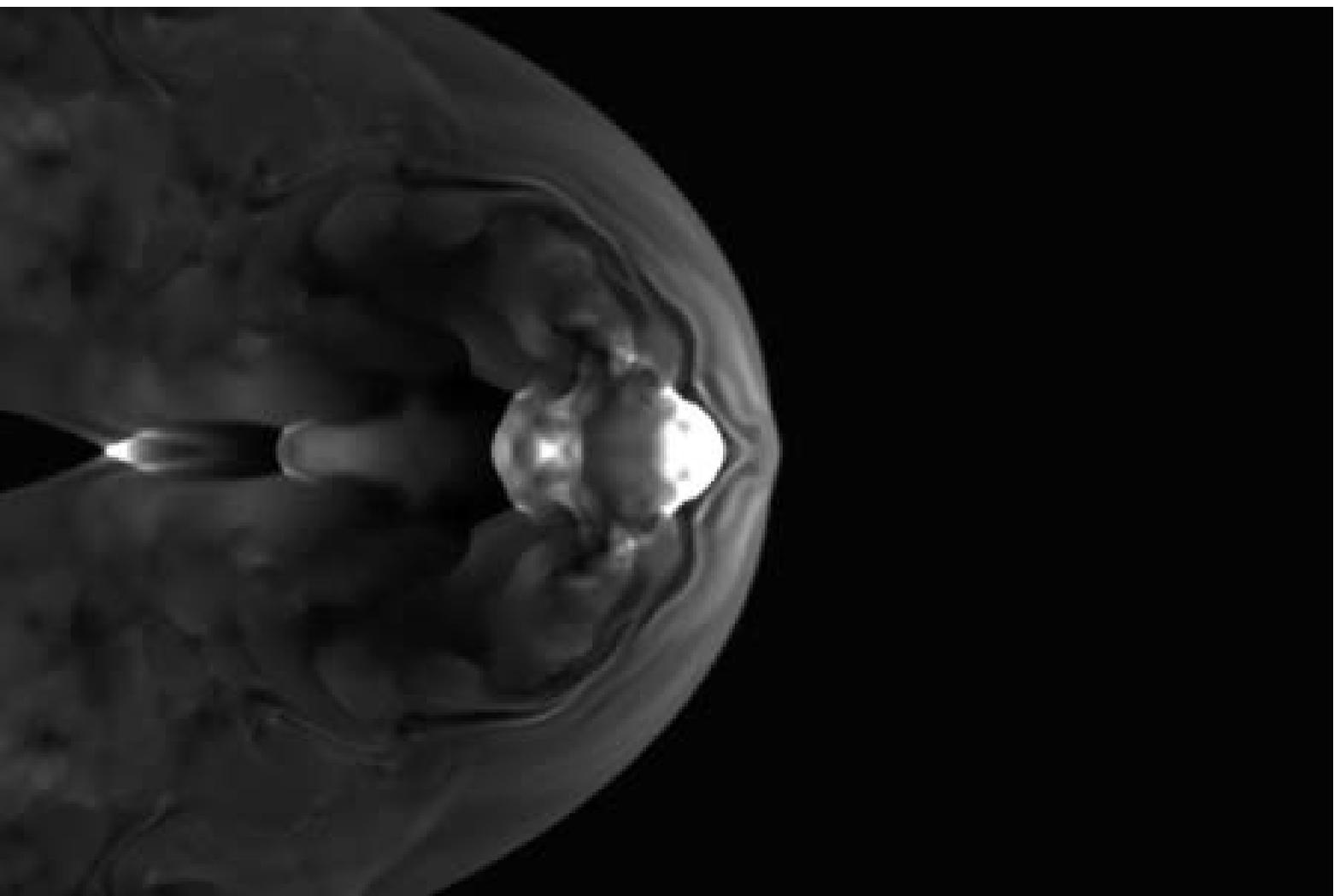}
&
\includegraphics[width=7cm]{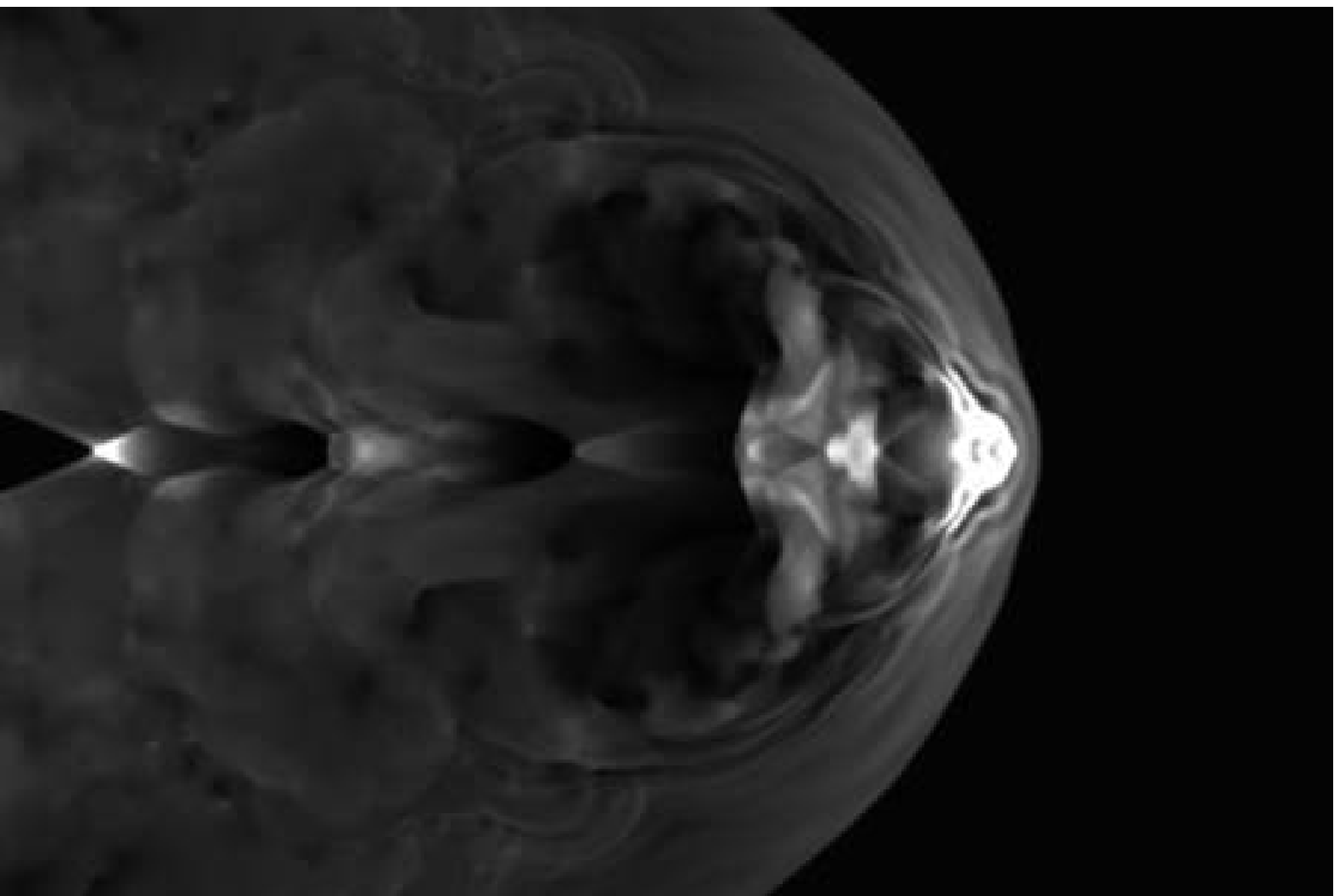}
\\
\includegraphics[width=7cm]{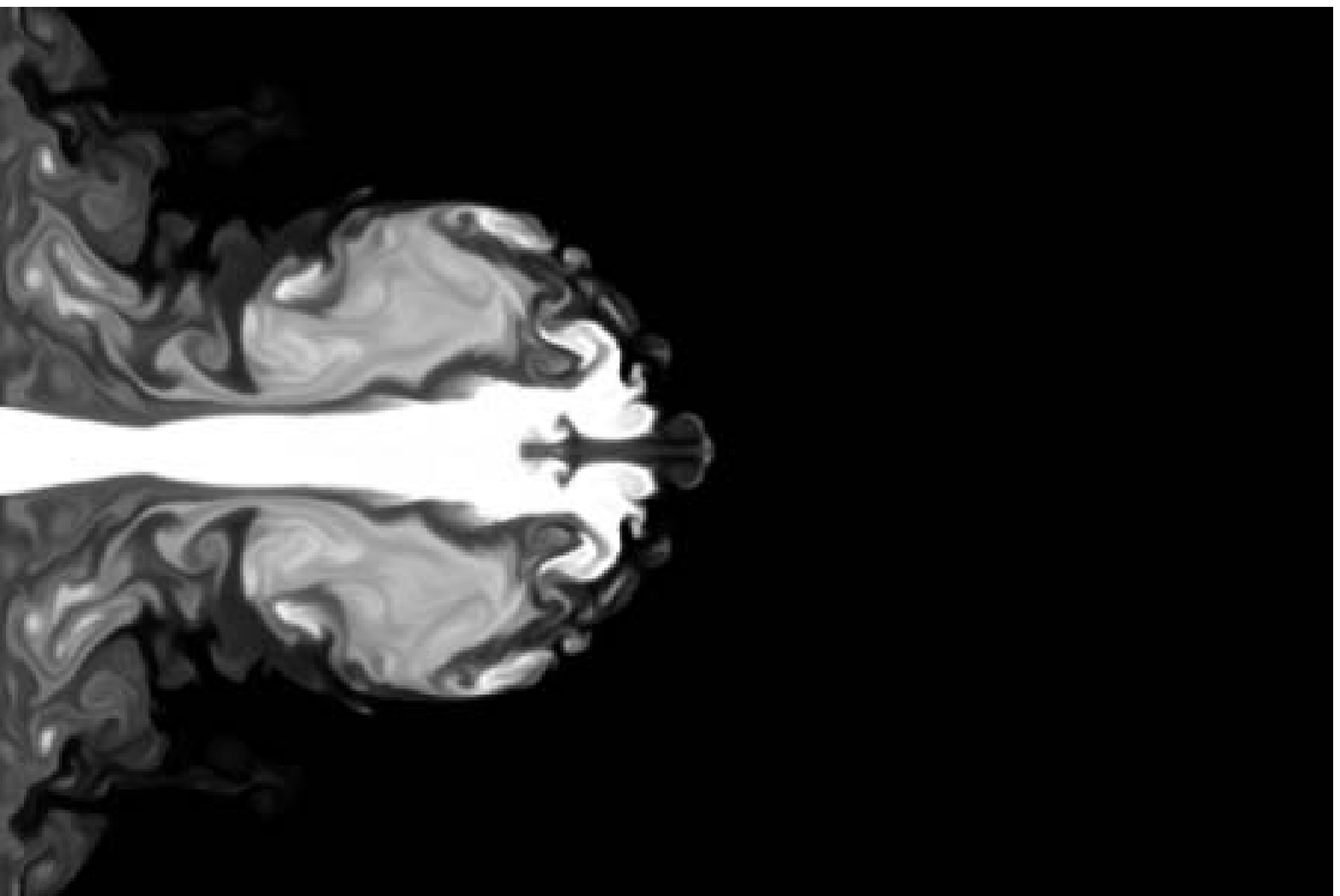}
&
\includegraphics[width=7cm]{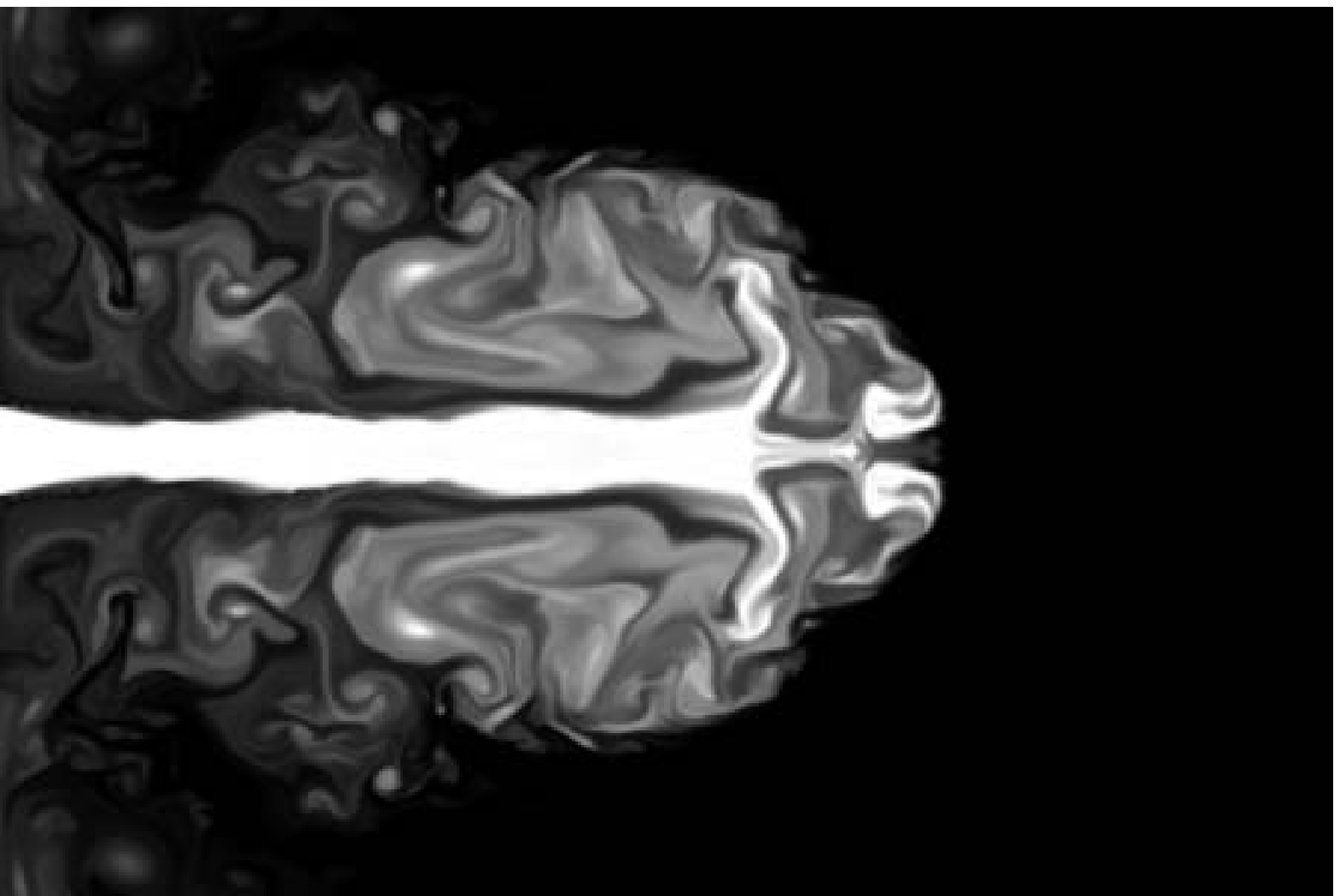}
\\
\includegraphics[width=7cm]{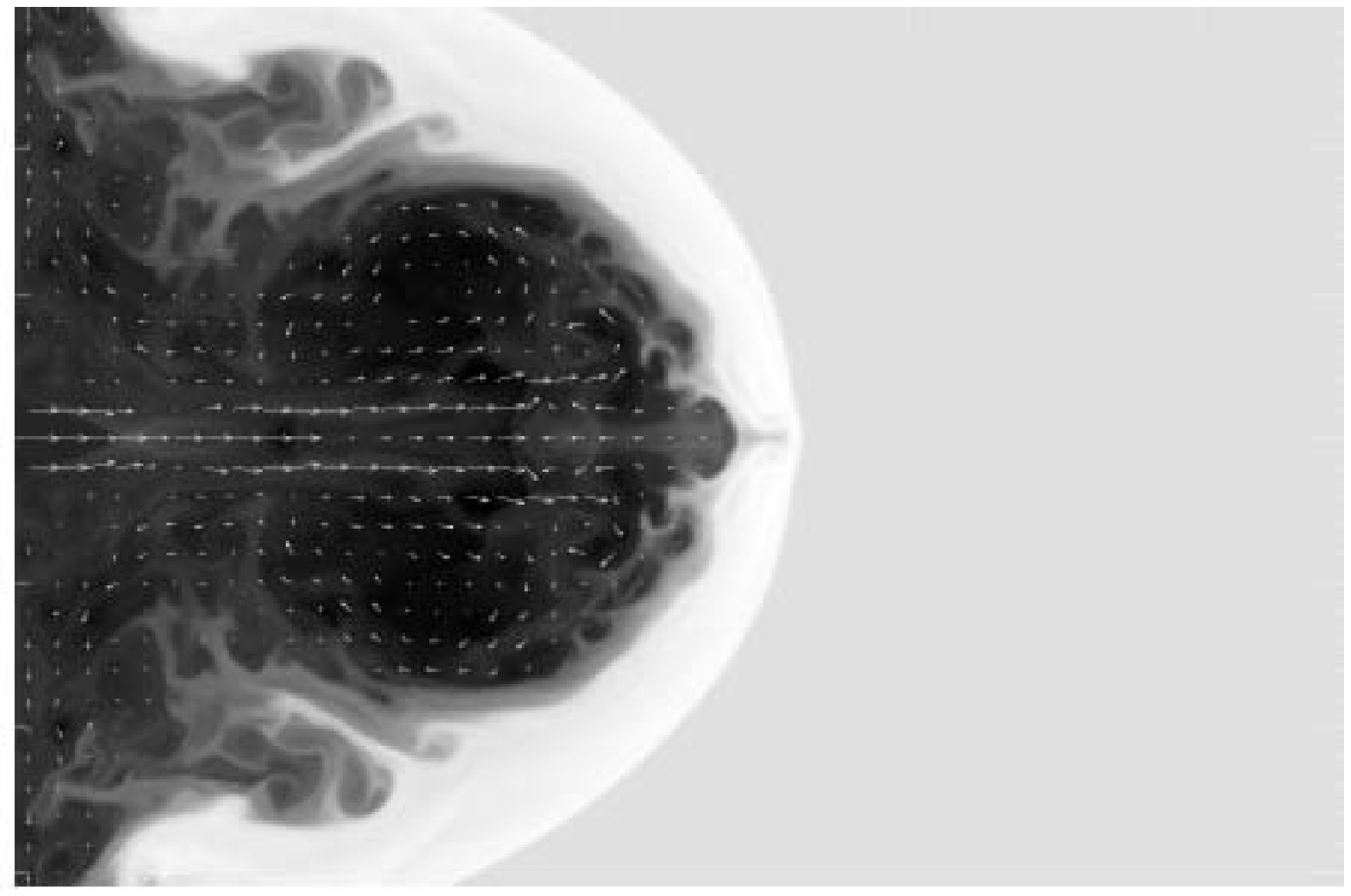}
&
\includegraphics[width=7cm]{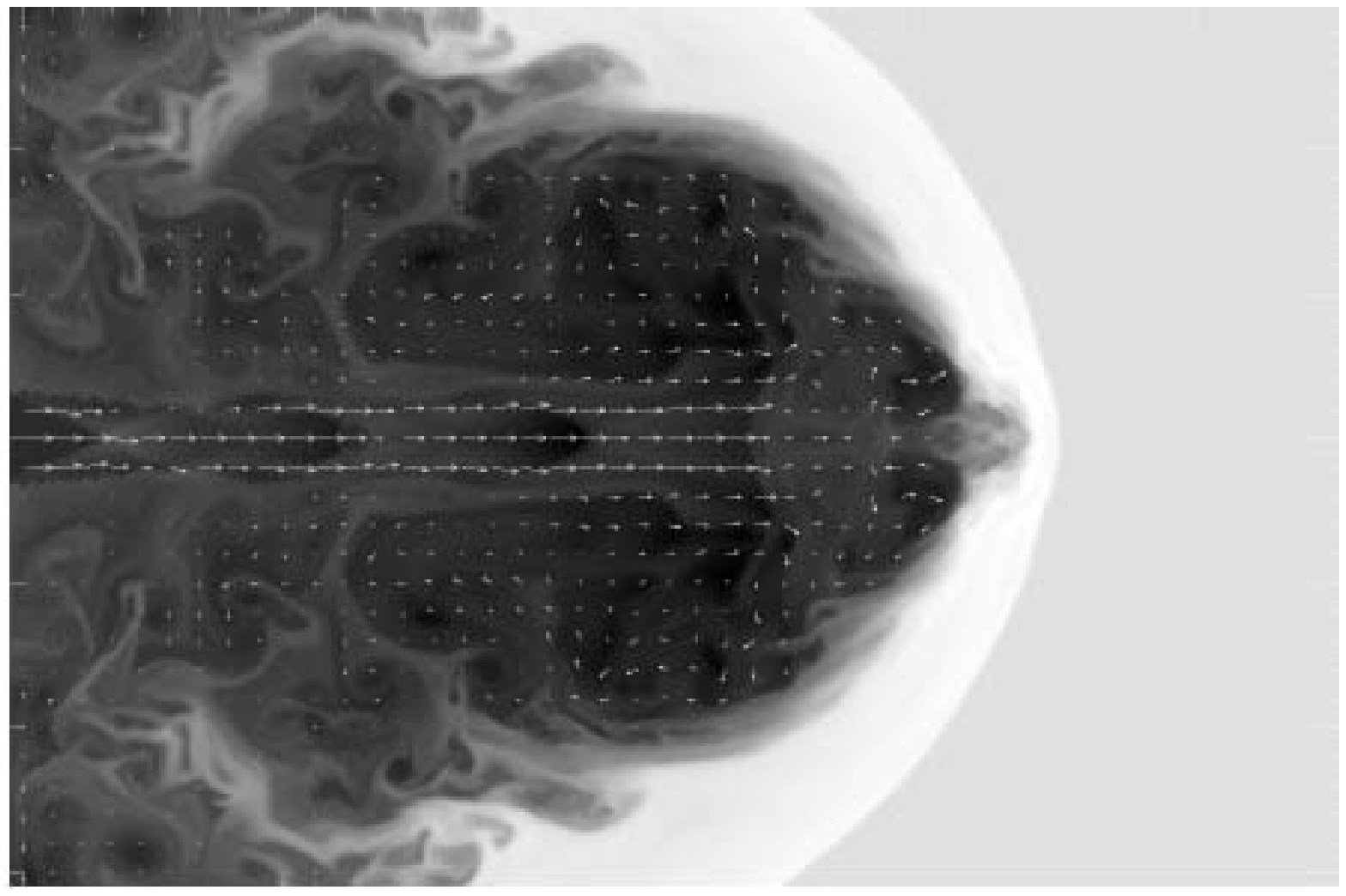}
\\
\end{array}$
\caption{
Further examples of ray-traced frames resembling Pictor~A,
with jet parameters 
$(\eta,M)=(10^{-4},10)$.
As in Figure~\ref{f:pageant-4m5.1},
the rows from top to bottom show
renderings, pressure, the tracer $\varphi$ and flow pattern.
}
\label{f:pageant-4mx.1}
\label{f:example.09}
\label{f:example.10}
\end{figure}

\begin{figure}[h]
\centering \leavevmode
$\begin{array}{cc}
\includegraphics[width=7cm]{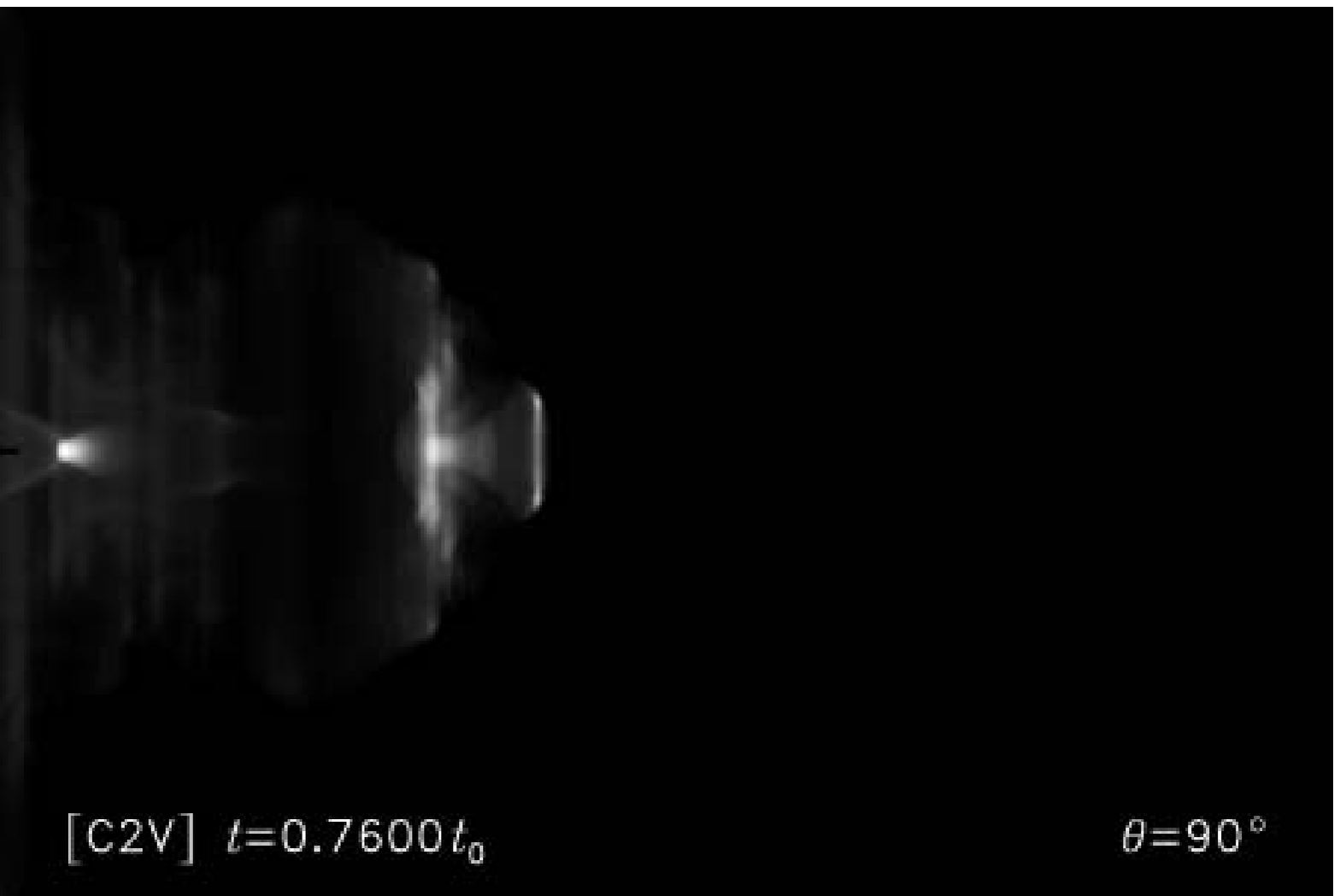}
&
\includegraphics[width=7cm]{jet-2m5_000f0226.eps}
\\
\includegraphics[width=7cm]{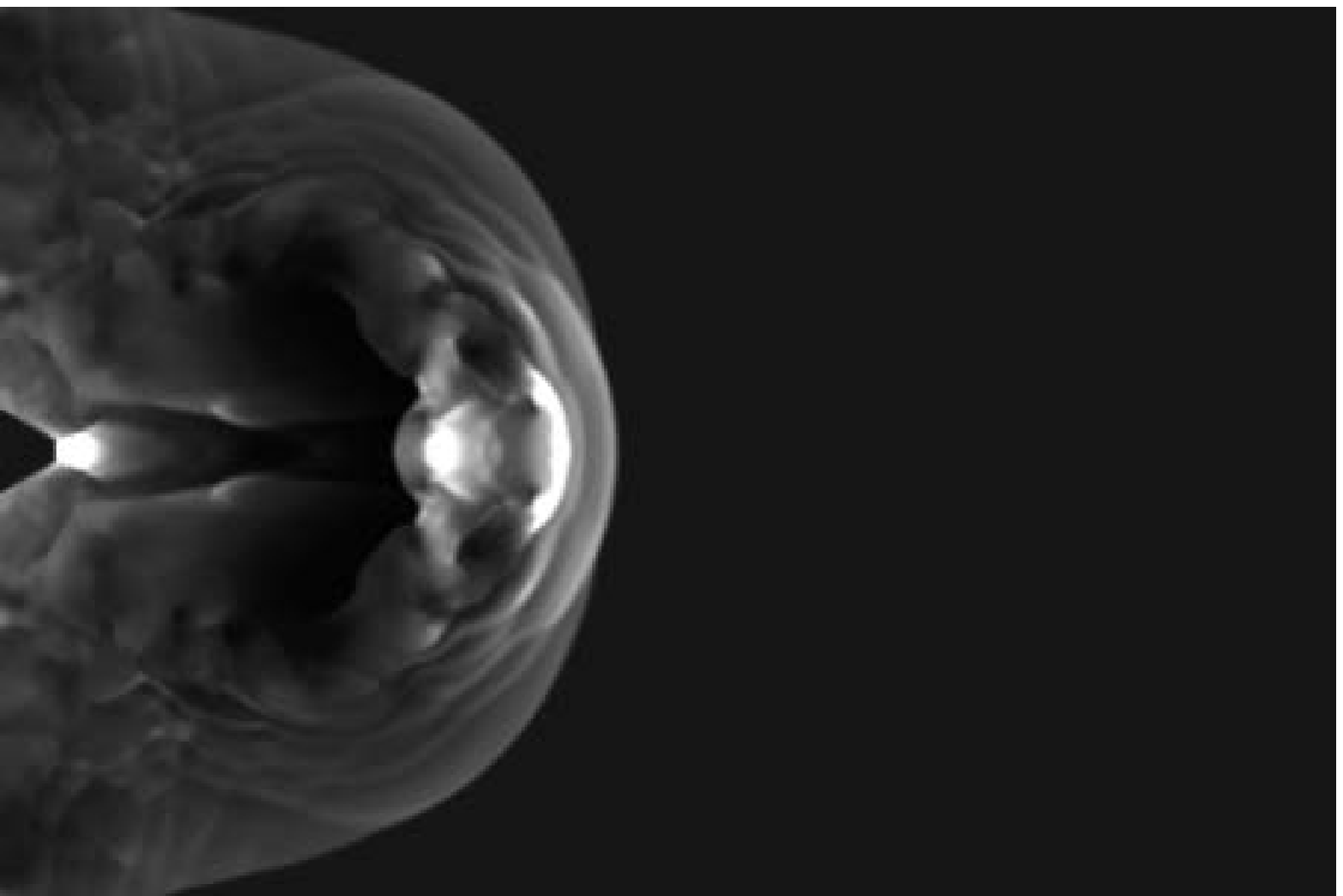}
&
\includegraphics[width=7cm]{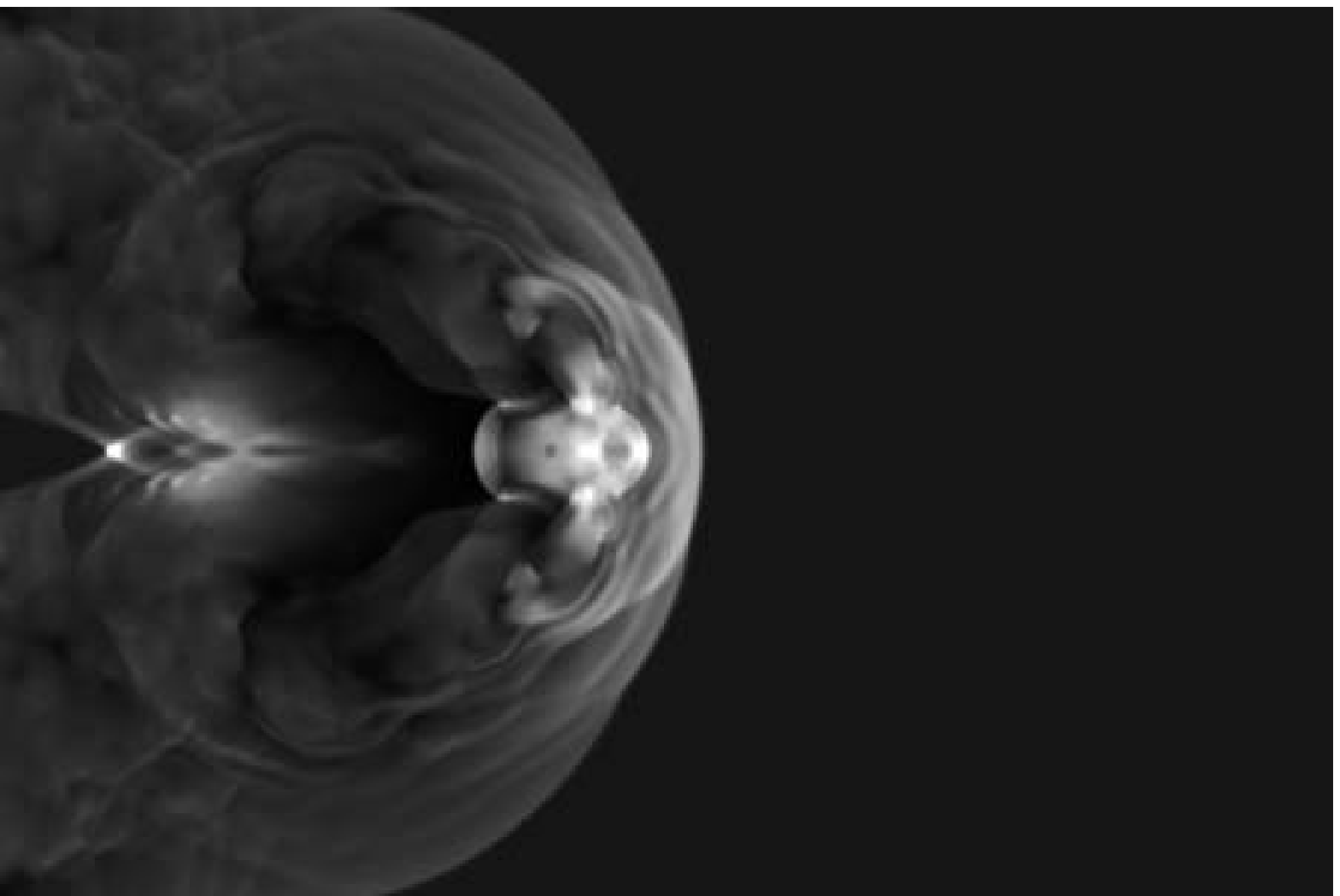}
\\
\includegraphics[width=7cm]{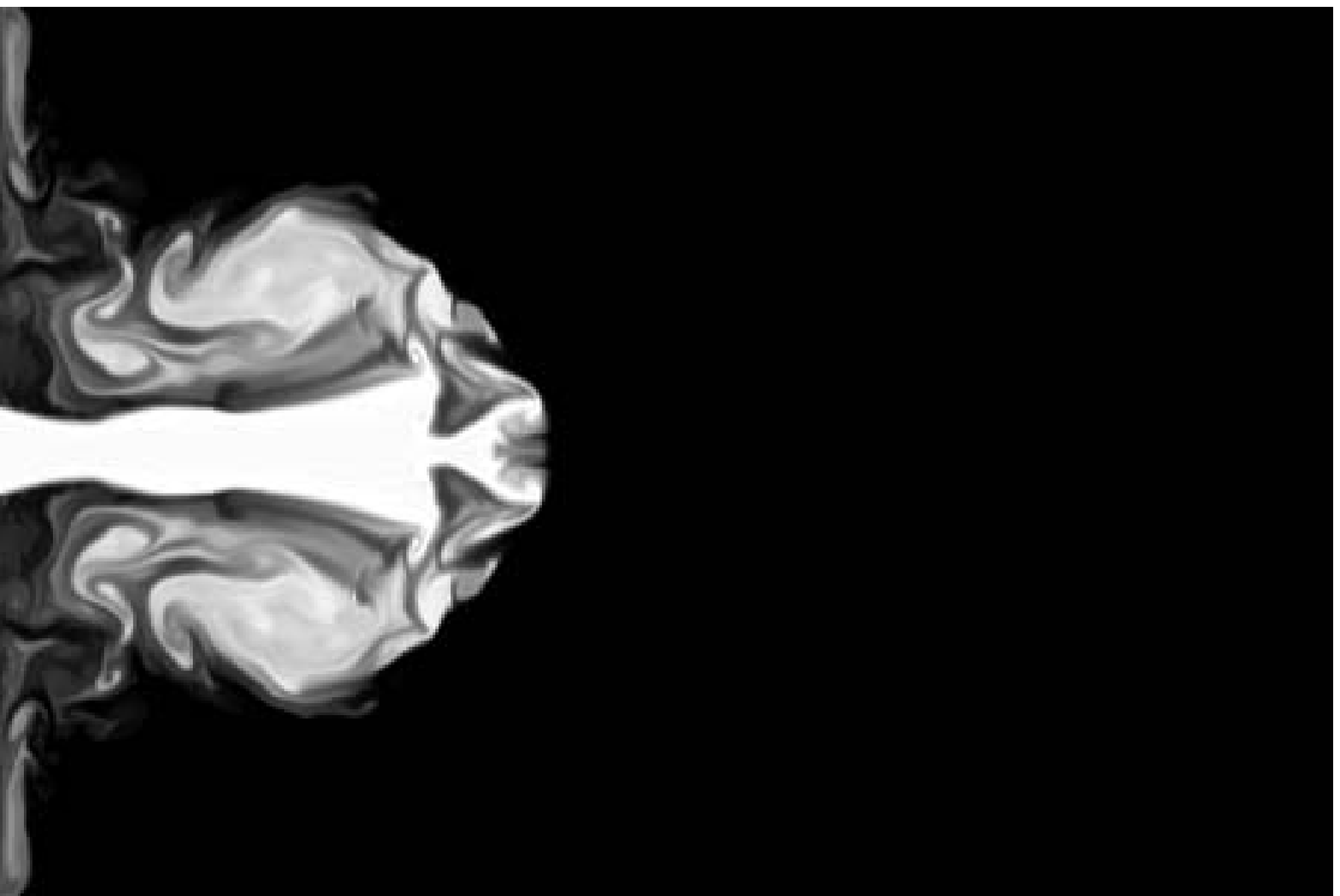}
&
\includegraphics[width=7cm]{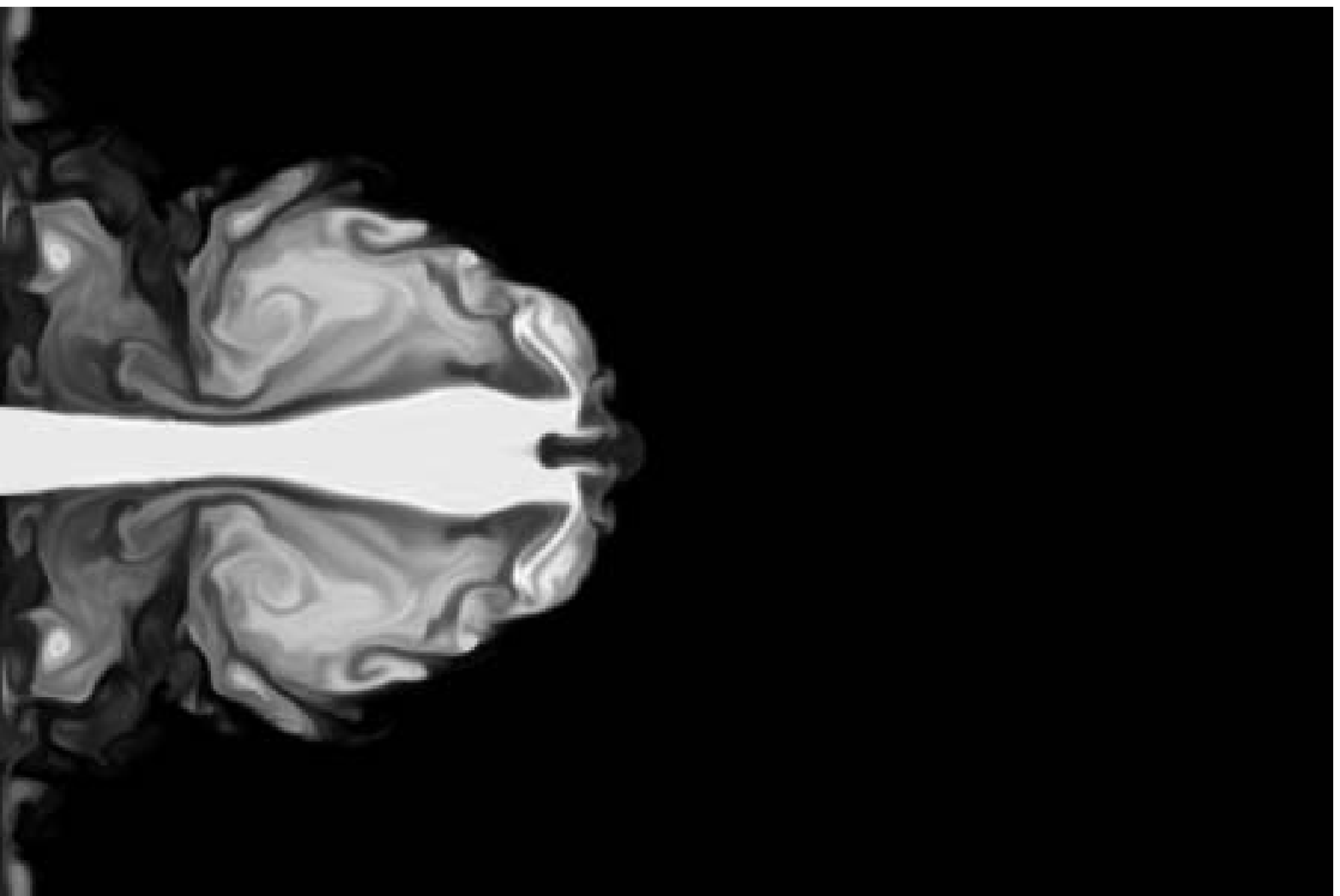}
\\
\includegraphics[width=7cm]{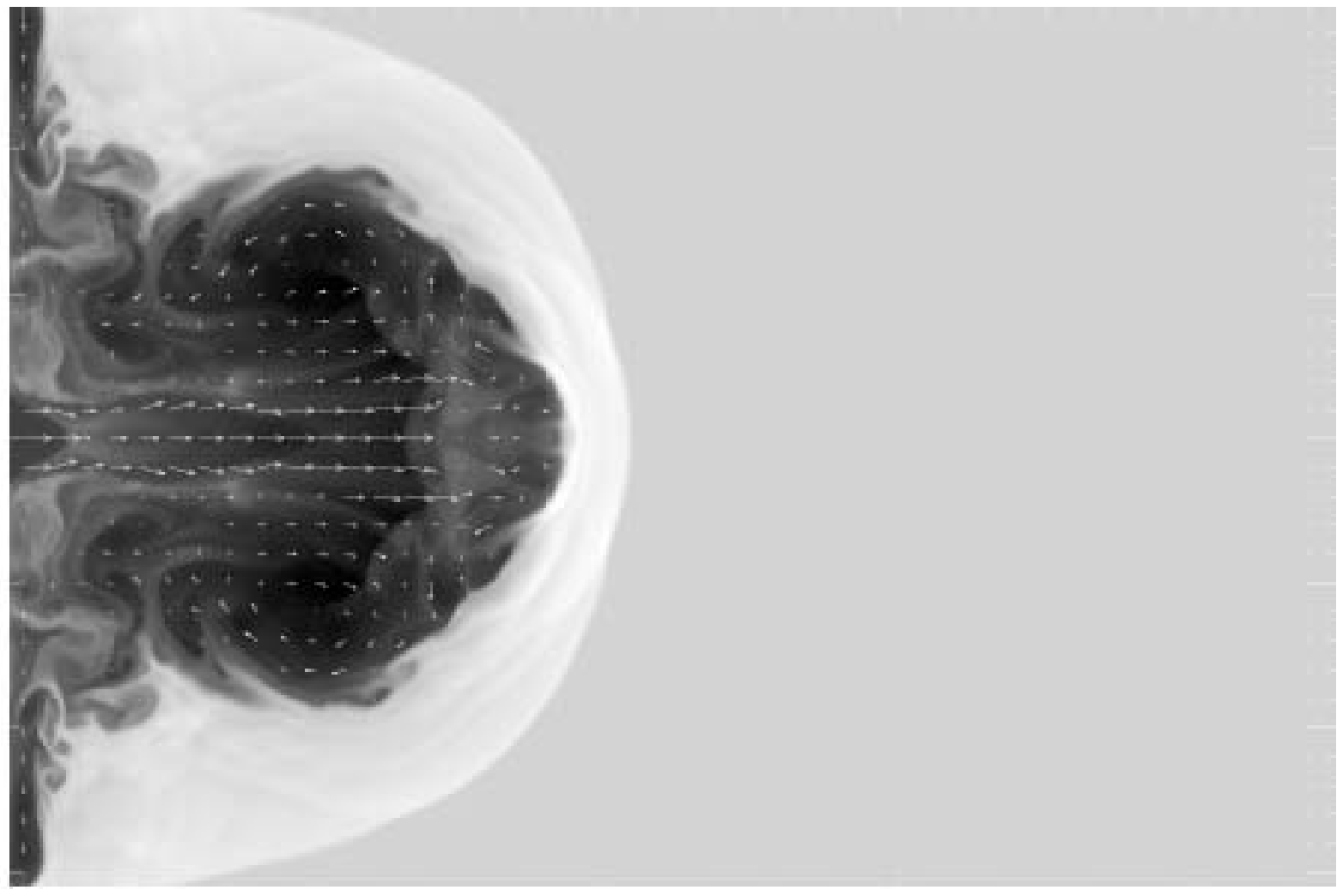}
&
\includegraphics[width=7cm]{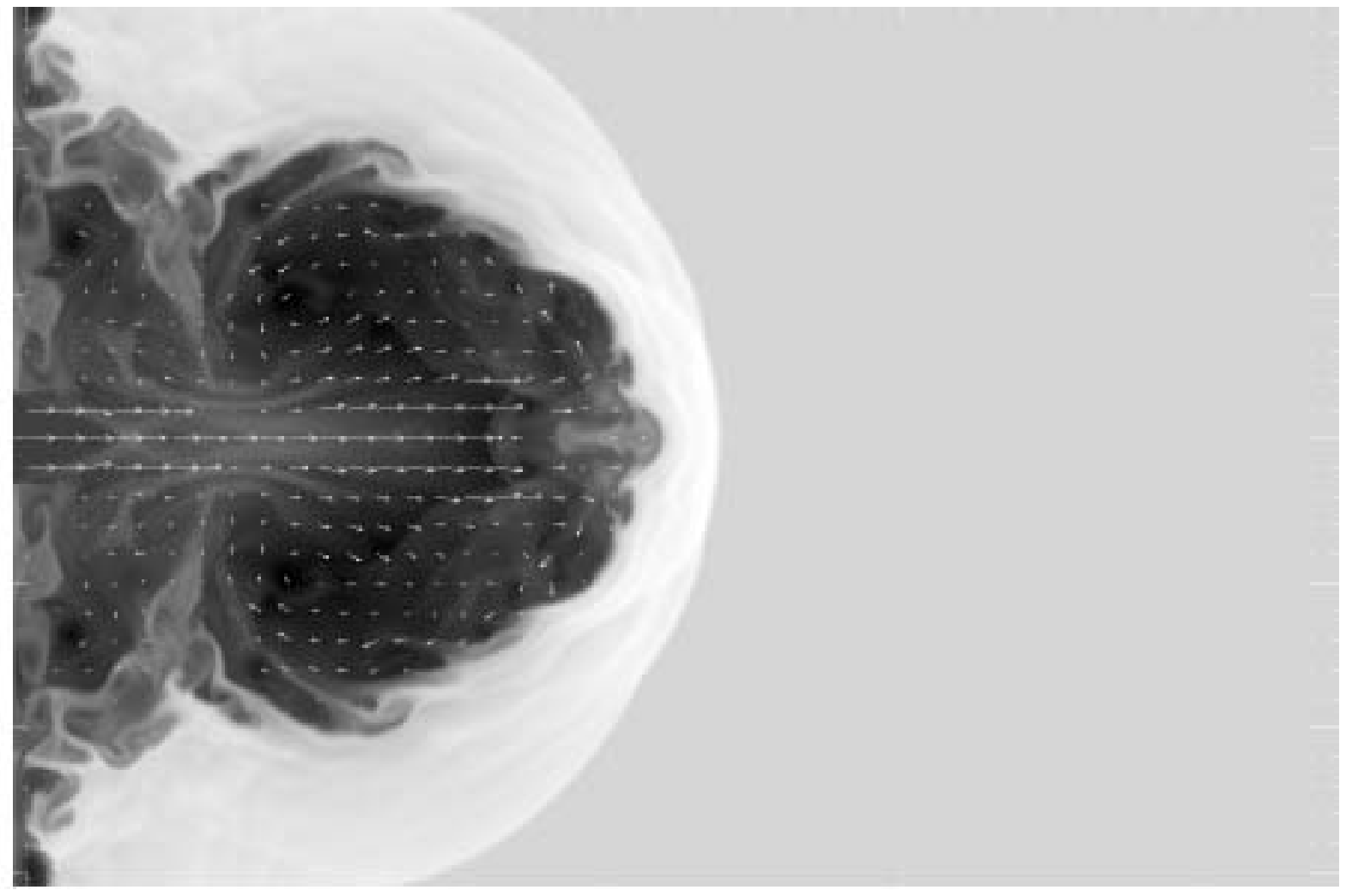}
\\
\end{array}$
\caption{
$450 \times 300$ pixel sub-region
of a jet with parameters 
$(\eta,M)=(10^{-2},5)$.
As in Figure~\ref{f:pageant-4m5.1},
the rows from top to bottom show
renderings, pressure, the tracer $\varphi$ and flow pattern.
}
\label{f:example.13}
\label{f:example.14}
\label{f:pageant-2m5}
\end{figure}

The reproduction of images resembling observations of
the western hot-spot of Pictor~A requires
a strong shock on the jet axis,
and some other, wider shock structure
in high-$\varphi$ regions of the backflow.
However more than one kind of physical situation
can produce an appropriate image.
Here we consider the particular
instantaneous arrangements of shocks and jet plasma
that yield some of the example images selected and displayed in
Figures~\ref{f:best_morphologies-4m5}-\ref{f:best_morphologies-2ml}.
Examples of the related dynamics are given in 
Figures~\ref{f:example.01}-\ref{f:example.14},
in which the panels show simulated intensity maps
along with corresponding snapshots of 
pressure, $\varphi$ and density with overlaid velocity vectors. 
(In referring to these figures 
we use L for left and R for right panels respectively.)

\begin{figure}
\centering \leavevmode
$\begin{array}{ccc}
\includegraphics[width=4.5cm]{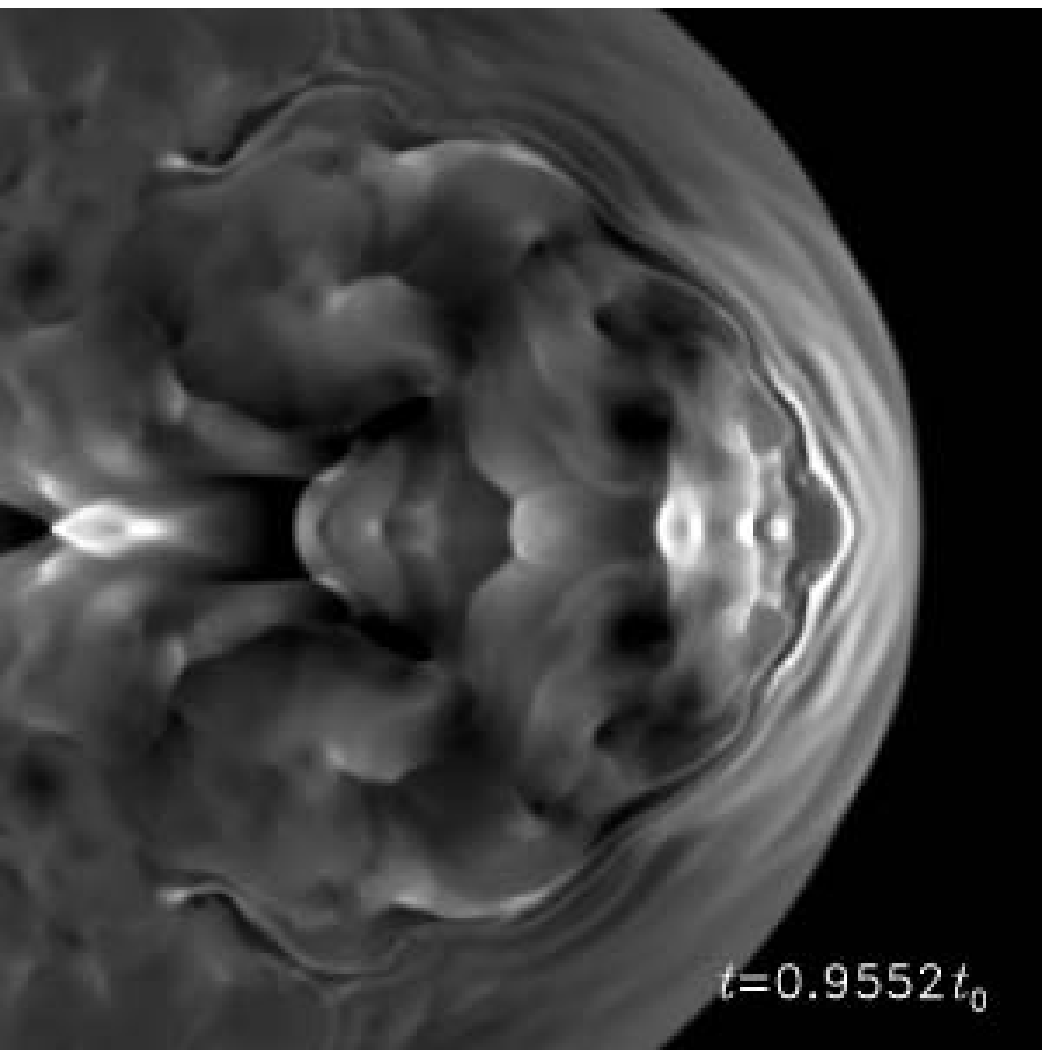}
&
\includegraphics[width=4.5cm]{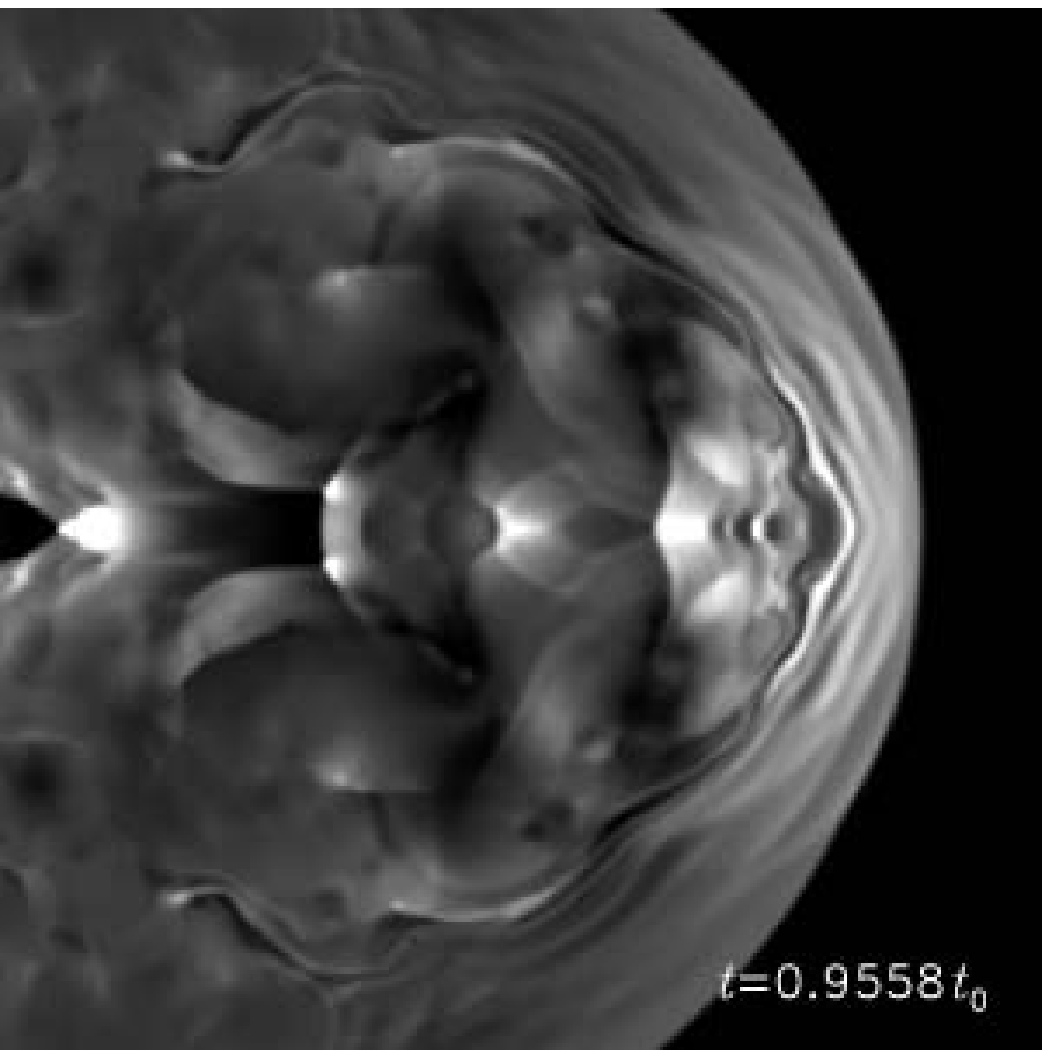}
&
\includegraphics[width=4.5cm]{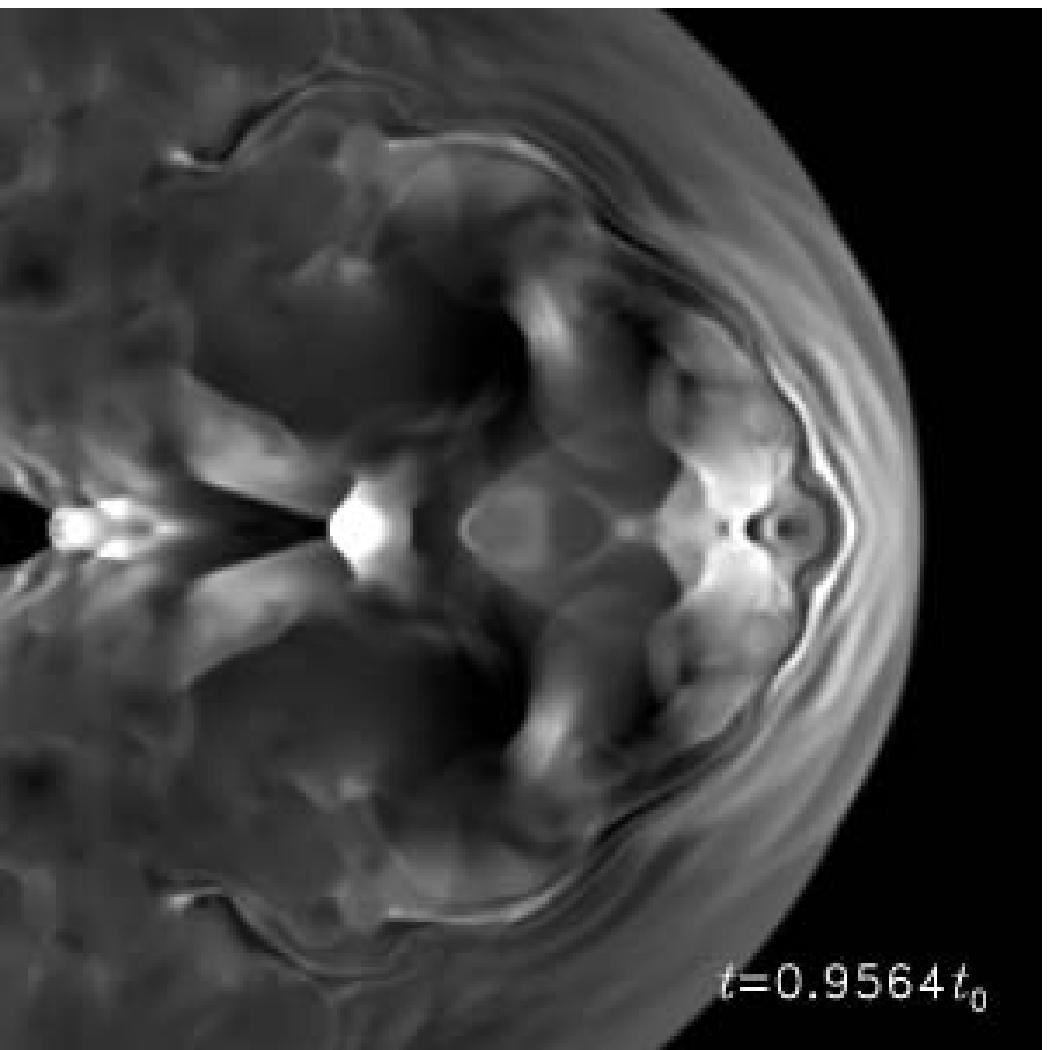}
\\
\includegraphics[width=4.5cm]{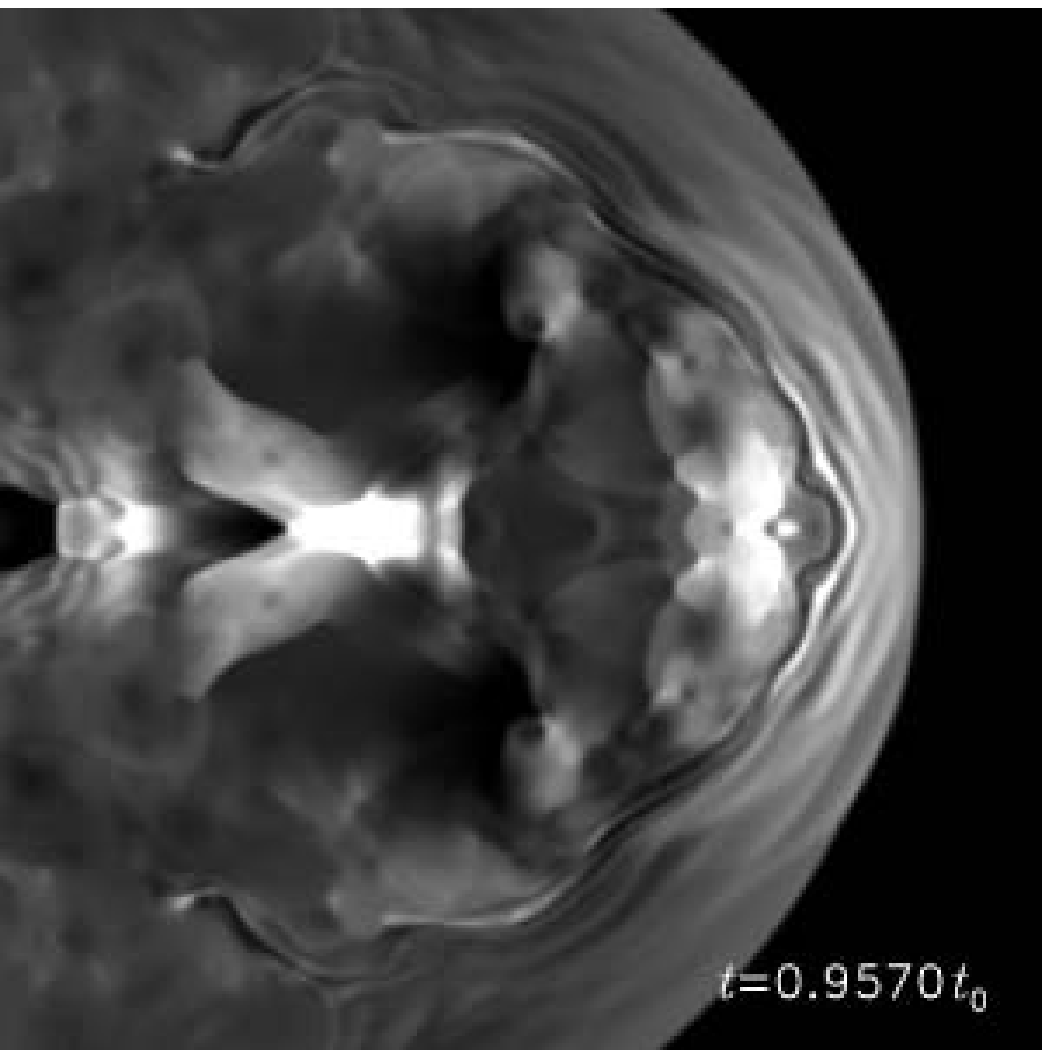}
&
\includegraphics[width=4.5cm]{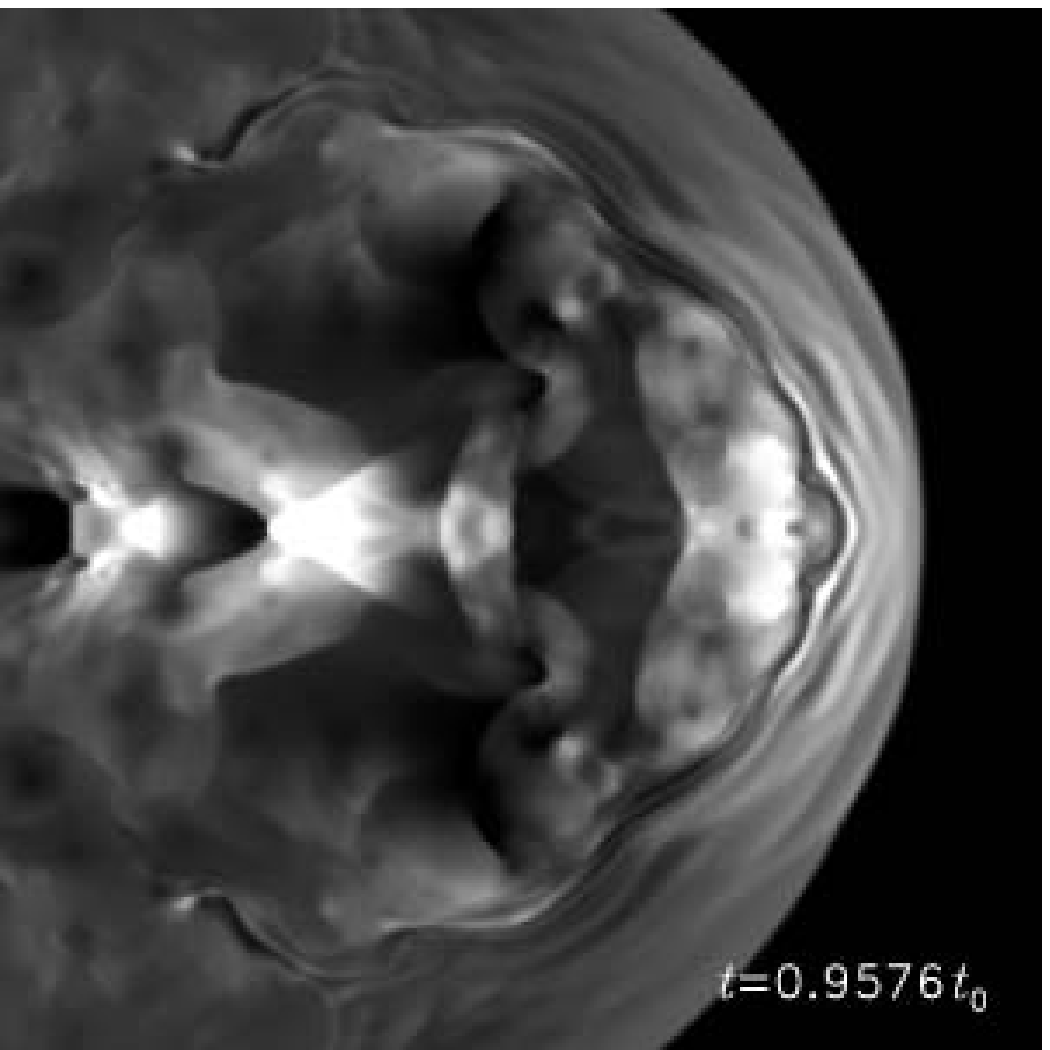}
&
\includegraphics[width=4.5cm]{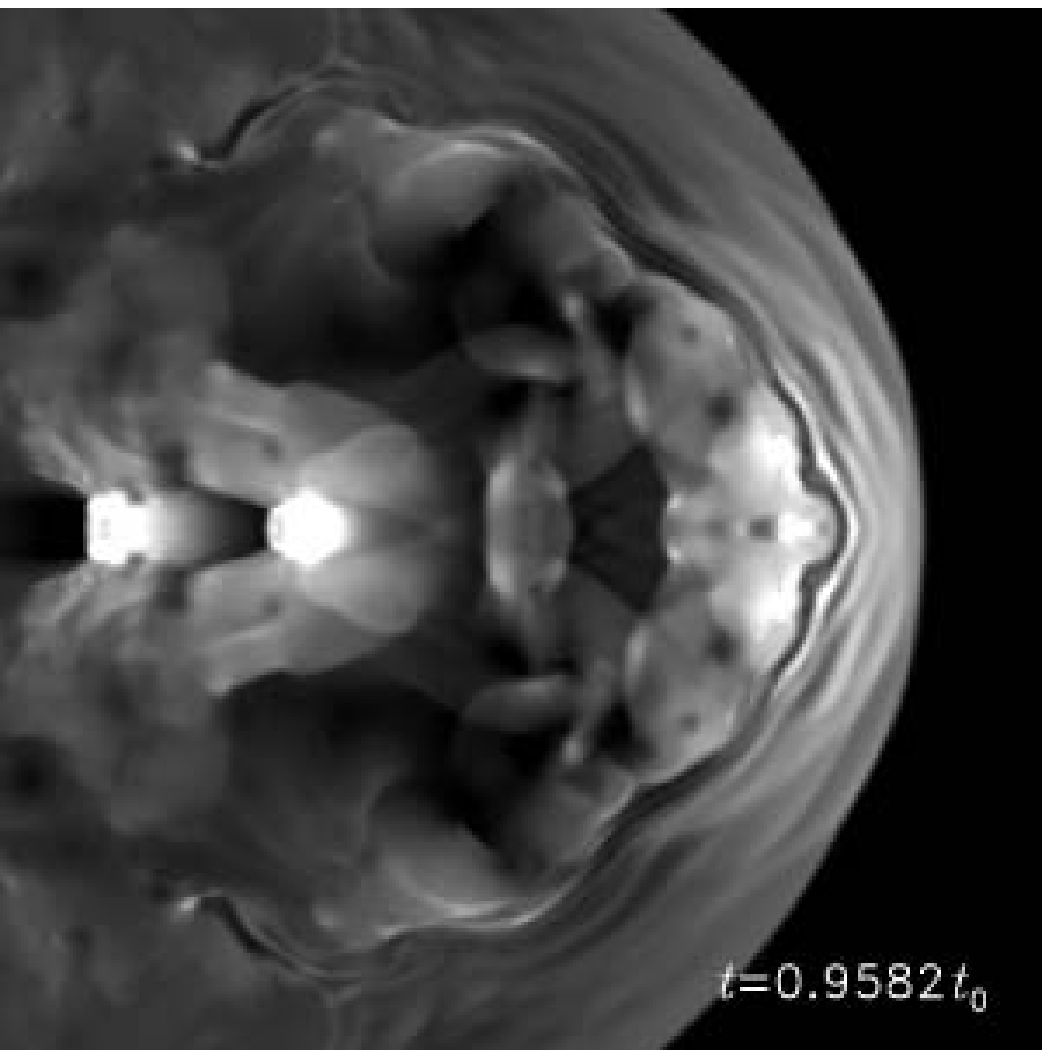}
\\
\end{array}$
\caption{
Sequence of pressure snapshots
during the emergence and disappearance of
a particular configuration of bar and hot-spot.
Jet parameters are $(10^{-4},5)$
and the left boundary is closed.
}
\label{f:sequence.pressure}
\end{figure}

\begin{figure}
\centering \leavevmode
$\begin{array}{ccc}
\includegraphics[width=4.5cm]{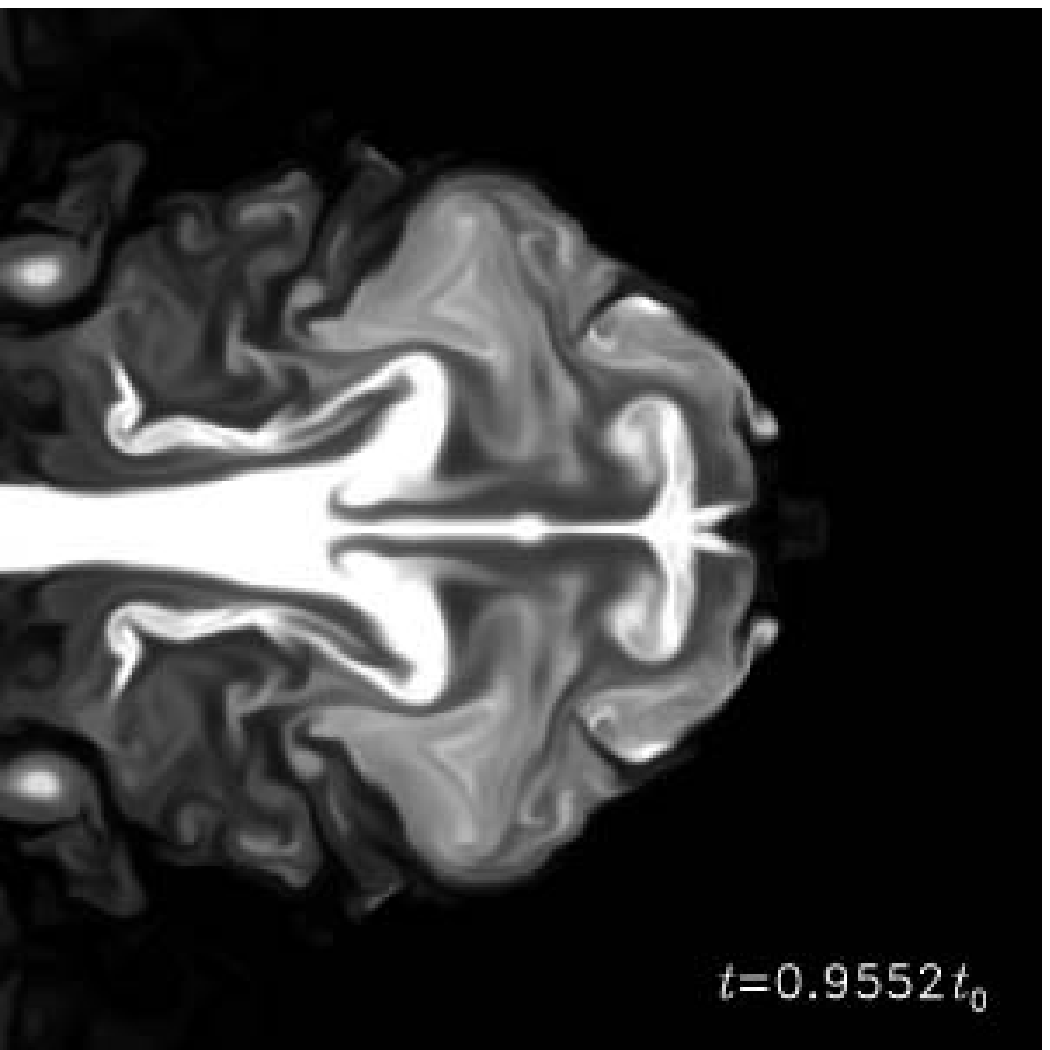}
&
\includegraphics[width=4.5cm]{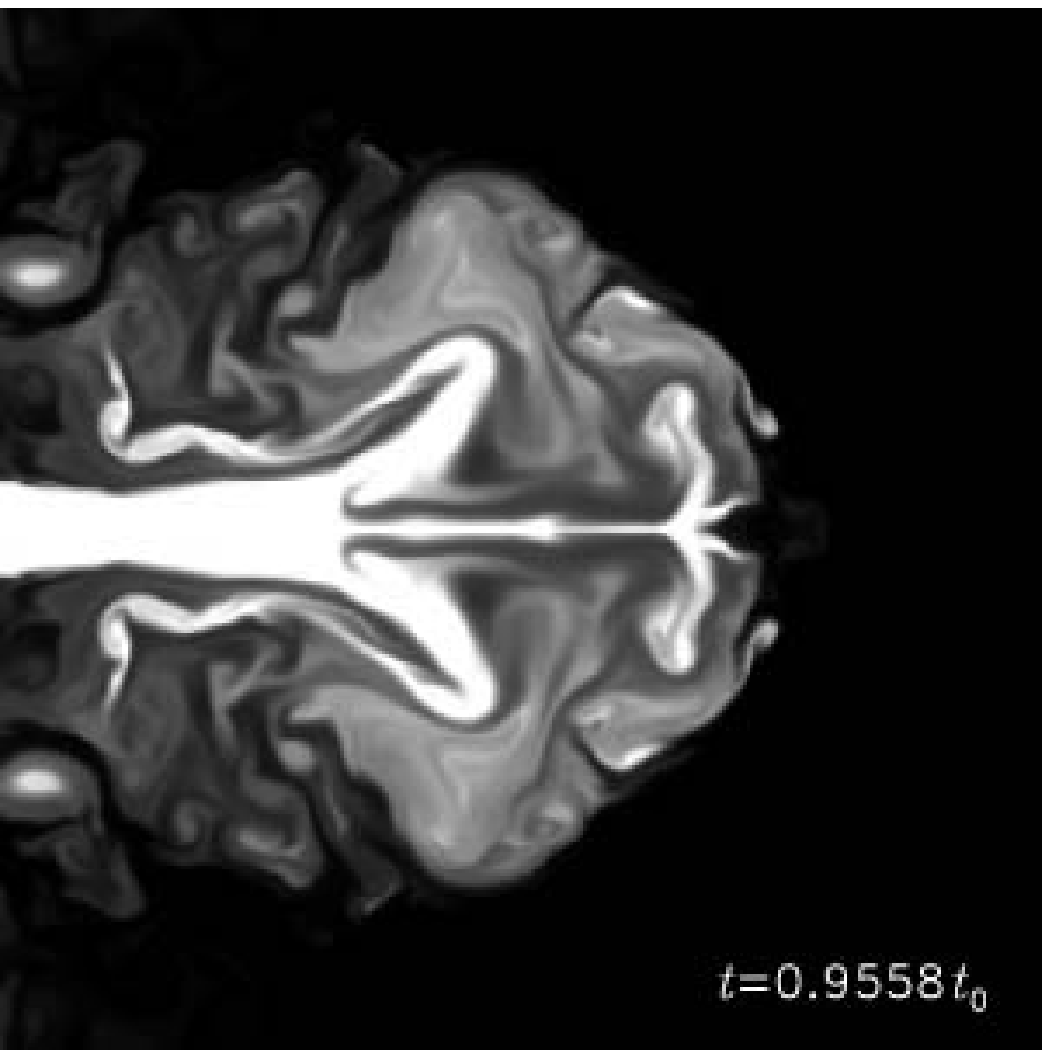}
&
\includegraphics[width=4.5cm]{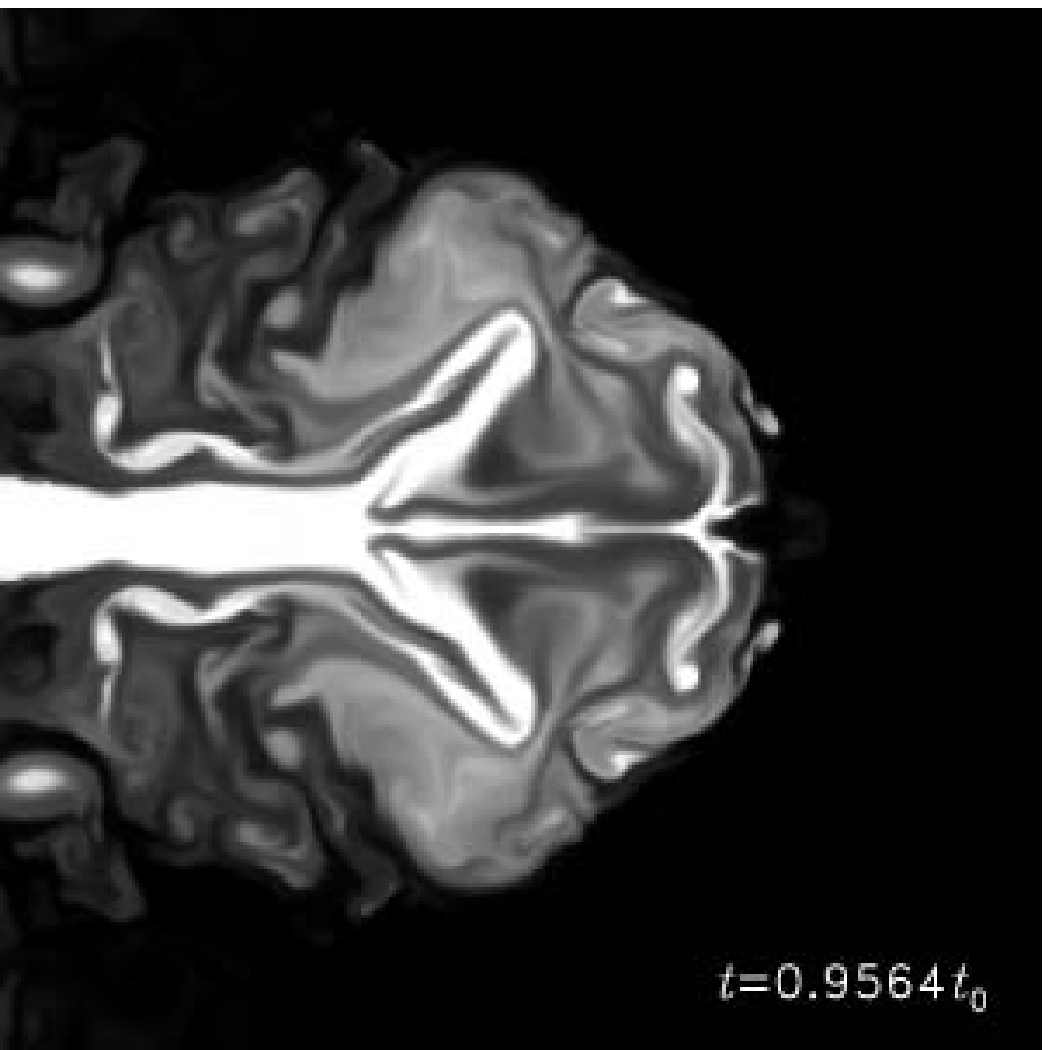}
\\
\includegraphics[width=4.5cm]{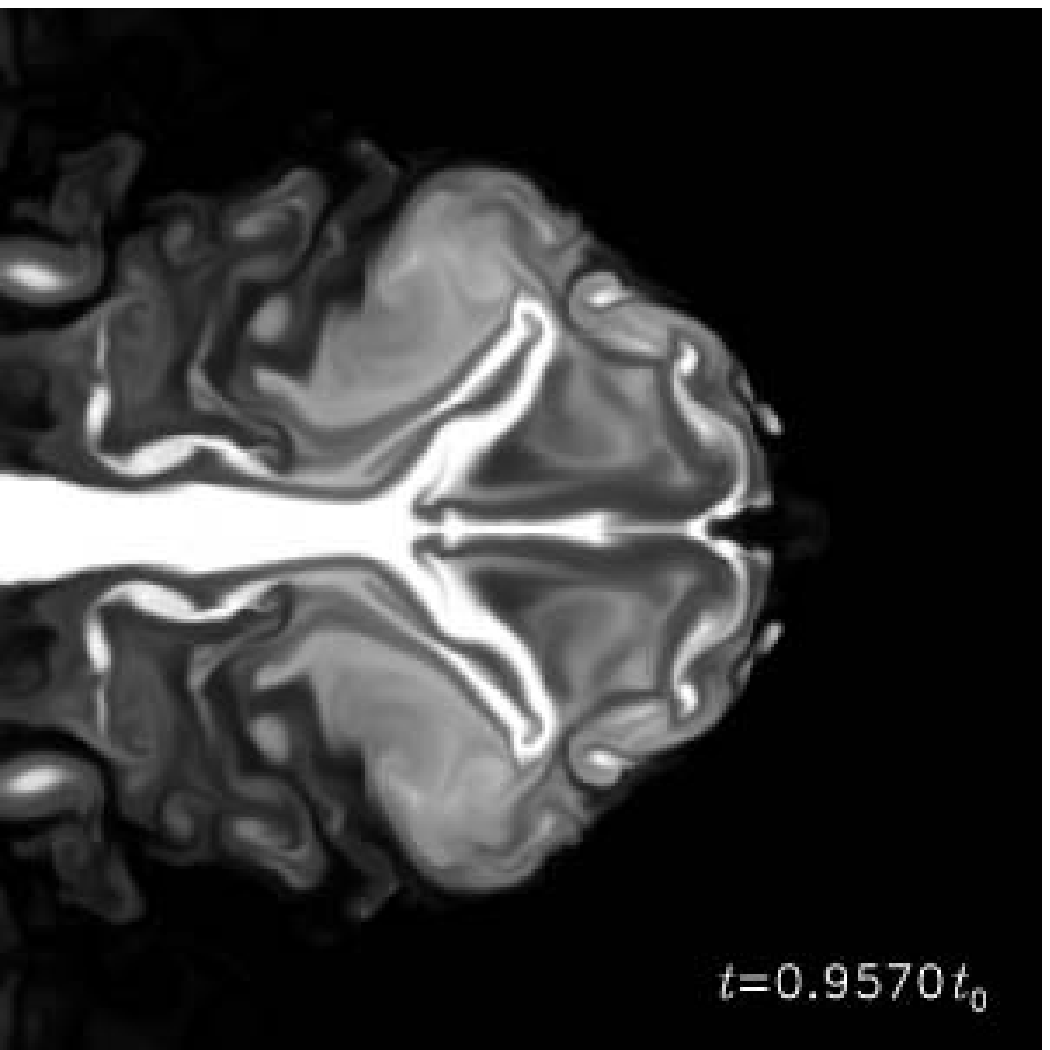}
&
\includegraphics[width=4.5cm]{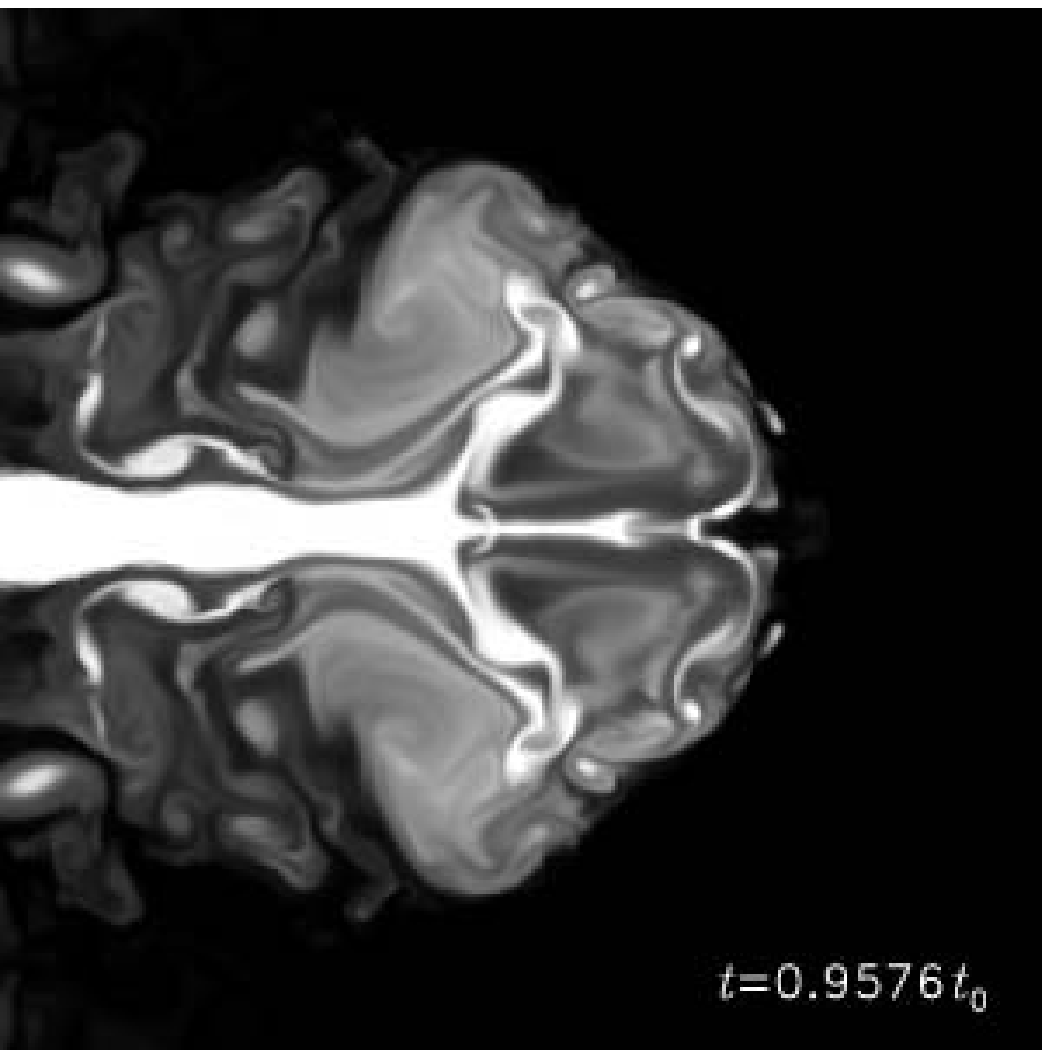}
&
\includegraphics[width=4.5cm]{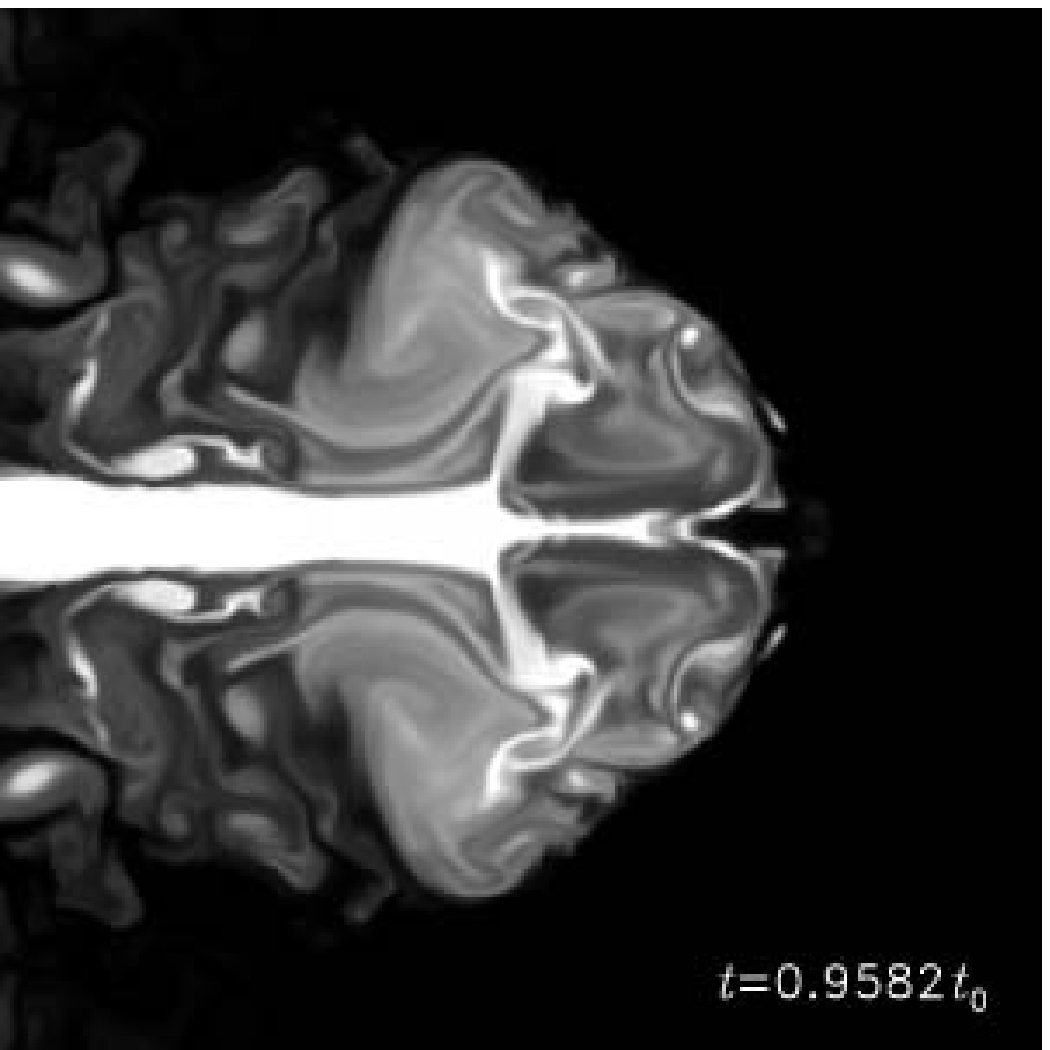}
\\
\end{array}$
\caption{
Sequence of scalar tracer images,
showing local relative concentration of jet plasma,
corresponding to the snapshots in figure~\ref{f:sequence.pressure}.
}
\label{f:sequence.tracer}
\end{figure}

\begin{figure}
\centering \leavevmode
$\begin{array}{ccc}
\includegraphics[width=4.5cm]{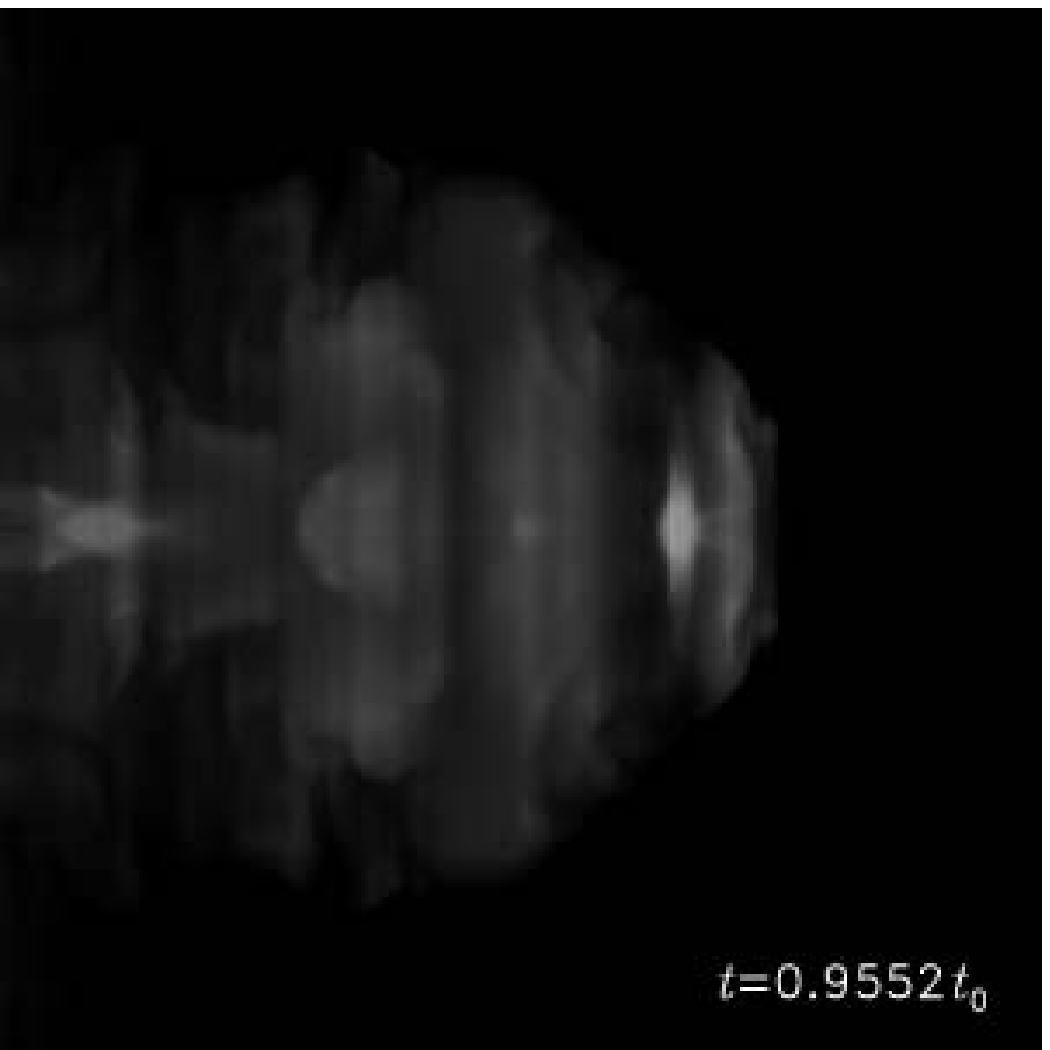}
&
\includegraphics[width=4.5cm]{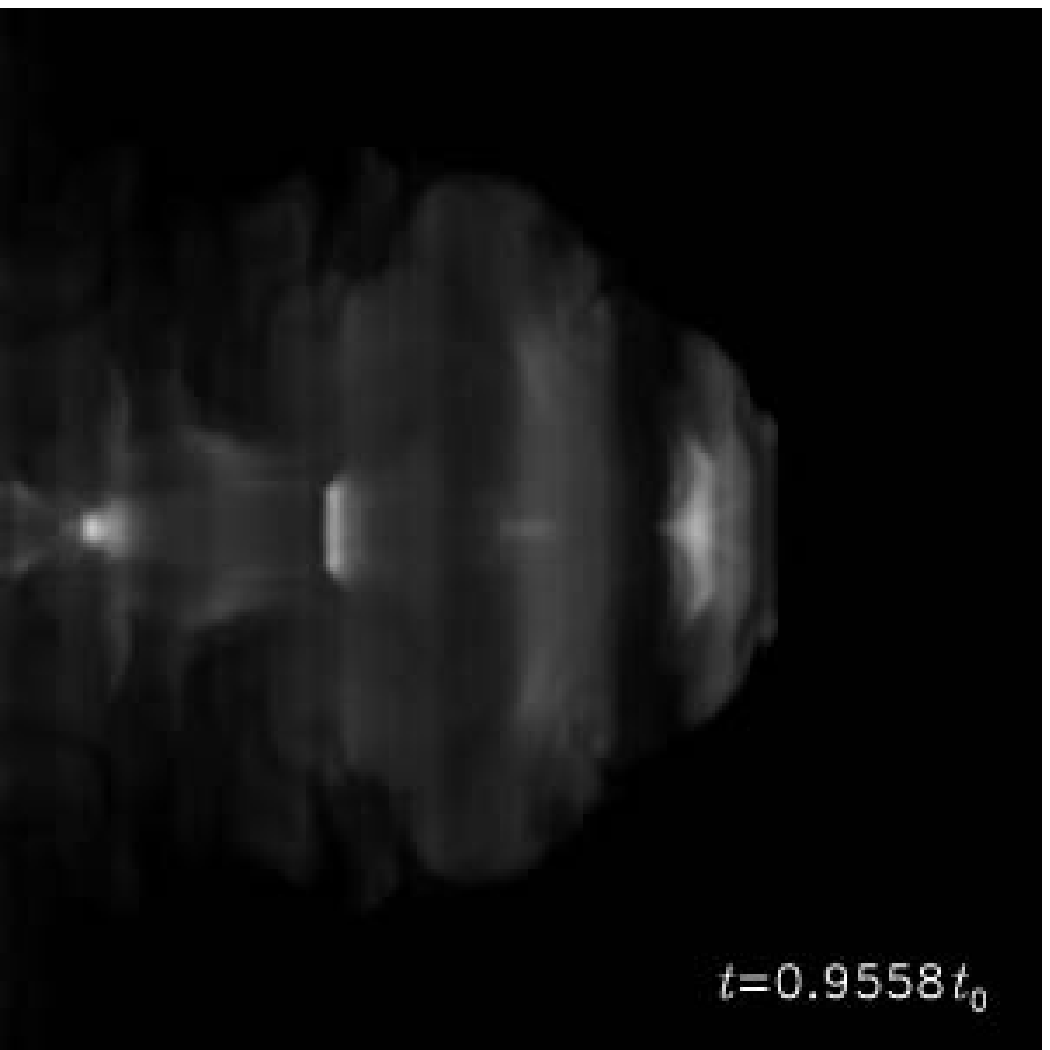}
&
\includegraphics[width=4.5cm]{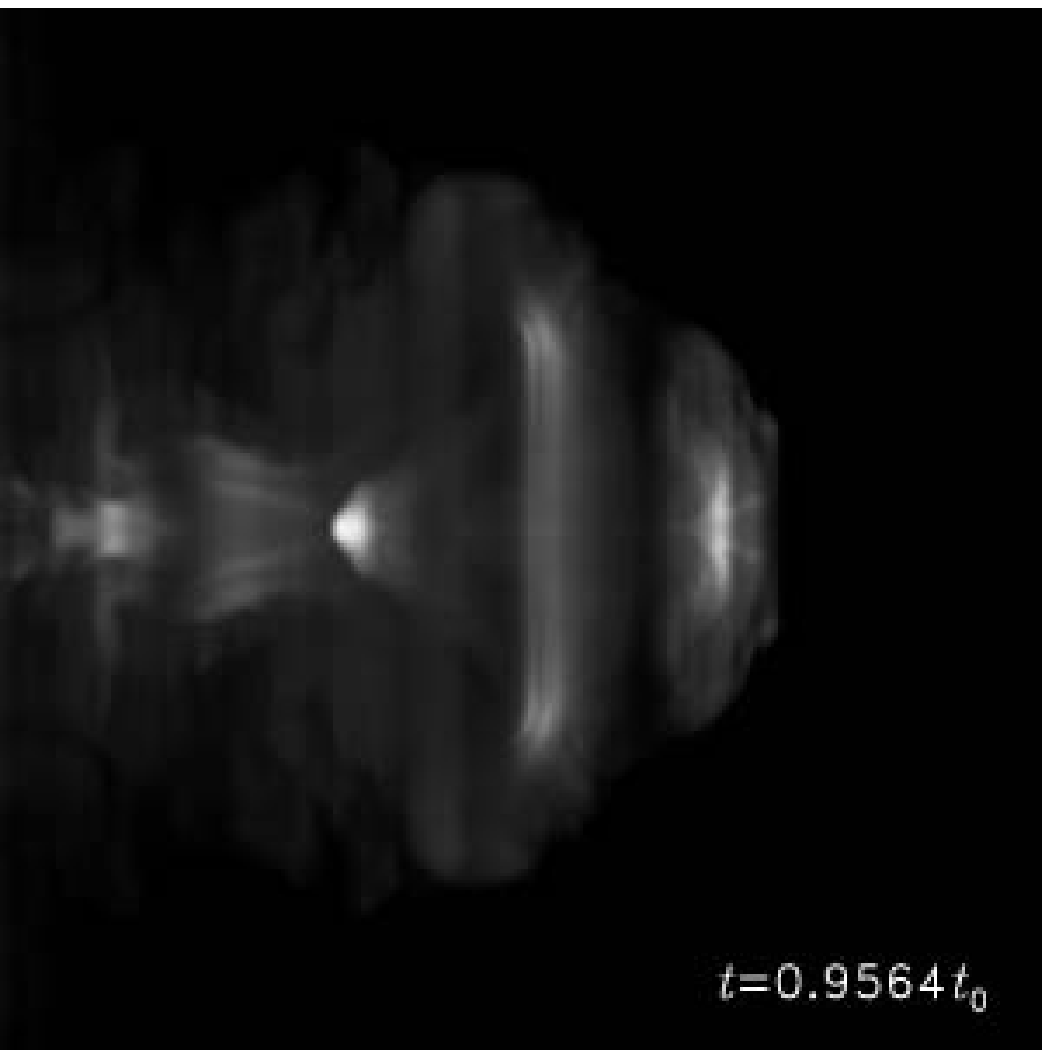}
\\
\includegraphics[width=4.5cm]{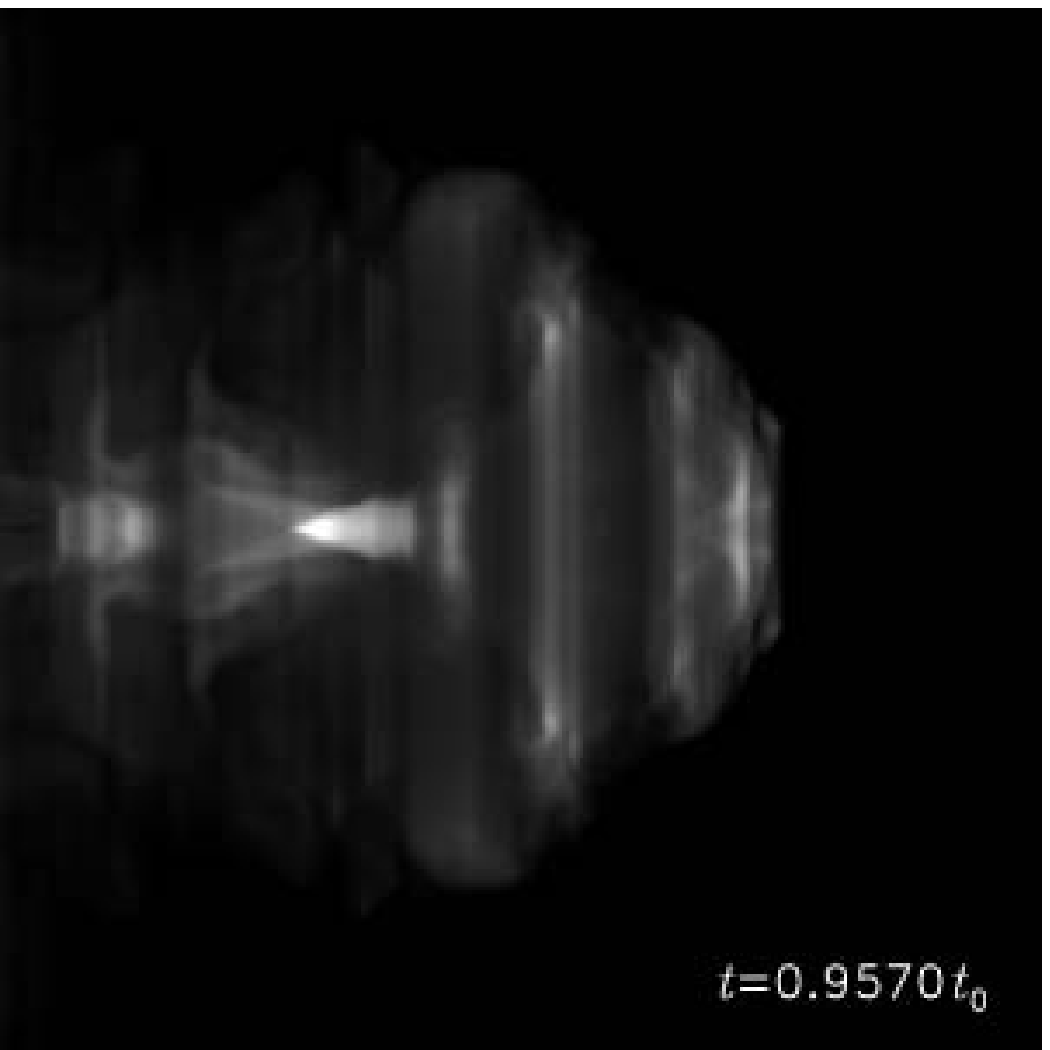}
&
\includegraphics[width=4.5cm]{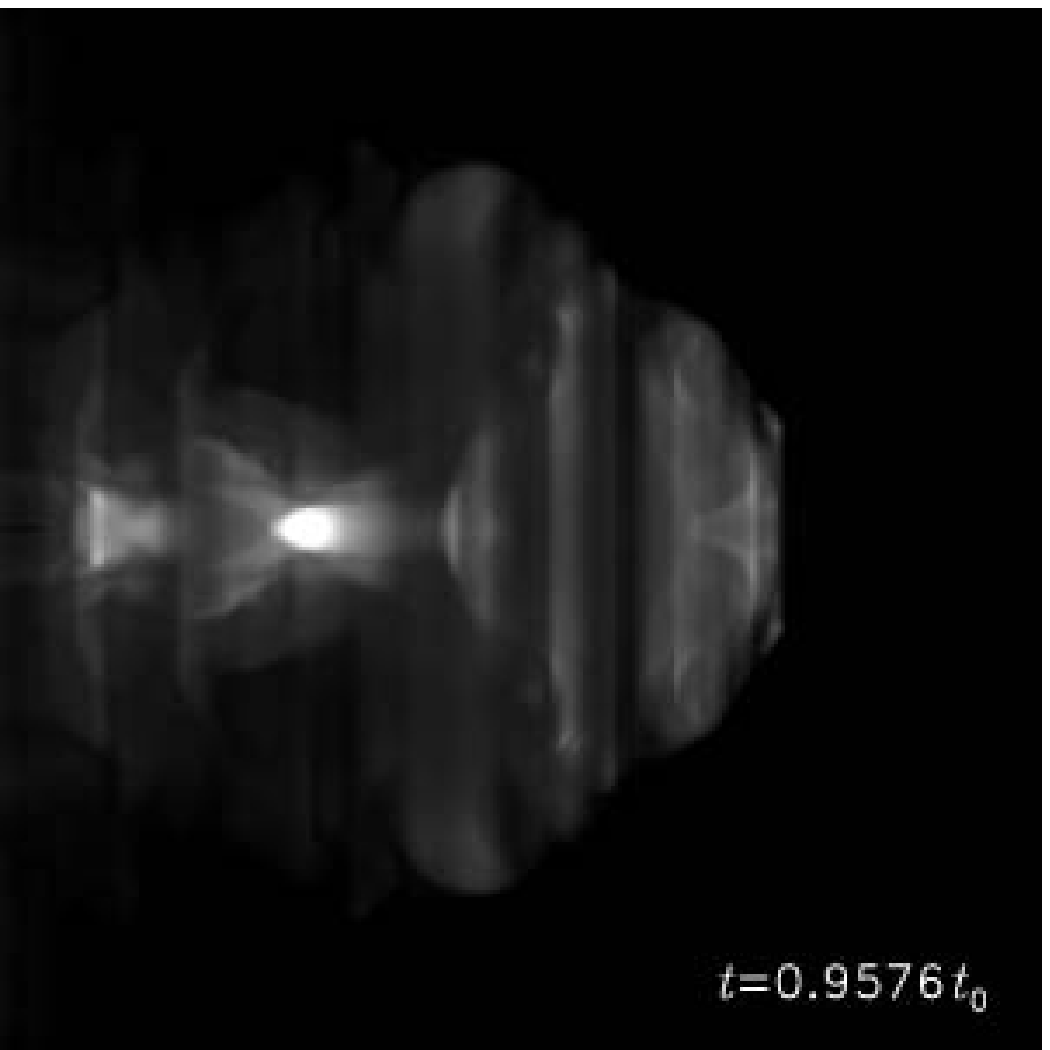}
&
\includegraphics[width=4.5cm]{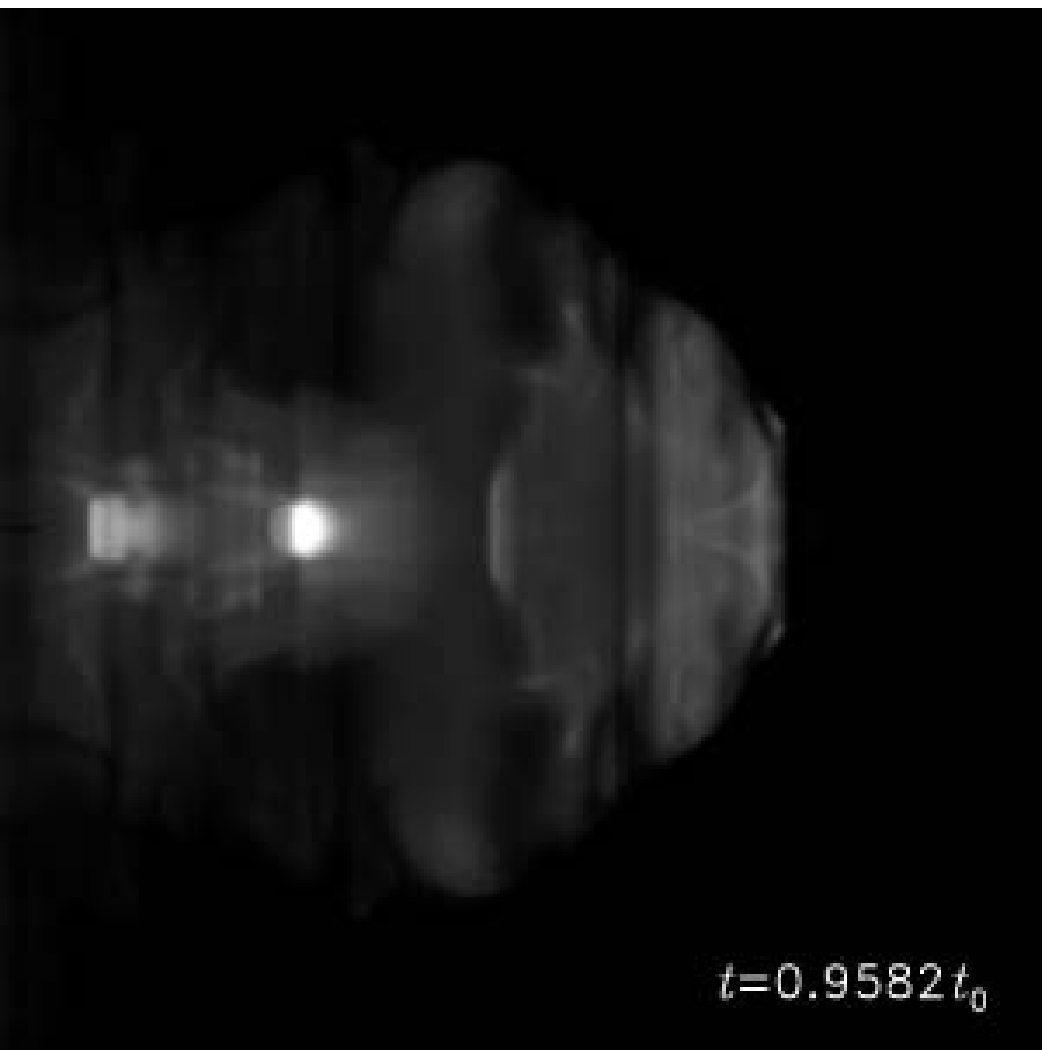}
\\
\end{array}$
\caption{
Sequence of surface brightness images rendered at $\theta=90^\circ$,
corresponding to the snapshots in figure~\ref{f:sequence.pressure}.
}
\label{f:sequence.raytraced}
\end{figure}

In practice, a bar and hot-spot may arise from different underlying
structures at any stage of the jet's surging behaviour.
A hot-spot is commonly the termination shock
at the head of the jet.
However amongst our $\eta=10^{-4}$ selections,
there are some cases
such as
Figures~\ref{f:example.01}L,
\ref{f:example.05}L,
\ref{f:example.06}R,
and the detailed case
Figures~\ref{f:sequence.pressure}-\ref{f:sequence.raytraced}
where the head of the jet is at the back of the leading cavity,
and not directly connected to the hot-spot.
In such cases the forward hot-spot consists of a high pressure region
of moderately mixed jet and thermal gas,
or is an isolated blob of high-$\varphi$ matter
previously separated from the jet.

A bar may result from a variety of different kinds
of annular or disk-like structures.
In some cases
(Figures~\ref{f:example.01}L,
\ref{f:example.02}R,
\ref{f:example.13}L)
the bar is a broad shock at the head of the jet,
or a body of jet plasma near the head.
In some cases,
it is the shocked rim of a funnel-shaped
or disk-shaped
volume of plasma that has split off the head of the jet
but yet remains largely unmixed with the thermal gas in the cocoon
(see Figures
\ref{f:example.06}R,
\ref{f:example.08}R,
\ref{f:example.10}R,
\ref{f:sequence.pressure}-\ref{f:sequence.raytraced}).
In other instances,
the bar is a wide annular region in a part of the backflow
well separated from the head of the jet,
when such a region either preserves a moderate or high
concentration of jet plasma,
or experiences a strong shock locally
(see Figures~\ref{f:example.04}R,
\ref{f:example.05}L
).

There are cases in which an ideal bar and hot-spot are formed,
but additional features spoil the resemblance to Pictor~A.
For example,
in one selection at $t=0.4085t_0$
(Figure~\ref{f:pageant-4mx.1}L;
or upper right panel of Figure~\ref{f:best_morphologies-4mx})
the bar and hot-spot are high-$\varphi$ shocks
in the vicinity of the jet's head,
but an additional strong reverse shock in the jet
appears as another bright feature to the left of the bar.
Such reverse shocks are an occasional by-product of the activity
that produces a hot-spot as a forward shock
and annular shocks propagating into the backflow.
Milder versions of the reverse shock may give the appearance
of the ``knot'' or ``pedestal'' observed on the jet axis.

\begin{figure}
\centering \leavevmode
$\begin{array}{ccc}
\includegraphics[width=4.5cm]{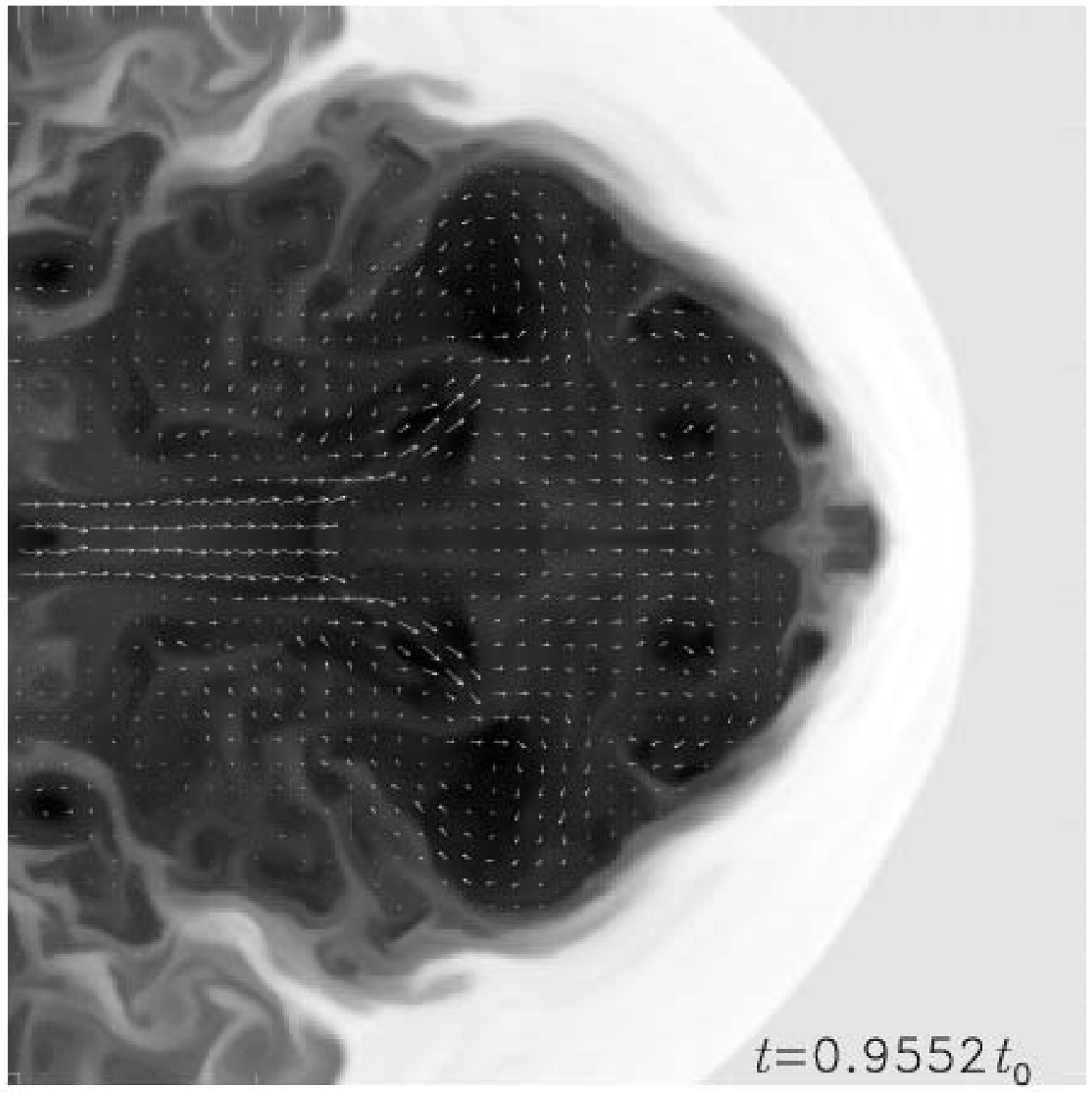}
&
\includegraphics[width=4.5cm]{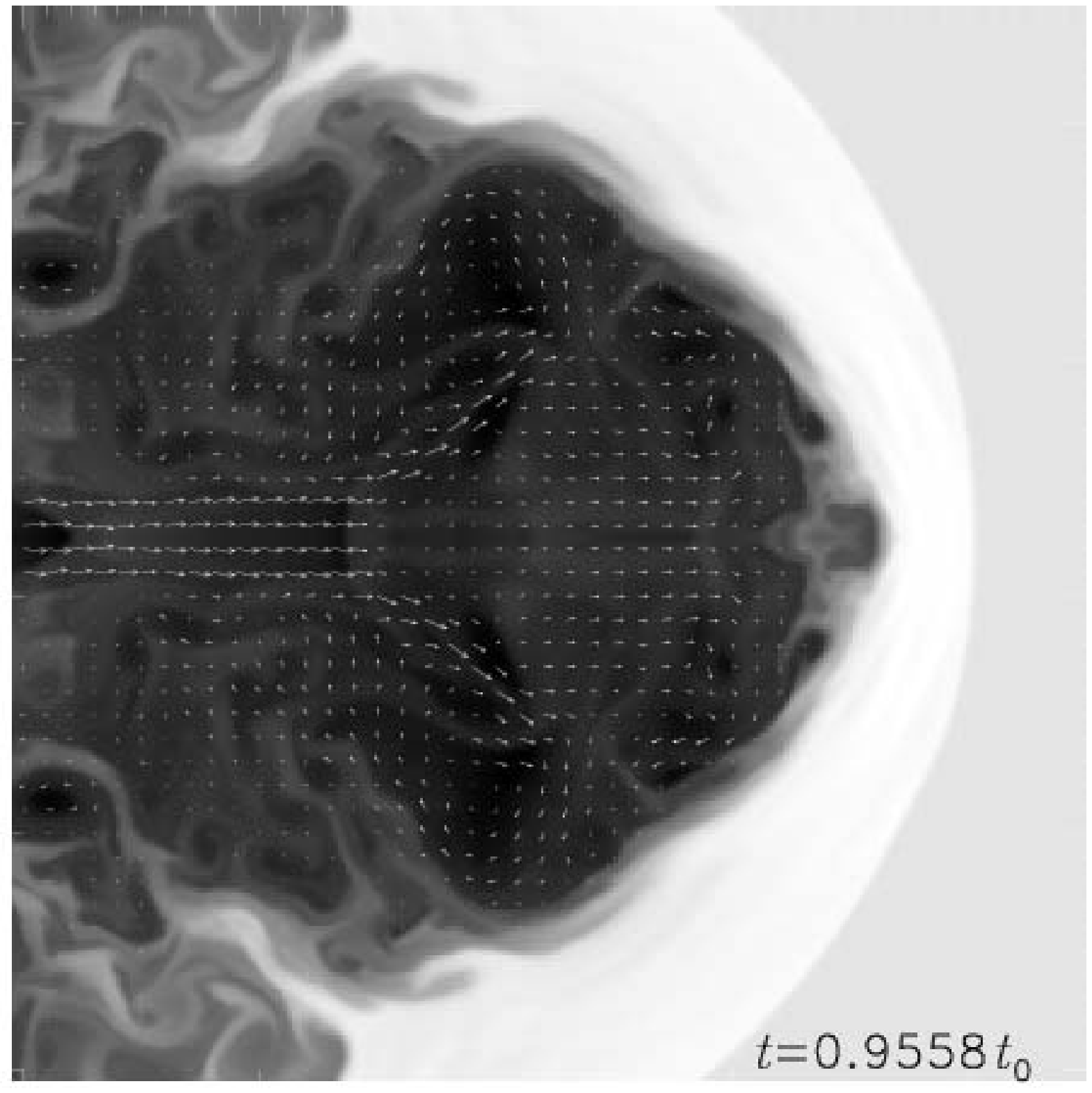}
&
\includegraphics[width=4.5cm]{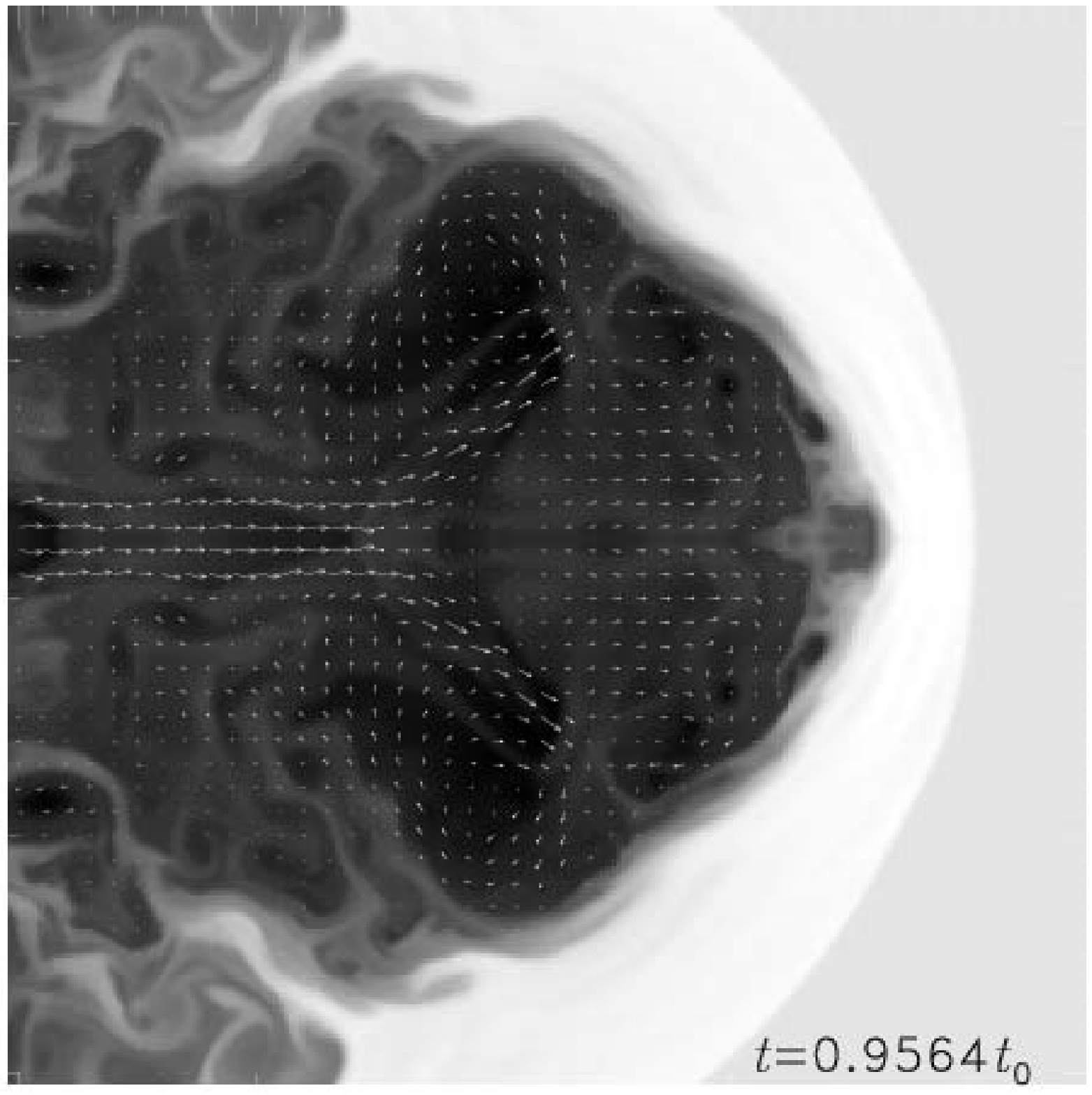}
\\
\includegraphics[width=4.5cm]{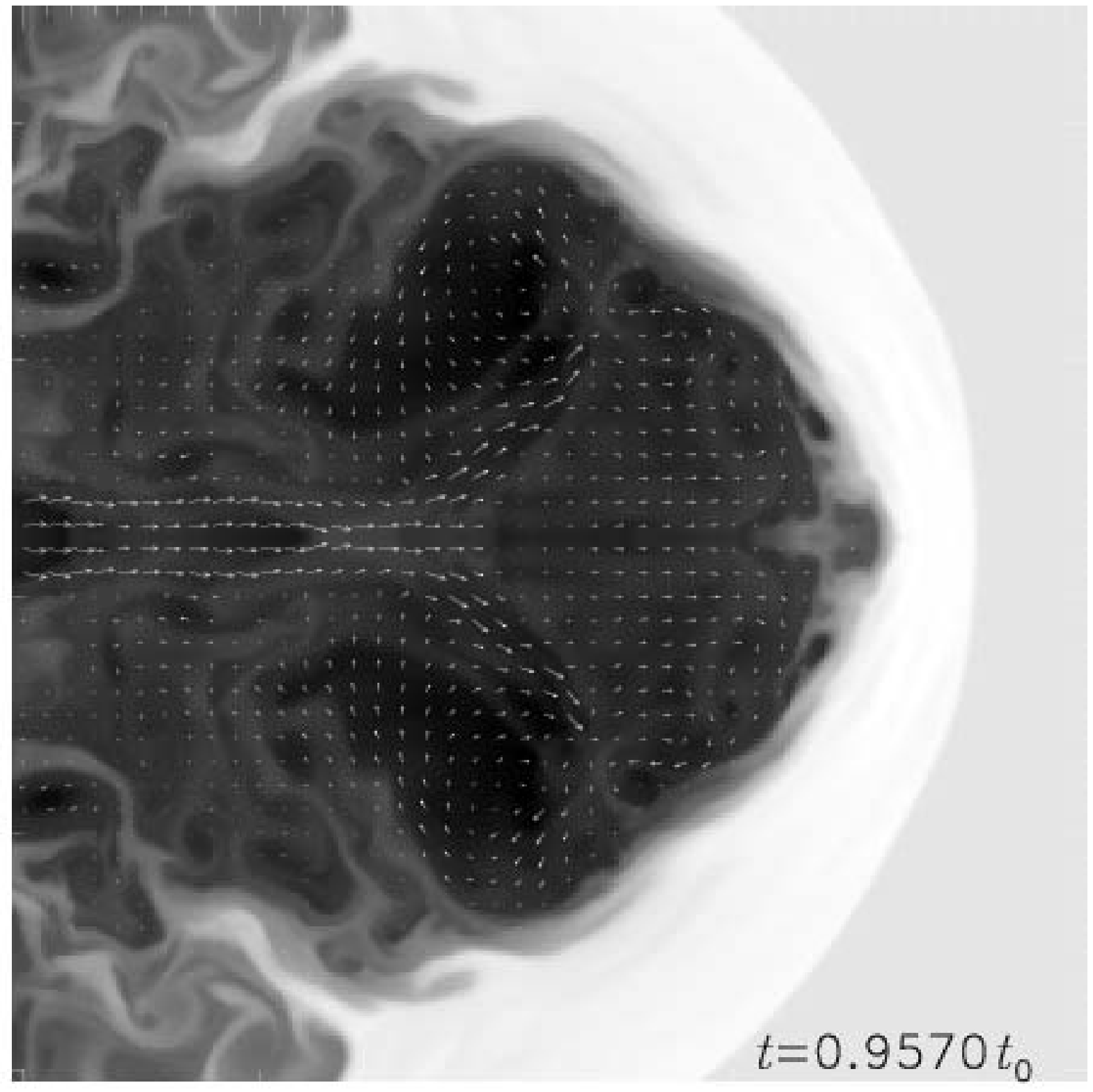}
&
\includegraphics[width=4.5cm]{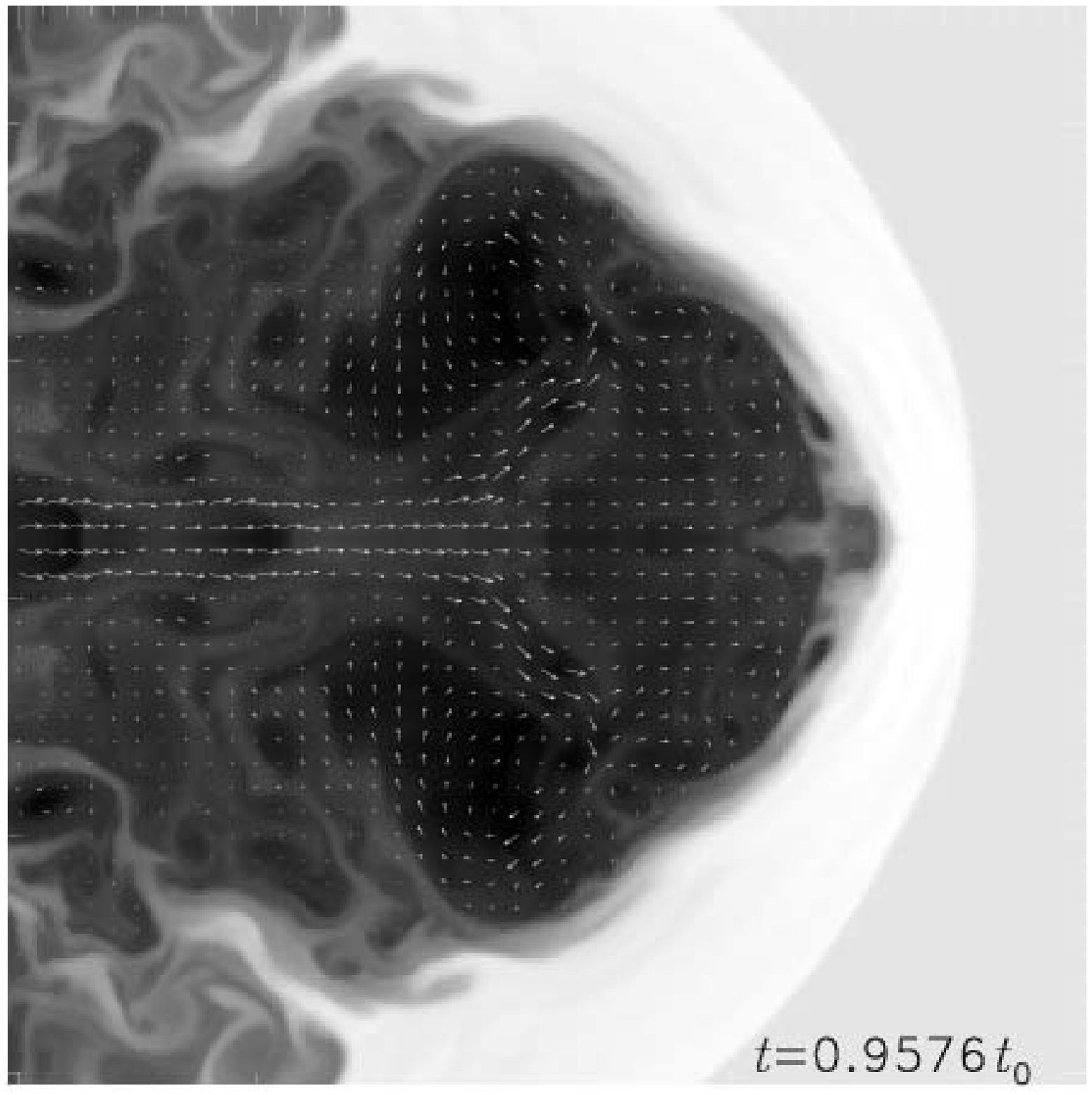}
&
\includegraphics[width=4.5cm]{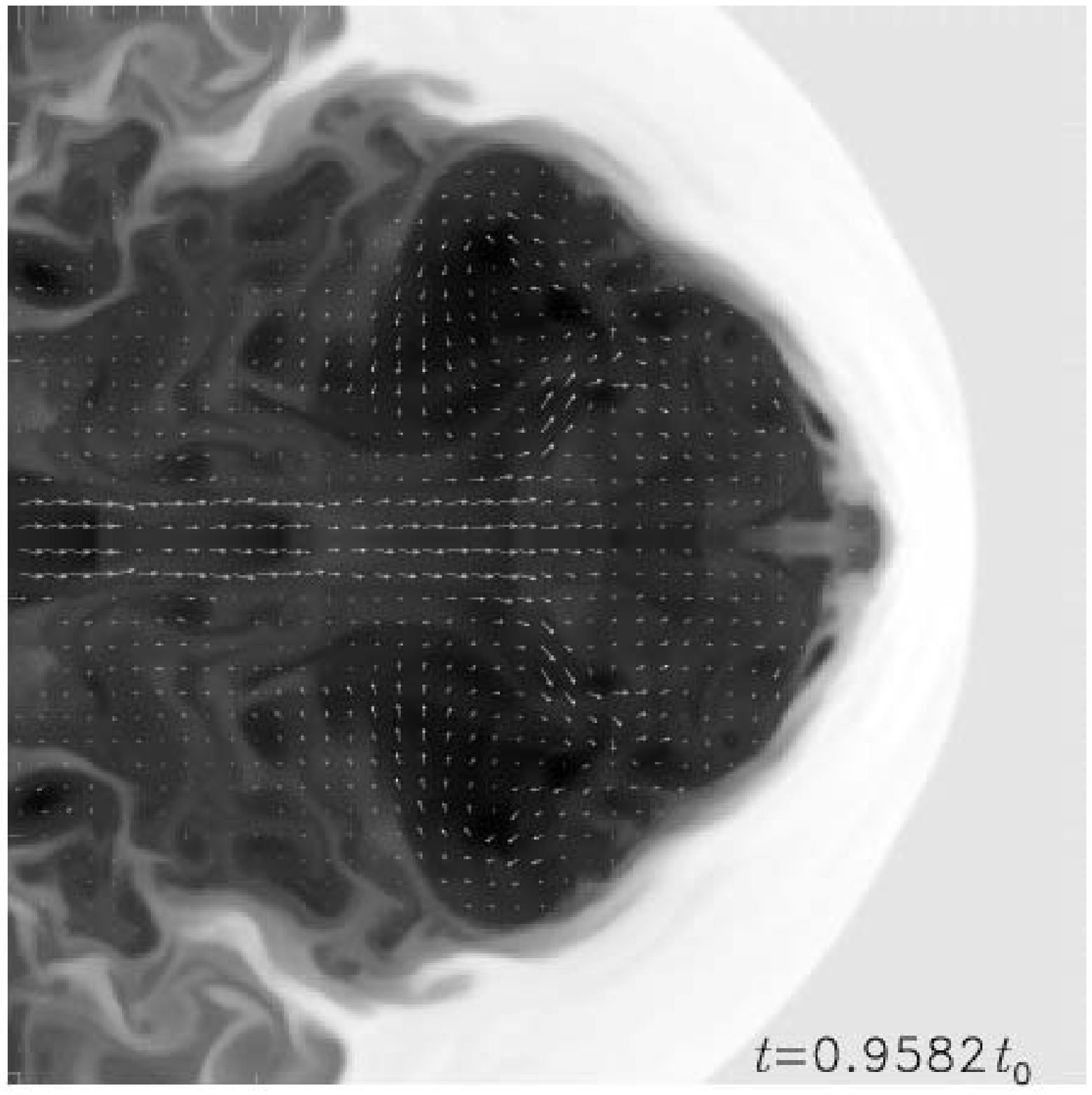}
\\
\end{array}$
\caption{
Sequence of snapshots of the flow vectors
superimposed on $\log\rho$ background,
corresponding to the snapshots in figure~\ref{f:sequence.pressure}.
}
\label{f:sequence.flow}
\end{figure}

It is not possible within the space of one paper to demonstrate the
physical details underlying the evolution
of all the physical configurations that
give the appearance of a bar and hot-spot.
In order to  describe
one particular process for the formation of a bright hot-spot and bar,
we show in
Figure~\ref{f:sequence.pressure},
a time sequence of linearly scaled pressure images
from the closed-boundary, $(\eta,M)=(10^{-4},5)$ simulation
with fine time resolution (simulation {\tt c4V}).
The first frame is at $t=0.9952t_0$ and subsequent frames
are at intervals of $0.0006t_0$ numerical time units.
Figure~\ref{f:sequence.tracer}
shows corresponding frames of the distribution of $\varphi$.
Figure~\ref{f:sequence.raytraced}
shows rendered images of the surface brightness at $\theta=90^\circ$.
Figure~\ref{f:sequence.flow} shows corresponding flow velocity fields
superimposed upon $\log\rho$ greyscale images.

At the beginning of the sequence (frame 1),
we see the usual haze of emission from weak shocks
in the $\varphi\approx0.5$ cavity at the head of the cocoon.
A bright hot-spot appears at the front of the cavity and this
is a high-pressure, shocked region with $\varphi\approx1$,
left behind by the head of the jet
when it was previously at the foremost extremity of its surging motion.
In this frame, the head of the jet is towards the rear of the cavity.
The jet splits at a terminal shock,
dividing into a funnel-shaped structure
consisting of unmixed jet plasma moving
in a radially outwards, positive-$z$ direction,
at speeds comparable to $v_{\rm j}$.
The outermost edge of this structure is curled back upon itself,
and trails off into the backflow closest to the surface of the jet.
The region between this backflow and
the inner surface of the funnel is underpressured
compared to average conditions in the cocoon,
however the outer lip of the funnel is a high pressure surface.

In the second and third frames,
the edge of the funnel has advanced outwards.
In the third frame it collides with a flow of $\varphi\approx0.5$ matter
moving in the negative-$z$ direction near the outer surface of the cocoon.
The flow of gas around the funnel causes a strong annular shock which also
results in low presures at smaller $r$ and $z$.
The shock produces significant emission,
and the result is a bright bar visible in images rendered at
$\theta \approx90^\circ$.

During the activity that creates the bright ring, the diamond and transverse
shocks moving in both directions within the jet
produce several brighter knots of emission.
The knots move and fluctuate in brightness,
but they maintain an approximately constant separation.
Frames~3 to 5 also show a hazy skirt of emission surrounding the jet
to the left of the brightest knot;
it is a backflowing, high-$\varphi$ body of fluid
that previously came off the edge of the funnel.

In frames~5 and 6
the front hot-spot fades,
as it spreads laterally, loses pressure
and undergoes a reduction in $\varphi$ as a result of mixing.
Meanwhile,
the bright bar that persisted from frames~3 to 4 has also faded.
The interaction between the funnel's edge
and the well-mixed, outer-front backflow
has progressed.
The shocks of the initial interaction
have separated and spread into
a wider and complicated shock structure.
The outer edge of the high-$\varphi$ funnel is swept
in the negative-$z$, radially outwards direction
by the well-mixed backflow from the head of the cocoon.
In later frames (not shown), the funnel structure is greatly disrupted.

\section{Discussion}

\subsection{Orientation}

In this paper, we have presented a detailed analysis of simulations of
light jets with a wide choice of parameters with the main aim of
determining the cause of the filament upstream of the western hot-spot in
Pictor~A. The dominant physical principle underlying this analysis is that
complex terminal shock structure is manifest in
simulations of light supersonic jets and that, in projection, such a shock
complex may resemble the Pictor~A radio and optical observations of the region
near the western hot-spot. Similar complexity has been realised in three dimensional
simulations of light supersonic jets by \citet{tregillis01a}. One of the advantges
of restrictio to a two dimensional simulation, at this stage, is the ability to map
out a larger region of parameter space and to utilise higher spatial resolution. In
particular, we have carried out simulations with a density ratio of $10^{-4}$; these
would be quite expensive in three dimensions.

One
of the most important constraints on the simulations is that the bar is a thin structure
and if it is three dimensional, it would appear to be almost edge-on.  Our rendering of
synthetic radio images at different orientations bears this out. Adequate morphologies
are not reproduced at low inclinations. Another decisive requirement that implies an
almost edge-on structure, is reproduction of a filament with low curvature.

At inclinations $\theta < 70^\circ$, geometric effects cannot convincingly produce a
long, straight bar frequently enough .
The projection of an annular structure with a ribbon-like cross-section
makes one side brighter than the other,
but the image is still significantly curved.
The accidental coincidence of the near and far sides of separate rings
in projection may give
the appearance of a straight feature,
but these tend to be narrower than the observed filament.
A structure with a very large radius of curvature
might appear straight,
but luminous structures of this type occur in our simulations
in regions where the coocon is no more than $\sim 2 r_{\rm j}$ in radius.
Three dimensional hydrodynamical effects
may admit more complicated explanations
for bar-like morphology. However
extreme departures from axial symmetry in the backflow seem unlikely,
given the strightness of the jet as indicated by the X-ray observations.
On the basis of non-relativistic hydrodynamics, we therefore think it likely
that the inclination angle, $\theta > 60^\circ$.

If there is any easy way to produce the appearance of a bar
in a basically axisymmetric backflow
viewed at $\theta < 70^\circ$,
it probably requires relativistic effects.
A ring or disk moving relativistically
appears more edge-on
than if it were at rest 
\citep[see ][]{bicknell96a}. 
This would require the
filament to be transient structure moving with a pattern speed larger than the
expected hot-spot speed $\sim 0.1-0.2 c$. Relativistic beaming could also
affect the brightness distribution if the matter passing through the filament
and hot-spot shocks moves relativistically, brightening an annular structure
on whichever side has a post-shock flow
directed towards our line of sight.
However the significance of these relativistic effects,
for hot-spots in general and for Pictor~A in particular,
remains to be demonstrated,
and is beyond the scope of our present study.

\subsection{Implications for physical parameters}

Although all choices of the jet parameters, jet density ratio, $\eta$, and
Mach number, $M$, yield instants with appropriate morphology,
the brightness characteristics of the resulting images
may distinguish the most plausible values of the Mach number.
Jets with larger Mach number typically produce terminal shock features
with higher peak intensity.
\citet{perley97a} had deduced a high Mach number ($M \approx 40$) for the jet
in Pictor A based on the large ratio of the minimum pressures in the hot-spot
and lobes. We have confirmed this in our simulations that show, for example,
a brightness contrast $\sim 100$ for $(\eta,M) = (10^{-2},50)$ (see
\S\ref{s:brightness_distribution}).  Nevertheless, the unusual
$(\eta,M)=(10^{-4},50)$ simulation
produces a lower contrast and widely distributed rings in the backflow,
and these features are clearly ruled out by observation.
This may mean that the real jet parameters are less extreme,
or else the powerfully turbulent backflow that occurs in this case
may be less disruptive in three dimensions.

We do not find such a clear diagnostic of the density contrast of the jet.
Cases with lower $\eta$ tend to produce broader cocoons,
especially the rounded cavity in $\eta=10^{-4}$ simulations.
For jets with larger $\eta$,
the jet plasma and shocked thermal gas mix more rapidly in the cocoon.
However all of the parameter sets we investigated permit
the appearance of ephemeral bar-like sructures with
a range of widths,
including some with $\approx 4r_{\rm j}$ radius as required by the
observations.

Many of our simulations
imply relativistic bulk flow,
$\beta \sim 1$ (see Table~\ref{table.jet.properties}). One aspect of the observations
where relativistic effects may be relevant concerns the large contrast between the
hot-spot and radio lobe empshasised by \cite{perley97a}. 
On the basis of hydrodyamics
alone, this would indicate a Mach number of approximately 40-50, or equivalently a
Lorentz factor of approximately 25. This has motivated the $M=50$ simulations described
above. 
Nevertheless, relativistic beaming may substantially contribute to the observed
hot-spot brightness. 
This aspect may need to be taken into account with future
simulations with a fully relativistic code.

\subsection{The orientation of the jet and filament
in relation to radio and X-ray observations}

In their analysis of the Chandra X-ray Observatory data on Pictor~A,
\citet{wilson01a} have suggested that
an inclination $\theta \approx 23^\circ$
is viable.
Larger angles are possible if jet magentic fields considerably less
than equipartition are allowed. An
inclination angle as small as
$23^\circ$ would be in direct contradiction to
our conclusion that the Pictor~A radio structure is viewed at
an angle of
$\theta \ga 70^\circ-90^\circ$.
It is therefore worthwhile to examine the question of orientation in
some detail.

The first part of the \citet{wilson01a} argument relates to the arm-length
ratio of the  radio source.
Assuming a velocity of advance for the hot-spot,
$\beta_{\rm HS} \approx 0.11$ \citep{arshakian00a} and the measured
arm-length ratio of the source,
\citet{wilson01a}
derive $\theta \approx 23^\circ$ but caution that this estimate
by itself  should not be taken too seriously for
an individual source.
The main limitation is probably the following:
The statistical result for hot-spot advance includes sources in which
the momentum of the jet is spread
over a wide area as a result of the ``Dentist Drill'' effect
\citep{scheuer82a}, thereby slowing its average advance.
However, as we have remarked in justifying an axisymmetric
simulation, the western jet in Pictor
A appears to be very straight.
That is, it does not appear to be
substantially deflected and therefore
the advance of the hot-spot may be faster than the average
\citet{arshakian00a} value.
Moreover, in contrast to the western lobe,
the eastern lobe contains two hot-spots
indicating some spreading of the
jet force over a larger area than each hot-spot and a consequent
slower expansion rate and larger
arm-length ratio.

Now consider the effect of orientation on the X-ray emission and jet
brightness ratio.
As \citet{wilson01a} remark, the X-ray emission resulting from  scattering of
the cosmic microwave background radiation by the electrons in the jet is
proportional to
$\delta^{4+2\alpha}$ where $\delta$
is the Doppler factor and $\alpha \approx 0.9 \pm 0.5$ is the
spectral index \citep{begelman87a,dermer95a}. 
The ratio of jet to counterjet X-ray brightnesses, $R \ge 10$
implies that $\beta \cos \theta \ga 0.2$ for $\alpha = 0.9$.
The radio estimate of $R \ga 3$ does not provide as stringent a limit --
$\beta \cos \theta \ga 0.14$.

The main constraint on the angle
$\theta$ is that it should not be so close to
$90^\circ$ that the western jet is Doppler dimmed, in absolute terms,
irrespective of the ratio of jet
to counterjet fluxes. That is,
$\delta$ should not be less than approximately $0.7$ for an order of
magnitude decrease in the X-ray flux.

\begin{figure}
\centering \leavevmode
$\begin{array}{c}
\includegraphics[width=6cm]{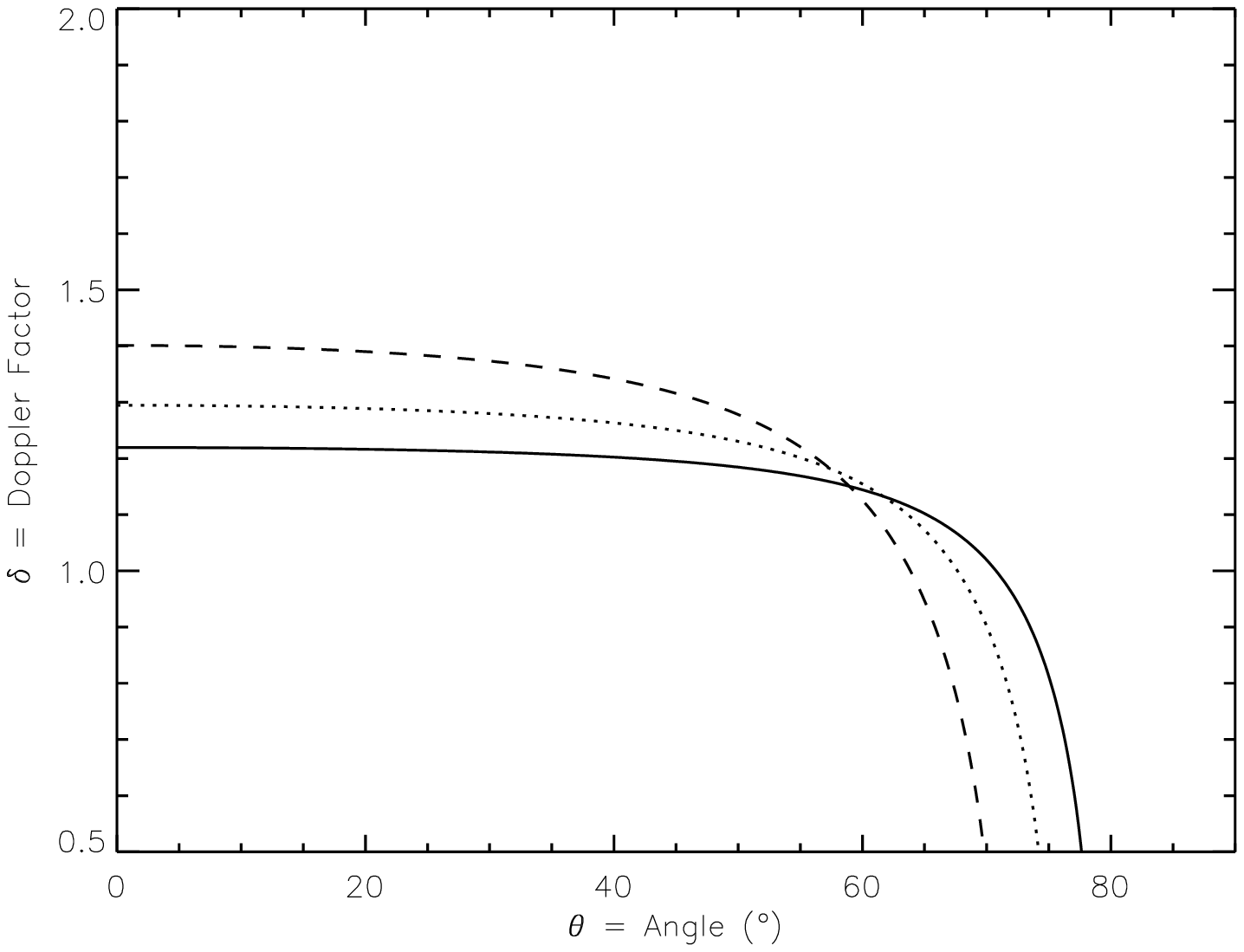}
\end{array}$
\caption{
The Doppler factor of the jet
for a given ratio of jet
to counter-jet surface brightness.
The cases $R=10, 20, 50$
are shown by solid, dotted and dashed lines respectively.
\citet{wilson01a} find $R>10$ for Pictor~A.
}
\label{f:delta}
\end{figure}

For a given ratio, $R$, of jet to counterjet X-ray fluxes, we have
\begin{equation}
\beta \cos \theta \approx \frac {R^{1/p} - 1}{R^{1/p}+1}
\end{equation}
where $p = 4+2\alpha$ for inverse Compton emission off the microwave
background. In
Figure~\ref{f:delta}, we show plots of the Doppler factor against
inclination for values of
$R=10, 20$ and $50$. An inclination of $70^\circ$ is consistent with
$\delta \ga 0.7$ for $R=10, 20$ but not for $R=50$. If $\theta =
70^\circ$ then \citet{wilson01a}
have shown that the magnetic field is about 0.03 times the minimum
energy value. This constitutes no
difficulty in principle since one does not necessarily expect minimum
energy conditions to apply in the
jet even though turbulent processes may be sufficient to produce
minimum energy conditions in the lobes. Moreover, one possible estimate of
the magnetic field in the hot-spot is an order of magnitude below
equipartition. For reference, a jet inclination of
$\theta = 70^\circ$ and
$R=10  (20)$ implies a jet velocity of $\beta
\approx 0.57 (0.74)$. These estimates of velocity are consistent with
other estimates in powerful
sources
\citep[e.g.][]{wardle97a}.
We conclude therefore that an orientation of
$\theta\approx 70^\circ$ is consistent with the X-ray and radio
constraints.
We note, however, that the steep decline in the Doppler factor at around
$\theta = 70^\circ-80^\circ$ indicates that the
constraints implied by the X-ray data are only {\em just} satisfied
if $\theta = 70^\circ$.

A jet velocity of $\beta \approx 0.6 - 0.7$ has other interesting
ramifications since gas would then emerge from oblique
shocks in the hot-spots with this velocity. Some of
this gas would be moving directly towards us with  $\delta \approx
3.1$ (for $\beta \approx 0.6$). The
flux from that region of the hot-spot would be enhanced by a factor
of 60 and this may be another explanation for the large
ratio of hot-spot surface brightness to lobe surface brightness
\citep{perley97a}.

If an almost side-on orientation proves to be untenable, then an
alternative is that the
limitations of an axisymmetric solution do not allow for a partial
ring. That is, the ring could be
brighter in one part that another as a result of three dimensional
hydrodynamic effects.
This may be consistent with
the faint structures north of the filament seen in the greyscale
image of the \citet{perley97a} radio data,
the \cite{roeser89a} optical image
\cite[see Figure~3 of ][]{wilson01a}
and the VLT optical image
(Figure~\ref{f:optical_image}).
These features {\em may} indicate a  non uniformly bright ring viewed at a
more acute angle. We do not favour this explanation because of the
straightness of the observed filament.

Another point in favour of $\theta \sim 90^\circ$ is that the
western hot-spot appears, in projection, at the edge of a wide lobe. If the
source were to viewed at, say $\theta \approx 20^\circ$, then the hot-spot
would most likely appear closer to the middle of the lobe. In addition such a
small viewing angle makes the total extent of the source quite large ($\approx
1.3 \> \rm Mpc$). Radio galaxies this large are not unknown, but are fairly
exceptional.

\subsection{Structure functions and hot-spot ratios}

The construction of cumulative distribution functions for hot-spot ratios
has been incidental to the main purpose of this paper. However, it is of
general interest and quantifies the role of intrinsic variability in using
such ratios for the estimation of say, relativistic effects.

The calculation
of structure functions in jet simulations could have wide applicability. For
example, the structure function of internal shocks could be applied to
blazars.

\section{Acknowledgements}

This work was supported by an Australian Research Council Large Grant,
A69905341 and grants of computing time from the ANU Supercomputer Facility.

\appendix
\section{Probability distribution of hot-spot ratio}
\label{appendix.hot-spot.ratio}

In each simulation the bright features in the vicinity of the hot-spot 
are unsteady, dynamic structures.
The intrinsic variability of the peak intensity, $I$, of the hot-spot
is easily quantified as a cumulative probability distribution function,
$G(I)$.
This function is numerically approximated
by sorting the instantaneous $I$ values 
from a representative period $t_0\leq t\leq t_0+\Delta t$ 
of the simulation,
in which $\Delta t \gg t_\mathrm{dyn}$.
The sample must be representative in the sense that 
it begins and ends during the time in which the cocoon is well
established and both the head of the jet and the bow shock
are not close to any of the boundaries of the computational grid.
If the intensity samples $\{I_n\}$ are ranked in ascending order,
$\mathcal{R}_n$,
for $n=1,\ldots,n_\mathrm{max}$,
then the cumulative distribution function of
the peak intensity is 
\begin{equation}
G(I)\equiv \mathrm{Pr}(I'\le I)
={{\mathcal{R}(I) }\over{n_\mathrm{max} }}
\end{equation}
where $I$ is one of the sample values.
If $I$ is not one of the sample values
but has neighbouring values are, $I_-<I<I_+$,
then $G(I)$ can be interpolated from $G(I_-)$ and $G(I_+)$.
Then we may define a probability density function,
\begin{equation}
g(I)={{dG}\over{dI}}
\ ,
\end{equation}
which may be evaluated for any $I$ by numerical differentiation
and/or interpolation as appropriate.

If the two jet and counter-jet have the same physical parameters
and the respective hot-spots are separated by a distance
exceeding the $t_\mathrm{dyn} c$
then the hot-spots are causally independent.
The probability distribution, $f(R)$, of the ratio, $R$, of hot-spot peak intensities
can be derived from the intrinsic distributions of one hot-spot.
\begin{equation}
f(R)\equiv\mathrm{Pr}\left({RI_2<I_1<(R+dR)I_2}\right)
\equiv {{dF}\over{dR}}
=\int_0^\infty g(RI_2) g(I_2) I_2 dI_2
\ .
\end{equation}
Alternatively, the cumulative distribution function $F(R)$
may be calculated by taking a double sum over
the sample of intensity values,
\begin{equation}
F(R)\equiv\mathrm{Pr}\left({R'\le R}\right)
= {1\over{{n_\mathrm{max}}^2}} \sum_{i,j}{
H(RI_j - I_i)
}
\ ,
\end{equation}
where $H$ is the Heavyside step function
($H(x)=1$ for $x\geq 1$ and $H(x)=0$ otherwise).


\end{document}